%% file: SmallReview.tex
\documentclass[preprint,authoryear,3p,12pt]{elsarticle}
\usepackage{amsfonts}%

\usepackage{graphicx}
\usepackage{dcolumn}
\usepackage{bm}
\usepackage[T1]{fontenc}
\usepackage{color}
\usepackage{comment}
\usepackage{braket} 
\usepackage{float}
\usepackage{amsmath}
\usepackage{xcolor}
\usepackage{amssymb}
\usepackage{mathtools} 
\usepackage{eqnarray,amsmath}
\usepackage{hyperref}
\newcommand{\abs}[1]{\left| #1 \right|} 
\usepackage{hyperref}
\usepackage{multirow}
\usepackage{ulem}

\makeatletter
\def\ps@pprintTitle{%
  \let\@oddhead\@empty
  \let\@evenhead\@empty
  \def\@oddfoot{\reset@font\hfil\thepage\hfil}
  \let\@evenfoot\@oddfoot
}
\makeatother

\DeclareMathOperator{\sech}{sech}
\DeclareUnicodeCharacter{0229}{\c{e}}

\journal{Physics Reports}
\begin{document}


\title{Few-body Bose gases in low dimensions - a laboratory for quantum dynamics}

\author[add1,add10]{S.~I.~Mistakidis}
\author[add2]{A.~G.~Volosniev}
\author[add3]{R.~E.~Barfknecht}
\author[add4]{T.~Fogarty}
\author[add4]{Th.~Busch}
\author[add3]{A.~Foerster}
\author[add7,add8]{P.~Schmelcher}
\author[add9]{N.~T.~Zinner}

\address[add1]{ITAMP, Center for Astrophysics $|$ Harvard $\&$ Smithsonian, Cambridge, MA 02138 US}
\address[add10]{ Department of Physics, Harvard University, Cambridge, Massachusetts 02138, USA}
\address[add2]{IST Austria (Institute of Science and Technology Austria), Am Campus 1, 3400 Klosterneuburg, Austria}
\address[add3]{Instituto de F{\'i}sica da UFRGS, Av. Bento Gon{\c c}alves 9500, Porto Alegre, RS, Brazil}
\address[add4]{Quantum Systems Unit, Okinawa Institute of Science and Technology Graduate University, Okinawa, Japan 904-0495}
\address[add7]{Center for Optical Quantum Technologies, University of Hamburg, Department of Physics, Luruper Chaussee 149, D-22761, Hamburg, Germany}
\address[add8]{The Hamburg Centre for Ultrafast Imaging, University of Hamburg, Luruper Chaussee 149, D-22761, Hamburg, Germany}
\address[add9]{Department of Physics and Astronomy, Aarhus University, Ny Munkegade 120, DK-8000 Aarhus C, Denmark}

\date{\today}

\begin{abstract}
Cold atomic gases have become a paradigmatic system for exploring
fundamental physics, which at the same time allows for applications in quantum technologies. The accelerating developments in the field have led to a highly advanced set of engineering techniques that, for example, can tune interactions, shape the external geometry, select among a large set of atomic species with different properties, or control the number of atoms. In particular, it is possible to operate in lower dimensions and drive atomic systems into the strongly correlated regime. In this review, we discuss recent 
advances in few-body cold atom systems confined in low dimensions from a theoretical 
viewpoint. We mainly focus on bosonic systems in one dimension and provide an introduction to the static properties before we review the state-of-the-art research into quantum dynamical processes stimulated by the presence of correlations. Besides discussing the fundamental physical phenomena arising in these systems, we also provide an overview of the calculational and numerical tools and methods that are commonly used, thus delivering a balanced and comprehensive overview of the  field. We conclude by giving an outlook on possible future directions that are interesting to explore in these correlated systems. 
\end{abstract}

\begin{keyword}
Cold atoms, one-dimensional systems, few-body physics, multicomponent bosonic settings, quantum dynamics, droplets, spinor gases, systems with impurities, quantum control, long-range physics, beyond mean-field approaches, strongly correlated gases
\end{keyword}


\maketitle

\tableofcontents

\section{Introduction and Motivation}

Over the last two decades ultracold atomic physics has become invaluable in fundamental research and quantum simulations of a number of physical models due to the existence of highly controllable experimental setups \citep{bloch2008,Zohar:2015,Saffman:2016,Gross:17,Szigeti:2021}. One particularly intriguing puzzle that is being addressed is the many-body problem, which has excited researchers and students since the dawn of quantum mechanics. 
It is, however, a distinctly difficult problem, as the complex dynamics of multi-particle systems is governed by an exponentially large number of degrees of freedom, which can rarely be fully accounted for. 

Trapped ultracold atoms are highly controllable and pure. This allows for tackling the many-body problem from the directions that are beyond solid state experimental set-ups, which traditionally provide the main motivation for studying many-body physics with cold atoms.  In particular, the high control of ultracold gases allow for an experimental realization of strongly interacting particles in small samples. This progress has made it feasible 
to study many-body correlations starting from few-body systems in a bottom-up approach, i.e., 
{\it  the crossover from few- to many-body physics}.
 In addition to their value for contributing to the understanding of the many-body problem, small samples have been recognised as promising systems by themselves, in which, for example, one can \textit{engineer quantum} properties on demand. The high level of control available in cold atom experiments can be used to explore a diverse range of few-body quantum effects, including the generation and control of entanglement, spin-dependent transport, the dynamical orthogonality catastrophe, few-body quantum `chaos', filamentation processes, dynamical phase separation, interaction-induced tunneling and Josephson-type dynamics, spontaneous droplet formation and many more. 

Here, we review the recent body of theoretical works on statics and dynamics of  few-body systems in one spatial dimension with an emphasis on bosonic particles. Our hope is that this review will stimulate present and near-future experimental and theoretical efforts. Therefore,
our emphasis is on setups and situations that can be addressed in modern cold atom experiments.
It is worth noting that other useful reviews exist on several topics related to the use of cold atoms as quantum simulators for many-body condensed-matter systems in low dimensions \citep{bloch2008,cazalilla2011,Guan2013Review,Langen2015}, and  the limit of strongly correlated systems has recently been reviewed by \cite{Minguzzi2022Review}. The reviews by \cite{Blume_2012} and by \cite{Sowinski2019Review} provide good introductions to the physics of ultracold few-body systems, but have limited overlap with what we cover here. In particular, they mainly concentrate on static problems.

The focus of the present review is on the physics of a few strongly interacting\footnote{In the literature, one can find various definitions of ``strongly''. In the review, this word will imply that beyond-mean-field theoretical methods should be employed. The interested reader can consult~\cite{RevModPhys.71.463,Bagnato2015,kevrekidis2015defocusing} for discussions of mean-field approaches, which are beyond the scope of the present review. In general, mean-field methods are of limited use in few-body physics.} cold bosons in one spatial dimension, assuming that the coordinate of each boson is given by a continuous variable. The particles are confined inside an external trap that will often be parametrized by a harmonic oscillator -- the typical situation in cold-atom research. We will review theoretical work and approaches to study the dynamical (and some appropriately chosen static) properties of such systems in a wide range of different situations. 

It is worth bearing  in mind that the continuum models considered in the review can be always connected to low-density lattice models~\cite{Essler2005HubbardBook}, such as a Bose-Hubbard chain~\citep{Haldane1980}. This connection allows one to study dozens of particles with powerful numerical approaches developed for lattice systems~\citep{Schmidt2007exact,Tezuka2008DMRG}. At the same time, exact solutions of the continuum models can be used to test the accuracy of these approaches, which is especially important for strong interactions in multicomponent systems~\cite{dehkharghani2017analytical}.

\subsection{Review outline}
\label{subsec:review_outline}

To set the scene, we will first briefly review selected and relevant aspects of the experimental progress made in cold-atom systems in recent years with respect to foundational many-body experiments (see Sec.~\ref{sec:exp}) and few-body systems (see Sec.~\ref{subsec:few_body_exp}). The progress in these areas has been the motivation for many of the theoretical developments that will be presented and it is therefore valuable to know the state of the experimental art to understand the developments in theory. 
After this we move to the main part of our review and start with a brief introduction to fundamental and exact methods for low-dimensional systems and introduce specific solutions for small systems in Section~\ref{sec:EarlyMod}. We then discuss single-component cold atomic systems in Section~\ref{sec:OneComp} and multi-component systems in Section~\ref{sec:spinor}
with a focus on collective modes, energy transfer, tunneling and quench dynamics in small systems. In Section~\ref{sec:polaron}, we give an introductory overview of the physics of impurities in trapped one-dimensional systems.
Each section will present fundamental examples that are targeted at new entrants to the field, and then provide a survey of more recent results for researchers looking to update their knowledge. 

The ensuing sections start with a discussion of spin-1 Bose gases in Section~\ref{spin_1_gases}. After introducing the state-of-the-art of the field, we analyze the role of the spin degree of freedom, which leads to a rich phase diagram and correlation dynamics in these systems. 
To ease the understanding of the respective spin  processes, we briefly discuss the corresponding mean-field approximation, which  sets the stage for studying complex many-body phenomena.
In 
Section~\ref{sec:Control}, we review proposals on driven dynamics and transport, as well as the emerging field of quantum control. The latter is particularly intricate for few-body systems and is a topic of extensive interest at present. We hope that the discussion here will give the reader a starting point (and motivation) for further exploration.

A very recent topic of great interest is that of quantum droplets. These  have
been discussed in many studies in the past few years using a variety of different methods, and the first experiments are now in place to gauge present theoretical understanding. Here, we review recent studies on droplet formation in Section~\ref{sec:Droplets_contact} with an initial focus on static properties in order to have a platform on which to discuss new proposals on how to investigate their dynamics.

Since the field of one-dimensional (1D) few-body systems heavily relies on a combination of analytical insights and advanced numerical techniques, we provide in Section~\ref{sec:Methods} an overview of some of the general techniques that are in use today. This includes exact diagonalization, Bethe ansatz, DMRG, ML-MCTDHX 
and mappings of strongly interacting systems to spin-chain descriptions. The goal of Section~\ref{sec:Methods} is to give the reader an idea of different options for quantitative studies, as well as to provide the right references to more in-depth technical discussions of each individual method and available resources for their implementation.

We move on to systems that are beyond the assumption of short-range interactions in Section~\ref{sec:beyond} and discuss the physics that long-range interactions can add to 
few-body systems. This includes $p$-wave interactions, particles with dipole moments and dipole-dipole couplings, as well as combinations of atoms and ions in traps. We believe that these research topics will strongly influence few-cold-atom physics in the coming years. Therefore, we find it necessary to conclude the main part of our review with a stepping stone into this research. 

Finally, we end our review by a summary and give our view of some of the exciting future perspective that this field holds, both in terms of the exploration of fundamental physics at the border of few- and many-body systems in low dimensions, as well as their potential use in quantum technologies in the future.

\subsection{Cold-atom experiments in one dimension}\label{sec:exp}

One advantage of having tunable trapping potentials in a cold-atom laboratory is the possibility to study low-dimensional systems by freezing out dynamics in certain spatial directions through strong confinement.
Some of the first experiments used a combination of optical 
and magnetic traps to restrict the motion of atoms in a weakly-interacting Bose-Einstein 
condensate along the transverse
directions  \citep{gorlitz2001,dettmer2001}.  In parallel, it was 
shown how a combined system of fermions and bosons can arrange 
itself in a magnetic trap so as to achieve an effective 1D confinement of the Bose gas component 
\citep{schreck2001}. Alternatively, geometric confinement 
was achieved using optical lattices \citep{greiner2001,moritz2003,tolra2004} but 
only intermediate 
confinement strengths could be achieved. 
Another direction to simulate low-dimensional systems 
is to use so-called atom chips \citep{fortagh1998,muller1999,reichel1999,dekker2000,folman2000}. 
These are surfaces with carefully 
crafted nano-wires that can trap atoms close
to the surface. Atom chips have allowed 
for a number of impressive experiments
with bosonic samples in particular the study of matter-wave interferometry
\citep{schumm2005,jo2007}, and  of
quantum and thermal noise \citep{hofferberth2007,hofferberth2008}. 
More information about the atom chip developments can be found in dedicated reviews \citep{fortagh2007,folman2008,Reichel2011}. 

Arguably, one of the main drivers behind the early pursuits of 1D systems was the seminal results of Girardeau \citep{girardeau1960}  who pointed out that strongly interacting Bose systems in 1D display  fermionic aspects in their behavior. In the early 2000s, this so-called Tonks-Girardeau regime became a topic of great interest due to the ability to produce controllable samples of cold atoms in lower-dimensional geometries, for review see \citep{bloch2008}. For the theoretical treatment of these systems, a key step was the work by Olshanii \citep{olshanii1998} that provided a comprehensive theoretical model of how to simulate 1D physics, including very strong interactions, by tuning the confining geometry of the system (see~\citep{Dunjko:01} for a discussion focused on the Tonks-Girardeau regime).

\begin{figure}[tb]
    \center
    \includegraphics[width=0.6\linewidth]{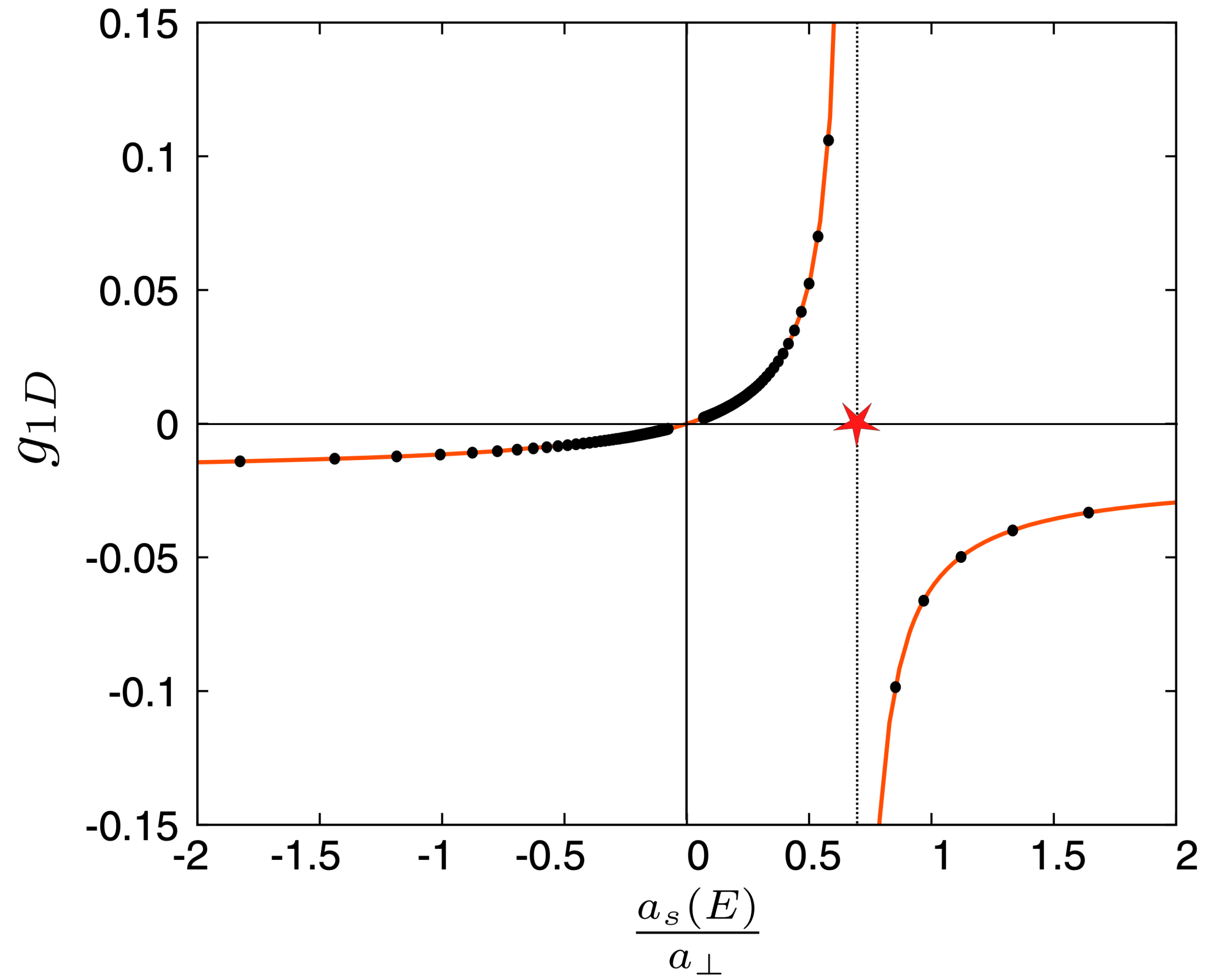}
    \caption{
    Effective 1D coupling strength $g_{1D}$ (in units of $\hbar/(\mu a_{\perp})$) with respect to $a_s(E)/a_{\perp}$ ($a_S(E)$ denotes the scattering length). 
    The location of the confinement-induced resonance is indicated by the red star. 
    The solid line depicts the analytical prediction of Eq.~(\ref{1D_coupling}), while the black dots refer to numerical calculations of the two boson problem with van-der-Waals two-body interactions. 
    Image from~\citep{greene2017universal}.
\label{fig:CIR_1D} }
\end{figure}

For quasi-1D systems, Olshanii argued that the pairwise interactions can be modeled via a 1D $\delta$-function interaction whose strength is given by
\begin{equation}
g_{1D}=\frac{2 \hbar^2 a_s}{\mu a_{\perp} (a_{\perp}-C a_s)},\label{1D_coupling}
\end{equation}
where $a_s$ is the three-dimensional (3D) scattering length, $C$ is a constant, $a_\perp$ denotes the harmonic oscillator length in the direction of the transverse confinement and $\mu$ is the reduced mass.  
Equation~(\ref{1D_coupling}) demonstrates that by changing $a$, one can adjust both the magnitude and the sign of the 1D interaction strength. 
The behavior of the effective coupling constant $g_{1D}$ as one varies $a_s/a_{\perp}$ is shown in Fig.~\ref{fig:CIR_1D}. The analytical result of Eq.~(\ref{1D_coupling}) (solid orange line) is in agreement with the numerical solution of a model two-body Hamiltonian with the potential $V_{vdw}=C_{10}/r^{10}-C_6/r^6$ (black dots). 
Notice also that for $a_{\perp}/a_s \to C$ the effective coupling constant diverges, $g_{1D}\to \infty$. This marks the position of the confinement-induced resonance (CIR) and is indicated by the red star\footnote{The physical origin of the CIRs is that they fulfill a Fano-Feshbach mechanism where all the energetically closed transversal modes of the trapping potential collectively support one bound state that lies in the energy interval of the open channel~\citep{bergeman2003,friedrich2006theoretical,greene2017universal}, i.e.~the ground state transverse mode. }. 
A more detailed discussion on this topic, including also the behavior of the bound states, can be found in~\citep{dunjko2011confinement,greene2017universal}.
To simulate the Tonks-Girardeau gas in a laboratory, the ratio $a_s/a_\perp$ should be tuned close to the position of CIR. One way to achieve it is to use the so-called Feshbach resonance technique~\citep{inouye1998}, see~\citep{chin2010feshbach} for review. 
This approach was utilized in one of the first experimental observation of CIR physics -- in two-fermion molecule formation studied using radio-frequency (RF) spectroscopy \citep{moritz2005}\footnote{
Corrections due to anharmonicity
of the confinement were also experimentally addressed by considering 
molecule formation with a sample of just two fermions \citep{sala2013}. This accurate analysis became possible in part due the calibration of the
scattering length performed in \citep{zurn2013cir}.}.

\begin{figure}[tb]
    \centering
    \includegraphics[width=\columnwidth]{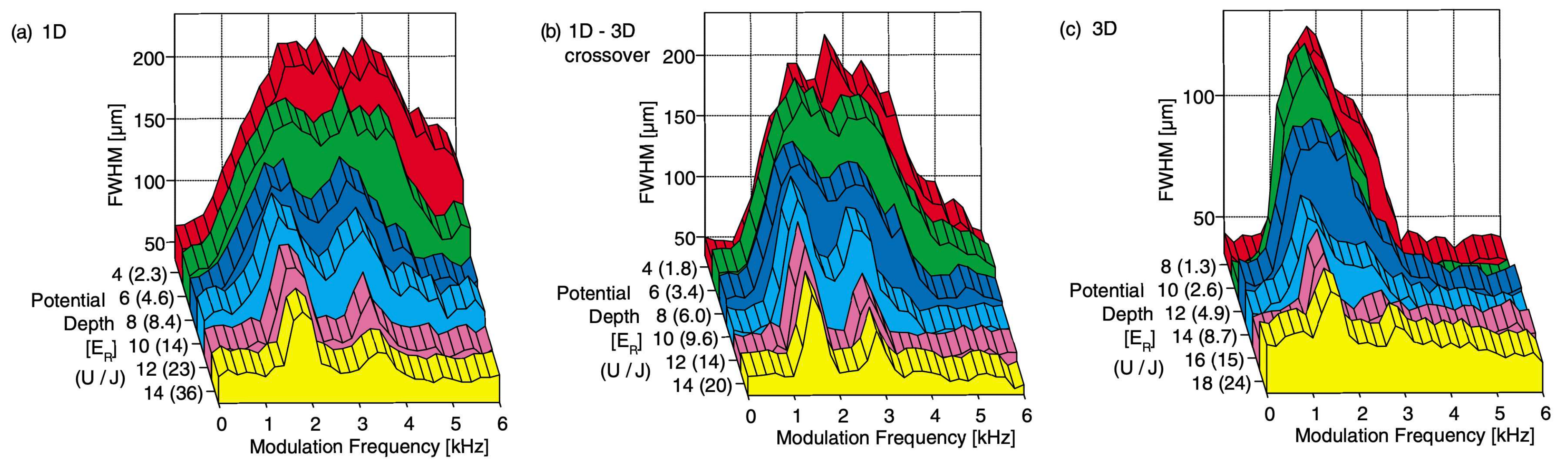}
    \caption{
    Measurement of a 3D to 1D crossover through the excitation spectrum of an array of 1D Bose gases as function of the depth of the trapping potential quantified through the ratio of interaction to hopping, $U/J$. The appearance of a discrete structure due to a vanishing superfluid component is clearly visible. The 3D case shows the superfluid to Mott insulator transition.
    Image from \citep{stoferle2004} }
    \label{fig:3Dto1DExcitationSpec}
\end{figure}

In 2004, two seminal experiments \citep{paredes2004, kinoshita2004} managed to reach the Tonks-Girardeau regime
by applying a (shallow) optical lattice
{\it along} the 1D spatial direction. The shallow longitudinal 
lattice  changes the kinetic energy of the particles but not 
enough for them to become trapped in the minima. Rather, the effect is to increase the effective mass of 
the bosons in the experiments, and this means that the relative
contribution of the interactions between the particles can dominate the kinetic term. This has made it possible to 
study the transition from 3D down to 1D physics with strong interactions \citep{stoferle2004} (see Fig.~\ref{fig:3Dto1DExcitationSpec}). 
These experiments provided the  
verification of the fact that 1D Bose 
gases do indeed adopt a fermionic nature in the presence of strongly 
(repulsive) short-range interactions as predicted by \citep{girardeau1960,Lieb1963GS}.  They probed different features of these {\it fermionized} systems
including the wave function overlap in the system \citep{kinoshita2005}, and 
the non-equilibrium dynamics in the 
Tonks-Girardeau regime~\citep{kinoshita2006}. 
The dynamics illustrated a quantum Newton's cradle (see also~\cite{Schemmer:2019,li2020} for recent realizations) and drove 
the research on the out-of-equilibrium dynamics, in particular, development of generalized hydrodynamics -- one of the main recent achievements in fundamental 1D physics. Note that this intrinsically many-body theory will be covered in this review only in passing; the interested reader should consult the recent pedagogical reviews ~\citep{Doyon:2020,Alba:2021}.

\begin{figure}[tb]
    \centering
    \includegraphics[width=\columnwidth]{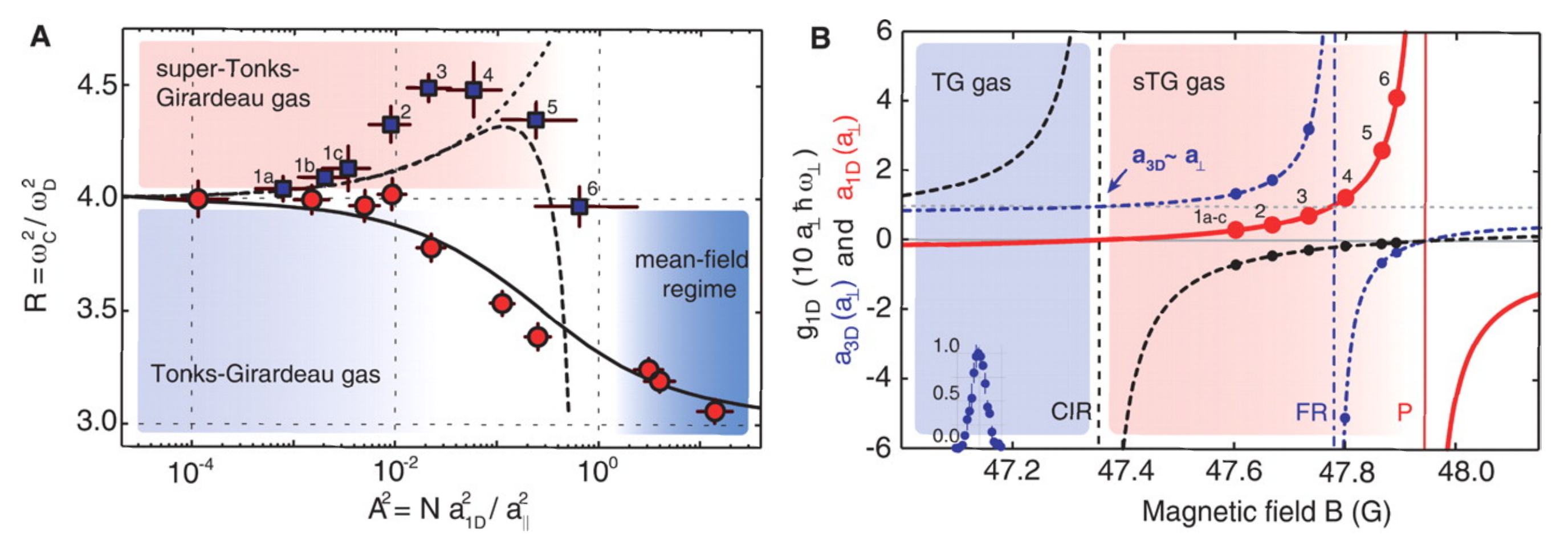}\hfill
    \caption{
    {\bf A}: Ratio of the frequency of the lowest axial compression mode over the lowest dipole mode, R, vs. the interaction strength. The blue squares show the measurements in the attractive regime and provide evidence for the super-Tonks-Girardeau gas. The red circles show the transition from the Thomas-Fermi to the Tonks-Girardeau regime for repulsive interactions.  {\bf B}: The 3D scattering length $a_\text{3D}$ (dashed-dotted), the 1D scattering length $a_\text{1D}$ (solid), and the 1D interaction parameter $g_\text{1D}$ (dashed) in the vicinity of the Feshbach resonance (FR) at 47.78 G. Image from \citep{haller2009} }
    \label{fig:haller}
\end{figure}

Soon after realization of fermionized Bose gases, tunability of $g_{1D}$ allowed experimentalists to transfer a strongly repulsive Bose 
gas into a regime with strong 
attractive interactions~\citep{haller2009}, see Fig.~\ref{fig:haller}. 
The latter has been denoted the super-Tonks-Girardeau regime \citep{Astrakharchik:05}. 
It is important to point out that a super-Tonks-Girardeau 
state prepared by moving slowly from strong repulsive interactions to strong attraction by crossing the CIR is a  highly excited state. Nevertheless, it lives long enough for it to be 
studied \citep{haller2009}. The nature of the CIRs was subsequently explored in greater detail \citep{haller2010}.

Observation of the Tonks-Girardeau gas and CIR implied that cold atomic gases could provide a universal platform for studying many celebrated 1D models~\citep{giamarchi2003,sutherland2004} in a very direct and tunable manner. Since then 1D Bose gas is a topic of continued pursuit with a number of milestone achievements. For example, the Yang-Yang (finite temperature) thermodynamics was observed in weakly and strongly interacting gases in~\citep{amerongen2008,Vogler2013}; Luttinger liquids were studied in~\citep{Haller2010_Luttinger,Gring2012}. The so-called Tomonaga-Luttinger liquid behavior in quantum 
critical systems was
engineered in~\citep{yang2017} (for observation of similar physics 
in a superconducting Josephson array, see \citep{cedergren2017}).
A recent study by \citep{wilson2020} showed how the 1D 
Bose gas in the strongly interacting regime will show dynamical 
fermionization in its momentum distribution  as it is released from the trap  (see \citep{rigol2005,minguzzi2005} for theoretical insight into this phenomenon). 

\subsection{Few-body regime}
\label{subsec:few_body_exp}

Experimental progress towards manipulating and controlling few-atom systems has a long history \citep{dumke2002,miroshnychenko2006,beugnon2007,karski2009,grunzweig2010,muldoon2012}.  
For bosonic systems, optical tweezer techniques have been a primary experimental tool, and they have seen a tremendous development in recent years. The seminal work of \citep{kaufman2015} showed how to use spin-exchange interactions in $^{87}$Rb atoms to achieve position-resolved coherence measurements and entangle two atom systems in a controlled manner. Furthermore, the same kind of  optical tweezer setup can be used to deterministically create entanglement even in systems where the atoms are non-interacting \citep{lester2018}. This was achieved using a beam-splitter and post-selection technique akin to the Hong-Ou-Mandel effect \citep{kaufman2014,lopes2015,preiss2015}, and can seen be as a step 
towards a protocol for measurement-based quantum information processing using atoms. 

The few-body results were soon followed by experiments that  showed how large-scale tweezer arrays can be experimentally realized. In \citep{endres2016} a total of 100 optical tweezers was used to obtain 1D arrays of up to 50 bosonic atoms with no defects in the system in a remarkably fast protocol, paving the way for  structured assembly of 1D atomic systems for quantum information and quantum simulation purposes. Moreoever, \citep{endres2016} also showed 
how one may replace lost atoms in the arrays. A similar feat was accomplished by \citep{barredo2016} who used up to 50 microtraps to assemble an array of bosonic atoms into a 2D lattice with full tunability. We note that it has even become possible recently to create an array of cold molecules using an optical tweezer setup \citep{anderegg2019} and this may open up the possibility to have longer-range interactions in such setups.

To fully understand the current experimental state of the art in trapping and manipulation of cold atoms, it is important to take a step back and consider the developments prior to the bosonic systems discussed above. Arguably, a large driver of these developments was the desire to trap both bosons and fermions with fermions getting a slightly earlier start as Pauli exclusion helps alleviate problems with loss and system collapse. To provide an adequate overview of the context of few-body experiments, it is therefore necessary to give a brief account of some of the seminal developments in that direction. Note also that in the strongly interacting regime in one dimension, there are large overlaps between the physics needed to describe both fermions and bosons as we will explicate later in the review. 

\begin{figure}[tb]
    \centering
    \includegraphics[width=0.7\linewidth]{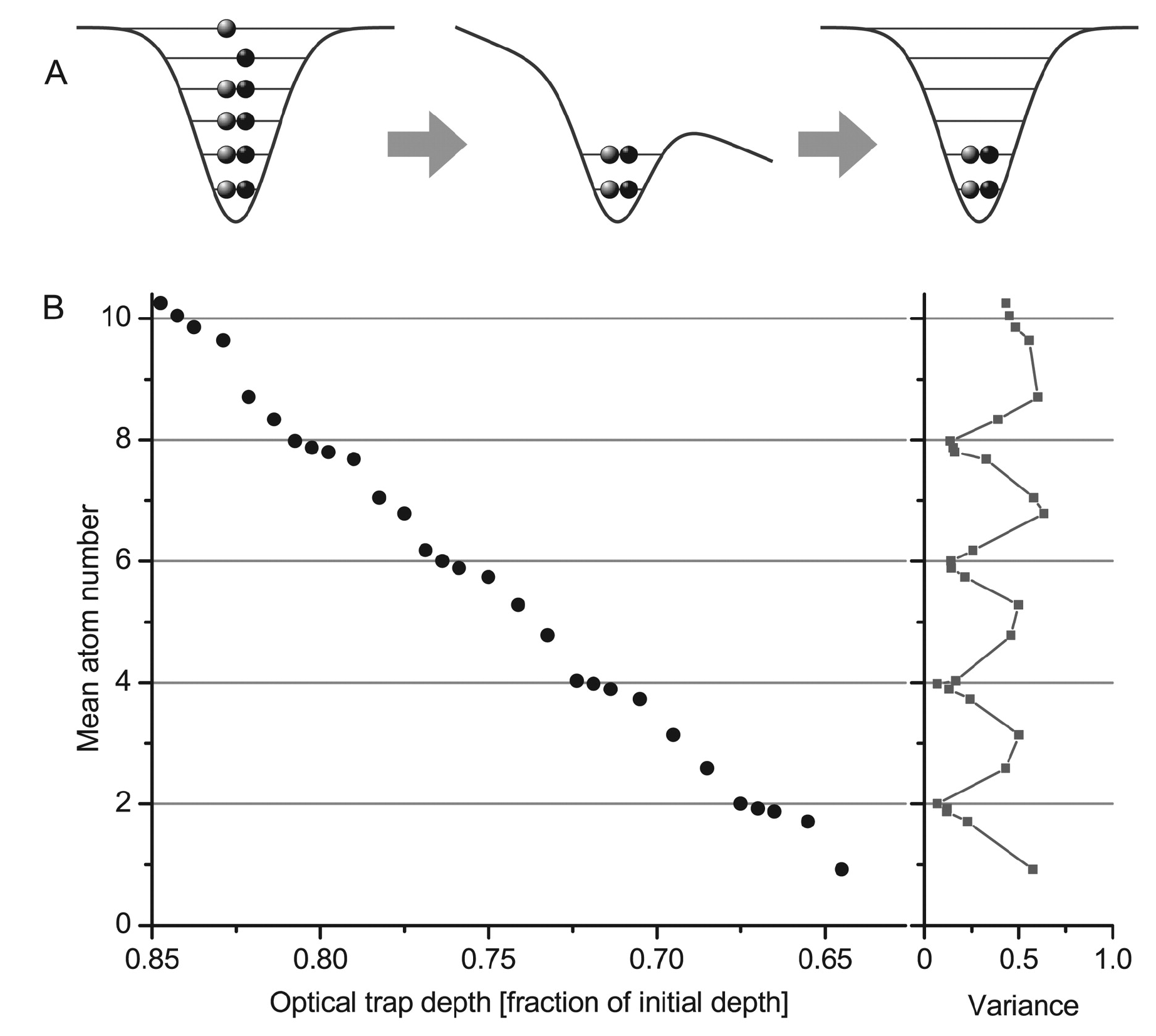}
    \caption{
    (A) Starting from a degenerate two-component Fermi gas of about 600 atoms in a microtrap, a few-particle system is created by adiabatically deforming the potential to allow atoms in higher levels to tunnel out. (B) The mean atom number decreases when the depth of the trap is reduced. This happens in step of two because each energy level in the trap is occupied with one atom per spin state. The variance is shown on the right and one can see that for even atom numbers, the number fluctuations are strongly suppressed. Image from \citep{serwane2011}}
    \label{fig:0d}
\end{figure}

Around a decade ago, a seminal work created samples with less than ten particles near zero temperature in a 1D trapping potential with fermionic $^{6}$Li atoms \citep{serwane2011}. This was done by loading a large number of atoms into a trap with a small `dimple' at the bottom and then `cutting' the trap to remove most of the atoms, see Fig.~\ref{fig:0d}. This experimental procedure paved the way for several studies of the basic properties of quantum systems starting from the few-body limit and building towards the collective states seen with many particles. In particular, this setup was used to create a simple two-body system in which the interaction strength could be varied from weak to strong  \citep{zurn2012}. This experimental work confirmed that indeed local properties of two particles with opposite internal (spin) states are identical to those of spinless fermions in the limit where the effective 1D interaction strength diverges  \citep{zurn2012}. 
This was shown by using a novel tunneling technique in which one particle could `spill' out of the trap with a tunneling rate that is dependent on the interaction strength of the pair. As a check on the setup, the experiment has been performed using two fermions in identical internal states, for which the tunneling rate is independent of the interaction strength as identical fermions cannot interact via a short-range $s$-wave interaction \citep{zurn2012}. 

These experimental techniques were later used to study the case of attractive interactions with more than two particles. It was shown that pairing occurs in system sizes down to six particles~\citep{zurn2013} with a clear signature in terms of an odd-even effect known from pairing studies in nuclei~\citep{ring1980,jensen2004,zinner2013}, quantum dots~\citep{kouwenhoven2001,reimann2002,hanson2007} and spectroscopy of small superconducting grains~\citep{Ralph1995,Black1996,Mastellone1998,Amico2002,VONDELFT2001}.  

The availability of tunable interactions and tunable particle numbers prompted an immediate experimental investigation into the emergence of many-body properties related to the impurity problem. The work of \citep{Jochim2013FewMany} showed how to study this problem starting from two particles and building it bottom-up. This effectively corresponded to constructing a Fermi sea atom by atom, and the energetics could be studied and compared to theoretical predictions. Remarkably, this experiment also found that already for the case of five or six identical fermions interacting with a single particle of opposite spin the system has an energy in agreement with many-body models for (repulsive) interaction strengths ranging from weak to strong. More specifically, the 1D Fermi impurity problem, or polaron problem, has a known solution without a trapping potential \citep{McGuire1965}. The experimental results approach the energy prediction for the polaron in the thermodynamic limit already for only moderate system sizes, indicating a fast convergence to a full Fermi sea interacting with an impurity in one spatial dimension.  
Having access to two-component fermions in tailored 1D traps in the strongly interacting regime enables the study of spin dynamics for few-body systems, and hence the study of magnetic correlations from the few-body point of view. In particular, the work of \citep{Murmann:15} showed how to use a tailored trap in 1D to build a Hubbard model starting from just two atoms. This means that the dynamics of the system as a function of the typical parameters can be explored. These parameters have typically been taken as fixed in theoretical Hubbard model studies of the many-body system, and it has remained uncertain to what extent they represent the true few-body parameters \citep{valiente2009scattering,buchler2010,valiente2015effective}. 
By using a double-well setup with a tunable barrier and a Feshbach resonance to tune the atom-atom coupling, \citep{Murmann:15} could 
explore the system going from the few-body equivalent of the Mott insulator to a superfluid phase, as well as enter the regime of attractive 
interactions. These kinds of experiments provide a starting point for building a quantum simulator of paradigmatic condensed-matter models from the bottom up. Furthermore, in the case of strong (repulsive) interactions, the Hubbard model becomes effectively a spin model as the motional degrees of freedom freeze out in 1D. Here the system should enter an antiferromagnetic regime. The work of \citep{Jochim2015SpinChain} showed that antiferromagnetic spin correlations could be detected for strong interactions in systems with three and four particles in a harmonic trap using the tunneling techniques mentioned above.  

\begin{figure}[tb]
    \centering
    \includegraphics[width=0.7\columnwidth]{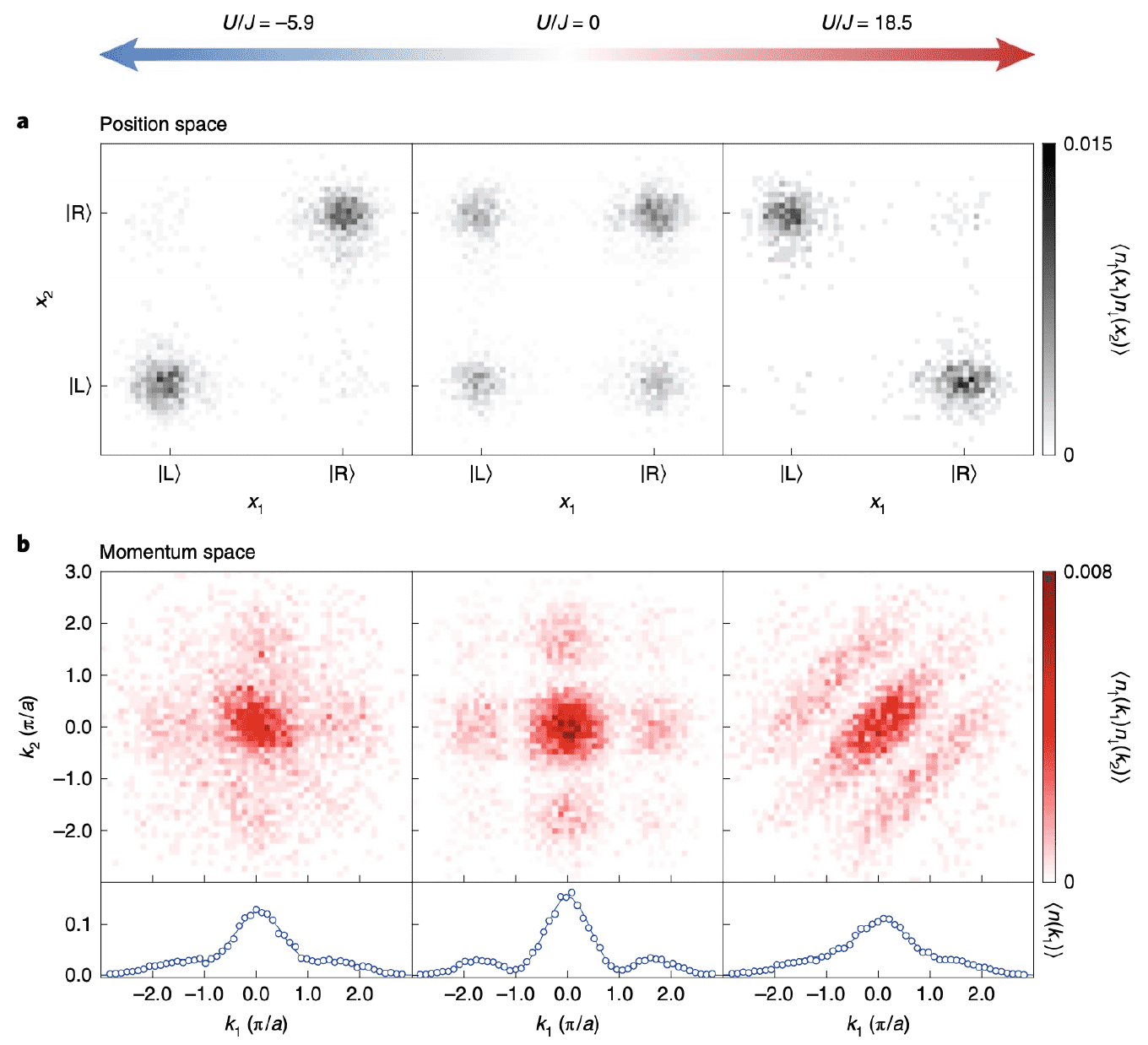}
    \caption{
    (a) Position and (b) momentum correlation functions of a Fermi-Hubbard dimer trapped in a double well potential for different interaction strengths, $U/J$, going from the attractive, through non-interacting and onto the 
    strongly repulsive regime. Image from \cite{bergschneider2019}.}
    \label{fig:SelimExp}
\end{figure}

In recent developments, it has become possible to image low-dimensional few-body Fermi systems \citep{bergschneider2019}\footnote{For the development of probing techniques for many-body systems, the interested reader can consult~\cite{Daniloff2021}.}. This was not only achieved in the traditional sense of knowing the spatial position of atoms, but also imaging in momentum space, see Fig.~\ref{fig:SelimExp}.
A double-well setup such as the one in \citep{Murmann:15} can then be combined with these new imaging techniques to study the correlations of as few as two particles and provide information about the level of entanglement in the system. These imaging techniques have also enabled the study of controlled few-body collisions in the system \citep{guan2019}. This can be used to study details of the dynamics in collisions and quench schemes, including the question of how the system may or may not enter relaxation, opening up a few-body angle on equilibrium, non-equilibrium and quantum chaotic behavior \citep{guan2019}. An extension of the manipulation and imaging techniques to 2D arrays of fermion pairs was recently reported
\citep{hartke2022quantum} and provides very long coherence times for potential use in cold atom-based quantum information and computing setups. 
In \citep{preiss2019}, a system of 1D fermions was captured in a set of optical tweezers and interference effects with high contrast were demonstrated. 
The key measurements involved the momentum distributions that could be used to reveal two- and three-body correlation functions 
for small system sizes \citep{becher2020}. This allows one to study Hanbury-Brown and Twiss (anti)-correlations for fermions using cold atomic 
gases in controlled setups. Furthermore, in a two-dimensional setting, the ability to probe the momentum distribution led to the first observation of a so-called `Pauli crystal' \citep{holten2020}. This is an elusive state that features higher-order density correlations between trapped fermions even without any interactions. It has also become possible to observe few-body shell effects in two-dimensional traps with fermions directly \citep{bayha2020}. For the two-dimensional setups it has been conjectured that by rotating the trap one may even access the exotic physics of fractional quantum Hall systems using either bosonic \citep{andrade2020,palm2020bosonic} or fermionic \citep{palm2020} few-body systems.
These new developments show how the imaging and tweezer techniques can be used also in low-dimensional systems for fermions and provide very 
interesting prospects for future work.

\input{Section02.tex}

\input{Section03.tex}

\input{Section04.tex}

\input{Section06.tex}

\input{Section05.tex}

\input{Section07.tex}

\input{section08.tex}

\input{section09.tex}

\input{Section10.tex}

\bibliographystyle{elsarticle-harv}

\bibliography{SmallReview}

\end{document}

%% file: Section02.tex
\section{Exact solutions}\label{sec:EarlyMod}

Long before 1D atomic quantum systems could be realized in the laboratory, theoretical models that would later become extremely relevant in the field of ultracold atoms were already well developed in the context of mathematical physics. This is due mainly to the fact that, given certain conditions, interacting 1D quantum systems can be exactly solved, in the sense that the wave functions and the entire energy spectrum can be obtained without approximations. One of the key ingredients for such an accomplishment is a mathematical tool developed by Bethe in 1931 \citep{Bethe1931Ansatz}, now called the coordinate {\it Bethe  ansatz} \citep{gaudin2014}: a trial wave function of a very general character upon which a set of constraints (given by the particular interacting model) is enforced.

Bethe's original work employed this approach to solve the Heisenberg chain, a quantum spin model of great importance in the field of condensed matter physics and quantum magnetism. Later developments brought about a series of works which made use of this same tool to exactly solve several other models. The ones that would turn out to be most valuable for experiments with atoms at low temperatures are models of particles in the continuum with short-range interactions that are modeled by a Dirac delta function, also known as contact interactions\footnote{See Sec.~\ref{sec:beyond} for a brief discussion of beyond-contact interactions.}. Among these models, the particular cases of the Lieb-Liniger model \citep{Lieb1963GS,Lieb1963LLExcited} for bosons and the Gaudin-Yang \citep{Gaudin1967Fermi, Yang1967Fermi} model for two-component fermions provide examples of the great impact that the applications of the Bethe ansatz would have on current developments in physics \citep{Batchelor2007After}. Comprehensive descriptions of both bosonic and fermionic models with contact interactions, including details on the Bethe ansatz solutions for each case, have been provided in \citep{cazalilla2011,Guan2013Review}.

Our goal in the next two sections is, therefore, to describe the building blocks for many of the problems explored in this review, specifically the solutions of a system with contact interactions for few-body systems and the limit of strong interactions. In Section \ref{subsec:BetheAnsatzMethods}, which is dedicated to a review of the underlying methods, we provide a formal description of the Bethe ansatz solutions for an arbitrary number of particles in the Lieb-Liniger model and other Bethe ansatz-solvable models.

\subsection{Two and three particles with contact interactions}

\begin{figure}[tb]
  \centering
\includegraphics[width=0.7\linewidth]{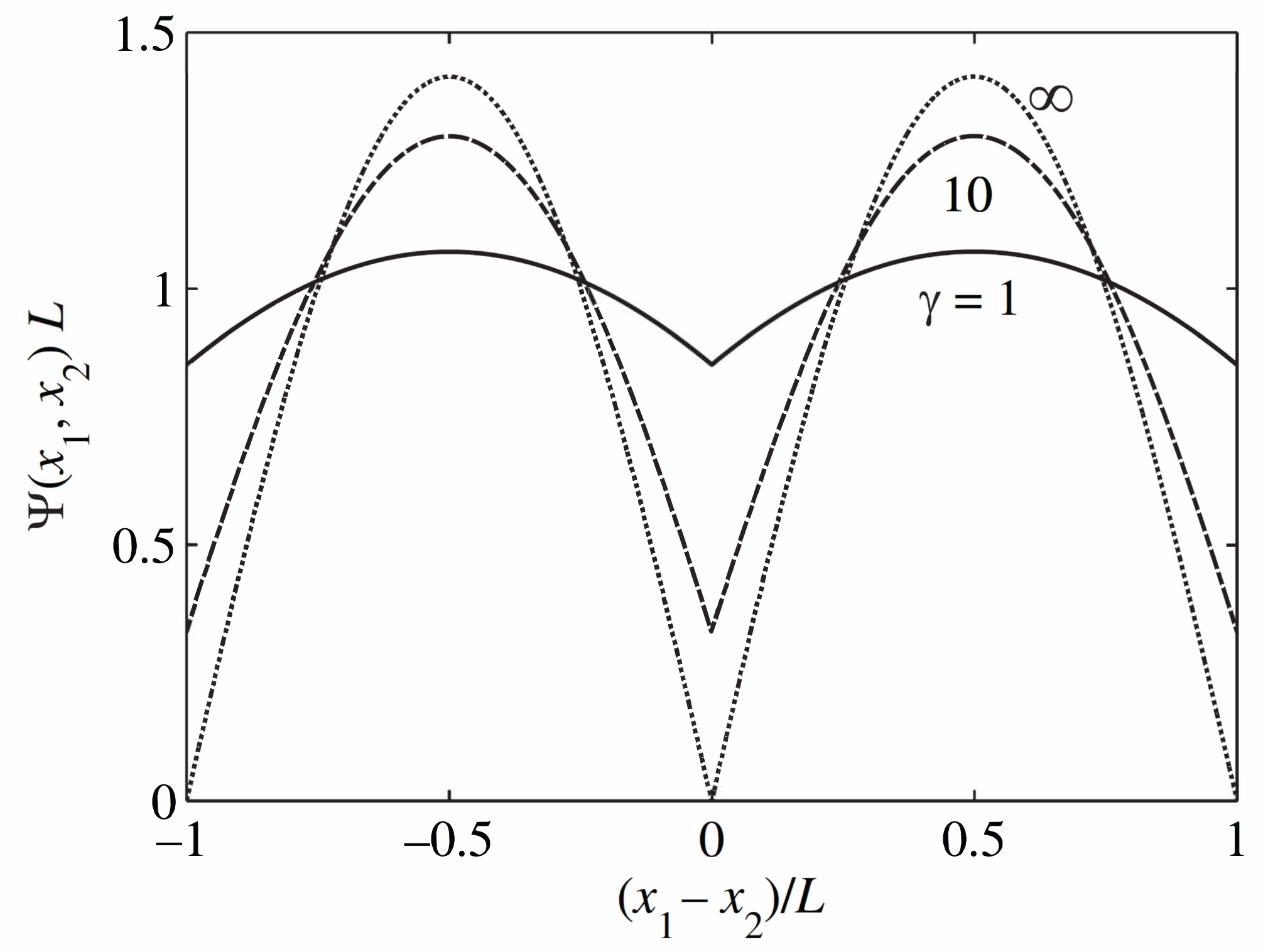}\hfill
\caption[fig2]{Ground state wave function of two interacting particles in a system of length $L$ with periodic boundary conditions. Each curve belongs to a different value of the interaction parameter $\gamma=c/\rho$, where $\rho$ denotes the particle density. At $\gamma \rightarrow \infty$, the wave function vanishes for $x_1=x_2$. Image from \citep{pethick2002}.}
\label{fig:2bodysol}
\end{figure}

The first relevant application of the Bethe ansatz is the case of two particles with contact interactions in a homogeneous potential and periodic boundary conditions. The Hamiltonian for this system, which is equivalent to the $N=2$ limit of the Lieb-Liniger model, reads:
\begin{equation}\label{llH_2p}
    H= - \left(\frac{\partial^2}{\partial x_1^2} + \frac{\partial^2}{\partial x_2^2}\right) +2c\delta(x_1-x_2),
\end{equation}
where $\hbar=m=1$ is assumed. The interaction strength is determined by the parameter $c$, which characterizes the repulsive ($c>0$) and attractive ($c<0$) regimes. 
The proposed Bethe ansatz wave function in this simple example reads \citep{eckle2019models}:
\begin{equation}
    \psi(x_1,x_2)= A_{12} e^{ik_1 x_1} e^{ik_2 x_2}+A_{21} e^{ik_1 x_2} e^{ik_2 x_1},
\end{equation}
where we sum over permutations of the {\it quasimomenta} $k_1$ and $k_2$, restricted to the region where $x_1<x_2$. Due to the bosonic symmetry of the wave function, it is enough to consider this domain only, since the amplitudes in the other region are the same. Assuming contact interactions leads to a relation for the coefficients $A_{12}$ and $A_{21}$ of the form 
\begin{equation}\label{n2conds}
    \frac{A_{21}}{A_{12}}=\frac{k_2-k_1+ic}{k_2-k_1-ic}.
\end{equation}
Minimizing the energy yields, for the ground state, $k_1=-k_2=k$, and the addition of periodic boundary conditions 
\begin{equation}\label{pb_condition}
\psi(x_1=0,x_2)=\psi(x_1=L,x_2)\,\,\,\,\text{and}\,\,\,\, \psi(x_1,x_2=0)=\psi(x_1,x_2=L)
\end{equation} 
allows us to write the resulting ground state wave function for $N=2$ as in a homogeneous potential as \citep{pethick2002}:
\begin{equation}\label{n2sol}
    \psi(x_1,x_2)=A \cos{\left[k\left(\vert x_1-x_2\vert -\frac{L}{2}\right)\right]},
\end{equation}
where $A$ is a normalization constant, and the quasimomentum $k$ and the interaction strength $c$ are related by $c=2k \tan{\left(kL/2\right)}$.
The wave function in Eq. \eqref{n2sol} respects bosonic symmetry when coordinates are exchanged and vanishes for $x_1=x_2$ when $c\rightarrow \infty$, corresponding to $k\to \pi/2$. Fig.~\ref{fig:2bodysol} shows an illustration of the wave function given in Eq. \eqref{n2sol}.

\begin{figure}[tb]
  \centering
\includegraphics[width=0.7\linewidth]{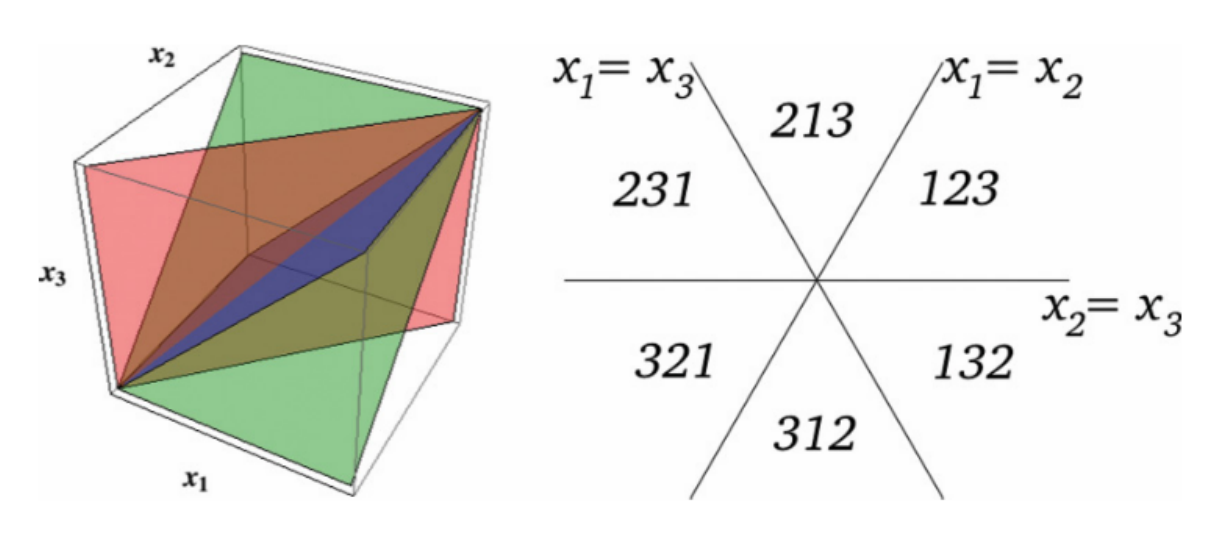}\hfill
\caption{ The two-body contact interactions of the three-body problem can be depicted as the intersections of the three planes where $x_i=x_j$ (left). The possible orderings of the particles describe 6 regions, with the lines representing the contact points. Image from \citep{lamacraft_2013} }
\label{fig:3bodyscattering}
\end{figure}

More generally, the Bethe ansatz wave function is written as a linear combination of $N!$ plane waves
\begin{equation}\label{coordBA}
    \psi(x_1,...,x_N)=\sum_P A(P)\exp{\left({i\sum_{j=1}^N k_{Pj}x_j}\right)},
\end{equation}
where $\{k_{P1},...,k_{PN}\}$ is a set of quasimomenta, $P$ describes a particular permutation of these numbers and $A(P)$ denotes the permutation-dependent coefficients. 
For $N=3$, for instance, the Bethe ansatz will be given by \citep{Takahashi1999OneDimensional}
\begin{eqnarray}
    \psi(x_1,x_2,x_3)=
    A_{123}e^{i(k_1x_1+k_2x_2+k_3x_3)} &+& A_{213}e^{i(k_2x_1+k_1x_2+k_3x_3)} + \nonumber \\
    A_{132}e^{i(k_1x_1+k_3x_2+k_2x_3)} &+& A_{312}e^{i(k_3x_1+k_1x_2+k_2x_3)} + \nonumber \\
    A_{231}e^{i(k_2x_1+k_3x_2+k_1x_3)} &+& A_{321}e^{i(k_3x_1+k_2x_2+k_1x_3)},
\end{eqnarray}
where again we are assuming a particular domain $x_1<x_2<x_3$. Formally, the three-particle bosonic problem in a homogeneous potential is simply an extension of the $N=2$ case described above. However, it is also an important example of a minimal model where integrability is explicitly manifested \citep{Jimbo1989YB}. In Fig. \ref{fig:3bodyscattering} we see a sketch of the planes that characterize the contact-interacting three-body problem in 1D, as well as the possible particle orderings on the ring.

The coefficients $A_{abc}$ corresponding to the wave function for $N=3$ respect six relations analogous to that of Eq. \eqref{n2conds}. They can be transformed into one another through the action of $K_{ab}=-(k_a-k_b+ic)/(k_a-k_b-ic)$, a set of objects that in turn obey
\begin{equation}\label{n3yb}
   K_{ab}K_{ac}K_{bc}=K_{bc}K_{ac}K_{ab}.
\end{equation}

Quite remarkably, this expression guarantees the existence of an exact solution and also holds for the $N$-body problem. In the three-body regime, it is the most elementary example of a Yang-Baxter equation \citep{Jimbo1989YB,Yang1991Braid,Batchelor2016YBExperiments}, a relation that can be physically understood as a pairwise factorization of interactions in the system. Since these operators permute the momentum values, the Yang-Baxter equation can be interpreted as a condition of non-diffraction, meaning that the sets of incoming and outgoing quasimomenta in scattering events are the same \citep{Mcguire1964Exactly,sutherland2004, lamacraft_2013}. The Yang-Baxter relations are also respected by the transfer matrices in two-dimensional lattice models \citep{Baxter1972Vertex}, and other interpretations of this equation exist in different domains (see, for instance \citep{Vind2016ExperimentYB,Batchelor2016YBExperiments}). A set of operators obeying the Yang-Baxter equation is generally a signature of integrability in many-body systems \citep{Batchelor2014Integrability}. A precise definition of the concept of quantum integrability, however, still remains a point of debate \citep{Caux2011Integrability}, and one typically needs to treat systems on a case 
by case basis with regard to questions of quantum integrability.

\subsection{The Tonks-Girardeau limit} 
\label{subsec:TGmodel}

A particular limit of the interacting Bose gas which has proven to be of great importance for both theory and experiment is the regime of infinitely repulsive interactions $(\gamma\rightarrow \infty)$. This regime is characterized by surprisingly simple solutions which are given by the so-called Bose-Fermi mapping, first proposed by Girardeau in 1960 \citep{girardeau1960}. It states that the wave function $\psi_B$ for ``impenetrable core" bosons can be written as
\begin{equation}
    \psi_B(x_1,...,x_N)=\prod_{i<j}\text{sgn}(x_i-x_j) \psi_F(x_1,...,x_N),
\end{equation}
where $\psi_F(x_1,...,x_N)$ is a wave function for identical spinless fermions. The main feature of the mapping is to identify the bosonic wave function as a symmetrized version of the fermionic wave function, which is antisymmetric under permutation of particle coordinates. The latter is obtained, for a system of $N$ particles, as the Slater determinant of single-particle solutions $\phi_n(x_i)$ for a given potential $V(x)$, where $n$ denotes the energy level. As a concrete example of this, the ground state solution for the bosonic problem in a system of length $L$ with periodic boundary conditions can be written, exploring the properties of the determinant as\footnote{Formally, this solution holds only for odd $N$ in the specific case of a ring geometry. For trapped systems, Girardeau's mapping holds for arbitrary $N$.} 

\begin{equation}
    \psi_B(x_1,...,x_N)=C\prod_{i<j}\Big| \sin{\left[\frac{\pi}{L}(x_j-x_i)\right]}\Big|
\end{equation}
where $C$ is a normalization constant. The corresponding expression for a harmonically trapped system has the form \citep{girardeau2001}
\begin{equation}
    \psi_B(x_1,...,x_N)=C'\prod_{i=1}^{N}\exp\left(-\frac{x_{i}^{2}}{2b^2}\right)\prod_{i<j}|x_j-x_i|,
\end{equation}
where $b$ is the corresponding oscillator length, and $C'$ is a normalization constant. Note the products of coordinate differences in both these cases. This structure is a result of the relation to Vandermonde determinants of the antisymmetric expressions, and is for instance also found in Laughlin wave functions at integer filling in the quantum Hall effect \citep{Scharf1994}.

The Bose-Fermi mapping, which is restricted to 1D geometries, also guarantees that the energies and spatial distributions between the bosonic wave function and its fermionic counterpart will be identical. Furthermore, it provides a way of understanding how the strong interactions between the bosons effectively mimic the Pauli exclusion principle, with the wave function vanishing for $x_i=x_j$. The mapping does not hold, however, for the momentum distribution, where the symmetry of the wave function plays a significant role \citep{Lenard1964MDist}. A comprehensive account of the exact solutions of strongly-interacting trapped gases is provided in \citep{Minguzzi2022Review}.

Finally, in the infinitely attractive limit ($c\rightarrow -\infty$), an excited state with a gas-like configuration can be found. In analogy to the repulsive case, it has been dubbed the Super Tonks-Girardeau regime, and it surprisingly features properties that are similar to a strongly repulsive 1D gas (see Fig.~\ref{fig:haller}), with even stronger correlations. This regime has been accessed theoretically with Monte-Carlo simulations where the many-body wave function is constructed from the exact solution of the two-body problem \citep{Astrakharchik:05}. 
Outside the Tonks-Girardeau limit, an exact solution for harmonically trapped systems is known only for two particles. For larger systems, however, useful approximations can be found. These solutions are the focus of our next Chapter.

%% file: Section03.tex
\section{Single component static and dynamics}
\label{sec:OneComp}

The ability to treat systems with small particle numbers exactly offers the possibility to study the dynamics of not only classical properties, such as density or momentum distributions, but also non-classical properties such as coherence and entanglement. To be able to do this in a manner relevant to currently existing experimental setups, we will in the following review different approaches for describing small systems, especially in inhomogeneous environments, and briefly discuss fundamental solutions and equilibrium properties as a basis for understanding the dynamical properties. 

\subsection{Two-particles in a harmonic trap}
\label{subsec:two_part}

As shown in the previous section already, the fundamental building block of many-particle systems is the two particle system and a lot of ideas and concepts already emerge at this level. We therefore begin with discussing the fundamental solution in 1D for two particles of mass $m$ confined by a harmonic trapping potential of frequency $\omega$ and interacting via a short-range delta-function interaction of strength $g_{\text{1D}}$\footnote{Note that in free space this quantity relates to the parameter $c$ in Eq. \eqref{llH} as $g_{\text{1D}} = c \hbar^2/m$.}  \citep{Busch1998TwoAtoms}. In 1D the Hamiltonian is given by
\begin{equation}
    \label{eq:TwoParticlesInTrap}
    H=\sum_{n=1}^2\left( -\frac{\hbar^2}{2m}\nabla_n^2+\frac{1}{2}m\omega^2 x_n^2\right) + g_{\text{1D}}\delta(x_1-x_2),
\end{equation}
and due to the quadratic form of both the kinetic energy and the particle-independent trapping potential it can be separated into a centre-of-mass and a relative part, $H=H_+ +H_-$, with
\begin{align}
    \label{eq:COM}
    H_+=&-\frac{1}{2}\nabla^2_+ + \frac{1}{2} x_+^2,\\
    \label{eq:REL}
    H_-=&-\frac{1}{2}\nabla^2_- + \frac{1}{2} x_-^2+g\delta(x_-).
\end{align}
Here $x_+=\frac{x_1+x_2}{\sqrt{2}}$ is the centre-of-mass coordinate and $x_-=\frac{x_1-x_2}{\sqrt{2}}$ is the relative coordinate. To achieve the simple forms for $H_+$ and $H_-$, we have used harmonic oscillator scaling where energies are given in units of $\hbar \omega$, lengths in units of the ground state width $a=\sqrt{\hbar/m\omega}$ and the interaction strength in units of $\sqrt{2\hbar^3\omega/m}$. 

\begin{figure}[tb]
    \centering
    \includegraphics[width=0.6\linewidth]{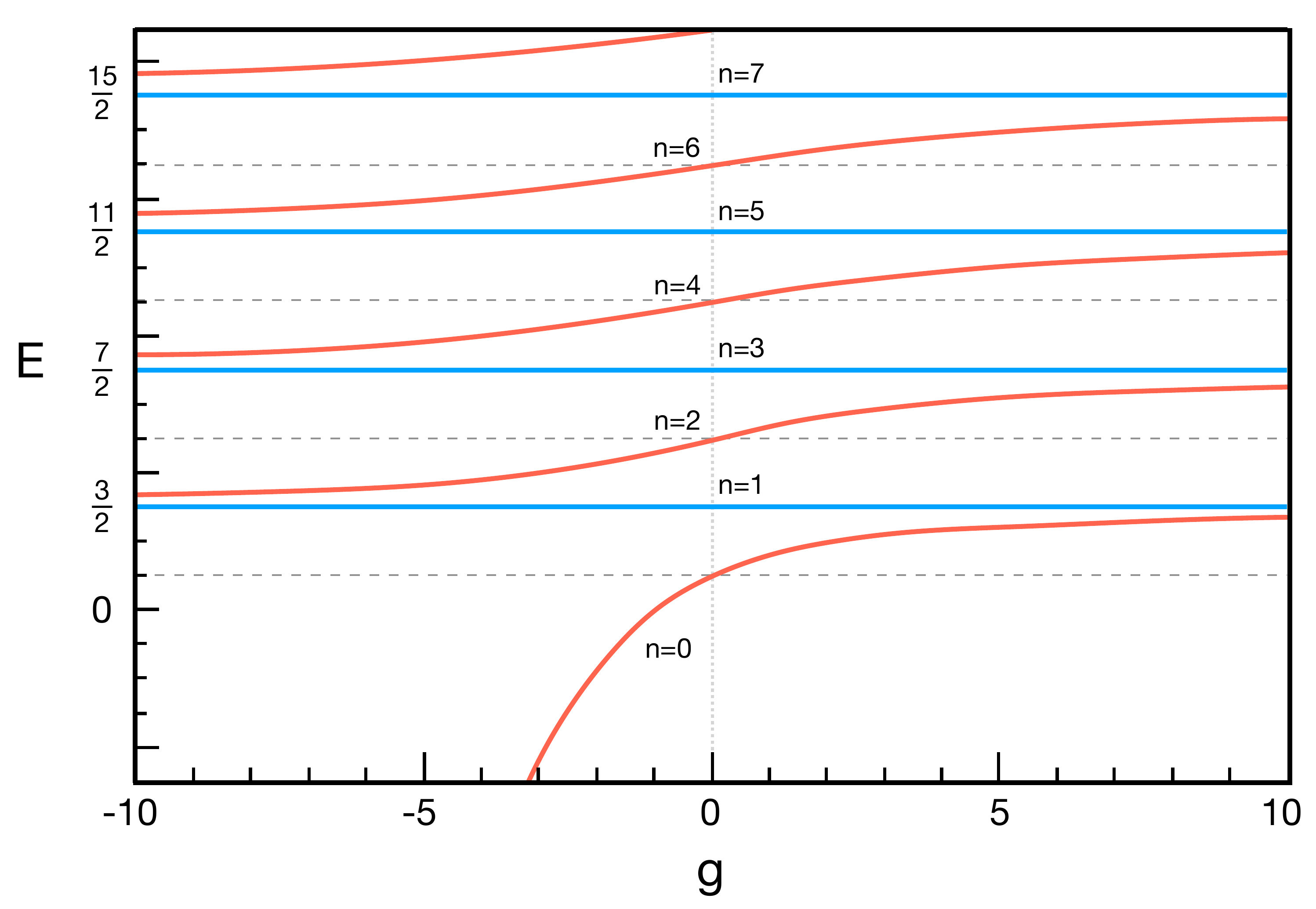}\hfill
    \caption{Spectrum of the relative part, $H_-$, of the two-particle Hamiltonian as a function of the interaction strength. The eigenvalues corresponding to odd eigenstates are not affected (blue lines), whereas the eigenvalue of the even eigenstates increase and decrease (red lines) for larger positive and negative values of $g$.}
    \label{fig:1D2PSpectrum}
\end{figure}

One can immediately see that the centre-of-mass Hamiltonian, $H_+$, simply describes a single particle harmonic oscillator with the well known solutions 
\begin{equation}
    \psi_n(x_+)= (2^n n!)^{-1/2} \pi^{-1/4} \mathcal{H}_n(x_+) e^{-\frac{x_+^2}{2}}, 
\end{equation}
where $\mathcal{H}_n$ is the $n$th order Hermite polynomial, and the energies are given by $E_n=(n+1/2)$. One is therefore left with solving the stationary Schr\"odinger equation for the relative Hamiltonian, which consists of a single particle harmonic oscillator with a contact potential at the origin. This immediately suggests that the odd eigenfunctions of the standard harmonic oscillator are still eigenfunctions of $H_-$, as they possess a node at $x_-=0$ and do not feel the additional contact potential. The even states, however, are affected and can be found to be
\begin{equation}
    \phi_n(x_-)=\mathcal{N}_n e^{-x_-^2/2} U\left(\frac{1}{4}-\frac{E_n}{2},\frac{1}{2},x_-^2\right),    
\end{equation}
where the  $U(a,b,z)$ are the Kummer functions and $\mathcal{N}_n$ is the normalization. The corresponding eigenenergies are given by the implicit equation 
\begin{equation}
    -g=2 \frac{\Gamma\left(-\frac{E_n}{2}+\frac{3}{4}\right)}{\Gamma\left(-\frac{E_n}{2}+\frac{1}{4}\right)}\;,
    \label{eq:two_body_energy}
\end{equation}
and the spectrum is shown in Fig.~\ref{fig:1D2PSpectrum}. One can see that, as one would expect, the odd eigenenergies are not affected by a change in $g$ and the even ones increase and decrease depending on the sign of $g$. In particular, it is worth noting that for $g\rightarrow\infty$ the energies of neighbouring odd and even states become degenerate and the form of each even eigenfunction approaches the absolute value of the next-higher-lying odd one (cf.~Section~\ref{subsec:TGmodel}).

While this model only contains essential elements, its applicability to ultracold atom experiments was demonstrated by the Esslinger group \citep{Stoferle:06} and a large number of subsequent works have taken advantage of it over the last two decades. More complex versions of the above model have also been developed, for example considering anisotropic traps \citep{Idziaszek:05,bougas2020stationary} or interactions via higher order partial waves \citep{Kanjilal:2004,Idziaszek:06,zinner2012}. A further extension to the situation where the two particles are trapped in differently centered harmonic oscillator potentials of the same frequency was analytically solved by \citep{Krych:2009}, who showed that in this case the relative Hamiltonian corresponds to a harmonic oscillator with an off-center delta function. It is worth noting that many solutions to the situations named above also exist in higher dimensions, but are not discussed in this review.

It is also possible to extend the Hamiltonian in Eq.~\eqref{eq:TwoParticlesInTrap} to include the effect of spin orbit coupling. The spectrum of such a system was discussed in \citep{Guan:2014}, the spin structure in \citep{Guan:2015} and a numerical treatment that included the Raman coupling term  was presented in \citep{Usui:2020}. Finally, the physics of two particles in an anharmonic potential, where the center-of-mass and relative degrees of freedom can couple, was explored in \citep{Matthies:2007}. 

The existence of analytical solutions allows one to have full access to all classical and quantum properties of the system, and we will discuss different stationary properties below. However, it is worth noting that the existing solutions can also be used to further applications in other areas. Examples of this are a more precise description of the interaction term of the Bose-Hubbard model \citep{buchler2010,Kremer:17}, the description of quantum gates \citep{Calarco2000, Negretti:2005}, or for discussing the measurement of spatial mode entanglement in ultracold bosons \citep{Goold:2009}.  Furthermore, the knowledge of the exact solutions for the delta-split harmonic oscillator has allowed one to study the many-body physics in the Tonks limit in such systems \citep{Busch:03,Goold:08,Goold:08b}.

Finally, the two-particle physics beyond the point-like interaction has been discussed in the literature (see also Section~\ref{Sec:LongRange}). In \citep{Oldziejewski:2016} the physics of two particles with dipolar interaction was treated (in full 3D) and the stationary and dynamical properties of two interacting molecules were determined in \citep{Dawid:2018,Dawid:2020,Sroczynska:2021}. Effects stemming from combining anisotropic traps and realistic interaction potentials were numerically discussed in \citep{Grishkevich:2009}.

\subsection{Three particles in a harmonic trap}

While systems with more than two particles in harmonic traps do not allow for analytically exact solutions in all generality\footnote{Consult~\citep{Beau2020,delCampo2020} for some unique examples of exactly solvable systems of a few equal-mass particles in a trap.}, considering the three particle problem can still be very helpful. The Hamiltonian for the most general system of two particles of species $A$ interacting with strength $g_A$, and one particle of species $B$ which interacts with the $A$ particles through $g_{AB}$, is given by 
\begin{equation}
    H=\sum_{n=1}^3\left( -\frac{1}{2}\nabla_n^2 + \frac{1}{2}x_n^2 \right) + g_A \delta(x_1-x_2)
    + g_{AB}\left[ \delta(x_1-x_3) +\delta (x_2-x_3) \right],
\end{equation}
assuming all masses to be equal and using harmonic oscillator scaling.
This Hamiltonian is separable when moving to Jacobi coordinates
\begin{eqnarray}
    X&=&(x_1-x_2)/\sqrt{2}\,,\\
    Y&=&(x_1+x_2)/\sqrt{6}-\sqrt{2/3}\,x_3\,,\\
    Z&=&(x_1+x_2+x_3)/\sqrt{3}\,,
\end{eqnarray}
which allow one to write the Hamiltonian as separate centre-of-mass and relative Hamiltonians, $H=H_{+}(Z)+H_{-}(X,Y)$. The center-of-mass part, $H_+(Z)=-\frac{1}{2}\nabla_Z^2+\frac{1}{2}Z^2$, again just describes a harmonically trapped single particle, while the relative part can be written as
\begin{align}
    H_-(X,Y)=&-\frac{1}{2}\left(\nabla_X^2+\nabla_Y^2 \right) + \frac{1}{2}\left(X^2 + Y^2 \right)\nonumber\\
     &+g_A\delta(X) + g_{AB} \left[ \delta \left(-\frac{1}{2}X+\frac{\sqrt{3}}{2}Y\right) 
     +\delta \left(-\frac{1}{2} X - \frac{\sqrt{3}}{2} Y \right) \right]\,.
\end{align}
For $g_A=g_{AB}=g$ the eigenstates of $H_{-}(X,Y)$ possess $C_{6\nu}$ symmetry, allowing one to write the interaction part of the Hamiltonian in cylindrical coordinates, $\rho=\sqrt{X^2+Y^2}$ and $\tan\phi=Y/X$, as 
\begin{equation}
    H_\text{int}=\frac{g}{|\rho|}\sum_{j=1}^6 \delta \left( \phi -\frac{2j-1}{6}\pi \right)\;.    
\end{equation}
This symmetry can significantly simplify the calculation of eigenstates when considering a truncated Hilbert space of non-interacting states or help in creating relevant trial wave-functions for which the conditions from particle exchange are simpler \citep{Harshman2012, GarciaMarch2014}. As a benchmark for approximate treatments, highly accurate spectra for three interacting bosons or fermions were determined in \citep{Amico2014}; see also the work of \citep{bougas2021few} using the hyperspherical formalism.
\begin{figure}
    \centering
    \includegraphics[width=0.7\linewidth]{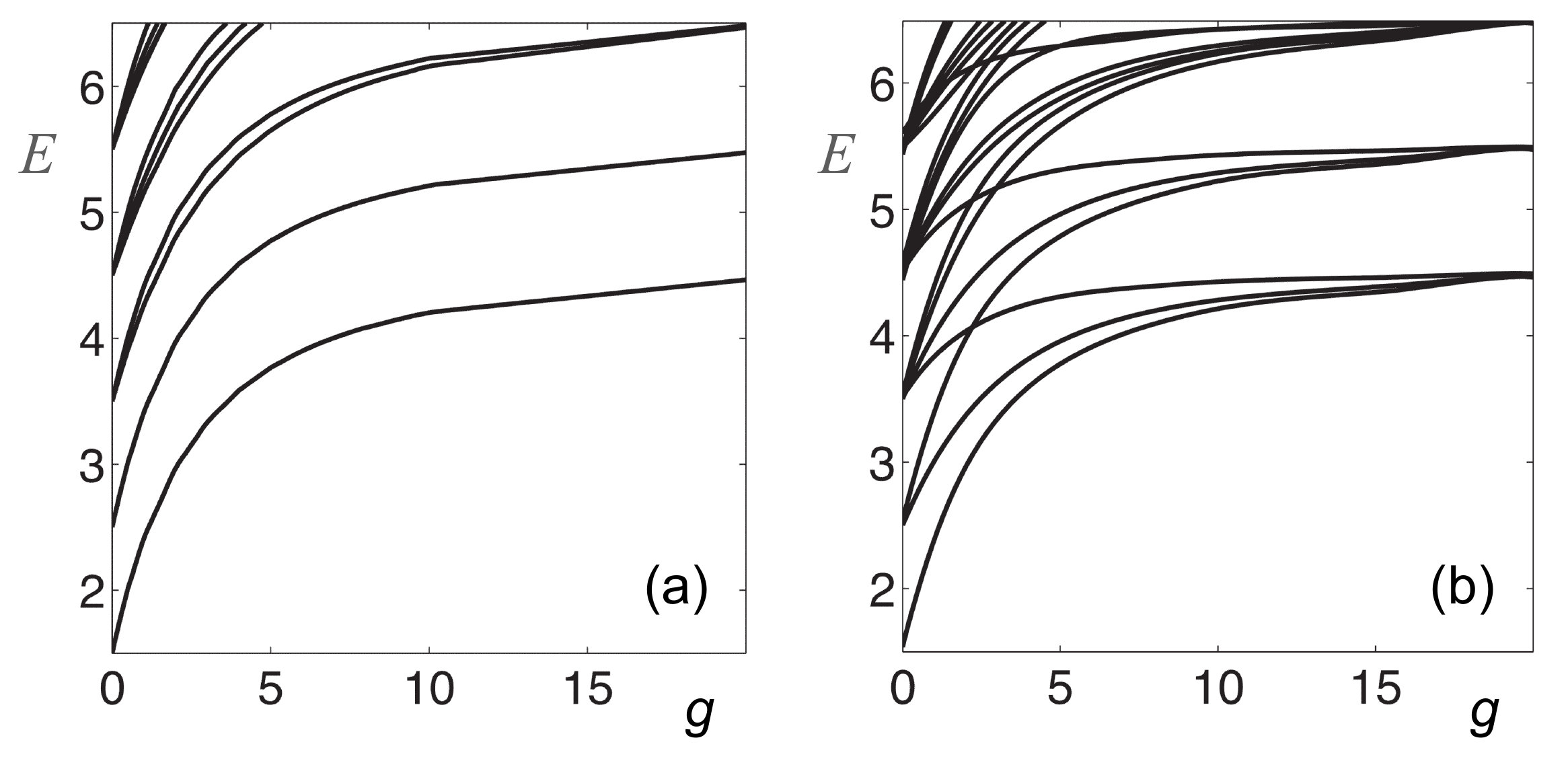}
    \caption{Spectrum as a function of the interaction strengths for (a) three indistinguishable bosons and (b) two indistinguishable bosons and a third distinguishable particle. The axes are scaled in harmonic oscillator units. Figure adapted from \citep{GarciaMarch:2016}.}
    \label{fig:three_part}
\end{figure}

In Fig.~\ref{fig:three_part}(a) the spectrum of three indistinguishable bosons is shown as a function of the interaction strength, $g$. Compared to the two particle case discussed above, the addition of an extra particle creates more degeneracies in the spectrum at $g=0$ and $g\rightarrow\infty$.  For the case of two indistinguishable bosons of species $A$ and a third distinguishable particle of species $B$ 
the spectrum is shown in Fig.~\ref{fig:three_part}(b). As there is no requirement for symmetric scattering between the $A$ and the $B$ particles, the spectrum for this situation contains more states compared to the panel~(a), which highlights the complexity that is introduced when multiple components are considered.

\subsection{Approximate approaches}

Even though a general solution for 1D gases with contact interactions for any number of particles exist in free space (see Section~\ref{subsec:BetheAnsatzMethods}), when going beyond three particles it is often helpful to consider approximate techniques that allow to include external trapping potentials or give easier access to specific quantities. In particular, for small systems it is possible to consider many-body variational approaches and in the following we will discuss a number of different ones. 

\begin{figure}[tb]
    \centering
    \includegraphics[width=\linewidth]{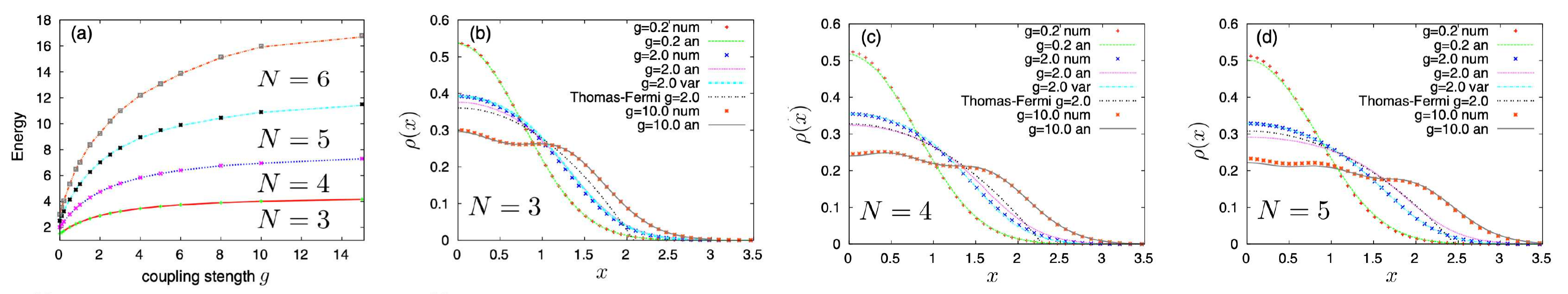}
    \caption{(a) Comparison of the total energies obtained numerically (marks) and analytically (lines) via the CPWF as a function of the coupling strength $g$. The numerical results for $N = 3,4$ and $5$ are obtained by exact diagonalization and for $N= 6$ by the multi-configuration time-dependent Hartree method. Panels (b)-(d) show the analytically and numerically obtained one-body densities for different particle numbers and coupling strengths. All quantities are in harmonic oscillator units. Figure adapted from \citep{Brouzos2012}.}
    \label{fig:CPWF}
\end{figure}

In \citep{Brouzos2012} this problem was approached for bosons by using the relative part of the Hamiltonian, after the center-of-mass part was decoupled (see Fig.~\ref{fig:CPWF}). All particle interactions are contained in the relative Hamiltonian and the authors constructed a set of $N$-body analytic wave functions, known as correlated pair wave functions (CPWF), from the exact two-body solutions in terms of parabolic cylinder functions $D_{\mu}$ as
\begin{equation}
    \Psi_\text{CP}=C \prod_{i<j}^{P} D_{\mu} \left(\beta |x_i-x_j|\right),
\end{equation}
where $P=\frac{N(N-1)}{2}$ is the number of particle pairs and $C$ is a normalization constant. Taking into account the boundary conditions from the point-like interactions yields a transcendental
equation for $\mu$ 
\begin{equation}
    \frac{g}{\beta}=-\frac{2^{3/2}\Gamma\left(\frac{1-\mu}{2}\right)}{\Gamma\left(-\frac{\mu}{2}\right)}\;,
\end{equation}
and while $\beta$ is known in the limiting cases of zero and infinite strengths interactions, it can be treated variationally for finite $g$. This approach leads to excellent agreement with exact numerical calculations for intermediate interaction strengths \citep{barfknecht2015,barfknecht2016} and it was later generalised to multi-component fermion systems \citep{Ioannis:2013} and also shown to match  quantum Monte-Carlo results for bosons in different potential settings \citep{Brouzos:2013}. 

While most variational ansatz approaches rely on the Jastrow split of the wave function into the non-interacting solution (center of mass part) and the interacting part (relative part), they normally require the  specific form of the pair-correlation function to be known. However, in \citep{Koscik:18} a general pair correlation function of the form 
\begin{equation}
    \varphi(x)=\frac{1}{\alpha}\left(1-\lambda e^{-\alpha|x|}\right)
\end{equation}
was suggested, which is independent of the exact form of the two-body interaction. Here $\alpha=\sqrt{2/N}$ and $\lambda$ is chosen to fulfill the discontinuity condition whenever two particles scatter. This is a significant simplification that still gives good results while being numerically inexpensive. A similar idea was used in \citep{Kocik_2020} to find the many-body ground-state of interacting p-wave fermions in a 1D harmonic trap. 

A different approach was presented in \citep{Wilson2014Geometric}, where a wave function for a gas in a harmonic trap was geometrically constructed from the known Bethe ansatz wave functions close to the trap centre where the interactions dominate ($\lambda<\Lambda$) and a Gaussian decay at the trap edges ($\lambda>\Lambda$)
\begin{equation}
\Psi(\lambda,\vec{\theta})=\left\{ \begin{array}{cc}
\psi_{B}\left(\vec{\kappa},\lambda,\vec{\theta}\right), & \lambda<\Lambda\\
A(\vec{\theta})\exp\left(-\alpha(\vec{\theta})(\lambda^{2}-\Lambda^{2})\right), & \lambda>\Lambda
\end{array}\right.,
\end{equation}
where $\lambda$ and $\vec{\theta}$ are the radial and angular components in hyperspherical coordinates and both, the decay parameter $\alpha(\vec{\theta})$ and the normalization $A(\vec{\theta})$, are fixed by the boundary conditions at $\lambda=\Lambda$. To find the ground state energy the position of the boundary $\Lambda$ and the quasi-momenta $k_i$ are used as variational parameters. The authors showed that this approach works in the attractive and the repulsive regime and can be highly accurate when compared to numerical and exact benchmarks for two- and three-particle systems. Related studies with few fermions can be found in \cite{Roditi2012}. 

A variational approach for systems consisting of two indistinguishable fermions and one distinguishable particle was presented in \citep{Loft:2015}. Since for the non-interacting limit ($g=0$) and for the impenetrable one ($1/g\rightarrow 0$) one can find exact solutions for the factorized wave-function in  Jacobi hypercylindrical coordinates \citep{volosniev2015EPJST}, the authors assumed that this factorisation approximately holds for finite interaction strengths as well. Based on this they constructed a variational ansatz and showed that such an approach can describe the energies and the spin-resolved densities of the system very well. 

A straightforward approach that uses an interpolatory ansatz for many-body systems in an inhomogeneous setting with finite strength interactions was presented in \citep{Andersen:16} and subsequently developed into a systematic framework for both static and dynamic few-body studies \citep{Lindgren2020Interpolatory}. Here the idea was to use a linear combination of known eigenstates in
the extreme limits of the interaction strength, $g\rightarrow 0$ and $g\rightarrow \infty$, as the system at intermediate-strength interactions contains a mixture of qualities from both limits. This technique was subsequently also applied to a systems of fermions with equal and unequal masses \citep{Pecak:17}, and for both cases good agreement to exact results for up to six particles was shown. The unequal mass few-body states in the limit $g\rightarrow \infty$ had been calculated around the same time 
using semi-analytical techniques \citep{dehkharghani2016JPhB}. This was later generalized to larger systems of contact interacting systems with unequal mass in harmonic traps, and used to infer the potential of both integrable, and ergodic quantum chaotic behavior in such strongly interacting systems \citep{Harshman2017}.

A model for few repulsively interacting particles trapped in a 1D harmonic well that included intrinsic $N$-body correlations was introduced in \citep{Andersen:17}. The Hamiltonian was a hybrid of the Calogero and the contact-interaction model and given by
\begin{equation}
    H_=\frac{1}{2}\sum_{i=1}^N\left(-\frac{\partial^2}{\partial x_i^2}+x_i^2\right)+\frac{\sqrt{2}g}{\rho}\sum_{\langle i,j\rangle}\delta(x_i-x_j),
\end{equation}
where $\rho$ is the relative hyper-radius for all $N$ particles. The two-body interactions strength therefore depends on the position of all other particles and the authors used this model to discuss insights into separability, symmetry, and integrability.

For strongly interacting systems one can construct a mapping onto Heisenberg spin chains \citep{Volosniev2014StrongInteractions,Deuretzbacher2014Mapping,Levinsen2015Ansatz, Yang:2015} (see also Section~\ref{sec09SC}). The basic idea is that for $g\rightarrow\infty$ the particles are arranged in a particular order, as the wave function vanishes whenever two approach each other. However, for small but finite values
of $1/g$, neighboring particles can exchange their positions, which effectively corresponds to nearest-neighbor spin interactions. One can therefore describe the system with an effective Hamiltonian that treats the particles as confined to a lattice for example as
\begin{equation}
    \mathcal{H}\simeq\frac{1}{g}\sum_{i=0}^{N-1}
    \left[J_i\mathbf{S}^i\mathbf{S}^{i+1}-\frac{1}{4}J_i \mathbf{I}\right]+E_0 \mathbf{I}.
\end{equation}
Here $J_i$ is the nearest neighbour exchange constant, $\mathbf{S}^i$ the spin operator at site $i$, and $\mathbf{I}$ is the unity operator. The rest of the Hamiltonian is constructed such that the strongly interacting state
has the energy $E_0$. This approach allows one to calculate analytically properties of general two-component systems in any external potential and can, in principle, be extended to systems with $SU(N)$ symmetry.  In Section~\ref{sec09SC} we discuss this technique in detail.

\subsection{Few-particle properties }

Using the models described above and others existing in the literature, a large amount of work has been carried out over the last two decades exploring properties in many different regimes. While densities and momentum distributions are often the first quantities to be analyzed, we focus in the following on properties stemming from the additional degrees of freedom in few-particle systems. 

For example, probing the correlations in few-body systems is not an easy task as it requires full reconstruction of the density matrix. However, recent theoretical progress has shown that such a task is possible through the measurement of position and momentum correlation functions when two atoms are trapped in a double well potential \citep{Bonneau2018}. This approach was successfully implemented for two interacting fermions in \citep{bergschneider2019} and the density and momentum correlation functions found are shown in Fig.~\ref{fig:SelimExp}. One can see that for attractive interactions double occupancy of each well is observed, while for repulsive interactions it is suppressed with each atom occupying different wells. In the non-interacting limit both particles are delocalised over both wells. In the momentum distribution one can see that the single particle coherence that is present in the non-interacting case is suppressed in the presence of strong interactions, both repulsive and attractive, while two-particle coherences are visible as interference patterns along the antidiagonal and diagonal respectively. The density matrix can be constructed from these momentum correlation functions allowing for the measurement of entanglement entropy and lower bounds on the concurrence. In \citep{becher2020} this technique was extended for systems with up to three particles, allowing for measurement of identical particle entanglement and the effect of antisymmetrization on correlations. 

The properties of systems with up to six particles in a double well geometry were explored in \citep{Zollner:2006} using a multiconfigurational approach over the whole range of interaction strength, and the same authors also subsequently studied the excitation spectrum \citep{Zollner:2007}.  Analytical results in the limit of two particles in a harmonic trap with a tunable zero-ranged barrier at the trap center were presented in \citep{Murphy:2007} and compared to a system with a finite width barrier in \citep{Murphy:2008}. The breakdown of a two-mode approximation for this situation was explored in \citep{Dobrzyniecki:2016}.

More detailed stationary states for a small number of harmonically trapped bosons with attractive interaction were explored in \citep{Tempfli:2008}, and  three distinct classes of states were found to emerge: (i) $N$-body bound states, (ii) bound states of smaller fragments and (iii) gas-like states that fermionize, that is, map to ideal fermions in the limit of infinite attraction (cf.~Section~\ref{subsec:TGmodel}). For two-boson systems in oscillator potentials with various powers of anharmonicity  several measures for the entanglement were numerically compared in \citep{Okopinska:2009,Okopinska:2010} and  the non-local properties of two interacting particles in separate traps at zero and at finite temperatures were calculated in \citep{Fogarty:2011}. 
The correlations for a larger gas of Lieb-Liniger type in a harmonic trap were shown to be accessible using the formalism of inhomogeneous Gaussian free fields (IGFF) \citep{Brun:18}, which was also compared to exact DMRG results and showed remarkable agreements for an approximate method. 

As briefly discussed in the introductory section, early landmark experiments on the behaviour of small systems have been a strong motivator for the theoretical works described here. These experiments include the fermionisation of two distinguishable fermions in a 1D harmonic potential \citep{zurn2012}, where for diverging interaction strength the energy and square modulus of the wave function for the two distinguishable particles becomes the same as for a system of two noninteracting identical fermions (cf.~Section~\ref{subsec:TGmodel}). This confirmed also the model of confinement-induced resonances predicted by \citep{olshanii1998} and the exact results of \citep{Busch1998TwoAtoms} in a new setting. The appearance of fermionisation was also confirmed to be compatible with the full quasi-particle theory of tunneling developed in \citep{Rontani:2012}. 
Additional physical insight into the role of statistics and interaction in many-body tunneling is provided in~\citep{delCampo2011decay,Gaston2011}.

\subsection{Dynamical properties, quenches and time-dependent modulations }\label{Dyn_prop_single_component}

Because the evolution equations of small systems can be integrated exactly, they are very interesting candidates to explore the non-equilibrium dynamics following a quench from a well-defined initial state.  With today's existing technologies, almost any part of the Hamiltonian can be quenched: kinetic energies using artificial gauge fields, external potentials using electromagnetic fields and interactions using Feshbach resonances. 

One of the experimentally most common non-equilibrium situations is the expansion of a quantum state into free space during a time-of-flight measurement. In this situation any external trapping potential is instantaneously switched off and the state is allowed to freely disperse. For a 1D Lieb-Liniger gas this situation was first discussed in \citep{Buljan:08}, where the authors assumed that the gas was initially localised in free space. Using exact solutions it was shown that the system enters a strongly correlated regime during the free expansion in the 1D direction, with the asymptotic form of the wave function having the characteristic of a TG gas. This is a result of the drop in density during the expansion, driving the system into the strongly interacting $\gamma\to \infty$ regime. The expansion of gases into a finite geometry (smaller box to bigger box) was studied in \citep{Grochowski:2020,Lebek:2021} and the appearance of structures due to de- and re-phasing was shown as a function of interaction strength.

One of the fundamental dynamical effects a quantum system can display is the non-equilibrium tunneling between two states. For single particles or non-interacting systems this is a textbook problem. However for interacting particle system the presence of scattering leads to interesting new effects and behaviours. An {\it ab initio} investigation based on the multi-configuration time-dependent Hartree method (see Section~\ref{subsec:ML-MCTDHA}) of the tunneling dynamics of a few-atom system in a double well trap was presented by \citep{Zollner2008}, covering the full range of interaction strengths. The authors found that for weak interactions approximate Rabi oscillations can be observed, whereas for larger interaction energies correlated pair tunneling emerges. Close to the fermionization limit these fragment and lead to a rather involved population dynamics in the two wells. More recently the behaviour of interacting two and three particle systems in a quenched double well potential was studied using a MCTDH-X method \citep{Schaefer:20}. Starting from the ground state in a simple harmonic oscillator, the authors studied the (quasi)-adiabatic vs.~diabatic switching of a central barrier. They calculated the evolution of the single-particle density matrix and the related von Neumann entropy and found that diabatic switching leads to efficient energy transfer between all particles across a large number of excited many-particle  states, while a quasi-adiabatic ramp only excites low lying states. Their results were confirmed for systems of up to 10 particles. In the strongly interacting limit, a spin model description of a few-body system with a central barrier varying in strength with time was discussed by \citep{Volosniev2016Heisenberg}, and a proposal for how to drive not just exchange of two particles but also exchange of two pairs of particles via an engineered barrier dynamics was presented.

As an alternative to the involved numerical solutions, \citep{Dobrzyniecki:18} treated the tunneling problem using a two-mode model that was not just based on the lowest single-particle eigenstates of the external potential, but used a basis of properly chosen effective wave functions. These were found by decomposing the initial many-body state into the basis of exact eigenstates of the many-body Hamiltonian, so that the interactions between particles and the specifics of the initial many-body quantum state were both taken into account. This approach was shown to significantly increase the accuracy of the evolution for intermediate interactions.

\begin{figure}[tb]
    \centering
    \includegraphics[width=0.75\linewidth]{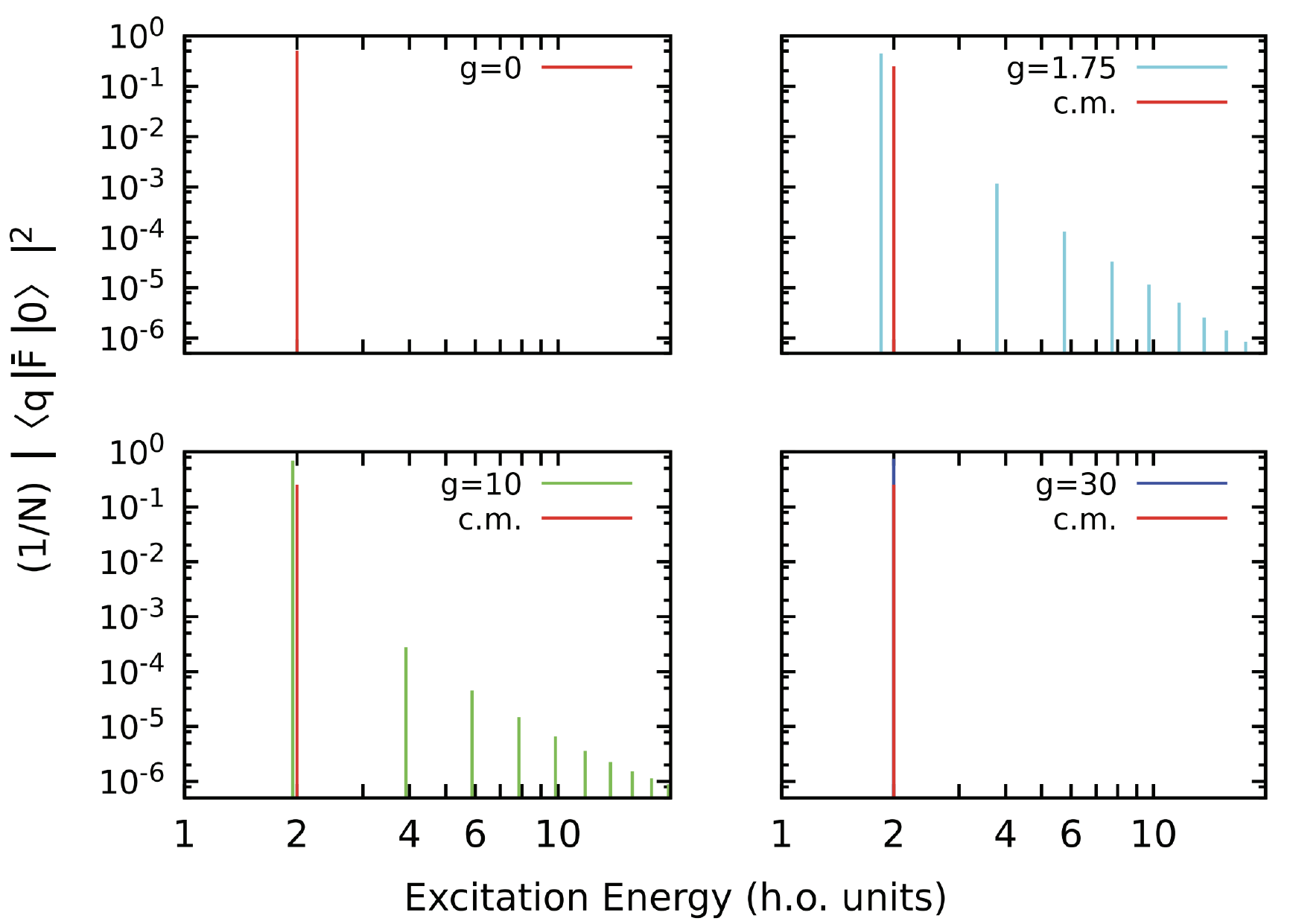}\hfill
    \caption{Strengths of the dynamic structure function (DSF) for the one body operator $\bar{F}=\sum_{i=1}^{N}x_i^2$ for a system of $2$ bosons with different contact interaction strengths $g = 0, 1.75, 10, 30$. The DSF is shown for the centre-of-mass (c.m.) component (red lines) which is independent of $g$, while the relative components (blue and green lines) are distributed over many excited states with its largest peak being the breathing mode. All quantities are in harmonic oscillator (h.o.) units. Image  from \citep{Ledesma:2019}.}
    \label{fig:DSF}
\end{figure}

In an effort to compare the exact dynamics of two-particles to the standard mean-field treatment of Bose-Einstein condensates, \citep{Sowinski:2010} compared the evolution of the largest eigenvalue of the reduced single particle density matrix (RSPDM) with the Gross-Pitaevskii equation (GPE) dynamics and found that the difference between them signified the presence of non-classical correlations in the system. A way to experimentally probe the correlations in an interacting two-particle systems was presented in \citep{Ledesma:2019}, where the response to a monopolar excitation was studied via the dynamical structure factor as a function of different values of the interaction strength (see Fig.~\ref{fig:DSF}). This allowed the authors to compute monopolar excitation energies, which are known to be directly related to the internal structure of the system \citep{pethick2002}.

\begin{figure}[tb]
    \centering
    \includegraphics[width=\linewidth]{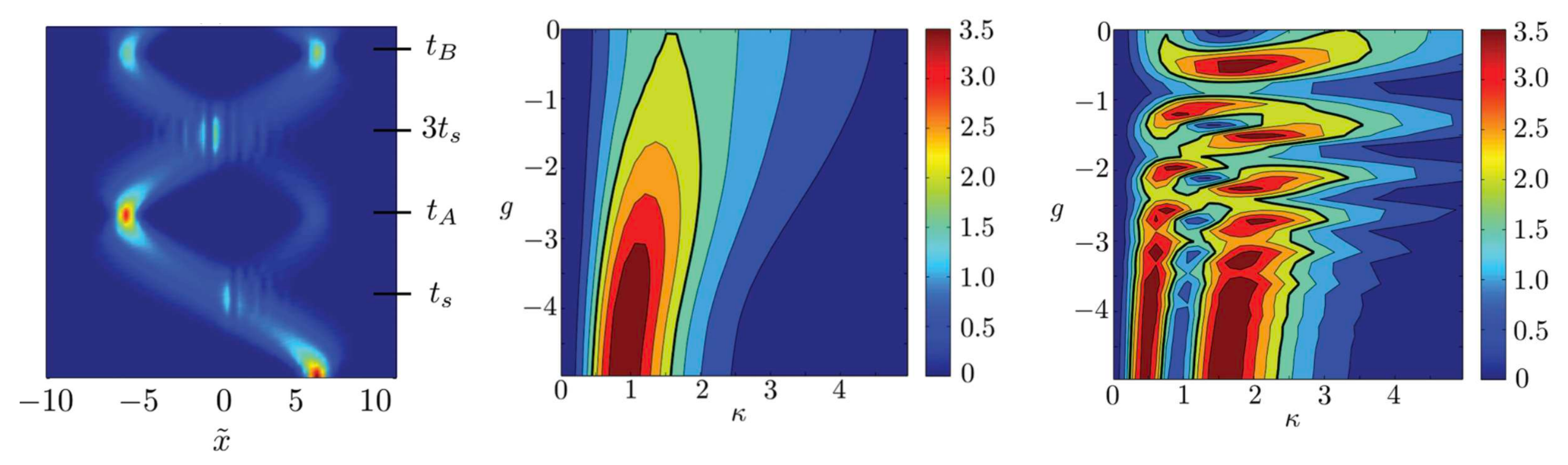}\hfill
    \caption{Left: Single-particle density versus time for two attractively interacting atoms. The barrier is point-like and positioned at $\tilde x = 0$ and the two particles are initially trapped at $\tilde x = 6$.  The times for scattering off the barrier ($t_s$), coming to rest at time at the classical turning point ($t_A$), recombining  and scattering a second time ($3t_s$), and come to rest again ($t_B$) are indicated. Also shown is the Quantum Fisher Information as a function of interaction strength $g$ and the barrier height $\kappa$ at $t=t_A$ (middle panel) and $t=t_B$ (right panel). All quantities are in harmonic oscillator units. Image adopted from \citep{Fogarty:2013}}
    \label{fig:BSNoon}
\end{figure}

Deeply connected to the question of coherence is the behaviour of two interacting atoms in an interferometer setting and in \citep{Fogarty:2013} it was shown that two interacting atoms impinging on a beam-splitter can evolve into states close to the NOON-state (see Fig.~\ref{fig:BSNoon}. 
It is worth noting that a similar situation for a gas of attractively interacting bosons scattering on a barrier was considered in \citep{Weiss:09,Streltsov:09} and the possibility for creating a many-body NOON state from quantum solitons was shown. Scattering of dimers made from short-range interacting atoms off a point-like barrier was explored in \citep{Gomez:19}, and it was shown that such processes can give insight into the integrability of the system. 
Creating entanglement by scattering small numbers of atoms off each other in a harmonic traps was investigated in \citep{Holdaway:2013}. For four interacting particles the authors were also able to identify parameter regimes in which attractive interactions lead to a quantum soliton-like state, for which the time average of the von Neumann entropy does not tend to a long-term mean value in contrast to the case of repulsive interactions of similar magnitude.  The idea to control the motional entanglement between two colliding atoms by dynamically varying the frequency of their common trap was explored in \citep{Haberle:2014} and it was shown that the amount of entanglement can be significantly increased with this dynamical tool.

Since for the two-particle system in a harmonic trap the exact eigenstates are known, it is interesting to see if analytical expressions can be obtained for non-equilibrium quench dynamics as well. \citep{Schmitz2013} explored the breathing dynamics following a trap quench for a two-particle system and, using the exact solution, calculated the full breathing mode spectrum. The authors found that it consists of many frequencies which are accompanied by a multiple of sidebands each. Using numerical methods for more particles, they were also able to show how this case evolves towards the single-frequency behavior that can be observed for larger particle numbers. The breathing mode for larger systems was explored in \citep{Gudyma:15} based on experimental results obtained in \citep{haller2009}, and the breakdown of scale invariance was shown by studying it close the the Tonks-Girardeau limit in \citep{Zhang:14}. A comprehensive study of the collective excitations in lower dimensions and for different correlation regimes was presented in \citep{DeRosi:2015,DeRosi:2016} and an anomaly in the temperature dependence of the specific heat of a 1D Bose gas due to the structure of the underlying excitation spectrum was recently reported in \citep{DeRosi:2021}.

For a two-particle system \citep{Budewig:2019} studied an interaction quench from repulsion to attraction and were able to derive analytical expressions for the expansion coefficients of the time-evolved two-body wave function. Quenching between zero to infinite interaction strength was discussed in \citep{Kehrberger:2018}, which is a situation in which analytical expressions for the dynamical evolution of the wave function can be derived. 
Quantifying the quench dynamics  for small bosonic systems after a trap or interaction change using thermodynamical quantities was studied in detail in \citep{GarciaMarch:2016} and a clear
qualitative link between the amount of irreversible work performed on the system and the increase in the degree of inter-atomic entanglement was found. Similarly, in \citep{Mikkelsen:20} it was shown that the variance of the work probability distribution after a trap quench is directly proportional to the infinite time-average of out of-time order correlations functions (OTOCs) but only when finite interactions are present between the particles.

A  quantum quench from an ideal Bose condensate to the Lieb-Liniger model with arbitrary attractive interactions was discussed in \citep{Piroli:16a}, and in more detail in \citep{Piroli:16b}, where it was found that the stationary state the system reaches at long evolution times after the quench shows a prominent hierarchy of multi-particle bound states. For a slow ramp in a spatially inhomogeneous system the formation of molecules was explored in \citep{Koch:2021} and the ensuing non-equilibrium dynamics in \citep{Koch:2022}. While in equilibrium the attractive Lieb-Liniger model is known to be thermodynamically unstable, the appearance of these multi-particle bound states after the quench ensures stability due to energy conservation. This is distinctively different to the situation of the Super Tonks-Girardeau gas (cf.~Section~\ref{subsec:TGmodel}), which is created by quenching from the strongly repulsive limit directly into the strongly attractive regime. This state was first described by \citep{Astrakharchik:05} using a diffusion Monte Carlo method and corresponds to a stable and highly excited Bethe state in the integrable Lieb-Liniger gas \citep{Batchelor2005SuperTonks,Chen:10}. It was observed by \citep{haller2009}, who confirmed its stability due to the presence of strong quantum correlations. As a step towards quantifying the stability of the Super Tonks state, which is related to the rates of inelastic processes such as photoassociation in pair collisions and three-body recombination, the local two- and three-body correlators at zero and finite temperature were analyzed in \citep{Kormos:11}. The Super-Tonks-Girardeau state in a 1D gas with attractive dipolar interactions was studied in \citep{Girardeau:12}.

Studies focusing on periodically modulating the trapping frequency in 1D for an interacting two- and few-particle systems were presented in \citep{Brouzos:2012,Ebert:2016} and a strong influence of the interactions on the behaviour near parametric resonances was found. The effect of a periodically oscillating magnetic field, which leads to a periodically oscillating scattering length, was investigated for small fermion systems in \citep{Yin:2016} and a technique to create controlled excitations based on this effect was presented. The breaking of the scale invariance in a system with smoothly varying external harmonic confinement in the nonadiabatic regime through finite interparticle interactions was discussed in \citep{Gharashi:2016}. 

\begin{figure}[tb]
    \centering
    \includegraphics[width=0.75\linewidth]{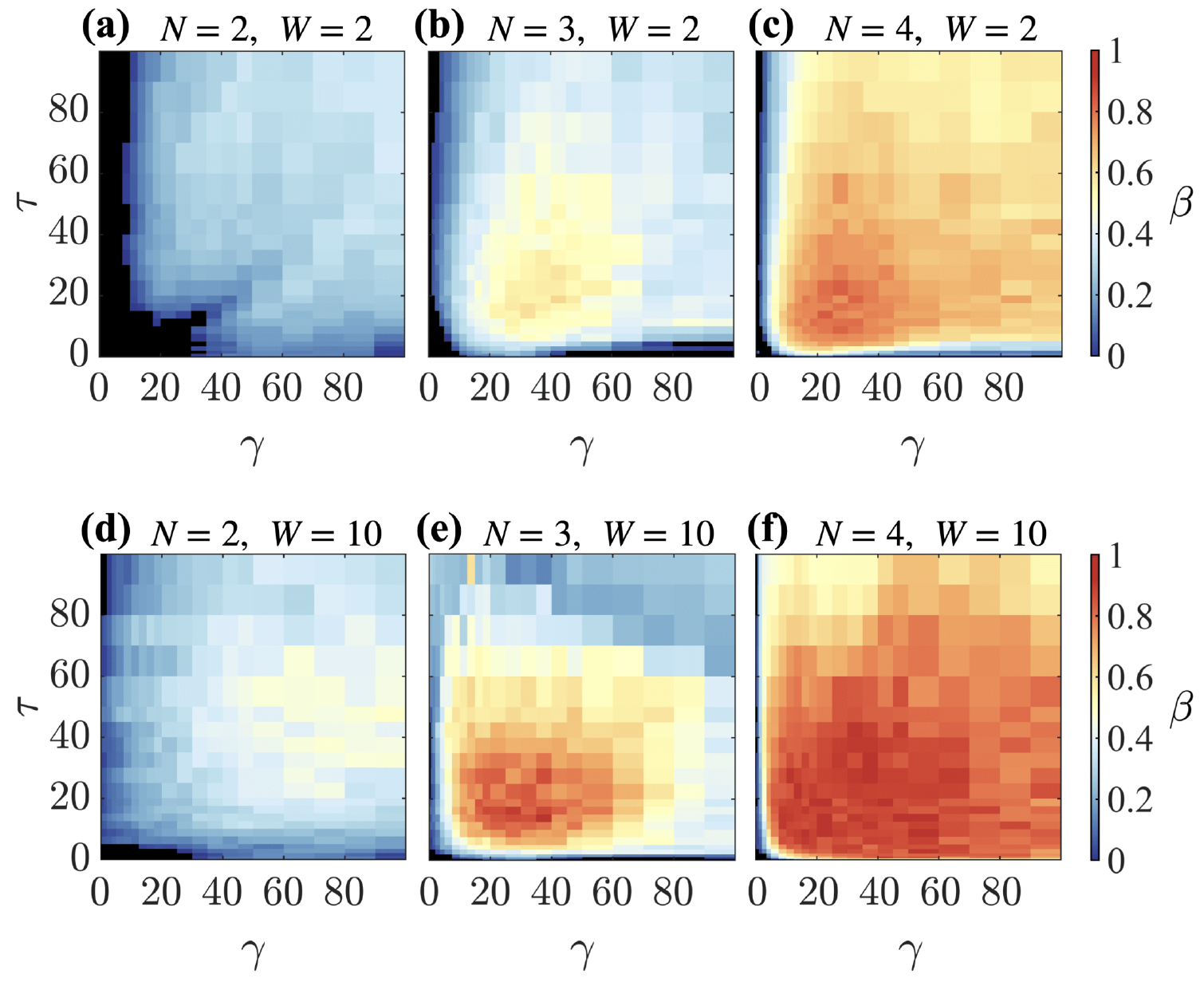}\hfill
    \caption{Brody distribution parameter $\beta$ for $N=2,3,4$ interacting bosons in a Kronig-Penney potential with $W$ wells as function of contact interaction strength $\gamma$ and delta function barrier height $\tau$ between adjacent wells. The Brody distribution is $P_\beta (s) = (\beta+1) b s^{\beta} \exp(-b s^{\beta+1})$ with $b =\left[ \Gamma \left( \frac{\beta+2}{\beta+1} \right) \right]^{\beta+1}$ and describes an integrable energy spectrum when $\beta\sim0$ and a chaotic one for $\beta\sim 1$. Black areas show $\beta<0$ indicating degeneracies and picket-fence spectra.  Image adopted from \citep{Fogarty2021}.}
    \label{fig:Qchaos}
\end{figure}

Similarly to quenching the interaction strength in a two-atom system, it is interesting to consider quenching the presence of an impurity (see also Section~\ref{sec:polaron}). In \citep{Keller:16} the non-equilibrium dynamics originating from such a process for an infinite mass impurity was considered, using an approximate variational Lagrange-mesh method.  Using the Loschmidt echo to quantify the irreversibility of the system following the quench and the probability distribution after long times
to identify distinct dynamical regimes, this work showed that the resulting dynamics depends strongly on the interaction strength between the atoms.  A similar situation, but allowing for kinetic energy for the impurity, was considered in \citep{Campbell2014Orthogonality}, where evidence of Anderson's orthogonality catastrophe was found even for small systems.

This situation was extended to a spinor impurity repulsively interacting with a Bose-Einstein condensate after a quench in \citep{Mistakidis2019QuenchBosePolarons}. Here the temporal evolution was done using ML-MCTDHX and the determination of the structure factor allowed the authors to identify three distinct dynamical regimes as a function of increasing interspecies interactions. These regions were found to be related to the segregated nature of the impurity and to the Ohmic character of the bath. In the miscible regime one can identify the polaron formation through the spectral response of the system, however for increasing interaction an orthogonality catastrophe occurs and the polaron picture breaks down. At this point the impurity transfers its energy to the environment, which signals the presence of entanglement in the many-body system. The situation for two interacting impurities was investigated in \citep{Mistakidis2020InducedBose2}, where the induced correlations between the polarons due to the background were examined, see Section~\ref{sec:polaron} for more details.

A further interesting question for quenched systems is the existence of dynamical phase transitions  \citep{Heyl2013,Heyl_2018}. 
For the Tonks Girardeau gas this was investigated in \citep{Fogarty:17}. Using the pinning transition for the situation where a commensurate lattice was instantaneously turned on, this work investigated the dynamics of the rate function and of local observables after the quench. Whenever the return probability (or Loschmidt echo) goes to zero, corresponding to singularities of the rate functions, the system is said to display dynamical criticality and this work addressed the question of whether this behaviour can be tied to the dynamics of physically relevant observables or order parameters in the systems. Comparing the periodicity of the collapse and revival cycle in the momentum distribution to that of the non-analyticities in the rate function shows that the two are only closely related for deep quenches.

Few-body systems have also been used to uncover the path from integrability to chaos, as they allow one to explore equilibration and thermalization in isolated systems \citep{ETH_review}. In \citep{Fogarty2021} integrability breaking was explored with interacting bosons trapped in multiwell potentials, and the presence of both, finite interactions and finite barrier heights between the wells, was shown to be able to result in strong signatures of quantum chaos, which intensifies with increasing particle number. In Fig.~\ref{fig:Qchaos} the integrable and chaotic regions are discerned by fitting the energy spectrum to the Brody distribution, with the fitting parameter $\beta \sim 0$ describing a Possonian spectrum and therefore integrability, while $\beta \sim 1$ describes a Wigner-Dyson distribution signifying strong energy level repulsion and therefore chaos. Recent work by \citep{Lydzba2021} has also investigated the emergence of chaos between two component interacting fermions in a double well potential and \citep{Yampolsky2021} used a quantum generalization of the Chirikov criterion of overlapping resonances to detect chaos in a two-particle system in the presence of a perturbation. Finally, \citep{Huber2021} used convolutional neural networks to identify chaotic eigenstates of two fermions and an impurity on a ring.

%% file: Section04.tex
\section{Two-component gases}\label{sec:spinor}

Multicomponent systems are not a straightforward extension of their single component counterparts. 
Even small samples can already hold a large amount of new physics, owing mostly to the presence of additional degrees of freedom, which result in novel properties of the corresponding mixture. While single component systems are defined by one scattering length, one external potential and one dispersion relation, two component systems can have species-selective trapping potentials, modified dispersion relations e.g. due to intercomponent coupling~\citep{pethick2002,pitaevskii2016bose}, and two additional scattering lengths, with one of them accounting for the inter-species scattering. 
They therefore allow one to explore various correlation regimes~\citep{tempfli2009binding,garcia2014quantum} and even feature an experimentally realizable phase-separation phase transition~\citep{timmermans1998phase,papp2008tunable}. 
Furthermore, one can also consider mixtures made from particles of different statistics (Bose-Fermi mixtures), which we will address below in Section~\ref{sec:Bose_Fermi}.

\begin{figure}[tb]
\centering
\includegraphics[trim={3cm 11cm 2cm 2cm},clip,width=0.5\textwidth]{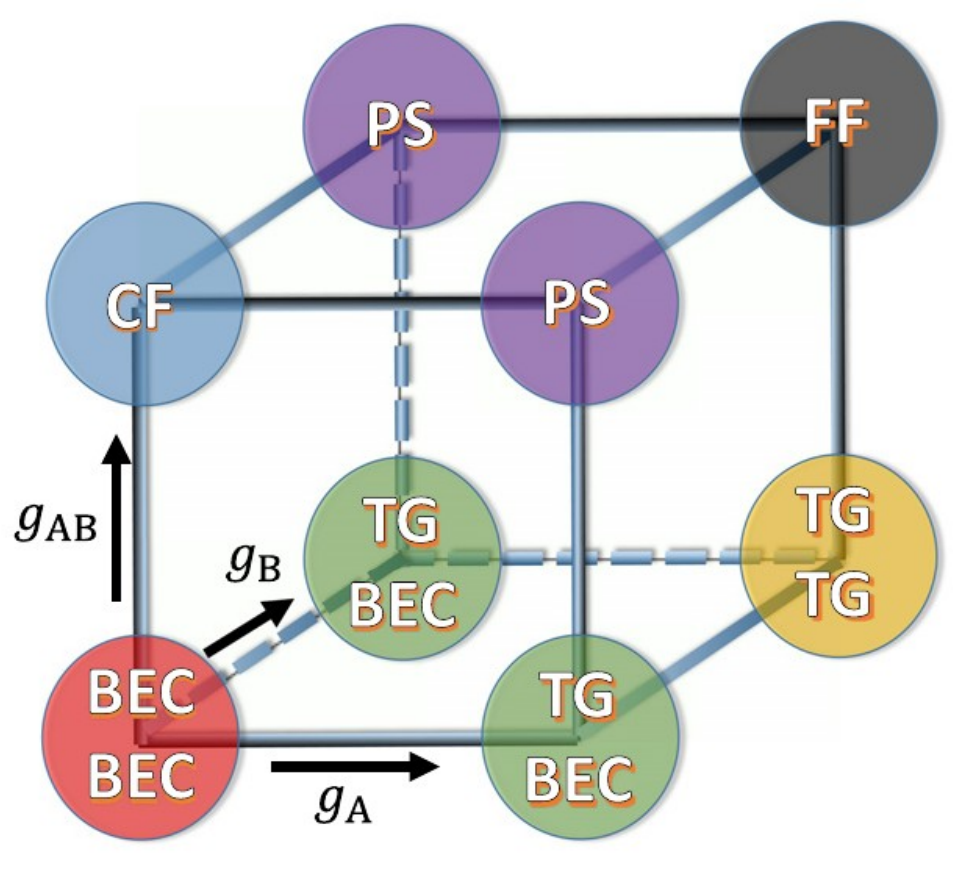}
\caption{Schematic representation (the {\it Miguel cube}) of the few-body ground states of a two-component bosonic mixture for different intra- and intercomponent interactions. CF stands for composite fermionisation, PS refers to phase-separation and FF stands for full fermionization. Figure from~\citep{garcia2014quantum}.}
\label{fig:few_body_cube} 
\end{figure}

Commonly, two-component bosonic settings consist of either two alkali metals e.g. $^{23}$Na-$^{87}$Rb, or two isotopes like $^{85}$Rb-$^{87}$Rb, or two hyperfine states of the same alkali metal~\citep{myatt1997production,hall1998measurements}. 
The interplay between the inter- and intraspecies scattering is a central issue in these systems since it determines the static and dynamical properties of the mixture and strikingly it allows for manipulating the degree of quantum correlations, see Fig.~\ref{fig:few_body_cube}.
The role of correlations has been extensively discussed in the literature.
Characteristic examples include, for instance, altered phase-separation regimes~\citep{garcia2013quantum,GarciaMarch2014,garcia2014quantum}, the generation of quantum droplets~\citep{petrov2015quantum,Petrov2016Liquids,cabrera2018quantum} (see in particular Section~\ref{sec:Droplets_contact}), noticeable modifications of the superfluid to Mott insulator phase transition~\citep{lkacki2013dynamics,dutta2015non}, composite fermionization processes~\citep{zollner2008composite,garcia2013sharp,Zinner2014Fractional,pyzh2018spectral}, quantum emulsive states~\citep{roscilde2007quantum}, emergence of instabilities~\citep{cazalilla2003instabilities,vidanovic2013spin} and intriguing phases such as paired or counterflow superfluidity~\citep{hu2009counterflow}. 

\subsection{Ground state properties of bosonic mixtures}\label{sec:ground_BB_mixture}

The basic model for a pseudospinor~\footnote{Notice that multicomponent setting containing $n$ individual components are often described as a fictitious spin-$(n-1)/2$ system. They preserve the atom number of each component prohibiting spin-mixing, while the internal states are not rotationally symmetric in spin space, see also~\citep{stamper2013spinor}.} system of ultracold atoms interacting via  point-like interactions is described by the Hamiltonian
\begin{align}
    H&=\sum_{n=1}^{N_A}\left[-\frac{\hbar^2}{2m_A}\frac{\partial^2}{\partial x_n^2}+V_A(x_n)\right]+g_A\sum_{n=1}^{N_A}\sum_{m=n+1}^{N_A}\delta(x_n-x_m)\nonumber+\sum_{n=1}^{N_B}\left[-\frac{\hbar^2}{2m_B}\frac{\partial^2}{\partial y_n^2}+V_B(y_n)\right]\\
    &+g_B\sum_{n=1}^{N_B}\sum_{m=n+1}^{N_B}\delta(y_n-y_m)+g_{AB}\sum_{n=1}^{N_A}\sum_{m=1}^{N_B}\delta(x_n-y_m). 
    \label{eq:Hamiltonian_pseudospinor}
\end{align}
Here, the components $A$ and $B$ are under the influence of the external trapping potentials $V_A(x)$ and $V_B(y)$ respectively. 
Such a setup can be experimentally realized either by considering two distinct hyperfine states of the same element (mass-balanced system, $m_A=m_B \equiv m$) or two different atomic isotopes (mass-imbalanced setting, $m_A \neq m_B$). 
In the former case the pseudospin degrees-of-freedom can correspond, for instance, to the hyperfine states $\ket{A} \equiv \ket{F=1,m_F=1}$ and $\ket{B}\equiv\ket{F=1,m_F=-1}$ of a $^{87}$Rb gas and the latter scenario describes a mixture of isotopes containing e.g. $^{87}$Rb and $^{7}$Li. 
The effective 1D intra-(intercomponent) coupling constants $g_{\sigma}$ ($g_{\sigma \sigma'}$) with $\sigma' \neq \sigma=A,B$ are related to the 3D $s$-wave scattering lengths, ${a^s_{\sigma \sigma'}}$~\citep{olshanii1998}. 
Experimentally, $g_{\sigma\sigma'}$ can be tuned either with the aid of ${a^s_{\sigma \sigma'}}$ utilizing Feshbach resonances \citep{kohler2006production,chin2010feshbach} or by means of the transversal confinement frequency ${{\omega _ \bot }}$ using the so-called confinement-induced resonances \citep{olshanii1998}, see also Section~\ref{fin_range_corrections}. 
Recent advances in optical trapping techniques allow one to imprint almost arbitrary shapes of the external trap and even create species-dependent potentials e.g.~by employing the so-called ``tune-out'' wavelength~\citep{leblanc2007species,henderson2009experimental}.

\begin{figure}[tb]
\centering
\includegraphics[width=\textwidth]{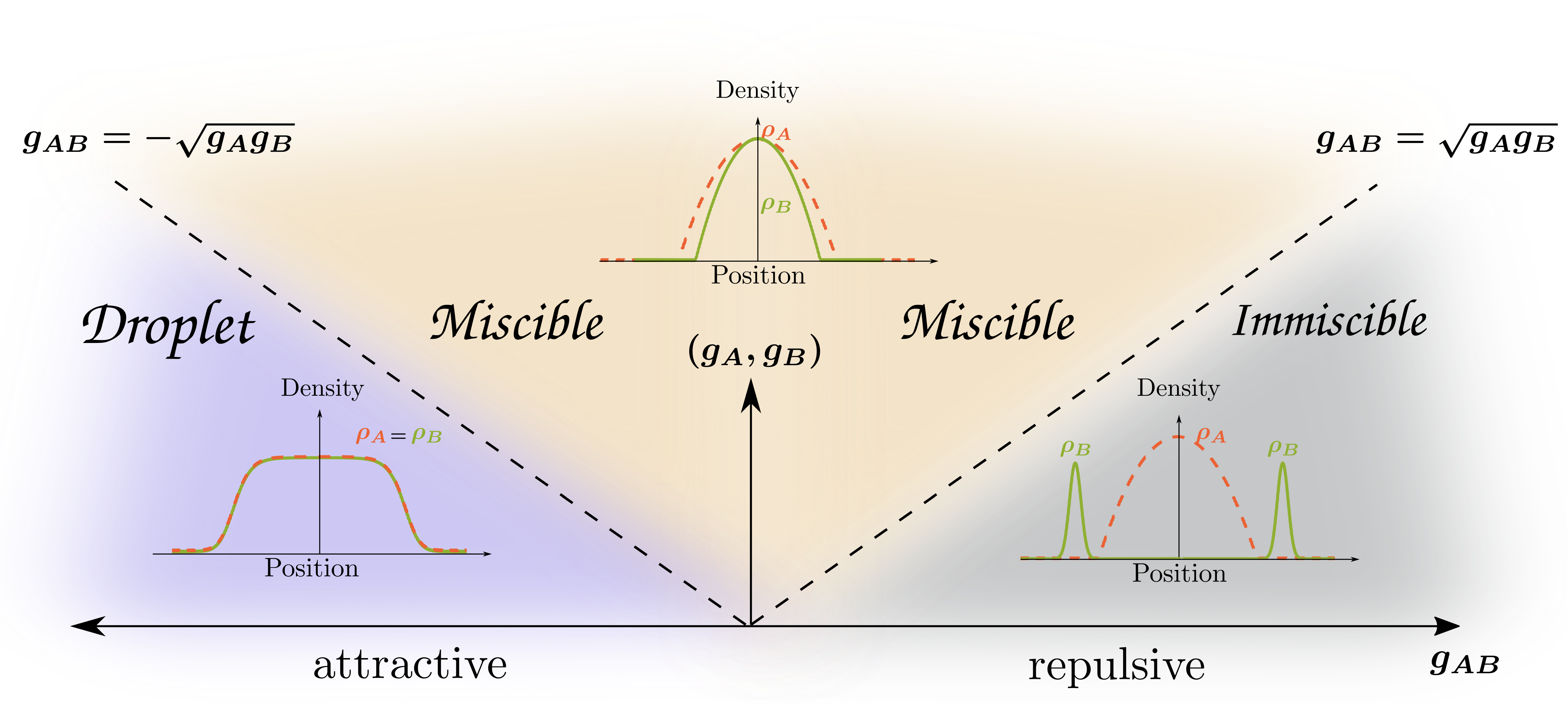}
\caption{Illustration of the ground state phase diagram of a two-component bosonic system with respect to the intra- ($g_{A}$, $g_{B}$) and intercomponent ($g_{AB}$) interactions. The density distributions ($\rho_A$, $\rho_B$) of each component are depicted together with the phase boundaries from the immiscible to the miscible state and then to the droplet regime. The droplet phase of matter arising in the attractive interaction regime and only in the presence of quantum fluctuations is analyzed in detail within Sec.~\ref{sec:Droplets_contact}.} 
\label{fig:miscible_transition} 
\end{figure}

Recently this topic has been comprehensively summarised in a review by Sowinski and García-March~\citep{Sowinski2019Review}, which mostly focused on the stationary properties of few-body mixtures in harmonic traps. 
A multitude of distinct phases can be formed stemming from the interplay between intra- and intercomponent couplings, see Fig.~\ref{fig:few_body_cube} for an overview. 
Indeed, there are six individual configurations for repulsive couplings~\citep{garcia2014quantum}.

The so-called \textit{BEC-BEC} limit appears when the particles are non-interacting and the
\textit{BEC-TG} phase occurs for infinite intracomponent coupling of one species and  all other interactions vanishing. 
Systems enter the \textit{TG-TG} phase if both intracomponent interactions are infinitely strong and the intercomponent coupling is zero. 
In the \textit{composite fermionization} regime the intracomponent couplings are suppressed, but strong intercomponent interactions exist. 
Finally, the so-called \textit{phase-separation} regime, is realized for infinite intercomponent repulsion, while one of the intracomponent strengths has a finite value and the other is zero. 
This state is also generated for finite interactions, see Fig.~\ref{fig:miscible_transition}, as long as the intercomponent coupling is larger than the mean of the intracomponent ones. 
In the finite interaction regime and in particular for two different hyperfine states of $^{87}$Rb phase-separation and associated mixing-demixing processes have been experimentally realized in quasi-1D traps~\citep{tojo2010controlling,eto2016nonequilibrium,eto2016bouncing}. 
Recently, another intriguing phase of matter has been theoretically predicted~\cite{naidon2021mixed,sturmer2022mixed} in the vicinity of the immiscibility threshold caused exclusively by the presence of quantum fluctuations and dubbed ``mixed-bubble". It refers to a configuration where one component acts as a pocket being trapped within the gaseous medium of the other one and it forms in intracomponent interaction imbalanced mixtures. Current investigations of this phase are mainly restricted to large atom numbers, while their persistence in few-body systems and dynamical properties remain elusive. 
It is worth mentioning that phase separation occurs under the same conditions in lattice two-component bosonic mixtures~\citep{mishra2007phase,alon2006phase,richaud2019mixing,altman2003phase,isacsson2005superfluid} which is a topic beyond the scope of the present review and has been summarized together with other intriguing phases of matter arising due to admixtures of energetically higher-bands e.g. in the review of~\cite{dutta2015non}. 

Coming back to Fig.~\ref{fig:few_body_cube}, the \textit{full fermionization} regime appears when all interaction strengths tend to infinity. 
In the case of intercomponent attractions and strong intracomponent repulsions it is possible to form a \textit{molecular TG} state among the species if the attraction is smaller or of the same order as the repulsion~\citep{tempfli2009binding}. 
A further increase of the attraction leads to a prominent bound state and eventually to the collapse of the gas. This is discussed in further detail in Section~\ref{sec:Droplets_contact}.
Finally, for strong attractive intracomponent and vanishing intercomponent couplings the so-called \textit{Super Tonks} (excited) state occurs (see also Section~\ref{subsec:TGmodel}), while tuning the intercomponent attraction to larger values a corresponding \textit{molecular Super Tonks} phase can appear.

There are several ways to induce nonequilibrium quantum dynamics of pseudospinor systems including, for instance, sudden changes or time-dependent variations of the interatomic interactions and the external confinement. 
Accordingly, it is possible to dynamically trigger complex tunneling mechanisms beyond the single component paradigm, create peculiar correlation patterns, realize few-body analogs of macroscopic phenomena, and trigger collective or local excitations of the binary system to name a few. 
Below, we review the observations of such processes in few-body and mesoscopic settings. 

Analysis of pseudospinor systems requires theoretical methods and numerical approaches that are able to account for both inter- and intracomponent correlations, see Section~\ref{sec:Methods}.
The latter are naturally enhanced for few-body systems especially in 1D. 
On the contrary, in the limit of large particle numbers and weak interactions, where quantum fluctuations can be neglected, the standard approach is the celebrated mean-field approximation dictated by the Gross-Pitaevskii equations of motion\citep{pethick2002,pitaevskii2016bose}. 
Within this framework each of the species, $A$ and $B$, is represented by only one macroscopic wave function. This way, the many-body problem reduces to an effective single-particle one, where all interaction effects solely manifest themselves in the deformation of the associated single-particle orbital. 
The Gross-Pitaevskii framework has been proven to accurately describe the physics of  weakly interacting single- and multicomponent bosonic gases. It can explain a plethora of stationary and nonequilibrium phenomena, such as the formation, interactions and dynamics of various types of nonlinear excitations~\citep{kevrekidis2007emergent,kevrekidis2015defocusing,kevrekidis2015solitons}. 
In this review, such processes will not be described, as they usually do not have a well-defined few-body counterpart and they have been summarized in other reviews, see e.g.~\cite{RevModPhys.71.463,Bagnato2015,edwards1996zero,kevrekidis2015defocusing}. Here, the focus is rather placed on processes where correlation-induced phenomena play a dominant role. 

In the so-called Thomas-Fermi limit where the interaction energy dominates, the kinetic energy of the mixture can be neglected \citep{pitaevskii2016bose,kevrekidis2007emergent}. The condition for phase-separation, which in general holds for 3D dilute Bose gases,  depends only on the relative interaction strengths between the atoms and in homogeneous systems the separation transition takes place at  $g_{AB}>\sqrt{g_Ag_B}$~\citep{ho1996binary,ao1998binary,pu1998properties,timmermans1998phase}, see also Fig.~\ref{fig:miscible_transition}.  
For condensates in inhomogeneous potentials usually regions exist where the kinetic energy of the atoms cannot be neglected and, as a consequence parameters such as the trap frequency, atom number and mass ratio can noticeably impact the miscible-immiscible phase boundary~\citep{wen2020effects,tanatar2000strongly,lee2016phase,gutierrez2021miscibility,pyzh2020phase}. As an example, for mass-imbalanced mixtures the degree of spatial localization of each component is naturally affected by the unavoidable change of the related confinement length scale~\citep{pflanzer2009material,pflanzer2010interspecies,cikojevic2018harmonically}, 
see also \citep{pkecak2019intercomponent,Pecak2016,cui2013phase,Harshman2017, Volosniev2017MassImbalance} for related effects in 1D fermionic mixtures. 
In the presence of a harmonic trap, the heavier component is usually stronger localised and has a ground state density that has a Gaussian distribution of width $\sqrt{\mathcal{\kappa}/2}$, where $\mathcal{\kappa}$ is the mass ratio. It  approaches a $\delta$-like shape for $\mathcal{\kappa}\to 0$. 
It is also noteworthy that keeping the interspecies repulsion fixed and increasing solely the mass ratio, one can also achieve a transition from a miscible to an immiscible phase. 
This transition behavior can be seen on the one-body density level and it is accompanied by a minimum in the two-body interspecies correlation function,  implying the formation of anti-ferromagnetic order~\citep{garcia2014quantum,garcia2013quantum}. 
Moreover, for strongly mass-imbalanced harmonically trapped few-boson mixtures and in the limit of large intercomponent repulsions it has been shown in~\citep{pflanzer2009material,pflanzer2010interspecies} that the heavier component acts as an effective ``material barrier'' for the lighter species, inducing an effective attraction between the atoms of the latter that increases with decreasing mass ratio.
This behavior can be explained by constructing an effective Bose-Hubbard model for the light atoms which contains a renormalized hopping and on-site interaction parameters.

\subsection{Energy transfer and breathing modes}\label{sec:energy_transfer} 

A fundamental mechanism that arises in multicomponent settings due to the intercomponent coupling is dynamical "correlation transfer" between the components. 
A detailed study of such process was presented in \citep{Kronke2015Dynamics} with the focus on quantum dynamics of a single atom of species $A$ in a finite bosonic environment made of $B$  atoms. 
Initially, both components are trapped in species-selective 1D harmonic traps such that their spatial overlap is rather small. 
Upon changing the initial displacement of the harmonic traps, the two species collide (see e.g.  \citep{vogels2002generation,perrin2007observation} for relevant experiments) and thereby interchange energy and correlations as well as build up interspecies correlations~\citep{deuar2007correlations,zin2018properties,ogren2009atom}. 
In particular, the atom $A$ oscillates in the harmonic trap and thereby periodically penetrates the $B$ species held in a displaced trap. 
It has been shown that the corresponding interspecies energy transfer accelerates with increasing the number of $B$ atoms which was related to a level splitting of the involved excited many-body states. 
The finite amount of interspecies correlations created stimulates coherent energy transfer processes which manifest in singlet and doublet excitations of the bosonic environment.

\begin{figure}[ht]
\includegraphics[width=\textwidth]{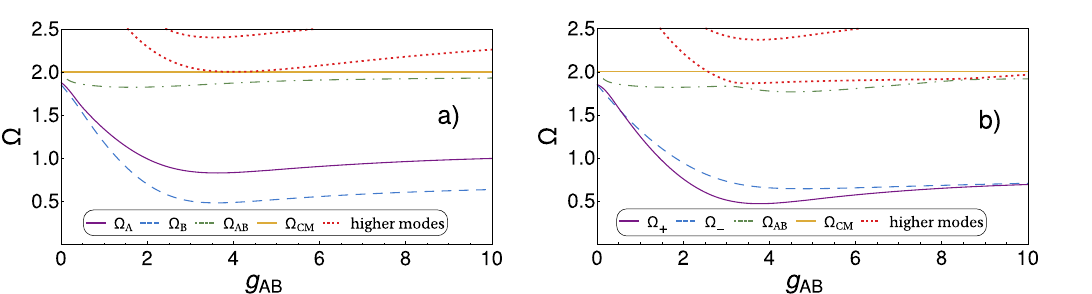}
\caption{Breathing mode frequencies, corresponding to energy differences between specific eigenstates of a 1D bosonic mixture of $N_A=N_B=2$ atoms as explained in~\citep{pyzh2018spectral}, with respect to the intercomponent interaction strength $g_{AB}$. The cases of interaction a) imbalance $g_A=1$, $g_B=2$ and b) balance $g_A=g_B=2$ setups are depicted. In particular, $\Omega_A$, $\Omega_B$ mark the breathing mode frequencies of the relative coordinate of each species, $\Omega_{AB}$ is the respective frequency of the relative coordinate 
between the individual center-of-masses of each species and $\Omega_{CM}$ refers to the total center-of-mass coordinate. 
$\Omega_+$ ($\Omega_-$) is the same with $\Omega_A$ ($\Omega_B$) but it is denoted differently since the involved excited eigenstates are symmetric and antisymmetric in terms of the particle exchange operator.  
Higher modes represent energetically excited breathing frequencies that posses a relatively small contribution.   
The collective breathing motion is initiated following a change of the trap frequency. 
All quantities are expressed in postquench harmonic oscillator units. 
Figure from~\citep{pyzh2018spectral}.} 
\label{fig:mixture_breathing} 
\end{figure} 

The correlated breathing mode dynamics of few-body binary bosonic mixtures confined in a 1D harmonic trap was investigated in \citep{pyzh2018spectral}. 
Using exact diagonalization with a correlated basis approach, the low-energy eigenspectrum was obtained for the complete range of repulsive intra- and intercomponent interaction strengths. 
In particular, the strongly coupled limiting cases of composite fermionization, full fermionization and phase-separation as well as all the intermediate symmetric and asymmetric intraspecies interaction regimes were examined in detail. 
Knowing the stationary spectrum of the system, one can see that the breathing dynamics in the linear-response regime can be triggered by slightly quenching the trap frequency symmetrically for both components. 
Depending on the choice of interaction strengths, the presence of one to three monopole modes was identified besides the breathing mode of the center-of-mass coordinate, see also Fig.~\ref{fig:mixture_breathing}. 
In particular, it was shown that for the uncoupled mixture each monopole mode corresponds to the breathing oscillation of a specific relative coordinate. 
Importantly, increasing the intercomponent coupling leads to multi-mode oscillations in each relative coordinate, which turn into single-mode oscillations of the same frequency in the composite-fermionization regime. 
For clarity the dependence of the breathing mode frequencies of the $N_A=N_B=2$ bosonic mixtures on the intercomponent coupling are depicted in Fig.~\ref{fig:mixture_breathing}. 
It is observed that for a mixture with distinct intracomponent interactions [Fig.~\ref{fig:mixture_breathing} a)] the breathing of the relative coordinate among the species center-of-masses, $\Omega_{AB}$, is degenerate with $\Omega_{CM}$ at $g_{AB}=0$ then decreases for larger $g_{AB}$ and revives back towards $\Omega_{CM}$ asymptotically. 
This is reminiscent of the interaction dependence of the relative coordinate breathing frequency of a single-component bosonic system~\cite{abraham2014quantum,Schmitz2013}. 
Turning to the relative coordinates of each species it is found that both $\Omega_A$ and $\Omega_B$ are substantially affected by $g_{AB}$ and always lie below the underlying center-of-mass dipole mode frequency which equals unity. 
The breathing mode frequencies of an interaction imbalanced mixture [Fig.~\ref{fig:mixture_breathing} b)] behave similarly to the balance scenario with the following two main differences: i) $\Omega_{AB}$ exhibits two minima and ii) the relative breathing modes of each species $\Omega_+$, $\Omega_-$ are the same in the decoupled case ($g_{AB}=0$), they separate at finite coupling and finally coincide asymptotically. As mentioned already above all these behavior of the breathing frequencies are traced back to the energy spectrum and the involved avoided-crossings between the eigenstates participating in the breathing mode. Additionally, as expected, in both cases the breathing frequency of the total center-of-mass $\Omega_{CM}$ is interaction-independent.

\subsection{Quench dynamics and pattern formation}\label{sec:quenches_BB}

\begin{figure}[tb]
\centering
\includegraphics[width=0.8\textwidth]{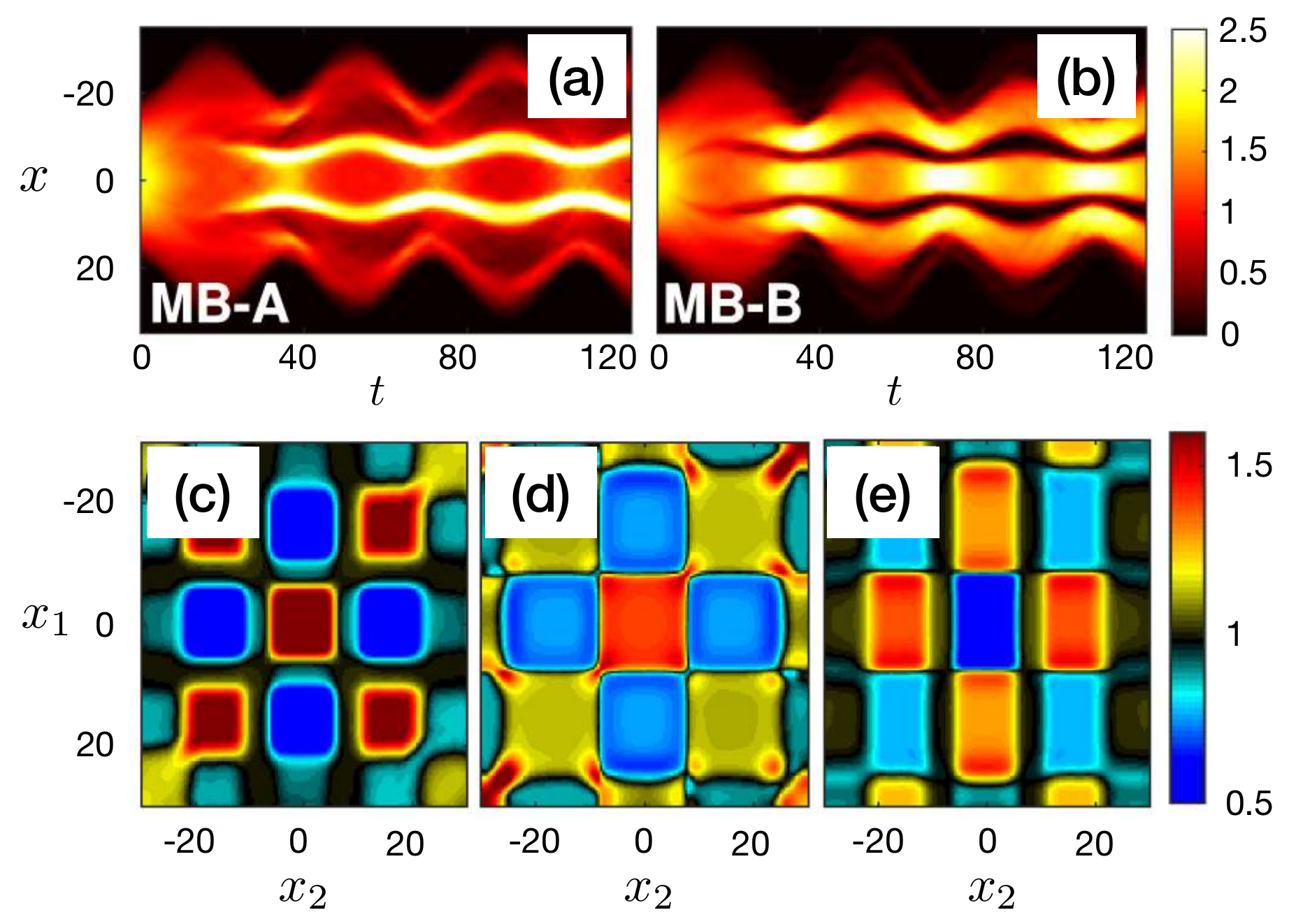}
\caption{(Upper panels) Density evolution of the (a) A and (b) B species following an interaction quench of a decoupled binary mixture into the immiscible phase. 
As time-evolves the originally smooth density profiles break into multiple filaments that are complementary between the species manifesting dynamical phase separation. 
(Lower panels) Snapshots of the two-body intracomponent correlation function of the (c) [(d)] A [B] species and (e) the interspecies two-body correlation function. A two-body intracomponent bunching (anti-bunching) occurs for parity symmetric outer (nearest neighbor) filaments explicating their domain-wall configuration. 
A similar correlation pattern takes place between the species evincing the entangled nature of the mixture and supporting the domain-wall structure between filaments of different components. 
Figures adapted from~\citep{mistakidis2018correlation}.}
\label{fig:quench_mixture_filaments} 
\end{figure}

The interspecies interaction quench-induced phase-separation dynamics of a harmonically trapped and repulsively interacting binary bosonic ensemble was discussed in \citep{mistakidis2018correlation}. 
It was shown that the increase of the interspecies repulsion results in a filamentation of the density of each species, see Fig.~\ref{fig:quench_mixture_filaments}(a), (b). The spontaneously generated filaments are strongly correlated and exhibit domain-wall structures at the two-body level as shown in Fig.~\ref{fig:quench_mixture_filaments}. 
Specifically, inspecting the off-diagonal elements of the intracomponent two-body correlations it is overall observed that nearest-neighbor (next-to-nearest-neighbor) filaments of the same component are anti-correlated (correlated).  
Recall that if the value of the two-body correlation function is larger (smaller) than unity then two-bosons exhibit a bunching (anti-bunching) tendency, while if it is unity then they are un-correlated. 
These correlation patterns imply the domain-wall formation between filaments of same component. 
On the other hand, the structure of the two-body intercompoent correlation function, besides showing the entangled character of the mixture, supports the fact that successive filaments of different components feature domain-wall configurations. 
Following the reverse quench protocol, i.e.~decreasing the interspecies interaction strength, the formation of multiple dark-antidark solitary waves was observed~\citep{katsimiga2020observation,kiehn2019spontaneous}. The experimental realization of such soliton configurations in a $^{87}$Rb BEC containing $~10^6$ atoms corroborated by an extensive theoretical analysis was also recently reported in \citep{danaila2016vector,katsimiga2020observation,mossman2022dense}. 
These solitary wave structures have been found to decay into the many-body environment soon after their generation, which is in sharp contrast to the predictions of the mean-field approximation. It constitutes a basic characteristic of many-body soliton entities whose properties have been extensively discussed, e.g.~in \citep{katsimiga2017dark,katsimiga2018many,dziarmaga2002depletion,mishmash2009quantum,delande2014many,syrwid2017quantum,syrwid2021quantum}. 
Moreover, experimental counterparts of these observations were reported by simulating single-shot images showcasing, among others, that the growth rate of the corresponding variance of a sample of single-shots\footnote{Detailed discussions regarding the theoretical implementation of the single-shot procedure in MCTDH type methods can be found in \citep{sakmann2016single,lode2017fragmented,katsimiga2017many,katsimiga2018many} for single component atomic ensembles and in \citep{mistakidis2018correlation,Mistakidis2019DissipationBosePolarons,erdmann2019phase,erdmann2018correlated,koutentakis2019probing} for multicomponent ones. Note also that in single component settings the behavior of the single-shot variance is related to the degree of fragmentation in the system.} probes  entanglement in the system~\citep{mistakidis2018correlation,katsimiga2017dark,katsimiga2017many}. 

The defect scaling and its relation to the celebrated Kibble-Zurek mechanism\footnote{The Kibble-Zurek mechanism~\citep{kibble1976topology,zurek1985cosmological} is based on the assumption that every system undergoing a phase transition experiences a phase of ``impulsive'' behavior where the topological defects  ``crystallize'' in the system. 
It predicts the scaling of the density of the generated topological defects with the rate at which the phase transition critical point is crossed.} when dynamically crossing the phase-separation boundary of binary Rabi-coupled bosonic mixtures was studied in~\citep{sabbatini2011phase,sabbatini2012causality} within a truncated Wigner approximation. 
The latter method accounts for quantum fluctuations around the mean-field solution~\citep{blakie2008dynamics}.  
Adding a Rabi coupling term with time-dependent strength $\Omega(t)$ to Eq.~(\ref{eq:Hamiltonian_pseudospinor}) will allow particle transfer among the components. 
Importantly, the miscible to immiscible phase transition can be realized by appropriately varying the Rabi amplitude above/below a critical strength $\Omega_{cr}$ in addition to choosing a combination of the interaction values. A detailed characterization of the ground state magnetic properties and excitation spectrum of Rabi-coupled two-component bosonic systems can be found in \citep{abad2013study,sartori2013dynamics,tommasini2003bogoliubov,recati2021coherently,sartori2015spin}. 
Notice that the presence of Rabi-coupling triggers spin-mixing processes among the internal states and favors their miscibility~\cite{recati2021coherently}. Exemplary, experimental demonstrations of such gases can be found in~\cite{williams2000excitation,zibold2010classical,nicklas2015observation,cominotti2022observation,shibata2019interaction,lavoine2021beyond} where, for instance,
the collective modes of these pseudospinor two-component sodium  gases~\cite{cominotti2022observation}, their miscible to immiscible phase transition through modulation of the interaction among the magnetic sublevels of $^{87}$Rb~\cite{shibata2019interaction} and measurement of beyond-mean-field effects in the expansion of a $^{39}$K two-component gas~\cite{lavoine2021beyond} were reported. 

Specifically, in~\citep{sabbatini2011phase,sabbatini2012causality}, a mixture of $N\approx 10^5$ atoms was initially prepared in the immiscible phase with $\Omega(0)>\Omega_{cr}$. 
Subsequently, $\Omega(t)$ was linearly ramped-down to zero with a finite ramp rate $\tau_Q$ giving rise to long-lived domain-wall configurations among the components. 
Remarkably it was shown that the number of spontaneously generated domains in a ring trap of length $L$ scales for varying quench rate with an exponent as predicted by the Kibble-Zurek mechanism. In particular the number of defects according to Kibble-Zurek theory is given by  
\begin{equation}
 N_d=\frac{L}{\xi_0} \left( \frac{\tau_0}{\tau_Q}\right)^{1/3},\label{eq:defect_number}
\end{equation}
where $\tau_0=\hbar/(2g_S \rho)$, $g_S=(g_A+g_B-2g_{AB})/2$ and $\xi_0=\xi_S/\sqrt{2}$ is related to the spin healing length $\xi_S=\hbar/\sqrt{2m\rho g_S}$, while $\rho$ denotes the uniform density of the gas. 
In the actual numerical simulations it was observed that for $\tau_Q>2$ms it holds that $N_d \propto \tau_Q^{-n}$ with $n=0.341 \pm 0.006$ which is in good agreement with the exponent in Eq.~(\ref{eq:defect_number}). 
Further deviations from this scaling law are present in elongated harmonic traps due to the spatial inhomogeneity induced by the external confinement. 
A comprehensive overview of various aspects of the Kibble-Zurek mechanism in different settings and dimensionalities can be found in the reviews ~\cite{del2014universality,polkovnikov2011colloquium,dziarmaga2010dynamics}. It should, however, be noted that all these studies operate in the macroscopic atom number regime where the interplay of quantum and thermal fluctuations is crucial and therefore lie beyond the scope of this review. For completeness, we rather mention, that notable manifestations of the Kibble-Zurek scaling in  elongated quenched two-component bosonic gases revealed the nucleation of dark-bright solitons~\cite{liu2016stochastic}, and exposed universal properties close to the quantum critical point and similarities with collective spin ensembles~\cite{nicklas2015observation}. Moreover, it was shown that the predicted dynamical critical exponent differs when the immiscibility threshold is crossed upon quenching the interparticle interactions~\cite{jiang2019universality} or the Rabi-coupling~\cite{sabbatini2011phase,lee2009universality}. Finally, observations of the Kibble-Zurek scaling in 1D inhomogeneous systems are summarized in~\cite{del2013causality}. 

\begin{figure}[tb]
    \center
    \includegraphics[width=0.9\textwidth]{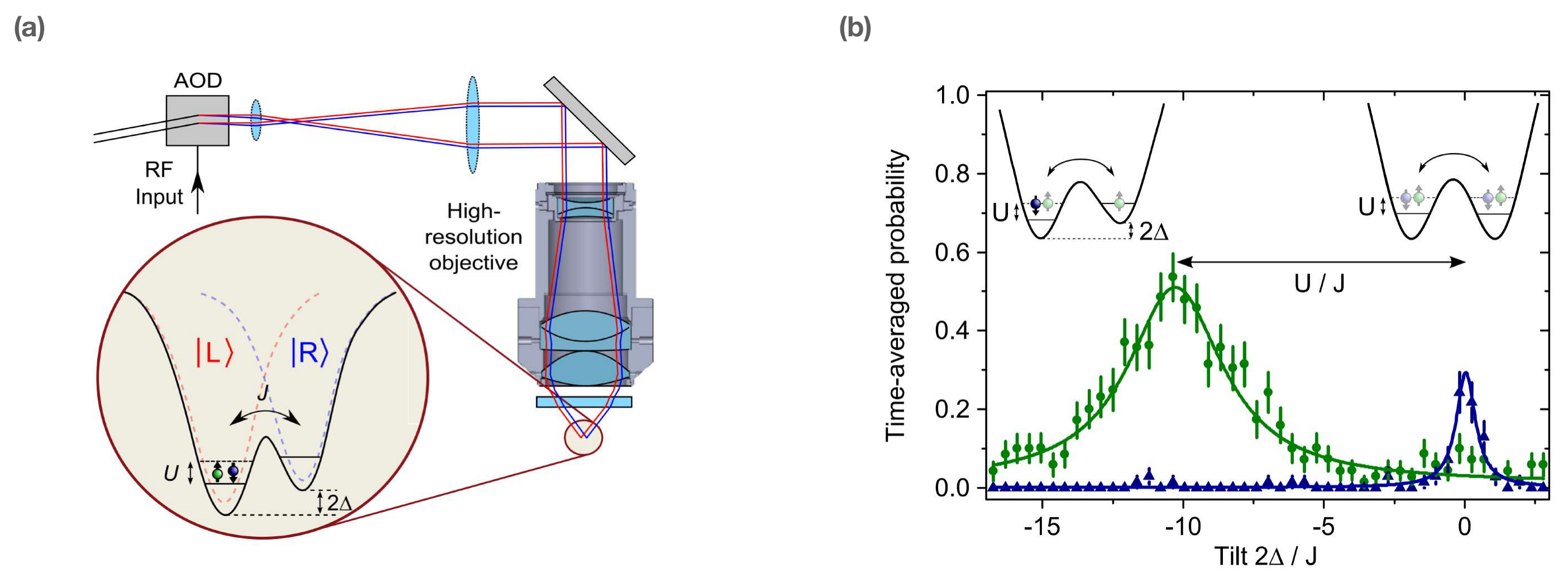}
    \caption{(a) Experimental setup for a 1D double-well potential created by focusing two laser beams with a high-resolution objective. An acousto-optic deflector (AOD) is utilized to independently control the intensity and position of the two laser beams, which allows one to adjust the tunnel coupling $J$ and the tilt $\Delta$ between the two wells. (b) Single-particle (green line) and pair tunneling (blue line) probabilities at the interaction energy $U/J=10.05\pm 0.19$ with respect to the tilt $\Delta$. The experimental data points have $1\sigma$ statistical uncertainty (error bars) and represent the time-averaged probability of detecting a single or a pair of particles in the right well. The system is initially prepared with two atoms in the left well and the the tunneling is switched on. As shown, pair tunneling is resonant in a symmetric double-well, while conditional single-particle tunneling takes place for $-2\Delta=U$. Figures adapted from~\citep{Murmann:15}}
\label{fig:tunneling_fermions_experiment} 
\end{figure}

\subsection{Tunneling dynamics}\label{sec:tunneling_BB}
As discussed in Section~\ref{sec:OneComp} the tunneling dynamics of single species bosons in the crossover from weak to strong interactions shows intriguing phenomena including common Josephson oscillations, pair tunneling, self-trapping, and fermionization~\citep{Zollner2008}. 
Extensions of these studies to binary bosonic mixtures were introduced in several works, e.g., in \citep{mazzarella2011rabi,mazzarella2009atomic,gillet2014tunneling}, demonstrating additional complexity due to the involvement of intra- and intercomponent interactions, as well as the mass-imbalance between the components. 
Indeed, in such non-linear systems time evolution cannot be simply explained by the knowledge of the single component system except for the two cases of either $g_{AB}=0$ and $g_A=g_B$ or $g_{AB}\to g_{A}$ with $g_{B}=0$ and $N_B=1$ in mass-balanced settings. Knowledge of the structure of the full few-body spectrum is required. 

Relevant dynamical aspects observed also for macroscopic bosonic ensembles include, for instance, \textit{macroscopic quantum self-trapping} and \textit{coherent quantum tunneling}~\citep{kuang2000macroscopic,smerzi1997quantum,raghavan1999coherent}, \textit{collapse and revival of population dynamics}~\citep{naddeo2010quantum} as well as \textit{symmetry breaking and restoring scenarios}~\citep{satija2009symmetry}. 
The Josephson or coherent quantum tunneling is a direct generalization of its single component counterpart (see also Section~\ref{Dyn_prop_single_component}), where the condensates oscillate symmetrically among the two potential wells, thus being reminiscent of the sinusoidal Josephson effect appearing in superconductors. 
Moreover, the macroscopic quantum self-trapping refers to an essentially ``locked'' population imbalance among the components. 
It is a quantum effect maintained in closed systems that originates from the presence of non-linear interactions.  However, it strongly depends on the system parameters such as the initial state and the number of atoms. 
Particularly, depending on the initial condition there are two different symmetry-broken phases of macroscopic quantum self-trapping in which the components reside either in distinct wells or at the same site. Moreover,  symmetry restoring tunneling modes exist, where the two species either avoid or follow each other in the course of the evolution. 
Along the same lines collapse and revival events which occur in the dynamics of the population imbalance between the two species are pure interaction effects\footnote{Precisely, both inter- and intracomponent interactions produce collapses and revivals of the population imbalance which counteract each other and are suppressed in the equal interaction limit.} and constitute a macroscopic quantum phenomenon. 
These phenomena are well-known in non-linear optics and are captured by the Jaynes-Cummings model~\citep{scully1999quantum}, which describes the interaction of a single-mode quantized radiation field and a two-level atom. 

The above tunneling processes have been observed in the large particle number limit and for weak interactions where the relevant dynamics can be described using the coupled set of mean-field Gross-Pitaevskii equations of motion~\cite{raghavan1999coherent,satija2009symmetry}, or using the lowest-band approximation and especially the Bose-Hubbard model~\cite{kuang2000macroscopic,smerzi1997quantum,naddeo2010quantum}. 
Note that the particular case of the two-site Bose Hubbard model, which is one of the simplest and most studied models that accounts for tunneling and self-trapping in double wells, can be solved exactly by Bethe ansatz methods \citep{Jon2006, Tonel2005}. 
However, for few-body systems and strong interactions, correlation phenomena become important and invalidate the aforementioned models. 
Indeed, the study of few-body systems is particularly useful, not only due to their numerical feasibility for \textit{ab-initio} calculations, but most importantly since they allow for an in depth understanding of the accompanying microscopic mechanisms which subsequently provide a bottom-up perspective to the processes occurring in larger systems. 
Additionally, recent experimental developments enable one to design and probe the tunneling properties of few-body systems. 
A characteristic example is the 
experimental realization of the tunneling of two distinguishable fermions where the interaction strength and the shape of the external potential were used to engineer the process~\citep{Murmann:15}, see Fig.~\ref{fig:tunneling_fermions_experiment}.  

The role of intra- and intercomponent interactions in the tunneling dynamics of few-body bosonic mixtures in a 1D double-well was discussed in~\citep{chatterjee2012few} for different initial configurations, namely a population imbalance and a phase-separated state. 
To elucidate the underlying few-body correlation processes a mixure of three $A$-species bosons and a single $B$-species atom was considered. 
Relying on the energy spectrum of the system, it was found that stronger interspecies repulsions result in a larger tunneling period. 
This mechanism is the few-body equivalent of \textit{quantum self-trapping} \citep{smerzi1997quantum,raghavan1999coherent,julia2010macroscopic,wuster2012macroscopic} which has been experimentally observed by~\citep{albiez2005direct} using $^{87}$Rb condensates, see also the review by~\citep{gati2007bosonic} and references therein. 
Therefore, the interspecies repulsion leads to particle ``binding'' and reduces the 
tunneling rate. 
Moreover, the tunneling behavior crucially depends on the initial configuration. 
Indeed, for complete population imbalance corresponding to either all atoms prepared in the same well or experiencing intercomponent separation, the individual components tunnel in- or out-of-phase. 
However, for ``partial'' population imbalance where one component is delocalized equally over the two wells and the other is localized a counterflow type dynamics takes place. We note that these oscillations among wells can also be observed in the limit of very strong interactions~\citep{Barfknecht2018Junction}. 

\begin{figure}[tb]
    \center
    \includegraphics[width=\textwidth]{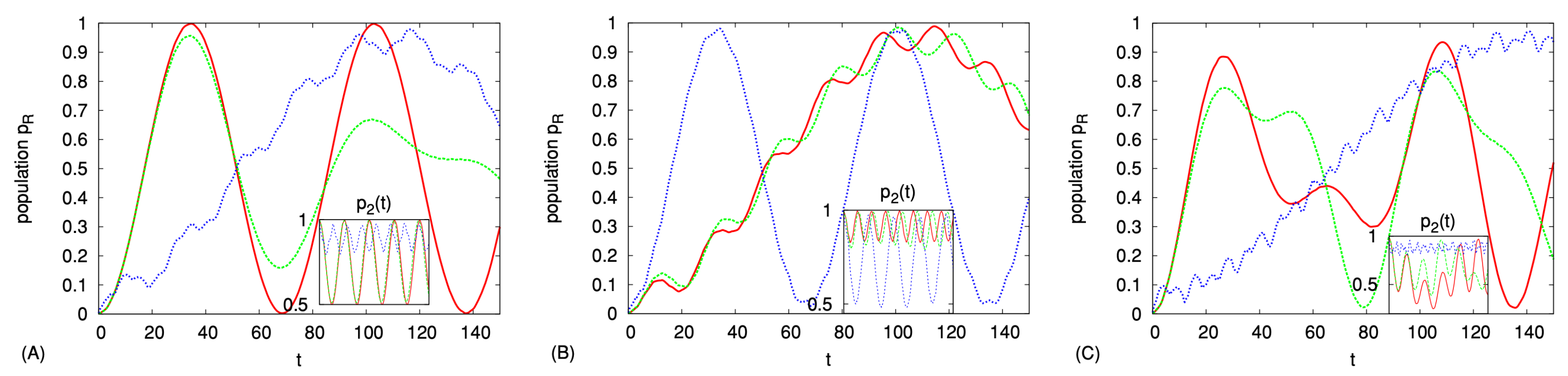}
    \caption{Characteristic tunneling regions of $N_A=2$ bosons through a single $B$ atom as captured by the population of the right well, see text for details. Shown are different mass ratios, namely $m_A/m_B=0.001$ (red line), $m_A/m_B=0.01$ (green dashed line) and $m_A/m_B=0.12$ (blue dashed-dotted line). The distinct panels correspond to intracomponent interactions (a) $g_{A}=0$, (b) $g_{A}=0.5$ and (c) $g_{A}=25$. The insets show the respective pair probabilities. In all cases $g_{AB}=8$. Figure adapted from \citep{pflanzer2009material}}
\label{fig:tunneling_mixture_material} 
\end{figure}

\subsection{Tunneling with mass imbalance}\label{sec:tunneling_massimb}
A typical observable for monitoring the tunneling behavior is the population imbalance among the left (L) and right (R) sites of the double-well in the course of the evolution. 
This is easily quantified by the expectation value of the one-body density given, for example, in the right well (0, $x_R$) by 
\begin{equation}
 p_R^{\sigma} (t)=\frac{1}{N_{\sigma}}\int_0^{x_R} dx \rho_1^{\sigma} (x,t).\label{prob_tunnel_1b}  
\end{equation} 
This provides a single-particle measure of the tunneling dynamics. 
The one-body density of the $\sigma=A,B$ component is given by $\rho_1^{\sigma} (x,t)=\braket{\Psi(t)|\hat{\Psi}_{\sigma}^{\dagger}(x) \hat{\Psi}_{\sigma}(x)|\Psi(t)}$, where $\ket{\Psi(t)}$ is the many-body wave function of the binary system and $\hat{\Psi}_{\sigma}(x)$ denotes the $\sigma$-component bosonic field operator acting at position $x$. 
Additionally, in order to infer the emergence of possible two-body mechanisms one can resort to the respective pair probability amplitude given by
\begin{equation}
p_2^{\sigma \sigma'}(t)= \frac{1}{\mathcal{K}}\int_{x_1 x_2>0} dx_1 dx_2 \rho_2^{\sigma \sigma'} (x_1, x_2,t).\label{prob_tunelpair_2b} 
\end{equation}
Here the normalization factor is $\mathcal{K}=N_{\sigma}(N_{\sigma}-1)/2$ for $\sigma = \sigma'$ and $\mathcal{K}=N_{\sigma} N_{\sigma'}$ if $\sigma' \neq \sigma$. 
Note that it is based on the reduced two-body density distribution 
\begin{equation}
\rho_2^{\sigma \sigma'} (x_1, x_2,t)=\braket{\Psi(t)|\hat{\Psi}_{\sigma}^{\dagger}(x_1)\hat{\Psi}_{\sigma'}^{\dagger}(x_2)\hat{\Psi}_{\sigma}(x_1)\hat{\Psi}_{\sigma'}(x_2)|\Psi(t)}. 
\end{equation}
The latter gives the probability of measuring simultaneously two bosons of the same ($\sigma=\sigma'$) or different species ($\sigma \neq \sigma'$) residing at positions $x_1$ and $x_2$, respectively.

The impact of the mass-imbalance ($m_A/m_B$) on the interspecies tunneling dynamics of few-body bosonic mixtures with $N_A=2$ light bosons and a heavy $B$ atom was investigated in \citep{pflanzer2009material,pflanzer2010interspecies,theel2020entanglement,theel2021many}. 
In~\citep{pflanzer2009material,pflanzer2010interspecies} 
the light $A$ bosons were originally loaded on the left-hand side of a harmonic trap and the $B$ atom was assumed to be tightly trapped at the center to provide a \textit{``material barrier''}. 
Then, upon release of the $A$ atoms the interspecies tunneling was initiated and monitored via the population of $A$ atoms in the right well given by Eq.~(\ref{prob_tunnel_1b}) and the corresponding pair probability for being in the same well given by Eq.~(\ref{prob_tunelpair_2b}). 
Focusing on relatively strong interspecies repulsion, $g_{AB}=8$, it was found that distinct interspecies tunneling regimes can occur depending on the intraspecies repulsion and the mass ratio, see Fig.~\ref{fig:tunneling_mixture_material}. 
In particular, in the non-interacting limit $g_{A}=0$ it is possible to realize Rabi oscillations for the smallest mass ratios, $m_A/m_B=0.001$, which turn into two-mode oscillations for increasing mass ratio, see the left panel in Fig.~\ref{fig:tunneling_mixture_material}. 

For a larger mass ratio ($m_A/m_B=0.012$) a second-order tunneling process of a stable atom-pair takes place and Rabi-oscillations are suppressed, a mechanism that is known and in fact realized to occur also for repulsive atom-pairs~\cite{winkler2006repulsively,folling2007direct}. 
This behavior is reversed for finite intraspecies repulsions, e.g.~$g_A=0.5$, where long period tunneling of the repulsive bound pair appears for $m_A/m_B<0.01$, while for even smaller $m_B$ the pair breaks and Rabi-type oscillations occur [middle panel of Fig.~\ref{fig:tunneling_mixture_material}]. 
When approaching the fermionization limit, $g_A\approx 25$, and for small mass ratios, $m_A/m_B=0.001$, the dynamics resembles the so-called \textit{fragmented pair tunneling} where the two non-interacting fermions (or equivalently infinitely strongly repulsive bosons according to the Bose-Fermi mapping~\citep{girardeau1960}) perform independently Rabi oscillations. 
However, for much smaller masses of the $B$ atom ($m_A/m_B=0.12$) a correlated pair tunneling behavior of the ``effectively attractive'' bound pair of identical fermions (Tonks regime) is observed as exemplarily shown in the right panel of Fig.~\ref{fig:tunneling_mixture_material}. 
All the processes can be assigned to specific transitions of the few-body spectrum as a function of the mass ratio. 

The dependence of the tunneling dynamics of few-boson mixtures in a 1D triple-well and in the presence of an additional harmonic trap on the intercomponent interaction strength was investigated in \citep{cao2012impact}, which considered
the mixture of two $A$  bosons and a single $B$ atom. The system is initially decoupled with the $A$ bosons residing in the ground state of the system and the $B$ atom being localized in the left well. 
Note that the ground state of the $A$-subsystem predominantly populates the middle well and that the outer wells (due to non-zero spreading of the bosonic ensemble) exhibit spatial correlations with one another. Such a configuration can be generalized to larger systems~\footnote{Recall that in the absence of the harmonic potential, lattice trapped single component bosons feature a superfluid to Mott-insulator phase transition~\citep{greiner2002quantum}.} where an additional trapping potential is present~\citep{batrouni2002mott,gemelke2009situ}.  
The dynamics is subsequently induced by switching on the intercomponent repulsion and allowing the $B$ atom to move through the lattice. 
For weak repulsions, it was shown that the aforementioned spatial correlations of the $A$-species particles triggers inter-site hopping through a mechanism called \textit{correlation-induced tunneling}, while the $B$-species atom remains practically localized. 
For intermediate interactions, a delayed tunneling process is observed due to destructive interference of the participating tunneling branches. 
Additionally, there are a few resonant tunneling interaction windows where "correlation transfer" from one species to the other takes place. 
Entering the strongly interaction regime an enhanced magnitude of the tunneling occurs, a phenomenon that is attributed to the contribution of higher-band states. 
Finally, it should be emphasized that the presence of interspecies correlations plays a crucial role for the appearance of next nearest neighbor tunneling channels of impurities in a lattice trapped majority component Bose gas as it has been demonstrated in \citep{keiler2020doping,keiler2019interaction,theel2020entanglement,theel2021many}.

\subsection{Bose-Fermi mixtures}\label{sec:Bose_Fermi}
In addition to the two-component Bose-Bose systems, degenerate mixtures where atoms obey different statistics, i.e., Bose-Fermi settings, have been experimentally realized, in particular, using sympathetic cooling~\citep{truscott2001observation,schreck2001,roati2002fermi,inouye2004observation,fukuhara2009all}. 
Recall, that $s$-wave interactions between spin-polarized fermions are suppressed due to the Pauli exclusion principle~\citep{pethick2002}. 
Thus, the corresponding 1D analogue of these mixtures is described by the Hamiltonian of Eq.~(\ref{eq:Hamiltonian_pseudospinor}) with zero intraspecies two-body interactions among the $N_F$ fermions\footnote{Here, the respective set of $N_F+1$ coupled Gross-Pitaevskii mean-field equations of motion, where the intraspecies correlations that are present in the Hartree-Fock approach are taken into account, can be found for instance in~\citep{karpiuk2004soliton,Mistakidis2019FermionsinBosons}}. 
Experimentally relevant mixtures of isotopes include $^{7}$Li and $^{6}$Li~\citep{truscott2001observation,schreck2001}, $^{6}$Li and $^{23}$Na~\citep{hadzibabic2002two} $^{40}$K and $^{87}$Rb~\citep{roati2002fermi,inouye2004observation}, or $^{173}$Yb and $^{174}$Yb~\citep{fukuhara2009all}. 

The interplay of Bose-Bose and Bose-Fermi interactions supports a plethora of phase configurations in the ground state that can be characterized by distinctive correlation properties. They also lead to peculiar dynamical features which differ from pure bosonic counterparts due to the fermionic nature of one of the components. 
These will be the main focus of our discussion below.
For completeness we note in passing that in higher dimensions several phenomena have been unveiled such as phase-separation processes~\citep{roth2002structure,lous2018probing}, non-trivial stability conditions~\citep{miyakawa2004peierls,cazalilla2003instabilities,belemuk2007stable}, interesting low-lying collective excitations~\citep{pu2002phonon,liu2003collisionless,rodriguez2004scissors}. For lattices, there are intriguing quantum phases like exotic Mott-insulator and superfluid states~\citep{zujev2008superfluid,dutta2010unconventional}, charge-density waves~\citep{lewenstein2004atomic,mathey2004luttinger} and supersolid phases~\citep{buchler2003supersolid,titvinidze2009generalized}.

The ground state phase structure of few-body Bose-Fermi mixtures in a 1D harmonic trap has been investigated in \citep{chen2018bunching,pkecak2019intercomponent} for varying interaction strengths and mass ratios\footnote{It is interesting to note that a system of three trapped ultracold and strongly interacting atoms or arbitrary statistics in 1D can actually be emulated using an optical fiber with a graded-index profile and thin metallic slabs \citep{GarciaMarch:2019gradedindexoptical}}. 
An important outcome was that the bosonic species exhibit a bunching-antibunching crossover for increasing interspecies repulsion, a behavior that originates from the presence of the entanglement among the two species. 
Furthermore, it was shown that the interspecies correlations create an induced bosonic interaction, which in turn elucidates the occurrence of the bosonic bunching. 
Moreover, as in the case of two-component Bose systems, an increasing interspecies repulsion for fixed intracomponent interaction leads to a transition from spatially miscible to immiscible Bose and Fermi components, whilst stronger attractions lead to a collapse behavior~\citep{cazalilla2003instabilities}.   
The impact of the particle imbalance on the structural deformations of the densities of each component in the strongly interacting regime and assuming equal boson-boson and boson-fermion couplings was considered in detail both in few-body~\citep{dehkharghani2017analytical}, and many-body homogeneous~\citep{cazalilla2003instabilities,Imambekov2006Mixture} settings. 
It has been argued that phase-separation is a sensitive phenomenon in Bose-Fermi mixtures and depends crucially on the system's composition such as the  even/odd number of atoms in each component.   

\begin{figure}[tb] 
    \center
    \includegraphics[width=0.9\textwidth]{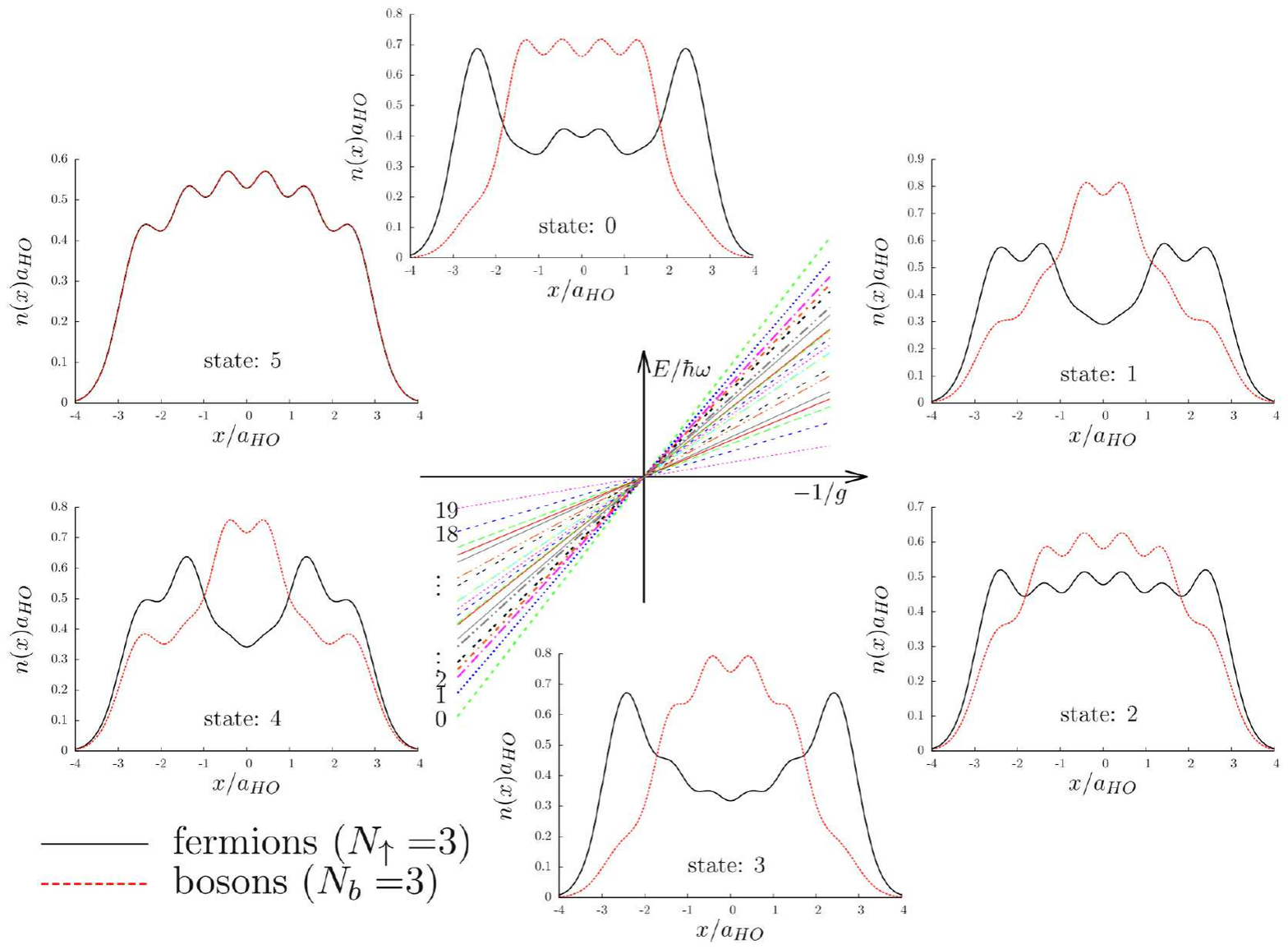}
    \caption{(Central panel) A sketch of the energy spectrum for a few-body Bose-Fermi mixture with three fermions and three bosons in the vicinity of $1/g=0$. The figure shows the first twenty states that become degenerate at $1/g=0$.
    The density profiles of the six energetically lowest-lying states (for $1/g=0+$) are shown in the other panels.
    The results were produced following the procedure of \citep{Volosniev2014StrongInteractions} (see also section~\ref{CloseToFermi}). 
    Figure from~\citep{dehkharghani2017analytical}.}
    \label{fig:ground_state_Bose_Fermi_few3} 
\end{figure}

Additionally, relevant exact stationary solutions of a mass-balanced Bose-Fermi mixture in the infinite repulsion limit and for equal intra- and intercomponent interactions have been constructed in the absence~\citep{hu2016strongly,batchelor2005exact} and in the presence~\citep{Massignan2015Magnetism,decamp2017strongly,dehkharghani2017analytical} of a 1D harmonic trap using the Bethe-ansatz and the Bose-Fermi mapping respectively~see~Fig.~\ref{fig:ground_state_Bose_Fermi_few3}. 
Degeneracies of the ground state energy, which are lifted away from the infinite interaction limit, were discussed and the universal energy relation for the homogeneous Bose-Fermi setting has been derived. 
Moreover, an adequate agreement in terms of the ensuing density and momentum distributions as captured by the analytical wave functions and predicted independently via  DMRG at strong repulsions has been reported in the trapped scenario \citep{Bellotti2017Comparing}.  
Also in this case, it was shown that an intercomponent phase-separation takes place for strong interactions, while the tail of the momentum distribution, defining the Tan contact\footnote{The Tan contact provides a measure of short-range two- and three-body correlations in single as well as multicomponent systems. 
This concept has been experimentally probed in higher-dimensional settings using radiofrequency spectroscopy and subsequent measurement of the structure factor via Bragg spectroscopy\citep{fletcher2017two,sagi2012measurement,wild2012measurements,stewart2010verification,hoinka2013precise}.}~\citep{tan2008energetics,tan2008large,tan2008generalized,barth2011,valiente2012PhRvA}, differs for bosons and fermions and assumes larger values when increasing the number of bosons or the number of fermionic species. 
A detailed analysis of the behavior of the Tan contact for Bose-Fermi mixtures in the thermodynamic limit at zero and finite temperatures can be found in \citep{pactu2019momentum}.  

For the ground state of a weakly interacting particle-imbalanced few-body Bose-Fermi mixture in a lattice, it was shown in~\citep{siegl2018many,wang2020emergent} that by varying the ratio between the inter- and intraspecies coupling strengths two distinct density configurations can form. 
Namely, when the interspecies coupling is larger (smaller) than the bosonic interaction, a phase immiscible (miscible) state is entered, which is characterized by a vanishing (complete) spatial overlap of the bosonic and fermionic single-particle density distributions, see also \citep{pollet2006phase,pollet2008mixture,barillier2008boson}.

\begin{figure}[tb]
    \center
    \includegraphics[width=0.9\textwidth]{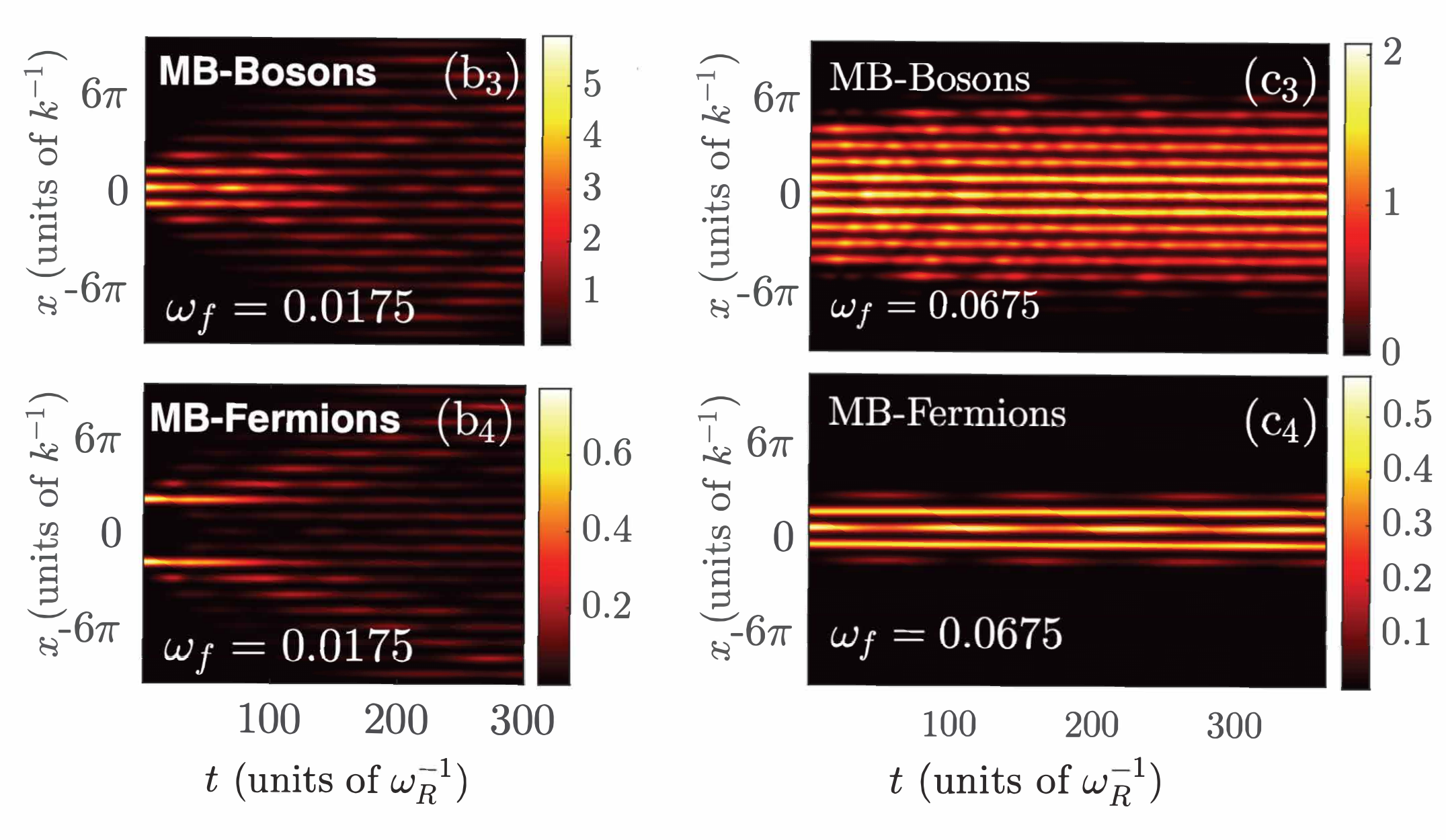}
    \caption{Resonant expansion dynamics of the density of a lattice trapped Bose-Fermi mixture following a quench from strong to weak confinement. The value of the postquench harmonic trap frequency ($\omega_f$) is provided in the legends and the prequench one is $\omega_i=0.1$. The initial phase of the mixture is either immiscible (left panels) or miscible (right panels). The mixture consists of $N_B=20$ bosons (upper panels) and $N_F=2$ fermions (lower panels) confined in a 19-well lattice with an imposed harmonic trap. The acronym MB stands for many-body designating that the dynamics is monitored within the variational ML-MCTDHX beyond mean-field approach. Time (length) is expressed in units of $\omega_R^{-1}=\hbar E_R^{-1}$ ($k^{-1}$), with $E_R=\hbar^2 k^2/(2m)$ being the recoil energy and $k$ the lattice wave vector. 
    Figures from~\citep{siegl2018many}. }
\label{fig:expansion_Bose_Fermi} 
\end{figure}

The behavior of the collective modes in a 1D harmonic trap of a mass-balanced Bose-Fermi mixture with equal intra- and interspecies repulsive couplings has been discussed in \citep{Imambekov2006Mixture,imambekov2006applications} employing a combination of the Bethe-ansatz technique and the local density approximation. 
Operating within the weakly interacting regime, it has been shown that for decoupled components the collective modes of the Bose-Fermi mixture are the same as the pure fermionic and bosonic modes, obeying the dispersion relations $\omega_f=n \omega_0$ and $\omega_b=\omega_0 \sqrt{n(n+1)/2}$ respectively. 
However, for finite boson-fermion interactions all collective modes besides the Kohn mode \citep{fetter1998,bijlsma1999,zaremba1999} are affected and change their value. 
Interestingly, in the strongly interacting regime the existence of low-lying collective oscillations leading to the counterflow of the two species was found~\citep{delehaye2015critical,chevy2015counterflow}. 
The properties of the low-lying collective excitations of Bose-Fermi mixtures have been extensively investigated, see for instance the works of~\citep{van2009trapped,banerjee2007collective,huang2019breathing,grochowski2020breathing,wen2019collective}.   
Lastly, it should be noted that the process of dynamical fermionization of strongly interacting Bose-Fermi mixtures subjected to quenches of the trap frequency has been reported in \citep{fang2009fermionization,patu2021dynamical}.

Turning to the dynamical response of few-body Bose-Fermi mixtures, the quench-induced expansion dynamics of such a particle imbalanced setting in an 1D optical lattice with an harmonic trap under quenches from strong-to-weak confinement was considered in \citep{siegl2018many}. 
It was shown that within the immiscible phase, a resonant-like response of both components occurs at moderate quench amplitudes [see the left hand side panels in Fig.~\ref{fig:expansion_Bose_Fermi}], which is a phenomenon that takes place also in the expansion dynamics of single component ensembles~\citep{koutentakis2017quench}. 
In particular, a decreasing confinement strength leads to various response regimes where the bosons either perform a breathing motion or only expand in the entire lattice. 
Simultaneously, the fermions undergo nearest-neighbor tunneling within the outer wells located at the edges of the bosonic cloud or exhibit a delocalized behavior over the lattice structure. 
A correlation analysis reveals that in the course of the evolution losses of coherence occur in the predominantly populated wells which are two-body anti-correlated among each other, while a correlated behavior for bosons and an anti-correlated one for fermions occurs within each well. 
The susceptibility of the mixture is significantly altered for an initial miscible state, see the right hand side panels in Fig.~\ref{fig:expansion_Bose_Fermi}. 
The bosons perform interwell tunneling reaching an almost steady state for long evolution times and the fermions expand toward the edges of the surrounding bosonic cloud where they are partly transmitted and partly reflected back to the central wells. 
It has been actually found experimentally~\citep{ronzheimer2013expansion} that the expansion dynamics of bosons indeed depends crucially on both the initial interacting state and the spatial dimension. 
For mass-imbalanced mixtures the heavy component remains almost unperturbed, whilst an increasing potential barrier height results in a suppressed expansion dynamics. 
The relaxation dynamics of doublons consisting of boson-fermion pairs in 1D lattices has been examined in~\citep{garttner2019doublon} and argued to stem from several processes including single-doublon decay, doublon-pair decay via triplon formation, and decay after doublon tunneling.

\subsection{Systems close to fermionization}\label{CloseToFermi} 
One interesting prospect of strongly interacting bosons is that their local properties can be studied using a non-interacting Fermi gas via the Fermi-Bose mapping~\citep{girardeau1960} (see also Section~\ref{subsec:TGmodel}).
This subsection discusses an extension to this mapping in the context of the multicomponent Hamiltonian from Eq.~(\ref{eq:Hamiltonian_pseudospinor})  with $V=V_A=V_B$, $m=m_A=m_B$, and $g_{A}, g_{B}, g_{AB}\to \infty$. For simplicity, it will be assumed that $g_A=g_{B}$. It is however straightforward to extend the discussion to the case with $g_A\neq g_B$. 

To set the stage, let us consider a two-body system that consists of an atom of type $A$, and an atom of type $B$. In the limit $g_{AB}\to \infty$, the ground state of the system is double degenerate. This degeneracy was already noticed for a harmonic trap, see Section~\ref{subsec:two_part}, but, in fact, it is a general feature independent of the external confinement $V$. The degeneracy occurs simply because the particles cannot exchange their positions when the interaction is infinitely strong, 
and the particle configuration $AB$ ($x_1<y_1$) is decoupled from the configuration $BA$ ($y_1<x_1$). The two orthogonal states $AB$ and $BA$ imply that the wave function at $1/g_{AB}=0$ reads as follows
\begin{equation}
  \Psi(x_1,y_1)=\begin{cases}
    a_1 \Phi_F(x_1,y_1), & \text{if $x_1<y_1$},\\
    a_2 \Phi_F(x_1,y_1), & \text{if $y_1<x_1$},
  \end{cases}
  \label{eq:Psi_Bose_Fermi}
\end{equation}
where $a_i$ is a number. The spatial profile, $\Phi_F$, of the state follows from the Fermi-Bose mapping, according to which, $\Phi_F(x_1,y_1)$ must be the wave function of spinless fermions, see~\citep{girardeau1960}. The existence of two independent coefficients $a_1$ and $a_2$ is a mathematical manifestation of the double degeneracy discussed above. 
It is worth noting that there is a special set of $\{a_1,a_2\}$ that is connected to eigenstates at finite values of $g_{AB}$. This subsection describes how to find this set, see also~\citep{Minguzzi2022Review}.

Let us consider the system where $g_{AB}/a$ is finite, but sets the largest energy scale of the problem. Here $a$ is a unit of length given by the trap\footnote{For example, for a harmonic trap $a=\sqrt{\hbar/m\omega}$, see Section~\ref{subsec:two_part}}. Under this condition, the energy of the system is given by 
\begin{equation}
    E\simeq E_F-\frac{K}{g_{AB}},
    \label{eq:energy_fermionization}
\end{equation}
where $K=-\lim_{g_{AB}\to\infty}\frac{\partial E}{\partial (1/g_{AB})}$, and $E_F$ is the energy of the corresponding ideal Fermi gas. For a harmonic trap, this form of $E$ can be motivated by the expansion of Eq.~(\ref{eq:two_body_energy}), see also Fig.~\ref{fig:ground_state_Bose_Fermi_few3} for larger systems. For a general case, the expansion of $E$ can be validated using the Hellmann–Feynman theorem.

Assuming that the form of the wave function is given by Eq.~(\ref{eq:Psi_Bose_Fermi}), the value of $K$ can be calculated using the Hellmann–Feynman theorem:
\begin{equation}
    K=\lim_{g_{AB}\to\infty}g_{AB}^2\frac{\langle\Psi|\delta(x_1-y_1)|\Psi\rangle}{\langle\Psi|\Psi\rangle}.
\end{equation}
The limit can be taken by using the Bethe-Peierls boundary condition for the delta-function interaction
\begin{align}
   \frac{2m}{\hbar^2}\lim_{g_{AB}\to\infty}g_{AB}\Psi(x_1=y_1)=\left(\frac{\partial \Psi}{\partial x_1}-\frac{\partial \Psi}{\partial y_1}\right)_{x_1=y_1+0}-
   \left(\frac{\partial \Psi}{\partial x_1}-\frac{\partial \Psi}{\partial y_1}\right)_{x_1=y_1-0}, 
   \label{eq:bethePeielrsBoundary}
\end{align}
which leads to the expression
\begin{equation}
    K=\frac{(a_1-a_2)^2}{a_1^2+a_2^2}\alpha; \qquad \alpha=2\frac{\hbar^4}{m^2}\int\mathrm{d}y_1\left|\frac{\partial \Phi_F(x_1,y_1)}{\partial x_1}\right|^2_{x_1=y_1},
    \label{eq:K_fermionization}
\end{equation}
where it has been assumed that $\Phi_F$ is normalized. Note that the value of $\alpha$ depends only on the shape of the external potential. 

Once $K$ is calculated, the spectrum at finite values of $g_{AB}$ can be found by minimizing the energy (maximizing $K$) with respect to the parameters $a_1$ and $a_2$:
\begin{equation}
    \frac{\partial}{\partial a_1}\frac{(a_1-a_2)^2}{a_1^2+a_2^2}\alpha=0, \qquad 
    \frac{\partial}{\partial a_2}\frac{(a_1-a_2)^2}{a_1^2+a_2^2}\alpha=0.
    \label{eq:min_alpha_K}
\end{equation}
This set of equations leads to the ground state of a two-body system having $a_1=-a_2$, which corresponds to a bosonic symmetry. The excited state is given by $a_1=a_2$, i.e., it is simply a fermionic state. This result is general for any external potential, because the shape of the potential appears in Eq.~(\ref{eq:min_alpha_K}) only via the overall pre-factor $\alpha$.  

The minimization problem given by Eqs.~(\ref{eq:K_fermionization}) and~(\ref{eq:min_alpha_K}) can also be formulated using the spin-chain Hamiltonian that acts in the Hilbert space of $|\uparrow\downarrow\rangle$ (cf. $AB$) and $|\downarrow\uparrow\rangle$ (cf. $BA$)
\begin{equation}
    H_s=E_F\mathbf{I}-J(\mathbf{P_{\uparrow\downarrow}-I}),
\end{equation}
where $\mathbf{I}$ is the identity operator, and $\mathbf{P_{\uparrow\downarrow}}$ is the swap operator ($\mathbf{P_{\uparrow\downarrow}}|\uparrow\downarrow\rangle=|\downarrow\uparrow\rangle$). Indeed, if  the expectation value of $H_s$ is written using the trial state $f_s=a_1|\uparrow\downarrow\rangle+a_2|\downarrow\uparrow\rangle$,
\begin{equation}
    \frac{\langle f_s| H_s| f_s\rangle}{\langle f_s| f_s\rangle}=E_F+J\frac{(a_1-a_2)^2}{a_1^2+a_2^2},
\end{equation}
then the connection to the minimization problem from Eq.~(\ref{eq:min_alpha_K}) is evident
assuming that $J=-\alpha/g_{AB}$.
\begin{figure}[t]
  \centering
    \includegraphics[width=\linewidth]{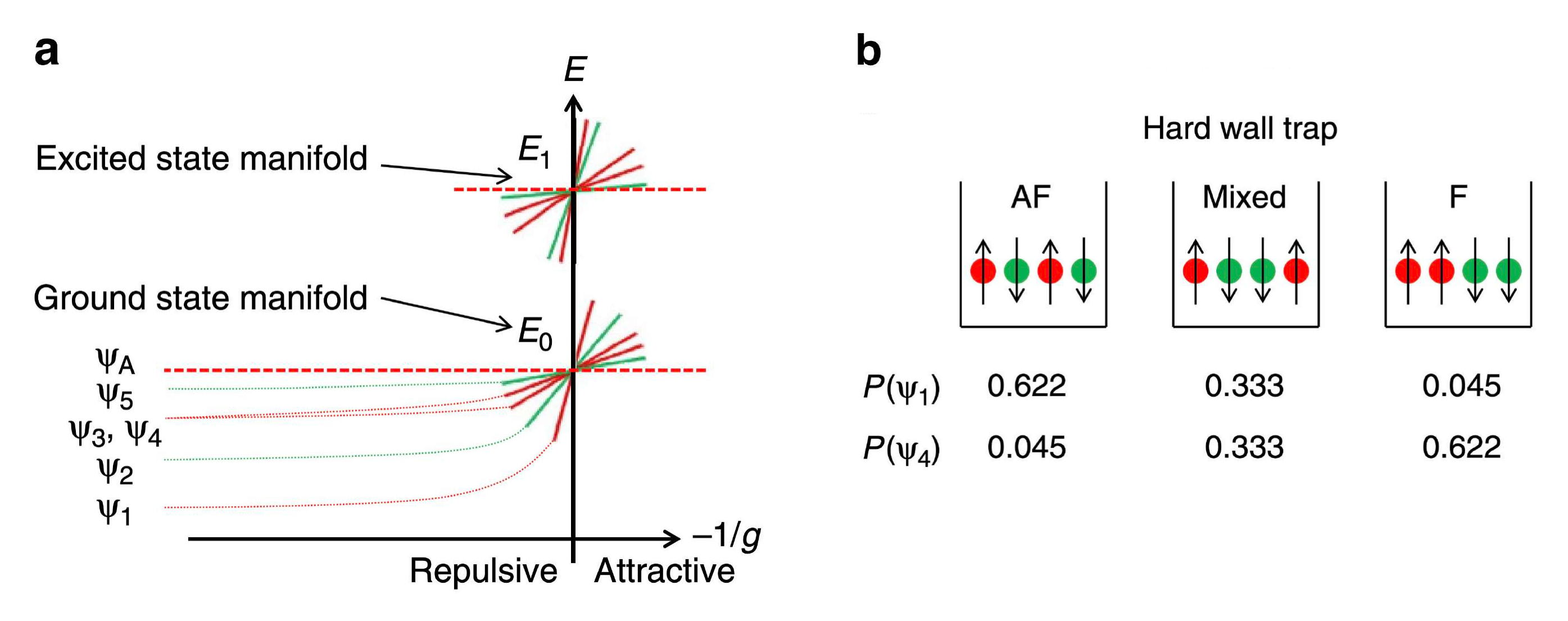}\hfill
    \caption{(a) Energies of a four-body system (2+2) close to fermionization ($1/g\to0$), with $1/\kappa=0$ for simplicity. Only the ground and first-excited state manifolds are shown.  
    (b) Relative populations of different configurations $\sim |a_i|^2$ for different states from the ground-state manifold. The ground state ($\Psi_1$) has the most probability in the antiferromagnetic configuration. Figure adapted from \citep{Volosniev2014StrongInteractions}.}
\label{fig:CloseToFermionization}
\end{figure}
 
The ideas above can be straightforwardly extended to larger systems. In fact, the hidden spin chain structure of a strongly interacting two-component system was also established in the context of the thermodynamic Bethe ansatz in \citep{Guan:2007}. A detailed discussion of a three-body system is presented in~\citep{Volosniev2014FewBody}; arbitrary particle numbers are discussed in~\citep{Deuretzbacher2014Mapping,Volosniev2014StrongInteractions,Volosniev2015Dynamics}. 
For all two-atom systems, and three-atom systems in parity-symmetric traps, the values of the $a_i$ can be deduced without maximising $E$ by considering symmetries only \citep{Harshman2016_symmetries}.
For three-atom systems in asymmetric traps, and for all larger systems there are not enough symmetries to fix $a_i$,~\citep{Harshman2016_symmetries,Harshman2016Symmetries2}, and $a_i$ will depend on the trap. At the moment, the spin-chain mapping is the simplest way to calculate the $a_i$. The mapping is applicable to all (fermionic, bosonic) strongly interacting systems, and can even be extended to systems with more than two components, see, for example,~\citep{Cui2016Spin}. The only assumption is that all masses are identical, although small mass imbalance can be included as a magnetic field term as discussed in~\citep{Volosniev2017MassImbalance}.

Let us briefly illustrate the spin-chain mapping for $g_{AB}=g$, and $g_{A}=g_{B}=\kappa g$. In this case, one derives the XXZ Hamiltonian (see Section~\ref{sec09SC} for more detail)
\begin{equation}
H_s =E_F\mathbf{I}-\sum_{i=1}^{N_A+N_B-1}J_i \left[\mathbf{P}_{i,i+1}-\mathbf{I}-\frac{1}{\kappa}(1+\sigma_z^{i}\sigma_z^{i+1})\right],
\label{eq:mapping_spin_chain}
\end{equation}
where $\mathbf{P}_{i,i+1}$ is the operator that swaps the spins at the $i$th and $(i+1)$th sites, and $\sigma_z^{i}$ is the third Pauli matrix that acts on the site $i$. The coupling coefficient $J_i$ is given by
\begin{equation}
J_i=-\frac{N!\hbar^4}{gm_A^2}\int_{x_1<x_2\cdots<x_N-1}dx_1...dx_{N-1}\Big|\frac{\partial \Phi_F}{\partial x_N}\Big|^2_{x_N=x_i},
\end{equation}
where $\Phi_F$ is a normalized wave function of the $N=N_A+N_B$ spinless fermions. 
For given values of $N_A$ and $N_B$, the spin-chain Hamiltonian operates in the sector whose total magnetization is $N_A-N_B$. For example, a balanced case has $M=0$. Experimentally, the mapping is most easily explored using systems with impurities~\citep{Jochim2015SpinChain}, see also~\citep{Meinert:17}, where it was observed that a strongly interacting Bose gas indeed creates a lattice for the impurity.

It is worth noting that specific properties of $\Phi_F$ have not been used so far. Therefore, the discussion above applies to excited states $\Phi_F$ as well as to the ground state, as long as $g_{AB}/a$ is the largest energy scale. For example, the energy spectrum for four particles in a box potential can be seen in Fig.~\ref{fig:CloseToFermionization}(a). Each function $\Phi_F$ gives rise to a manifold of states, which become degenerate in the limit $g_{AB}\to \infty$. The number of states in each manifold follows from the number of distinct particle configurations at $1/g_{AB}=0$, i.e., $(N_A+N_B)!/(N_A!N_B!)$. For completeness, the figure also illustrates the probabilities of these configurations in the ground-state manifold, see panel (b). The ground state, $\psi_1$, is mainly ``anti-ferromagnetic'' (AF) which follows from the structure of the Hamiltonian. However, the excited states may have a ``ferromagnetic'' configuration as a dominant one.    

The correspondence between a strongly interacting system and a spin chain turns out to be very useful. In particular because there exist powerful tools (e.g., DMRG, see Section~\ref{subsec:DMRG}) to solve problems formulated in a spin-chain formalism for a large number of particles. This allows one to study many-body systems in a continuum that cannot be addressed otherwise. As such, the mapping provides important reference points for benchmarking of standard numerical methods for diagonalizing the Hamiltonian $H$, and an example of this is given in \citep{Bellotti2017Comparing}. Finally, the spin-chain formulation allows one to propose new simulators for studying spin dynamics and magnetic correlations without the need for an external lattice potential. A brief review of some of the relevant in this context results is given in the next two paragraphs. 

\begin{figure}[t]
  \centering
\includegraphics[width=1\linewidth]{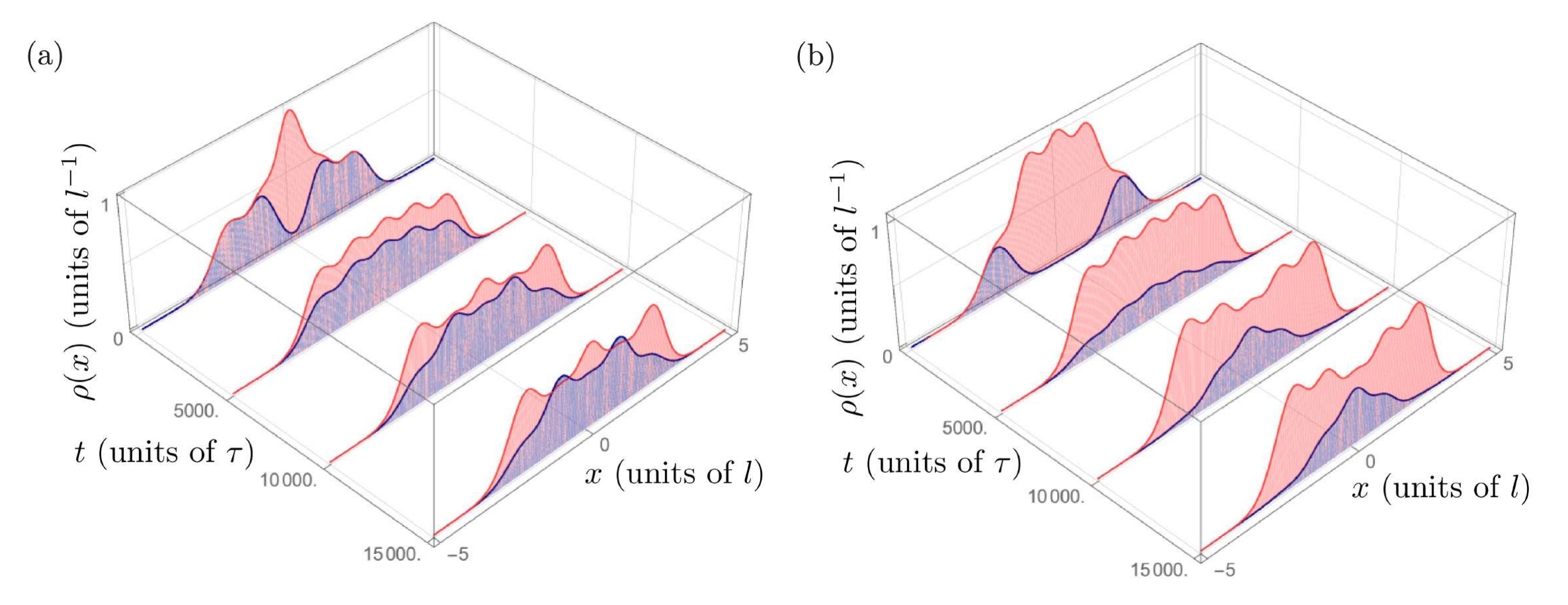}\hfill
\caption{Dynamical evolution of the spin densities in a two-component harmonically trapped bosonic system under an adiabatic modification of the intraspecies interaction parameter $\kappa$. The particle number in each case is (a) $N_\uparrow=3, N_\downarrow=2$ and (b) $N_\uparrow=4, N_\downarrow=1$. Figure adapted from \citep{Barfknecht2017Magnetic}, where the parameters $l$ and $\tau$ were used as harmonic-oscillator units of length and time, respectively.}
\label{fig:magneticbosons}
\end{figure}

The exchange coefficients of the Heisenberg model in Eq.~(\ref{eq:mapping_spin_chain}) depend on the shape of the external trap, which provides an opportunity to engineer systems with desired dynamical properties. One aim here can be a design of an inhomogeneous spin chain that allows for perfect short-distance quantum communication, which is one of the key ingredients for achieving scalable quantum information processing in lattice-based schemes.
The idea to use spin chains for such state transfer was introduced in~\citep{Bose2003communication} and later it was shown that inhomogeneity might be used to control the quality of quantum communication~\citep{Kay2010Transfer}.  
The mapping of Eq.~(\ref{eq:mapping_spin_chain}) provides a simple way for constructing inhomogeneous spin chains that allow for perfect state transfer as discussed in~\citep{Volosniev2015Dynamics,Loft2016Transfer,Barfknecht2018Imbalance,Barfknecht2018Junction,Iversen2020Transfer}.
One possible technical advantage of those approaches in comparison to previous studies is the use of a global potential, which bypasses the need to address exchange coefficients locally. Further modifications of the trap may even allow for engineering systems that have conditional state transfer properties for use in coherent quantum spin transistors~\citep{Marchukov2016Transistor}.

Depending on the parameter regime strongly interacting systems can exhibit ferromagnetic or antiferromagnetic correlations \citep{Volosniev2014StrongInteractions,Deuretzbacher2014Mapping,Massignan2015Magnetism,Cui2019magnetism}. The transition between different regimes can be achieved by changing the parameter $\kappa$, as illustrated in Fig.~\ref{fig:magneticbosons}. The figure shows the spin densities of a two-component harmonically trapped bosonic system, where the change in magnetic correlations is realized by an adiabatic modification of the intraspecies interaction parameter. At $t=0$, the system with $\kappa=0.1$ is initialized in the ground state. The parameter $\kappa$ is then adiabatically increased, and at around $\kappa\simeq 1$, which corresponds to $t=5000/\omega$, the system changes its structure from ``ferromagnetic'' to ``antiferromagnetic''.  The existence of this transition motivates one to study spin dynamics within and across different phases. An interesting question in this respect is the relaxation dynamics starting from a spin-separated state. For small system sizes, the effective Hamiltonian allows one to investigate long-time dynamics that may feature traces of universality and thermalization~\citep{Pecci2021} (see also~\citep{Lebek2022Dynamics} where the corresponding dynamics of strongly attractive 1D quantum gases is studied without employing the effective Hamiltonian). Besides the time dynamics driven by a not-eigen initial state, one can study time evolution of strongly interacting systems with time-dependent trap potentials~\citep{Volosniev2016Heisenberg,Barfknecht2019Spindensity}, time-dependent interparticle interactions~\citep{Barfknecht2017Magnetic}, and time-dependent spin-flip operations~\citep{Barfknecht2019}.

%% file: Section06.tex
\section{Gases with impurities}\label{sec:polaron}

The spinor systems from Section~\ref{sec:spinor} in the limit of high polarization ($N_A\gg N_B$) make a self-sufficient topic, important in its own right. The component with $N_A$ atoms is often called an environment, and $N_B$ atoms are referred to as impurities. Realization of such systems with cold atoms sheds new light onto long-standing problems in condensed matter physics, and motivates new investigations disconnected from balanced systems. 
In particular, {\it ``quantum systems with impurities''} 
serve as a testbed for studying static and dynamic properties of quasiparticles (e.g., polarons\footnote{The polaron concept first introduced to study the motion of an electron in dielectric lattices (see~\citep{Frohlich1954} and references therein) is now central in the physics of degenerate gases. If the host medium is fermionic (bosonic), then the problem is usually called the Fermi (Bose) polaron.}) in the regimes (and with the accuracy) that are beyond standard solid-state set-ups, see, e.g., research papers~\citep{Zwierlein2009FermiPolaron,Grimm2012FermiPolaron,spethmann2012,Cetina2016,Jin2016,Arlt2016,Skou2021Polaron}, and reviews~\citep{Chevy_2010review,Massignan2014Review,Schmidt2018Review,Scazza2022Review}.

The concept of a quasiparticle reduces a many-body system with an impurity to a one-body model with effective parameters. 
Unlike the situation in 3D, the residue of a quasiparticle\footnote{The residue is an overlap between a non-interacting system (with $g_{AB}=0$) and an interacting one.} can vanish in the thermodynamic limit in 1D for a mobile impurity. For homogeneous Fermi gases, this is a manifestation of the Anderson orthogonality catastrophe that can be studied accurately using the Bethe ansatz~\citep{Castella1993Exact} or numerically~\citep{Castella1996Orthogonality}. A similar behavior occurs in trapped systems. For example, for a strongly interacting system in a trap\footnote{For a weakly interacting trapped Bose gas, the impurity can be pushed to the edge of the trap if the impurity-boson interaction is sufficiently strong (see the discussion in Section~\ref{subsec:single_impurity}) which complicates the comparison to the classic Anderson orthogonality catastrophe.}, the solution of~\citep{Levinsen2015Ansatz} demonstrates that the overlap of a system with and without the impurity scales as $\sim N_A^{-1/2}$. Although, this scaling is derived for large systems, one may expect to see a manifestation of this result already with a handful of fermionized (strongly interacting, cf.~Sec.~\ref{subsec:TGmodel}) bosons~\citep{Campbell2014Orthogonality}. 
For weakly-interacting Bose gases in the thermodynamic limit, infrared divergences are also common. For example, they appear in perturbative calculations~\citep{Pastukhov2017Impurity} and mean-field calculations (in momentum space)~\citep{Grusdt2017BosePolaron}\footnote{Note that beyond mean-field calculations can yield polaron energies free of this divergence~\citep{Kain2018Static}.} of the energy spectrum. This anomalous behavior affects the applicability of the quasiparticle concept, in particular, it might imply that correlations spread much slower than what is expected in the quasiparticle picture~\citep{Kantian2014Competing}. 
In spite of this, one-body effective models might be useful in 1D (at least for certain observables). This statement will be illustrated in this section with a few simple cases.

The focus of the present section is on 1D systems with a finite number of particles. The open question touched upon here is defined as: ``When and how can one use an effective one-body description to study time evolution of such systems?''.
From the point of view of this review, this question is connected to an exploration of the crossover from few- to many-body physics. It is addressed here by considering quench dynamics of impurities in a weakly-interacting harmonically-trapped Bose gas. 
For the sake of clarity, the Hamiltonian of Eq.~(\ref{eq:Hamiltonian_pseudospinor}) can be conveniently re-written in the form ($m_A=m_B=1$, $\hbar=1$, and $\omega=1$)
\begin{equation}
H=H_{A}+\sum_{n=1}^{N_B}\left[-\frac{1}{2}\frac{\partial^2}{\partial y_n^2}+\frac{y_n^2}{2}+g_{AB}\sum_{m=1}^{N_B}\delta(y_n-x_m)\right],
\label{eq:ham_imp}
\end{equation} 
where $H_{A}$ describes the bath of $N_A$ majority particles (cf.~Eq.~(\ref{eq:Hamiltonian_pseudospinor})).
For simplicity, it is assumed that there are no fundamental impurity-impurity interactions, i.e., $g_{BB}=0$; the system is somewhere in between few- and many-body worlds ($N_A\lesssim 100$); the bath (Bose gas) is weakly-interacting, so that without impurities it can be described using the mean-field approximation. 
The standard way to analyze the Hamiltonian in Eq.~(\ref{eq:ham_imp}) is to consider first the case with $N_B=1$ and then, for cases with $N_B\geq 2$, study impurity-impurity interactions mediated by the environment. This approach will  be utilized below.

\begin{figure}[tb]
  \centering
    \includegraphics[width=\linewidth]{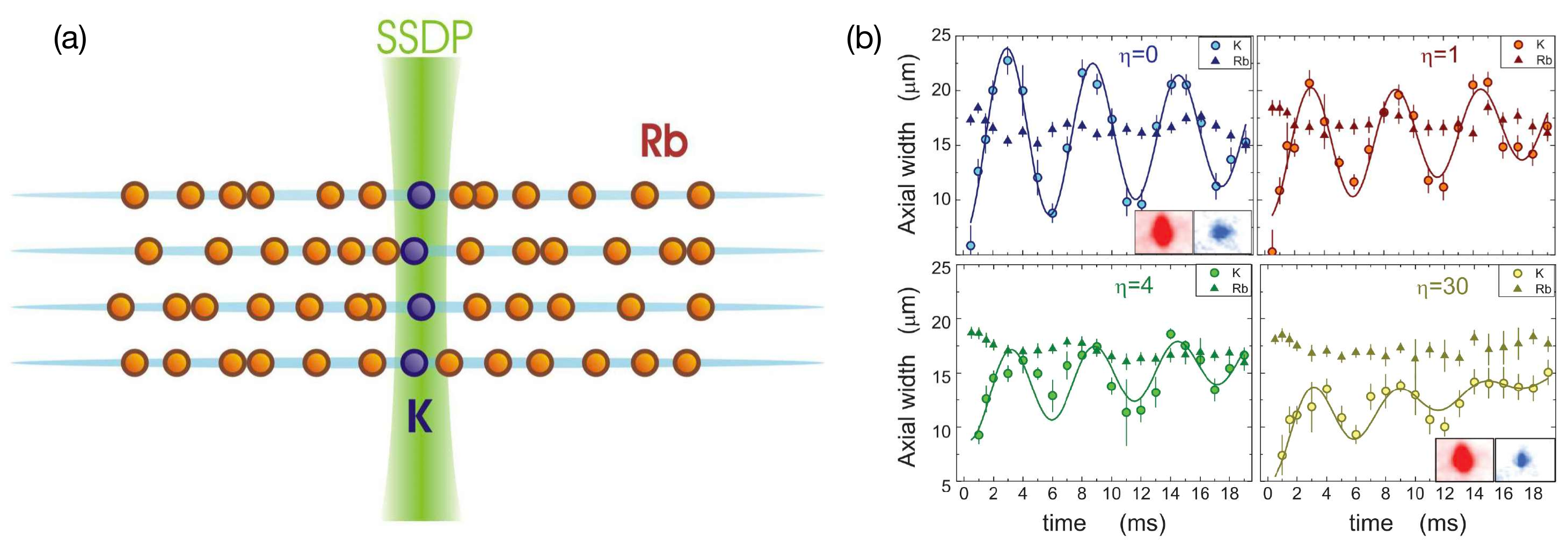}\hfill
    \caption{(a) Ultracold Rb atoms and K impurities are loaded into an array of 1D systems and a species-selective dipole potential (SSDP) light blade spatially localizes the impurities into the center of the tubes. (b) Oscillations of the K impurities axial width after switching off the SSDP for different interaction strengths with the surrounding Rb bulk, $\eta$. Triangles indicate the axial size of the Rb cloud.  Figure adapted from \citep{Catani2012Experiment}.}
    \label{fig:Catani_experiment}
\end{figure}

It is worth noting that theoretical studies of the Hamiltonian~(\ref{eq:ham_imp}) are motivated (in part) by cold-atom experiments with highly polarized mixtures~\citep{Catani2012Experiment,Fukuhara2013SpinImpurity, Fukuhara2013BoundStates, Meinert:17}. The experiment described in~\citep{Catani2012Experiment} is (arguably) the one to have in mind while reading this section. Figure~\ref{fig:Catani_experiment} illustrates this experiment. First (at $t=0$, $t$ for time), a cloud of bosonic rubidium atoms is prepared together with a few potassium atoms. The impurities are initially tightly confined by a `light blade'. Then, the dynamics is initiated by extinguishing this light blade, and the size of the impurity cloud is measured. In \citep{Catani2012Experiment}, the resulting data were analyzed using a one-body effective model, see~\citep{Grusdt2017BosePolaron,Ghazaryan2022} for a follow-up analysis of the data.

\subsection{Single impurity}
\label{subsec:single_impurity}

The simplest one-body model, which provides a stepping stone to more complicated constructions, can be derived within the mean-field approximation. Assuming that the impurity does not affect the bosonic cloud,  the wave function of the system is written as a product state $\phi_A\phi_B$, where $\phi_B$ describes the impurity atom, and $\phi_A$ describes the bosons. The effective Hamiltonian for the motion of the impurity then follows from Eq.~(\ref{eq:ham_imp})
\begin{equation}
    h_{\mathrm{MF}}=-\frac{1}{2}\frac{\partial^2}{\partial y^2}+\frac{y^2}{2}+g_{AB}n_A(y),
\end{equation}
where the subscript of $y$ has been omitted, for simplicity. The density of bosons $n_A$ can be estimated using the Thomas-Fermi approximation (see, e.g.,~\citep{pethick2002}) to be $n_A(|x|<R)=R^2/(2g_{AA})(1-x^2/R^2)$, where $R=(3g_{AA}N_A/2)^{1/3}$. The subscript $\mathrm{MF}$ indicates that the effective model is derived in the mean-field approximation. Up to an irrelevant energy shift, $h_{\mathrm{MF}}$ reads for $|y|<R$   
\begin{equation}
    h_{\mathrm{MF}}=-\frac{1}{2}\frac{\partial^2}{\partial y^2}+\left(1-\frac{g_{AB}}{g_{AA}}\right)\frac{y^2}{2}.
    \label{eq:h_MF}
\end{equation}
The effect of the bath is taken here into account by renormalizing the frequency of the external trap. The reader is referred to~\citep{Volosniev2015RealTime,Schecter2016Depletons} for complementary derivations of $h_{\mathrm{MF}}$. Equation~(\ref{eq:h_MF}) is simple, still, it allows one to make a number of reasonable predictions for the behavior of the impurity. For the ground state, the impurity is pushed to the edge of the trap as the interaction increases\footnote{In contrast,  an impurity in a Fermi gas resides in the center of the trap, see, e.g.,~\citep{Lindgren2014Fermionization, Levinsen2015Ansatz}.}, minimizing the interaction energy and leading to a smaller depletion of the condensate. This prediction of the mean-field approximation is in accord with more elaborate theoretical models of~\citep{Dehkharghani2015Impurity,Dehkharghani2015Bosons,GarciaMarch2016Entanglement}. Below, it will be shown that $h_{\mathrm{MF}}$ leads to qualitatively correct results also for the corresponding quench dynamics. 

A step beyond $h_{\mathrm{MF}}$ is the effective Hamiltonian
\begin{equation}
    h_{\mathrm{eff}}=-\frac{1}{2m_{\mathrm{eff}}}\frac{\partial^2}{\partial y^2}+\frac{k_{\mathrm{eff}}y^2}{2},
    \label{eq:h_eff}
\end{equation}
which includes a renormalized effective mass, $m_{\mathrm{eff}}$, together with a renormalized spring constant, $k_{\mathrm{eff}}$. For weak interactions, one can expect that $m_{\mathrm{eff}}\simeq 1$, and $k_{\mathrm{eff}}=1-g_{AB}/g_{AA}$ in agreement with mean-field calculations. The beyond mean-field corrections to $m_{\mathrm{eff}}$ and $k_{\mathrm{eff}}$ can be estimated using the local density approximation and knowledge of the polaron properties in homogeneous media acquired, e.g., from~\citep{Grusdt2017BosePolaron,Parisi2017BosePolaron,Volosniev2017BosePolaron,Panochko2019MeanField,Jager2020Deformation}. To address a more general question of the applicability of Eq.~(\ref{eq:h_eff}), the effective model should be benchmarked against ab-initio numerical data. 

\begin{figure}[tb]
  \centering
    \includegraphics[width=\linewidth]{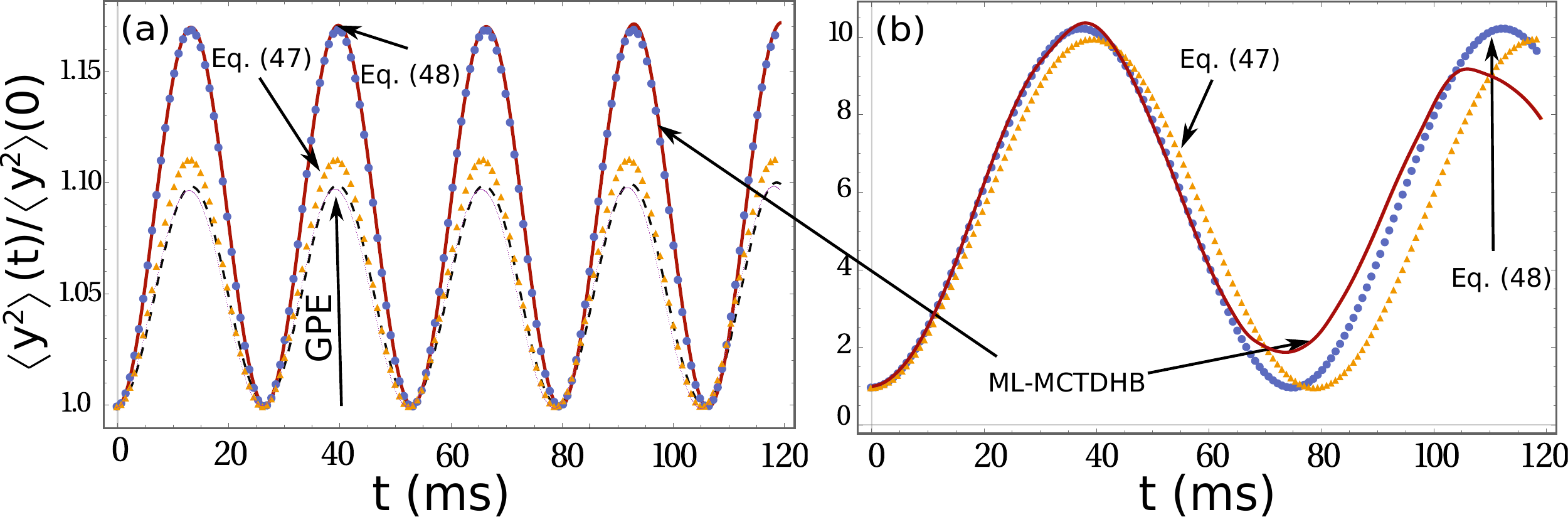}\hfill
    \caption{The size of the impurity cloud $\langle y^2 \rangle (t)/\langle y^2 \rangle (0)$ as a function of time. The (red) solid curves show the ML-MCTDHB results. The blue dots illustrate the fits with
    the effective Hamiltonian~\ref{eq:h_eff}. The yellow triangles are the results of the Hamiltonian~\ref{eq:h_MF} (no fit parameters). The (black) dashed curve (in (a)) shows the result obtained using the two coupled Gross-Pitaevskii equations (GPE). Panel (a) is for $g_{AB} = 0.1g_{AA}$, the parameters fitted to the ML-MCTDHB results are $m_{\mathrm{eff}}/m = 0.983$, $k_{\mathrm{eff}} /k = 0.87$. Panel (b) is for $g_{AB} = 0.9g_{AA}$, the fit parameters are $m_{\mathrm{eff}}/m = 0.937$ and $k_{\mathrm{eff}} /k = 0.104$. The other parameters are $N_A = 100$, $\omega= 2\pi \times 20$Hz, $g_{AA} = 10^{-37}$Jm and $m_A=m_B = m(^{87}\mathrm{Rb})$. Figure is adapted from~\citep{Mistakidis2019Effective}.}
    \label{fig:Figure_impurity}
\end{figure}

Figure~\ref{fig:Figure_impurity} illustrates such data obtained with the ML-MCTDHB method (see Section~\ref{subsec:ML-MCTDHA}) for the dynamics following a sudden change of $g_{AB}$ at $t=0$. It is assumed that the system was in the ground state with $g_{AB}=0$ at $t<0$. The figure demonstrates time evolution of the width of the impurity cloud $\langle y^2 \rangle$ for $t>0$. One can see that for small values of the final $g_{AB}/g_{AA}$, the effective description of $h_{\mathrm{eff}}$ from Eq.~(\ref{eq:h_eff}) is nearly perfect, see  panel~a). 
This panel also shows $\langle y^2 \rangle$ calculated using the coupled Gross-Pitaevski equations. The conclusion here is that beyond-mean-field effects must be taken into account even for weak interactions\footnote{For strong interactions, beyond-mean-field effects dominate; results obtained with the coupled Gross-Pitaevskii equations are not accurate, and we do not show them in Fig.~\ref{fig:Figure_impurity}b).}. 
For stronger interactions, the agreement between ML-MCTDHB and the effective one-body model is good at short time scales, however, $h_{\mathrm{eff}}$ becomes less adequate at longer evolution times, see  panel~b). This is expected. For strong interactions, the impurity can move outside of the Bose gas, and the effective description must fail. This regime is characterized by large energy exchange~\citep{Peotta2013Breathing, Kronke2015Dynamics}, which is not captured by the simple effective Hamiltonian presented here.

The Hamiltonian $h_{\mathrm{eff}}$ can also describe the dynamics of impurities in the bulk of a Bose gas for other quench protocols, see, e.g.,~\citep{Mistakidis2019QuenchBosePolarons,Mistakidis2019DissipationBosePolarons}. Those studies further support the use of effective one-body models for the description of the dynamics, at least for certain observables. They also establish a connection between an impurity in a finite trapped Bose gas and the traditional polaron picture valid for a homogeneous infinite system.   
For example, the parameter
$m_{\mathrm{eff}}$ that enters $h_{\mathrm{eff}}$ was connected to the effective mass of the traditional polaron in a weakly interacting case~\citep{Mistakidis2019DissipationBosePolarons}. This was done by comparing to the results of the Fr{\"o}hlich Hamiltonian (for the mapping onto the Fr{\"o}hlich Hamiltonian, see~\citep{Casteels2012Impurity}). In spite of this success, further work is needed to establish the limits of applicability of the effective model in Eq.~(\ref{eq:h_eff}). In particular, it is important to consider other observables. 

In this section, the impurity problem is considered solely from the point of view of the impurity, reducing the role of the environment to the renormalization of the mass and the spring constant.  Naturally, the environment should be considered in more detail if one aims to understand the limits of applicability of quasiparticle models and learn about the energy exchange between the impurity and the bath, see, e.g.,~\citep{Lampo2017BosePolarons, Ghazaryan2022} for a relevant discussion based upon quantum Brownian motion. Such studies are also needed for  possible applications of impurities in thermometry ~\citep{Mehboudi2019polarons,Khan2022thermometry}.

\subsection{Two impurities}

Mobile impurities attract each other when immersed in a Bose gas even if $g_{BB}=0$, because the local deformation of the medium caused by the impurities is minimized when the impurities are close to each other~\citep{Recati2005Casimir,Bruderer2007Polaron}\footnote{Note that if the impurities are not identical, then it is possible that the induced impurity-impurity interaction is repulsive~\citep{Schecter2014Casimir,Reichert2019SemiInfinite,Brauneis2021FlowEquations}. This situation is beyond the scope of this review.}. This attraction is dubbed impurity-impurity interaction induced (or mediated) by the environment. It can be strong enough to be observed in state-of-the-art cold-atom experiments~\citep{Fukuhara2013BoundStates,DeSalvo2019Mediated,Edri2020Mediated}.

\begin{figure}[tb]
    \centering
    \includegraphics[width=\linewidth]{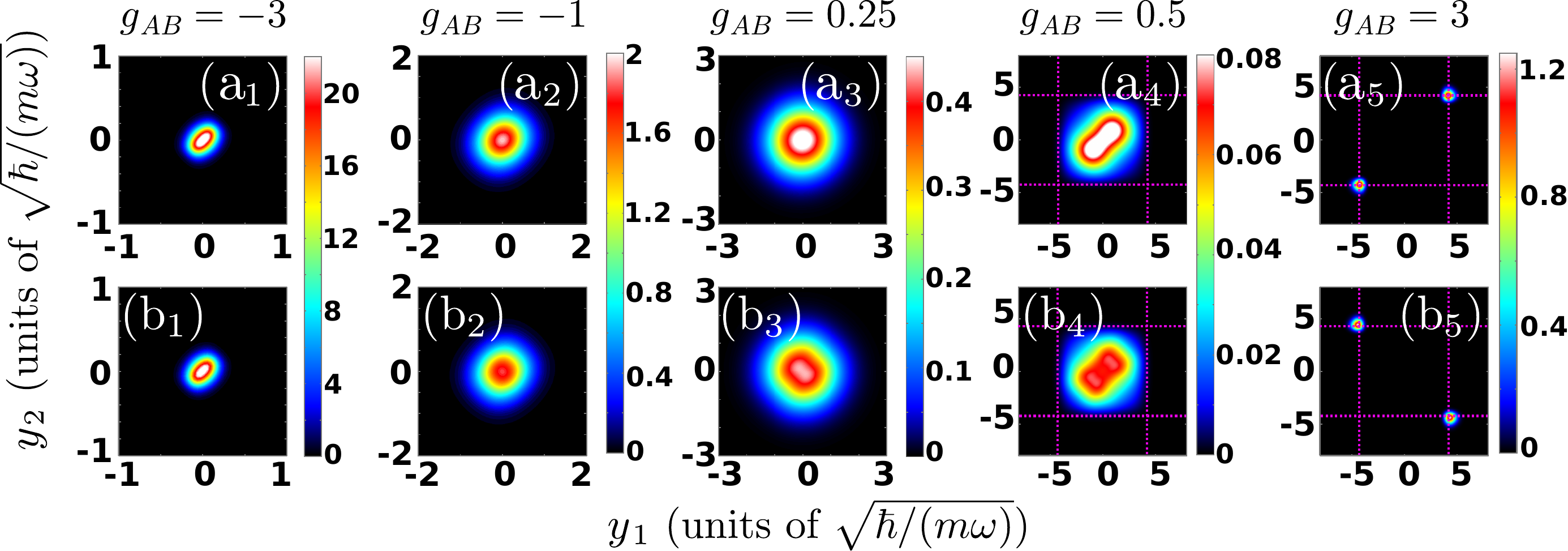}\hfill
    \caption{Two-body reduced density matrix for two bosonic impurities immersed in a harmonically trapped Bose gas. The system is in the ground state. (a$_1$)-(a$_5$) are for non-interacting ($g_{BB}=0$) and (b$_1$)-(b$_5$) are for interacting ($g_{BB} = 0.2$) impurities.  The panels showcase the density matrix for different impurity-boson couplings (see the legend). The parameters of the gas are $N_A = 100$ and $g_{AA}=0.5$. 
    Harmonic oscillator units are used. The dashed (magenta) lines mark the Thomas-Fermi radius of the gas. Figure is adapted from~\citep{Mistakidis2020InducedBose2}.}
    \label{fig:Figure_induced_ground}
\end{figure}

Induced interactions can lead to in-medium bound states -- `Bose bipolarons' -- whose investigation is motivated in part by their analogy\footnote{To put the analogy on solid ground, one will need to estimate the importance of the long-range Coulomb repulsion, which strongly modifies properties of weakly-bound states, in particular, their universal behavior~\citep{Schmickler2019Coulomb,Luna2019Papenbrock}. In any case, Bose bipolarons provide an excellent toy model for studying the origin of short-range attraction mediated by the environment and its relation to electronic correlations in standard condensed matter set-ups. A particularly exciting direction is establishing a connection to high-temperature superconductors.} to the electronic bipolarons, see, e.g.,~\citep{Devreese2009Bipolarons}. It is worth noting that even though the presence of the environment modifies few-impurity physics in all spatial dimensions~\citep{zinner2014FBS,Naidon2018Yukawa,Camacho2018Landau,Camacho2018Bipolarons}, one expects most prominent effects in 1D where correlations are in general stronger. In particular, any attractive interaction leads to a bound state in the 1D world~\citep{LandauQM}. 

The existence of the in-medium bound states can be seen in the two-body reduced density matrix for the two impurities as illustrated in Fig.~\ref{fig:Figure_induced_ground}, see also~\citep{Dehkharghani2018Coalescence}. For any non-vanishing value of $g_{AB}$, two impurities without bare interactions ($g_{BB}=0$) correlate as if they were interacting attractively. This can be inferred from the shape of their two-body density matrix, which is elongated along its anti-diagonal, see Figs.~\ref{fig:Figure_induced_ground}~(a1)-(a5). 
Note that the impurities are repelled from the center of the trap if impurity-medium interactions are strong and repulsive (cf. Eq.~(\ref{eq:h_MF}) for $g_{AB}>g_{AA}$). In this case, the induced impurity-impurity interaction leads to two clusters made of two impurities outside the Bose gas, see Fig.~\ref{fig:Figure_induced_ground}~(a5). Induced interactions can compete with the bare interactions as demonstrated in Figs.~\ref{fig:Figure_induced_ground}~(b1)-(b5). For example, for small values of $|g_{AB}|$, the induced interaction is weak and the impurity-impurity correlations are determined by the value of $g_{BB}$. This is also the case when the impurities are pushed out of the Bose gas, at least for the considered parameters, see Fig.~\ref{fig:Figure_induced_ground}~(b5). However, the induced and bare interaction are of equal importance if $g_{AB}\sim 0.5$, see Fig.~\ref{fig:Figure_induced_ground}~(b4).

An effective model for two impurities in a Bose gas can be constructed as a straightforward extension of Eq.~(\ref{eq:h_eff}) 
\begin{equation}
    h^{(2)}_{\mathrm{eff}}=h_{\mathrm{eff}}(y_1)+h_{\mathrm{eff}}(y_2)+g_{BB}^{I}\delta(y_1-y_2),
    \label{eq:h_eff_2}
\end{equation}
where $g_{BB}^{I}$ is the parameter that characterizes the strength of the induced interactions, and $y_1$ and $y_2$ are the positions of impurity atoms. 
The interaction is modelled here by a zero-range induced potential, which is equivalent to a low-energy approximation. Since the short-range part of the induced potential is attractive for two identical impurities~\citep{Recati2005Casimir,Bruderer2007Polaron,Schecter2014Casimir,Reichert2019Casimir,Dehkharghani2018Coalescence}, one expects that $g_{BB}^{I}<0$.

The simplest approach to estimate $g_{BB}^{I}$ is based on the Born-Oppenheimer approximation in which the impurities are immobile. The induced potential is determined from the ground-state energy of the Hamiltonian
\begin{equation}
H(y_1,y_2)=H_{A}+g_{AB}\sum_{m=1}^{N_A}\left(\delta(y_1-x_m)-\delta(y_2-x_m)\right),
\end{equation}
where, as before, $H_{A}$ is the Hamiltonian for bosons. For the weakly-interacting Bose gas,
the ground state of $H(y_1,y_2)$ is captured by the mean-field approximation as can be shown by comparing to ab-initio calculations~\citep{Brauneis2021FlowEquations,Will2021Bipolarons}.

It is straightforward to derive the ground-state energy, $E(y_1,y_2)$, of $H(y_1,y_2)$ for weak interactions, i.e., small values of $g_{AB}$. 
Indeed, if there is a known solution to a system with a single impurity placed at $y_1$
\begin{equation}
\left(H_{\mathrm{A}}+g_{AB}\sum_{m=1}^{N_A}\delta(y_1-x_m)\right)\phi_{y_1}=\epsilon \phi_{y_1}, 
\end{equation}
then $E(y_1,y_2)$ follows from first-order perturbation theory
\begin{equation}
    E(y_1,y_2)=\epsilon + g_{AB} n_{y_1}(y_2),
    \label{eq:Ex1x2}
\end{equation}
where $n_{y_1}(y_2)=N_{A}|\phi_{y_1}(y_2)|^2$ is the density of bosons at the position of the second impurity. 
The density $n_{y_1}(y_2)$ can be approximated by the mean-field result (see, e.g.,~\citep{Volosniev2017BosePolaron})
\begin{equation}
    n_{y_1}(y_2)=\rho\mathrm{tanh}^2(\sqrt{\gamma}\rho |y_2-y_1|+d), \quad d=\frac{1}{2}\mathrm{asinh}\left(\frac{2\rho\sqrt{\gamma}}{g_{AB}}\right),
    \label{eq:density_mean_field}
\end{equation}
where $\rho$ is the density of the Bose gas with $g_{AB}=0$, and $\gamma=g_{AA}/\rho$ is the Lieb-Liniger parameter. For the sake of discussion, we assume that $g_{AB}>0$.
The impurity-impurity interaction in the Born-Oppenheimer approximation for $g_{AB}\to0$ follows now
from Eqs.~(\ref{eq:Ex1x2}) and~(\ref{eq:density_mean_field})
\begin{equation}
    V_{\mathrm{eff}}(|y_1-y_2|)=g_{AB}\rho\left(\mathrm{tanh}^2(\sqrt{\gamma}\rho |y_2-y_1|+d)-1\right), 
    \label{eq:Veff_exp}
\end{equation}
which is always attractive. In the limit $|y_1-y_2|\to\infty$, it decays as 
\begin{equation}
\label{eq:Veff_BO}
    V_{\mathrm{eff}}(|y_1-y_2|)\simeq -\frac{g_{AB}^2}{\sqrt{\gamma}} e^{-2\sqrt{\gamma}\rho|y_2-y_1|},
\end{equation}
in agreement with calculations based upon the Bogoliobov approximation~\citep{Recati2005Casimir}. 
The value of $g_{BB}^I$ is computed from the net volume of the potential $V_{\mathrm{eff}}(|y_1-y_2|)$, which leads to
\begin{equation}
    g_{BB}^I=-\frac{g_{AB}^2}{g_{AA}}.
\end{equation}
Within first order perturbation theory, this expression can be obtained also without the Born-Oppenheimer approximation~\citep{Mistakidis2020InducedBose1}. 

One comment is in order here. According to Eq.~(\ref{eq:Veff_BO}), the induced impurity-impurity interactions decay exponentially. This is a mean-field result, which is modified for large values of $|y_1-y_2|$ due to quantum fluctuations. In particular, for finite values of $g_{AB}$ and $g_{AA}$\footnote{The potential changes if the impurities are impenetrable as discussed in~\citep{Reichert2019SemiInfinite}. If $1/g_{AA}=0$, the system fermionizes, which also leads to the change of the long-range physics, see, e.g.,~\citep{Recati2005Casimir}.}, the tail of the potential is $\sim 1/|y_1-y_2|^3$~\citep{Schecter2014Casimir,Reichert2019Casimir}.  This long-range behavior is not relevant here, since its effect is too weak for the considered systems, see, e.g., the discussion in~\citep{Mistakidis2020InducedBose1,Will2021Bipolarons, petkovic2021}, which show that the mean-field calculation of the induced interaction is adequate for trapped Bose gases with $N_A\simeq 100$.

\begin{figure}[tb]
  \centering
    \includegraphics[width=\linewidth]{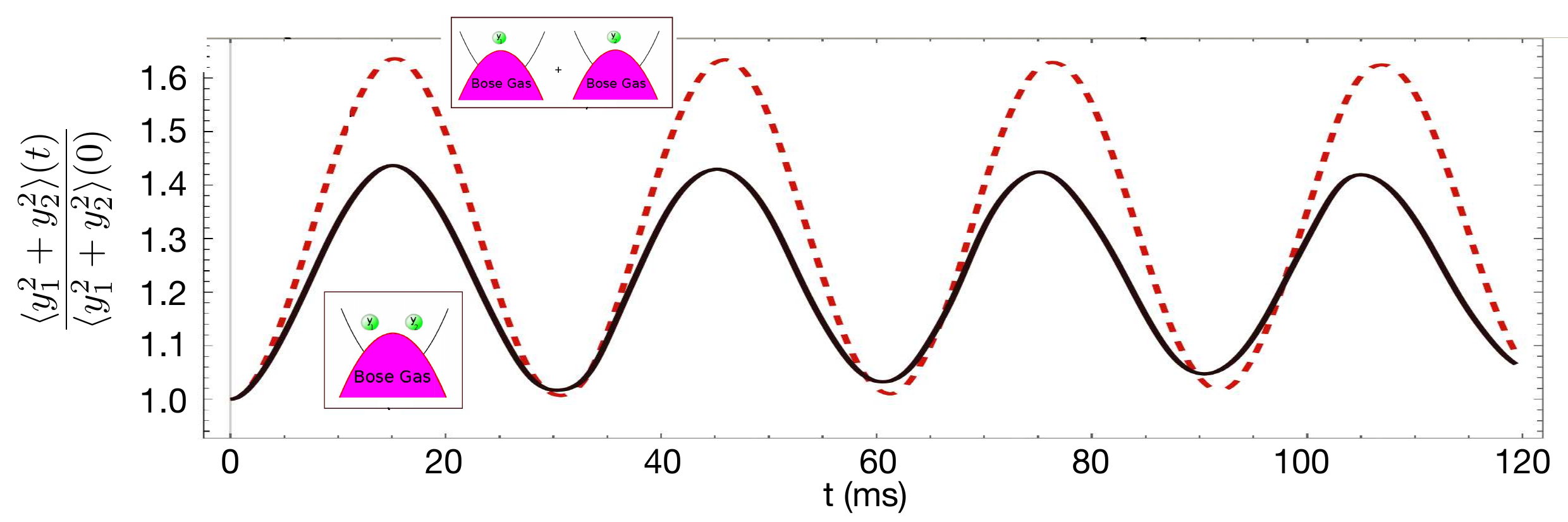}\hfill
    \caption{Time evolution of $\langle y^2_1 +y^2_2\rangle(t)/\langle y^2_1 + y^2_2\rangle(0)$ after a rapid change of the boson-impurity interaction strength, $g_{AB}$. The data are calculated using the ML-MCTDHB method. The (red) dashed curve describes the situation where the impurities are in different traps, which corresponds to the single impurity case, cf. Fig.~\ref{fig:Figure_impurity}. The (black) solid curve presents the dynamics of two impurities in the same trap. There is no free-space impurity-impurity interaction, i.e., $g_{BB}=0$. Therefore, the difference between the two curves can be interpreted as a manifestation of attractive  impurity-impurity interactions induced by the environment. The interaction strength at $t>0$ is $g_{AB}/g_{AA} = 0.3$, all other parameters are as in Fig.~\ref{fig:Figure_impurity}. Figure adapted from~\citep{Mistakidis2020InducedBose1}.}
    \label{fig:Figure_two_impurities}
\end{figure}

Having determined $g_{BB}^I$, one can compare the dynamics governed by $h^{(2)}_{\mathrm{eff}}$ to numerical simulations of the Bose gas with two impurities. This was done using the  ML-MCTDHB method in \citep{Mistakidis2020InducedBose1} for quench dynamics that follows a sudden change of $g_{AB}$ from zero to some finite value. First, that work demonstrated that the induced impurity-impurity interaction can be observed in the dynamics of the impurities, see Fig.~\ref{fig:Figure_two_impurities}. Second, it was found that the effective Hamiltonian $h^{(2)}_{\mathrm{eff}}$ captures the dynamics for weak couplings well. For stronger couplings, however, the analysis showed that the effective description based upon $h^{(2)}_{\mathrm{eff}}$ fails. It is worth noting that the effective description for two impurities fails for the parameters for which the effective model for a single impurity still works.

All in all, one can conclude that the potential $g_{AA}^{I}\delta(x_1-x_2)$ is sufficient to describe weak induced interactions. However, it should be modified for stronger interactions. 
In particular, one expects that at stronger interactions an external trap may play an important role, as it will induce interactions that depend not only on the distance between the impurities, but also on the position of the impurities~\citep{Dehkharghani2018Coalescence,Chen2018Entanglement}. 

This section presented observables for which effective models can describe the dynamics accurately in the weakly interacting regime. However, there are observables for which such models do not suffice. For example, the effective models fail to predict the entanglement between two impurities induced by the environment~\citep{Mistakidis2020InducedBose1}, implying that further work is needed to establish the limits of applicability of $h^{(2)}_{\mathrm{eff}}$.

\subsection{Further directions}
  
{\it More than two impurities}. To account for $N_B$ impurities in a Bose gas, it is natural to use the effective Hamiltonian 
  \begin{equation}
    h^{(N_B)}_{\mathrm{eff}}=\sum_{i=1}^{N_B} h_{\mathrm{eff}}(y_i)+g_{BB}^{I}\sum_{i>j}\delta(x_i-x_j),
    \label{eq:h_eff_N}
\end{equation}
which assumes that many-body (three-body and beyond) induced interactions are weak. This is a typical approximation for dilute gases. The effective Hamiltonian $h^{(N_B)}_{\mathrm{eff}}$ implies that multi-polaron bound states can be formed, which was used in~\citep{Will2021Bipolarons} to explain the ground-state energy of a Bose gas with five impurities computed using the Quantum Monte Carlo method. Comparing general predictions of $h^{(N_B)}_{\mathrm{eff}}$  to exact numerical data is still outstanding. Furthermore, three-body forces must be explicitly calculated at least for static impurities in order to understand when two-body induced interactions are dominant.

{\it Non-vanishing bare interactions and finite-range effects.} If a non-negligible interaction between the bare impurities exists, the effective Hamiltonians needs to be modified. The simplest way to do that is to add $g_{BB}$ to $g_{BB}^I$, which is a reasonable approach for weak couplings. Its accuracy was tested by comparing to the Bethe ansatz results for two bosonic impurities in a Fermi gas, see~\citep{Huber2019BoundStates}. 

For fermionic impurities (or when $1/g_{BB}=0$) and weak impurity-boson interactions, one can still use the effective Hamiltonian $h^{(N_B)}_{\mathrm{eff}}$. In this case, the zero-range interaction is not felt by fermions, and can be removed from the Hamiltonian. Uncorrelated dynamics of fermionic impurities was indeed observed for weak interactions in ML-MCTDHB simulations~\citep{Mistakidis2019FermionsinBosons}. For strong boson-impurity interactions, one needs to include finite-range effects of the induced impurity-impurity interactions. They may lead to impurity-impurity correlations in homogeneous systems~\citep{Pasek2019Induced}. Note that in the trap the impurities are pushed outside a weakly interacting Bose gas before those correlations become strong~\citep{Dehkharghani2018Coalescence,Mistakidis2019FermionsinBosons}.

 Finite-range effects can also lead to a situation where  a three-body bound state exists, but no two-body bound states, i.e., to the so-called Borromean binding, see~\citep{jensen2004,BraatenHammer2006} and references therein. Note that the Borromean systems were observed in low spatial dimensions only for two-particle interactions with a repulsive barrier~\citep{Nielsen19992D,Volosniev2013Borromean}. Surprisingly, \citep{Will2021Bipolarons} suggests a barrier in the interaction between mobile polarons, which may allow for Borromean binding of polarons.  
 
Further work is also needed to understand finite-range effects associated with the bare interactions. These are particularly important in systems with dipolar interaction~\citep{Ardila2018DipolarPolaron,Kain2014Polarons} and ionic impurities~\citep{Astrakharchik2021Ion}.

{\it Tonks-Girardeau gas.} For strong boson-boson interactions, one deals with the Tonks-Girardeau gas discussed in Section~\ref{subsec:TGmodel}. The Tonks-Girardeau gas can be mapped onto the Fermi gas, which is outside the scope of the present section. However, for the sake of completeness, we briefly discuss this important limiting case here focusing on the time dynamics. 

It is possible to investigate time dynamics of small harmonically trapped Fermi gases ($N_A\simeq 5$) with impurities using {\it ab initio} methods. For example, it is feasible to calculate the dynamical evolution of spectroscopic observables and densities after a quench~\citep{Mistakidis2019FermiPolarons,Kwasniok2020}.  Note that for fermionic (as well as bosonic) impurities, the energy shift due to the presence of the additional impurity is small and hard to observe in the spectrum~\citep{Mistakidis2019FermiPolarons,Mistakidis2020InducedBose2}. 
Instead, \citep{Mistakidis2019FermiPolarons,Kwasniok2020} propose to use the relative distance between impurities as a measure of impurity-impurity interactions mediated by the environment.

{\it Other trapping potentials.} It is worth noting that induced interactions in harmonically trapped Bose gases do not appear substantially stronger than interactions mediated by Fermi gases. This happens in particular because the impurities are pushed to the edge of the trap before the induced interaction can become large. However, the situation might be different for homogeneous systems. For example, two fermionic impurities do not form a `bipolaron' in a fermionic medium~\citep{Flicker1967TwoFermions}, but they do so in a bosonic environment~\citep{Pasek2019Induced}. This motivates further studies of few-body systems in other trapping potentials (in particular in a ring and a box), see for example the recent investigation by~\citep{Theel2021} that explores the role of the confining potential in the collisional dynamics of two correlated impurities.

%% file: Section05.tex
\section{Spin-1 bosonic gases}\label{spin_1_gases}

\subsection{Overview and state-of-the-art}

Atomic BECs, in which the involved magnetic sublevels lead to spin degrees of freedom are often called spinor or pseudo-spinor condensates.
They provide ideal candidates for investigating quantum magnetic order in few- and many-body settings~\citep{kawaguchi2012spinor,stamper2013spinor} and
can be described by a $2F+1$ component, spin-$F$,  vectorial order parameter. 
The spin-1 case in this class of systems is the most studied and will be also the focus of our discussion. Various experimental demonstrations found the ground state of $^{23}$Na to be anti-ferromagnetic \citep{stamper1998optical,stenger1998spin} and the ground state of $^{87}$Rb to be ferromagnetic~\citep{chang2005coherent,widera2006precision}. 
Recently, also strongly ferromagnetic $F=1$ spinor condensates have been realized in $^{7}$Li~\citep{huh2020observation,kim2021emission}. 
Importantly, spin relaxation collision processes in the $F=1$ hyperfine manifold 
can give rise to spin-mixing i.e. coherent and reversible population transfer between the distinct hyperfine states, but they can also limit the lifetime of spinor BECs to about a second e.g.~for $^{87}$Rb. 
It is worth mentioning that the condensates with $F=2$ exhibit a more complex phase diagram~\citep{saito2005diagnostics} but are far less explored experimentally~\citep{kronjager2006magnetically,kronjager2005evolution}. 
They provide fruitful platforms for future investigations. 

Remarkably, spin-1 systems host a plethora of different magnetic ground states and phase transitions between them, originating from the competition between interparticle interactions and external fields~\citep{ho1998spinor,ohmi1998bose}, as well as a wealth of nonequilibrium phenomena. 
A short list of the latter includes spin-mixing dynamics~\citep{pu1999spin,law1998quantum}, the formation of metastable domain structures~\citep{miesner1999observation} and textures~\citep{leanhardt2003coreless,guzman2011long}, quantum spin tunneling~\citep{stamper1999quantum}, spinor non-linear excitations such as half quantum vortices~\citep{seo2015half,kim2021probing} and skyrmions~\citep{leslie2009creation,choi2012observation} in higher dimensions and solitons~\citep{lannig2020collisions,bersano2018three,katsimiga2021phase} in 1D as well as entanglement generation for creating metrologically valuable states~\citep{strobel2014fisher,feldmann2018interferometric,pezze2018quantum}. 

These systems are often investigated with the goal to understand the underlying correlation effects~\citep{cui2008quantum,leslie2009amplification,konig2018quantum,saha2020strongly}. 
Notable correlation phenomena include, in particular, alterations of the system ground state magnetic properties~\citep{mittal2020many,hao2017ground,hao2006density,hao2016weakening} and the study of universality in the spin dynamics of spinor ${}^{87}$Rb BECs~ \citep{prufer2018observation,prufer2020experimental}. 
Strikingly, it is nowdays experimentally feasible to extract and monitor the correlation functions in 1D spin-1 Bose gases~\cite{prufer2020experimental}. 
In the following we will describe, in a pedagogical manner, the behaviour of spin-1 Bose gases in 1D and analyze their correlation-induced phenomena.

\subsection{One-dimensional setting and basic definitions}

The many-body Hamiltonian describing an 1D ultracold spin-1 Bose gas with $N$ atoms of mass $M$ experiencing an external confinement $V(x)$ is given by 
\begin{align}
    H=& \int dx \sum_{\alpha,\beta=-1}^{1} \hat{\psi}_{\alpha} ^ {\dagger}(x) \Big[ -\frac{\hbar^2}{2M}\frac{\partial^2}{\partial x^2} \delta_{\alpha \beta} + V(x)\delta_{\alpha \beta} -p(f_z)_{\alpha\beta}+q(f_z^2)_{\alpha\beta}\Big] \hat{\psi}_{\beta}(x)\nonumber\\
    &+ \frac{1}{2}\int dx \Big[ c_0 \sum_{\alpha,\beta=-1}^{1}\hat{\psi}^{\dagger}_\alpha (x)\hat{\psi}^{\dagger}_\beta (x) \hat{\psi}_\beta (x)\hat{\psi}_\alpha (x)\nonumber\\
    &+ c_1 \sum_{\alpha,\beta,\gamma,\delta=-1}^{1} \sum_{i \in \{x,y,z\}}^{}(f_i)_{\alpha\beta}(f_i)_{\gamma\delta} 
   \hat{\psi}^{\dagger}_\alpha (x)\hat{\psi}^{\dagger}_\gamma (x)\hat{\psi}_\delta (x)\hat{\psi}_\beta (x)\Big]. \label{spin1_Hamiltonian}
\end{align} 
The bosonic field operator $\hat{\psi}_\alpha(x)$ acts on the involved magnetic sublevels (components) having a spin-$z$ projection $m_F \equiv \alpha= \{-1,0,1 \}$ in the $F=1$ hyperfine manifold. 
Also, the Pauli-$x$, $y$ and $z$ matrix elements are $(f_x)_{\alpha\beta}=\delta_{\alpha,\beta+1}+\delta_{\alpha,\beta-1}$, $(f_y)_{\gamma\delta}=-i\delta_{\gamma,\delta+1}+i\delta_{\gamma,\delta-1}$ and $(f_z)_{\alpha\beta}=\alpha\delta_{\alpha\beta}$ respectively where the indices $\alpha,\beta,\gamma,\delta \in \{ -1,0,1 \}$ refer to the individual spin components along a particular ($x$, $y$, $z$) spin direction. 
The second line of Eq.~(\ref{spin1_Hamiltonian}) represents the atom-atom interactions describing the intra- and intercomponent effective coupling strengths, while the last term refers to spin-mixing processes of the spinor system~\citep{kawaguchi2012spinor}. 
The remaining contributions constitute the non-interacting part of the  Hamiltonian.

The parameters $q$ and $p$ describe the quadratic and linear Zeeman energy shifts \citep{kawaguchi2012spinor,zhang2003mean}. They lead to an effective detuning of the $m_F=\pm 1$ components with respect to the $m_F=0$ one and are experimentally tunable by either adjusting the applied magnetic field~\citep{santos2007spinor} or employing a microwave dressing field~\citep{leslie2009amplification,bookjans2011quantum}. 
Also, $c_0=\frac{2\hbar^2(a_0+2a_2)}{3Ma_\perp^2}$ and $c_1=\frac{2\hbar^2(a_2-a_0)}{3Ma_\perp^2}$ correspond to the spin-independent and spin-dependent effective interaction coefficients~\citep{stamper2013spinor,kawaguchi2012spinor}. Here $a_0$ and $a_2$ are the 3D $s$-wave scattering lengths of the atoms in the scattering channels with the total spin $F=0$ and $F=2$, respectively~\citep{klausen2001nature}; $a_\perp=\sqrt{\hbar/(M\omega_\perp)}$ sets the transversal confinement length scale assuming the trapping frequency $\omega_\perp$. Importantly, a positive (negative) $c_{1}$ accounts for anti-ferromagnetic (ferromagnetic) spin interactions~\citep{ho1998spinor,ohmi1998bose}, whereas $c_0>0$ ($c_0<0$)  accounts for repulsive (attractive) interparticle interactions. 
For anti-ferromagnetic (ferromagnetic) couplings the condensate lowers its energy by minimizing (maximizing) its average spin. 
The values of the 3D scattering lengths $a_0$ and $a_2$ for $^{87}$Rb and $^{23}$Na can be found in~\citep{kawaguchi2012spinor,stamper2013spinor} while for $^{7}$Li in~\citep{huh2020observation}. 
Experimentally, $c_1$ is adjustable by means of the microwave-induced Feshbach resonance technique~\citep{papoular2010microwave}, whilst $\omega_{\perp}$ can be manipulated using confinement induced resonances~\citep{haller2010}.

Using the spinor systems one can study the competition between the involved spin (magnetic) interactions and an external magnetic field~\citep{he2009strong}. 
Naturally, an external magnetic field breaks the spin rotational symmetry~\citep{stamper2013spinor,stamper2001peeking} and a spatially homogeneous external field alters the total Zeeman energy of the spinor system and thus the ground state spin configuration. It also leads to Larmor precessions of the components, e.g., around the $z$-axis. 
In contrast, a spatially inhomogeneous magnetic field exerts magnetic forces on the components and can result in phase separation processes between the spin states~\citep{ho1998spinor}. 
Such fields allow also for spatial dephasing of the otherwise coherent spinor dynamics and can even cause diabatic spin flips~\citep{kawaguchi2012spinor,stamper2013spinor}. 

Before delving into quantum fluctuation driven processes and few-body mechanisms it is instructive for these spinor settings to initially discuss the somewhat simpler mean-field limit. 
The latter provides a pedagogical description of the involved magnetic processes, including spin-mixing and spin domain formation, which will serve as a starting point for analyzing more complex correlation mechanisms. 

\subsection{Mean-field picture}\label{mean_field_spin1}

Spinor bosonic systems have been widely studied within the mean-field approximation in the past decades, which has enabled, among others, researchers to interpret relevant experiments in both 1D~\citep{lannig2020collisions,bersano2018three} and higher spatial dimensions~\citep{weiss2019controlled}. 
Since our focus is the impact of quantum fluctuations we do not aim  
to provide a comprehensive description of the mean-field studies as many of them were already summarized e.g. in the review articles~\citep{kawaguchi2012spinor,stamper2013spinor}. 
Rather, we shall discuss the mean-field approximation in passing, which allows us to provide a simple picture of the emergent spin-mixing processes. 

Mean-field theory assumes that the ground state is macroscopically occupied and this allows one to replace the three spinor field operators by a three component vectorial order parameter, namely $(\psi_1,\psi_0,\psi_{-1})^{\intercal}=\braket{(\hat{\Psi}_1,\hat{\Psi}_0,\hat{\Psi}_{-1})^{\intercal}}$.  
This assumption accounts for the hybridization of the spin and spatial degrees-of-freedom and consequently ignores the contribution of excited states or effects stemming from quantum fluctuations which will be discussed later on. 
In particular, following a variational principle for the resulting product mean-field ansatz one can derive a coupled system of Gross-Pitaevskii equations of motion for the different spin components~\citep{ho1998spinor,ohmi1998bose,kawaguchi2012spinor,stamper2013spinor}. 
They are described by the individual spin wave functions $\Psi_{\pm 1}(x;t)$, $\Psi_0(x;t)$ for the $\ket{F=1, m_F=\pm1}$ and $\ket{F=1, m_F=0}$ components respectively. 
Particularly, the $m_F=\pm1$ spin components obey 
\begin{align}
i \hbar \partial_t \Psi_{\pm 1} &=
(\mathcal{H}_{0}+q\mp p)\Psi_{\pm 1}  
+ c_{0} \big( |\Psi_{+1}|^2 + |\Psi_{0}|^2 + |\Psi_{-1}|^2 \big)\Psi_{\pm 1}\nonumber \\&+c_{1}\big( |\Psi_{\pm1}|^2 + |\Psi_{0}|^2 - |\Psi_{\mp 1}|^2 \big)\Psi_{\pm 1}+c_{1}\Psi^{*}_{\mp 1}\Psi^{2}_{0},
\label{GPE_spin1_symcomp} 
\end{align}
while the $m_F=0$ magnetic sublevel satisfies 
\begin{eqnarray}
i \hbar \partial_t \Psi_{0} =
\mathcal{H}_{0}\Psi_{0}  
+ c_{0} \left( |\Psi_{+1}|^2 + |\Psi_{0}|^2 + |\Psi_{-1}|^2 \right)\Psi_{0}
&+c_{1}\left( |\Psi_{+1}|^2 + |\Psi_{-1}|^2 \right)\Psi_{0}\nonumber
\\&+ 2c_{1}\Psi_{+1}\Psi^{*}_{0} \Psi_{-1}. \nonumber\\
\label{GPE_spin1_0comp} 
\end{eqnarray} 
Note that in the above equations the single-particle Hamiltonian term is given by $\mathcal{H}_0\equiv -\hbar^2/(2M) \partial_{x}^2 +V(x)$, 
with $V(x)$ being the external confining potential. 
Interestingly, the penultimate terms of Eqs.~(\ref{GPE_spin1_symcomp})-(\ref{GPE_spin1_0comp}) are related to the spatial dynamics of the system and can lead to the formation of spin domains~\citep{miesner1999observation}. 
The last term of these equations gives rise to coherent spin-mixing processes driving particle transfer between the spinor components~\citep{law1998quantum,pu1999spin,zhang2005coherent}. 
The aforementioned mechanisms take place on similar timescales, i.e. they cannot be separated and their interplay results in the coupling between the spin and the spatial degrees-of-freedom of the condensate. 
It is also worth mentioning that both processes are influenced by the presence of external magnetic fields. 
The dispersion relations of spin-1 Bose gases within the Bogoliubov approximation are discussed in detail in \citep{kawaguchi2012spinor,stamper2013spinor}, and
they commonly consist of a phononic and two degenerate magnonic branches. 

\begin{figure}[tb]
    \center
    \includegraphics[width=0.9\linewidth]{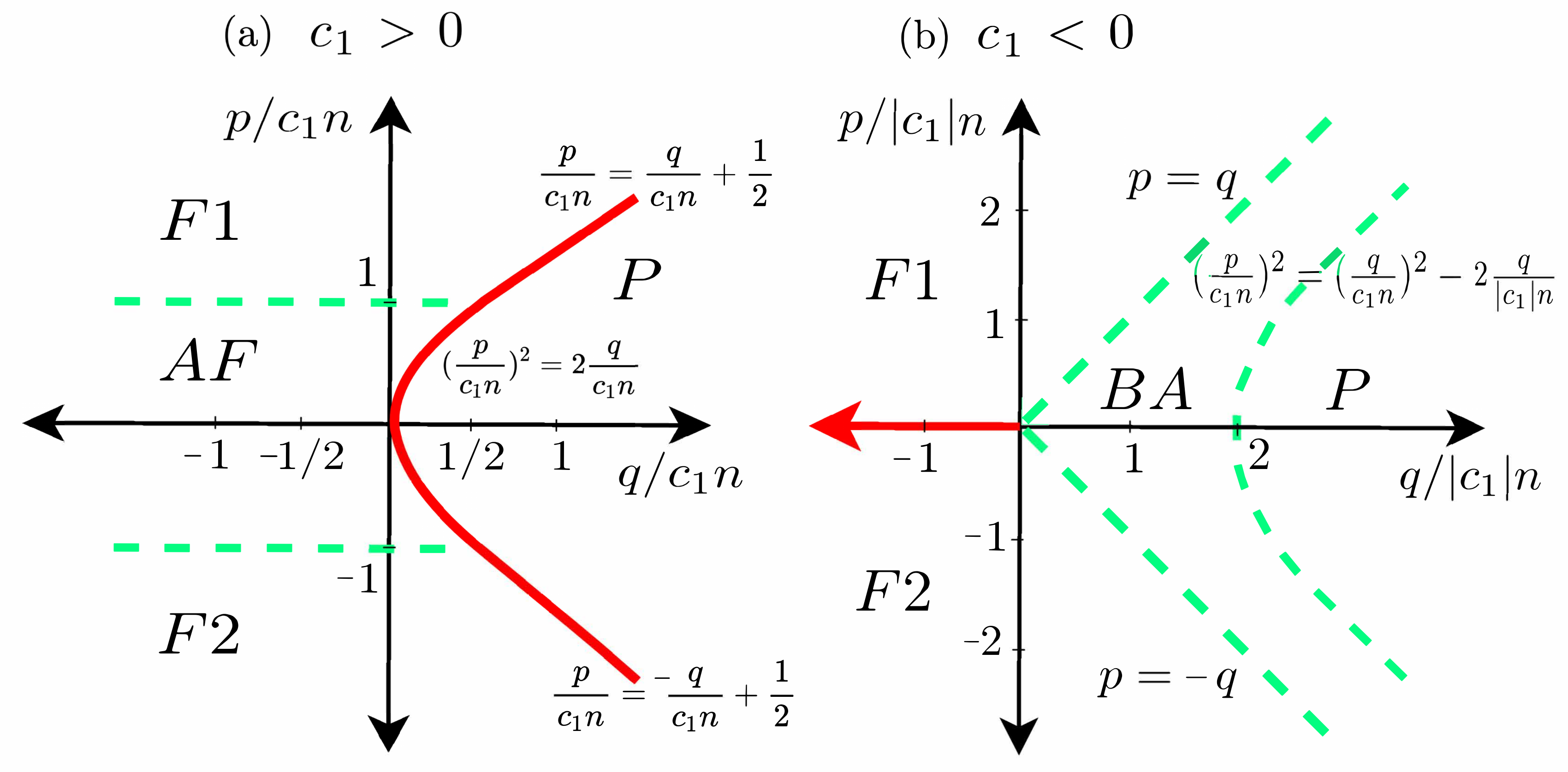}
    \caption{Ground state phase diagram of the spin-1 Bose gas for (a) anti-ferromagnetic $c_1>0$ and (b) ferromagnetic $c_1<0$ spin-dependent interactions and varying linear $p$ and quadratic $q$ Zeeman parameters. Red solid (green dashed) curves indicate the boundaries of the first- (second-) order quantum phase transitions as predicted in the thermodynamic limit. The total density of the gas is given by $n$. Depending on the sign of $c_1$ the phase diagram features two ferromagnetic phases (F1), (F2), an anti-ferromagnetic one (AF), a polar one (P), and a broken axisymmetry (BA) phase. Figure  from~\citep{mittal2020many}.}
\label{fig:spin1_phasediag} 
\end{figure}

\subsection{Phase diagram}\label{phases_spin1} 

The features of spin-1 Bose gases that unveil their richness compared to binary mixtures become evident when inspecting the ground-state phase diagram. 
We remark that another exciting direction refers to the so-called $SU(N)$ models that have been mainly analyzed for spin chain fermionic systems, see e.g. the review by~\cite{Guan2013Review} and a relevant brief discussion in Sec.~\ref{sec09SC}. This topic although interesting since it reveals exotic phases~\cite{taie2022observation,nataf2014exact,yamamoto2020quantum,wu2006hidden,bauer2012three,choudhury2020collective} and for large $N$ the $SU(N)$ symmetry enforces the presence of quantum fluctuations certainly lies  beyond the scope of the present review. 
The interplay between the sign of the spin-dependent interactions, $c_1$, and the strength of the linear, $p$, and quadratic, $q$, Zeeman terms gives rise to distinct ground states characterized by specific magnetic properties~\citep{kawaguchi2012spinor,jacob2012phase,stamper2013spinor} and first- or second-order phase transitions\footnote{For first- (second) order phase transitions the ground state of the system characterized by the population of a single spin state deforms abruptly (smoothly) to another spin state (a superposition where a second spin-state acquires finite population) across the underlying phase boundary.} between them~\citep{carr2010understanding}. 
The corresponding phase diagram is schematically depicted in Fig.~\ref{fig:spin1_phasediag}. 
The locations of the phase transition boundaries are determined in the thermodynamic limit $N\to \infty$ and within the mean-field realm where quantum fluctuation effects are ignored. However, it should be noted that \textit{"few-body"} analogues of these phases have been shown to occur e.g. in~\cite{mittal2020many} but featuring only small shifts of the boundaries due to the involvement of correlations. 

To quantify the emergent first- and second-order phase transitions between the individual phases the magnetization $\mathcal{M}$ along the spin-$z$-axis and the polarization $\mathcal{P}$ can be used 
\begin{equation}
\mathcal{M}=n_{1}-n_{-1}~~~{\rm{and}}~~~\mathcal{P}=n_0-(n_1+n_{-1}), \label{Magnetization}
\end{equation}
where $n_{m_F}=(1/N)\int dx \abs{\Psi_{m_F}}^2$ and $m_F=0,\pm1$. 
Note that $-1\leq \mathcal{M}\leq 1$ and thus $\mathcal{M}=\pm1$ refers to a fully magnetized state along the $\pm z$ direction; $-1\leq \mathcal{P}\leq 1$, which allows one to distinguish between the fully magnetized ($\mathcal{M}=1$) and un-magnetized ($\mathcal{M}=0$) configurations when varying the Zeeman parameters. 
Figure~\ref{fig:spin1_phasediag}(a) demonstrates that for anti-ferromagnetic interactions $c_1>0$, two ferromagnetic phases occur with the atoms populating either the $m_F=+1$ (F1) or the $m_F=-1$ (F2) component. 
Also an anti-ferromagnetic phase (AF) is present, where the bosons reside in a superposition of the $m_F=+1$ and the $m_F=-1$ components while the $m_F=0$ state remains un-occupied. 
In all cases $\mathcal{P}=-1$ while $\mathcal{M}=+1$ [$\mathcal{M}=-1$] for the F1 [F2] phase and $-1<\mathcal{M}<1$ in the AF state. 
Additionally, there is the Polar phase (P) in which all particles are in the $m_F=0$ state and therefore $\mathcal{P}=1$, $\mathcal{M}=0$. 
Among the aforementioned phases different types of phase transitions take place. For instance, at $c_1>0$ upon crossing the F1 to the P phase for increasing $q$ the spin-state contributing to the ground state of the system changes without accessing a superposition spin-state of the $m_F=0$ and the $m_F=1$ components.  
As such, a first-order transition takes place. 
However, when transitioning from the F2 phase to the AF phase for larger $p$ values, the system features a second-order transition. 
Indeed, if the F2 phase is given by $m_F=-1$, it   gradually evolves into a superposition of the $m_F=\pm1$ components (AF phase) by increasingly populating the $m_F=1$ component. 
Turning to ferromagnetic interactions, characterized by $c_1<0$, besides the above-described phases an additional one emerges, the so-called broken-axisymmetry phase (BA) where the three spin-states $m_F=\pm1, 0$ are populated in a non-equal fashion. 
For this reason, $-1<\mathcal{P}<1$ and $-1<\mathcal{M}<1$ in the BA phase. 
The corresponding layout of the ensuing quantum phase transitions as a function of the linear and the quadratic Zeeman coefficients as well as the boundaries of the individual states are shown in Fig.~\ref{fig:spin1_phasediag}(b). 

\begin{figure}[tb]
    \center
    \includegraphics[,width=0.8\textwidth]{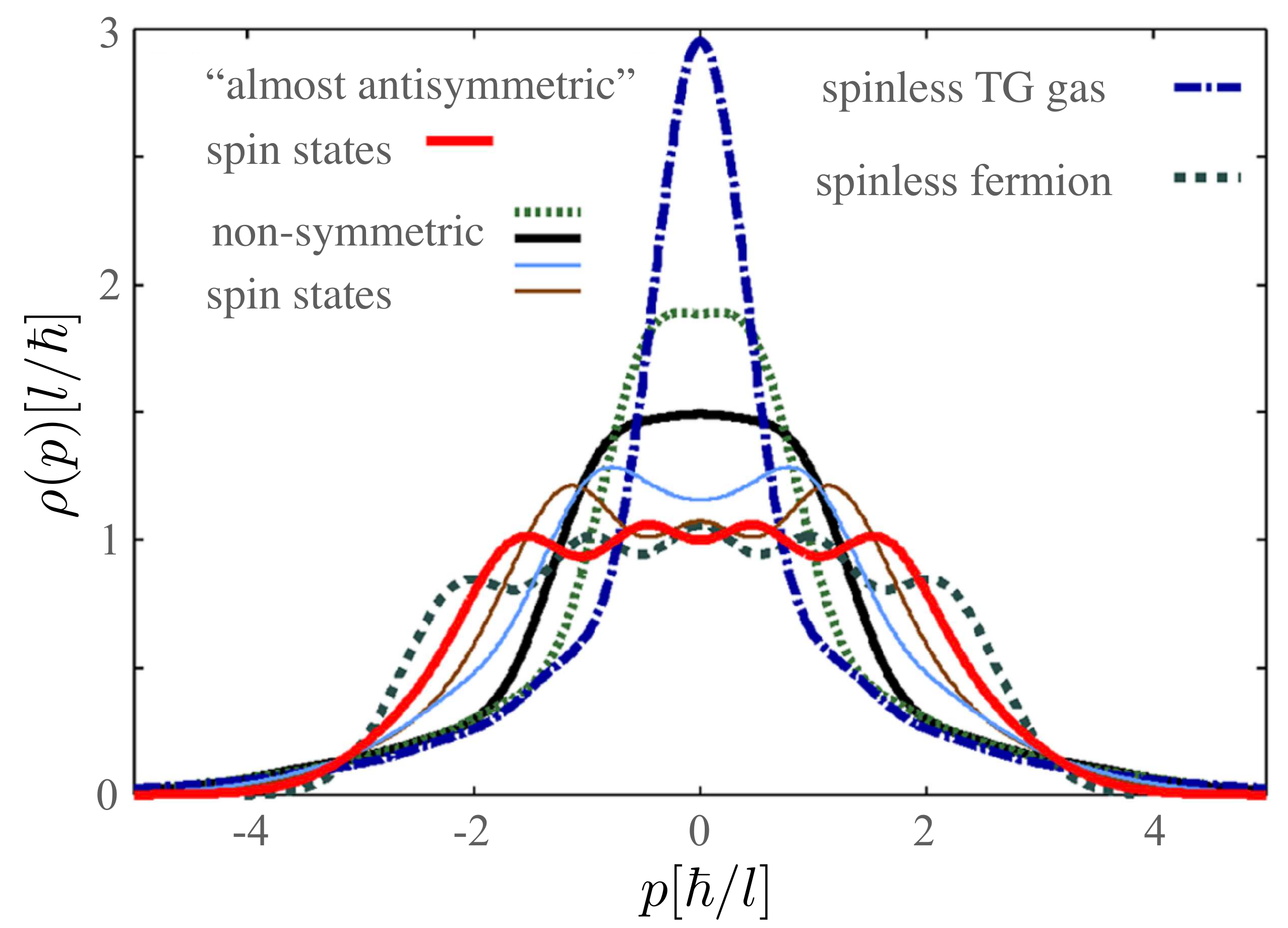}
    \caption{Momentum distribution of five spin-1 bosons in distinct degenerate ground states obtained through exact diagonalization. The gray dashed line presents the momentum distribution of five noninteracting fermions. It becomes evident that the shape of the momentum distribution depends on the symmetry of the spin function. Figure from~\citep{deuretzbacher2008exact}. }
    \label{fig:spin1_momentum} 
\end{figure} 

Having briefly analyzed the mean-field ground state phase diagram, we next focus on the impact of the intra- and intercomponent correlations on the magnetic phases of harmonically trapped spin-1 Bose gases that have been discussed in \citep{mittal2020many,kanjilal2021variational}.  The validity of the mean-field approximation in 1D trapped systems for large particle numbers was investigated, e.g., in \citep{gautam2015analytic,matuszewski2010ground, matuszewski2010magnetic,chern2019ground,lin2011characterization}.   
It was argued that correlation-induced phenomena manifest across the second-order phase transitions and become suppressed for the first-order ones. 
More precisely, for both ferromagnetic ($c_1<0$) and anti-ferromagnetic ($c_1>0$) spin-spin interactions the corresponding transition boundaries are shifted compared to the mean-field approximation resulting in a relatively smaller ($p$, $q$) interval where the AF and the BA states occur. 
This means that correlations favor states where the atoms are polarized in a single-spin component, a phenomenon that is more prominent in the few-body limit. 
Finally, spin-orbit coupled 1D lattice trapped  strongly interacting spin-1 Bose gases have been studied using a density matrix renormalization group technique~\citep{saha2020strongly} and spin-chain methods~\citep{Cui2016Spin} explicating the emergence of additional phases or altered properties of the existing ones. 
For instance, they have shown that the presence of spin-orbit coupling gives rise to a charge density wave state and a tendency for suppression of (anti-)ferromagnetic correlations.    

Exact solutions of the spin-1 gas consisting of hard-core bosons, in the absence of an external magnetic field, have been constructed through an extension of Girardeau's Bose-Fermi mapping theorem \citep{deuretzbacher2008exact}. Namely, in the limit of $c_0\to \infty$ and $c_1\to 0$ where the bosons behave as non-interacting spinless fermions with zero spin interactions. 
The eigenspectrum of the corresponding three particle system in the limit of large but finite repulsion and in the presence of spin interactions has also been analyzed within exact diagonalization revealing the structure and symmetry properties of the underlying wave functions. 
Inspecting the corresponding spin densities, various structures have been shown to occur ranging from phase miscible and immiscible to magnetic domain-wall like configurations among the individual components. 
Moreover, the momentum distribution of the eigenstates turned out to be a promising diagnostic of the symmetry of the spin wave function having a wider shape for non-symmetric spin functions than the one of a spinless Tonks-Girardeau gas, see Fig.~\ref{fig:spin1_momentum}. 
The properties of the one-body momentum distributions 
in few-boson spin-1 gases in the Tonks-Girardeau limit and under the constraint that the energy splitting between different spin states is smaller than $k_B T$, thus entering the so-called spin-incoherent Luttinger Liquid regime~\citep{fiete2007colloquium}, have been discussed in \citep{jen2016spin,jen2017spin}. 
A broadening of the individual component momentum distributions and a tendency to acquire the same shape for increasing particle number was found along with an analytical prediction of universal $1/p^4$ dependence for high momenta. 

\begin{figure}[tb]
    \center
    \includegraphics[trim={0cm 14cm 0cm 0cm},clip,width=0.9\textwidth]{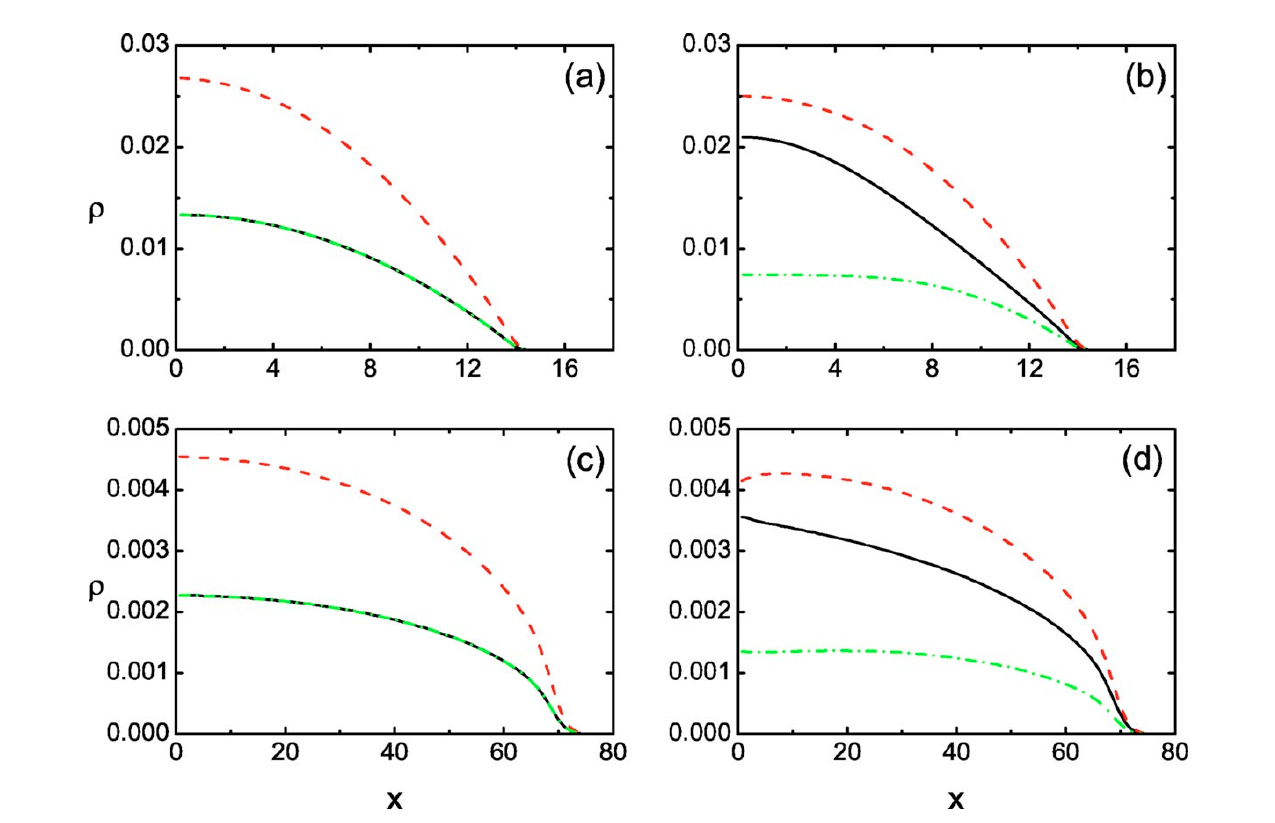}
    \caption{Ground state density distributions of the individual components (see legend) of a 1D harmonically trapped (a), (b) weakly (Thomas-Fermi regime) and (c), (d) strongly interacting (TG limit) ferromagnetic spin-1 Bose gas. Solid (dashed-doted) curves represent the density of the $m_F=1$ ($m_F=-1$) state; the dashed curve shows the density of the $m_F=0$ state. The magnetization is (a), (c) $\mathcal{M}/N=0$, and (b), (d) $\mathcal{M}/N=0.2$. Length and density profiles are given in units of 1.2$\mu m$ and $N\sqrt{\hbar/(m \omega)}$ respectively with $\omega$ being the harmonic trap frequency. Figure from~\citep{hao2006density}.}
\label{fig:spin1_spatial_configurations_new} 
\end{figure} 

For anti-ferromagnetic gases~\citep{hao2017ground} with fixed magnetization a transition from a Gaussian to a fermionized multi-hump density structure was found to take place upon increasing $s$-wave interactions. In the fermionized limit phase separation emerges for  $\mathcal{M}=0.25$, while for $\mathcal{M}=0.5$ the $m_F=0$ component is depopulated and the others fermionize. 
Off-diagonal long-range order\footnote{1D quantum phases cannot exhibit true long-range order. Rather quasi-long-range order appears meaning that the correlation function $\braket{\Lambda^{\dagger}(x)\Lambda(0)}$ of an operator $\Lambda(x)$ falls of algebraically $\sim \abs{x}^{\beta}$ as $\abs{x}\to \infty$.} is evident up to intermediate $s$-wave interactions and becomes suppressed in the strong interaction regime. 
Recall that off-diagonal long-range order\footnote{
It is worth noting that off-diagonal long-range order (ODLRO) is associated with the BEC phase but not with superfluidity. Indeed, superfluid systems where BEC is not possible lack true ODLRO. Moreover, in finite systems no true ODLRO can exist and therefore the term coherence refers to the off-diagonal of the single-particle reduced density matrix quantifying short- and long-range one-particle correlations, see also Refs.~\citep{penrose1956bose,yang1962concept,colcelli2020finite} for the scaling behavior of the largest eigenvalue of the single-particle density matrix with respect to the atom number.} is a measure of coherence which can be experimentally measured~\citep{bloch2000measurement} and it is related to the fact that the off-diagonal elements of the system's single-particle reduced density matrix remain finite for long distances~\citep{hohenberg1967existence,yang1962concept,penrose1956bose}. 
Independently, such crossover behavior towards fermionization has been discussed in~\citep{hao2006density} for both ferromagnetic and anti-ferromagnetic gases using a modified Gross-Pitaevskii framework based on the local density approximation and the analytical solution of the homogeneous system.
As a case example, Fig.~\ref{fig:spin1_spatial_configurations_new} (a) [(b)] depicts the corresponding density profiles of a ferromagnetic $^{87}$Rb weakly interacting gas for $\mathcal{M}/N=0$ [$\mathcal{M}/N=0.2$]. 
As expected, the $m_F=\pm 1$ components are discernible only in the case of nonzero magnetization. 
The respective density configurations in the Tonks limit are provided in Figs.~\ref{fig:spin1_spatial_configurations_new} (c), (d). 
They become wider due to the strong interactions and feature a sharp decrease at the edges. This sharpness of the density edges is reminiscent of the Fermi-Dirac distribution. 
Additionally, it has been argued that an increase of the spin-dependent interactions results in the weakening of the fermionization meaning that less than $N$ peaks appear in the density profiles and all components show the same behavior~\citep{hao2016weakening}. 
This phenomenon is attributed to the formation of composite pairs due to the strong spin-exchange interactions. 
Similar density distributions were found independently using a density functional approach~\citep{wang2013density}. 

\subsection{Spin-mixing dynamics and the single mode approximation}\label{spin_mixing} 

To gain a better intuition into emergent spin-mixing processes, we first rely on a reduction of the mean-field description before reporting on recent advances on the arguably more complex many-body problem. 
As explained in Section~\ref{mean_field_spin1} the characteristic timescales of the spin and the spatial dynamics are generically inseparable. 
However, under certain conditions they can be decoupled. 
This occurs when the spin healing length $\xi_s=\hbar/\sqrt{2Mn\abs{c_1}}$ becomes larger than the size of the condensate, where $n$ is the peak density of the system. 
Then, $\Psi_1(x)\approx \Psi_{-1}(x)\approx \Psi_0(x)$ which allows for a separation of the spatial and spin dynamics, a situation that is known as the single-mode approximation (SMA)~\citep{ho1998spinor,yi2002single}. 
Here, it is assumed that the order parameter possesses the form $\Psi_i(x)=\sqrt{N}\phi(x)e^{-i\mu t/\hbar}\zeta_i(t)$ with $\phi(x)$ representing the common spatial mode normalized to unity and $\mu$ being the chemical potential. 
Also, $\mathbf{\zeta}=(\zeta_1,\zeta_0,\zeta_{-1})^{\intercal}$ is the spin state vector satisfying $\abs{\mathbf{\zeta}}=1$. 

Substituting the SMA ansatz into the full system of Gross-Pitaevskii Eqs.~(\ref{GPE_spin1_symcomp})-(\ref{GPE_spin1_0comp}) it is possible to show that the spatial dynamics separates from the one related to the spin degrees-of-freedom~\citep{yi2002single} and is governed by the spin-independent part of the Hamiltonian of Eq.~(\ref{spin1_Hamiltonian}). 
On the other hand, the spin states obey a set of coupled equations of motion \citep{kawaguchi2012spinor,stamper2013spinor} which can be further simplified by using the normalization condition $\sum_i \rho_i=1$ with $\rho_i=\abs{\zeta_i}^2$, the conservation of the magnetization $\mathcal{M}$ and assuming that $\zeta_i=\sqrt{\rho_i}e^{-i\theta_i}$. 
It can be showcased that under the previous conditions the spin dynamics is determined solely by the fractional population $\rho_0(t)$ of the $m_F=0$ spin state and the relative phase $\theta=\theta_1+\theta_{-1}-2\theta_0$ of the spin components:
\begin{align} 
\dot{\rho}_{0}&=\frac{2c_1 N}{\hbar} \rho_0 \sqrt{(1-\rho_0)^2-\mathcal{M}^2} \sin \theta, \label{SMA1}\\
\dot{\theta}&=-\frac{2\delta}{\hbar}+ \frac{2c_1N}{\hbar} \bigg[(1-2\rho_0)+\frac{(1-\rho_0)(1-2\rho_0)-\mathcal{M}^2}{\sqrt{(1-\rho_0)^2-\mathcal{M}^2}} \cos\theta \bigg]. \label{SMA2}
\end{align} 
According to the Breit-Rabi formula~\citep{breit1931measurement,stamper2013spinor} $\delta=(E_1+E_{-1}-2E_0)/2$ is the quadratic Zeeman shift and $E_i$ with $i=-1,0,1$ denotes the energy of the respective hyperfine state. 
These two equations represent a classical nonlinear pendulum with total energy 
$E=c_1N\rho_0\big[ (1-\rho_0)+ \sqrt{(1-\rho_0)^2-\mathcal{M}^2} \cos \theta \big]+\delta (1-\rho_0)$. Interestingly the set of Eqs.~(\ref{SMA1}) and (\ref{SMA2}) can also be derived via $\dot{\rho}_0=-(2/\hbar)\partial E/\partial \theta$ and $\dot{\theta}=(2/\hbar) \partial E/\partial \rho_0$. 
The latter set of coupled equations represents a nonlinear Josephson oscillator revealing an equivalence between spin-mixing processes and the Josephson dynamics in the context of superconductors~\citep{barone1982physics} and superfluids~\citep{smerzi1997quantum,cataliotti2001josephson}. 
A proper choice of the initial populations and phases of the spin components as well as of the strength of the external magnetic field allows the creation of a manifold of dynamical trajectories that have been realized experimentally~\citep{black2007spinor,zhao2014dynamics} and are manifested due to the inherent nonlinearity of these equations.

In this context, efforts towards appreciating the impact of correlations in spin-mixing processes have also been put forth. Indeed, the role of quantum fluctuations in the spin dynamics of $F=1$ sodium trapped atoms with $N<10^3$ by expanding the full Hamiltonian within the SMA around the mean-field ground state has been discussed in \citep{cui2008quantum}. 
It was argued that quantum fluctuations are dominant only for small values of the quadratic Zeeman field, in particular $q \ll 4 c_1$, and that time evolution of the particle number of the participating spin states provides a probe for beyond mean-field dynamics. 
A promising candidate for demonstrating these effects experimentally is a multicomponent condensate coupled to a single cavity mode where the eigenfrequencies of the system depend crucially on the population of the individual hyperfine states, see~\citep{mivehvar2021cavity} and references therein.

Additionally, the number conserving Bogoliubov theory for finite uniform spin-1 anti-ferromagnetic condensates allows one to explore the instability of the polar state~\citep{kawaguchi2014goldstone}; see also \citep{bookjans2011quantum} for an experimental investigation of the dependence of this instability on the quadratic Zeeman energy. 
The theory framework is based on an ansatz where all atom pairs reside in the same two-particle state while the origin of this process is traced back to the unstable Goldstone magnons at $q=0$\footnote{For $q=0$ the magnon modes of the spinor BEC become unstable and correspond to the Goldstone modes associated with the breaking of the $SO(3)$ rotation symmetry in the spin space in the Hamiltonian of the system~\citep{kawaguchi2012spinor,stamper2013spinor}}. 
The latter show an algebraic growth and lead to the fragmentation of the system. 
It is worth mentioning that in the thermodynamic limit the relevant timescale of this instability diverges and as a consequence the respective mean-field state is stable. 

\begin{figure}[tb]
    \center
    \includegraphics[width=0.9\textwidth]{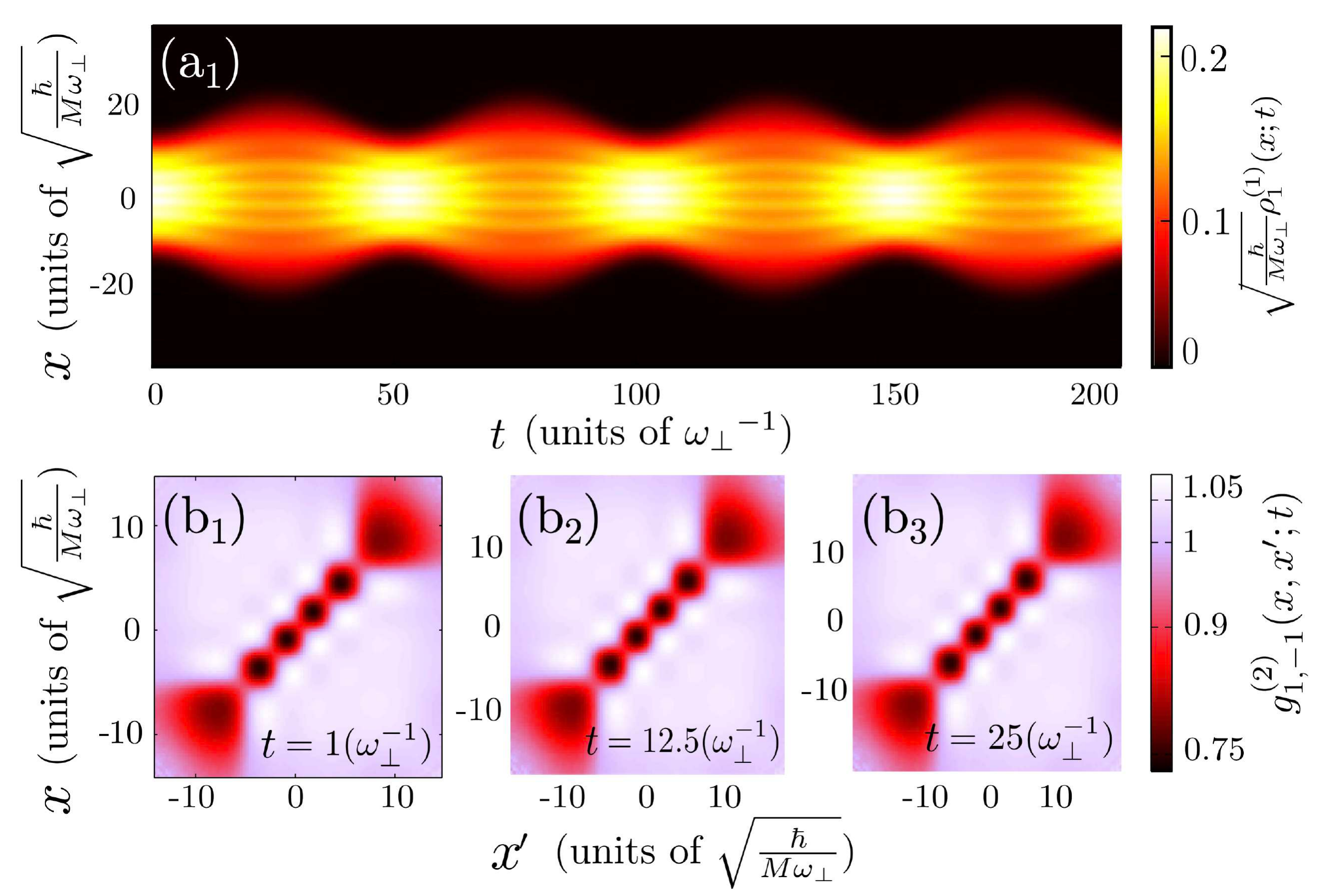}
    \caption{($a_1$) Density evolution of the $m_F=+1$ component of the 1D spin-1 gas with $N=20$ bosons initialized in the AF phase and subjected to a quench of the trap frequency from $\omega=0.1 \omega_{\perp}$ to $\omega=0.07 \omega_{\perp}$. ($b_1$), ($b_2$), and ($b_3$) Profiles of the two-body inter-component correlation function among the $m_F=+1$ and $m_F=-1$ components. The gas is characterized by spin-independent $c_0=0.5 \sqrt{\hbar^3 \omega_{\perp}/M}$ and spin-dependent $c_1=0.018\sqrt{\hbar^3 \omega_{\perp}/M}$ interaction strengths. The Zeeman coefficients are $p/(c_1n)=0.04$ and $q/(c_1n)=-0.44$. Figure from~\citep{mittal2020many}.}
\label{fig:spin1_breathing} 
\end{figure} 

The correlation effects in the quench induced breathing dynamics of an 1D harmonically trapped spin-1 Bose gas using an \textit{ab-initio} approach have been explored in \citep{mittal2020many}. 
The emergent collective oscillation frequency is almost the same for all participating hyperfine states and for different initial phases. 
Spin-mixing processes take place only in the BA phase. 
Interestingly, the presence of interparticle correlations leads to filamentary-like patterns as shown in Fig.~\ref{fig:spin1_breathing}. These multihump structures build upon the background density of each component and become more prominent in the few-body limit or for larger spin-independent interactions. 
Each filament is a coherent structure showing two-body anti-correlations while neighboring ones feature substantial coherence losses and are correlated with one another.

%% file: Section07.tex
\section{\label{sec:Control}
Control: driven dynamics and transport}

Since ultracold atom systems are usually of finite size and due to their ultracold nature disconnected from any external environment, studies of driven dynamics and transport have to be realised in a closed system setting. This is therefore different from low-dimensional settings in, for example, condensed matter, where usually leads are attached to systems and open system approaches need to be employed. However, it is worth noting that two-terminal transport measurements are in principle possible in ultracold atom systems and have been realised in cold Fermi gases \citep{Krinner:2017}. 

A more popular way to study transport in finite sized system is to use geometries that are periodic \citep{Amico_2005}. Experimentally ring-like traps have seen significant development over the last decade \citep{ramanathan2011,moulder2012,wright2013,ryu2013,Herve:2021}, and while no experimental setting has been created that strictly has a one-dimensional azimuthal degree of freedom only, there is no technical issue that would prevent this.
Nevertheless the stationary properties in such periodic systems for small and strongly correlated gases have been theoretically explored \citep{Alon:04,Sakmann:05}.

\subsection{Driven dynamics}

Systems in which the atoms are not stationary but move under a controlled external force are interesting from a fundamental point of view, but also have applications in areas such as atomtronics. 
The latter is driven by a highly active community and their progres is well summarised in a number recent reviews and roadmaps \citep{Amico:2017,atomtronics,Amico:21}. 
We will therefore in the following only mention applications briefly when relevant.

To study the effects of correlations and coherences on emerging, non-trivial dynamics, 
\citep{Cominotti2014} explored the control of persistent currents over a wide range of interactions for bosons trapped in a 1D ring of length $L$.  The flux was created by a rotating barrier of strength $U_0$ and the Hamiltonian in the co-rotating frame was given by
\begin{equation}
    H=\sum_{j=1}^{N} \frac{\hbar^2}{2m}\left( -i\frac{\partial}{\partial x_j} -\frac{2\pi}{L} \Omega \right)^2 + U_0 \delta(x_j) + \frac{g}{2} \sum_{j,l=1}^{N}\delta(x_l-x_j),
\end{equation}
where the Coriolis flux $\Omega=MVL/2\pi\hbar$ depends on the speed of the barrier $V$. The presence of the barrier breaks the integrability of the model, and allows one to explore the interplay between the interactions and the quantum fluctuations as a function of the barrier height. The authors showed that finite interactions and barrier heights yield a maximal current amplitude due to the competition between correlations and fluctuations. This can be understood by realising that increasing interactions weaken the importance of the barrier. However, going towards the strongly correlated Tonks-Girardeau regime the increasingly fast spatial decay of phase correlations (and therefore increased quantum fluctuations) screen the barrier less.

Including a lattice potential along the ring geometry allows one to explore the effect of lattice filling on the ability to create persistent currents. \citep{Arwas2017} showed that at unit filling such currents are suppressed as the Mott-insulator phase restricts the flow of particles around the ring. However, tuning the barrier energy to be on the order of the interaction energy creates a resonant behaviour in the current which coincides with sudden changes of the particle occupation at the position of the barrier.  

Exact results for the persistent currents of two interacting bosons in a ring lattice were obtained in \citep{Polo2020}, using a plane-wave ansatz for the Bose-Hubbard model subject to a synthetic magnetic field that drives the current. In the presence of the lattice, interactions can couple the center-of-mass and relative coordinates of the particles, which is contrary to what is seen in the continuous system described by the Lieb-Liniger model. While for repulsive interactions this only has a marginal effect on the persistent current, for attractive interactions the current dynamics does depend on the interactions due to nontrivial dynamics in the relative coordinate.
In fact, in systems with attractive interactions solitonic solutions can appear and have interesting applications in quantum metrology \citep{Polo:2021,Naldesi:22}.

\begin{figure}[tb]
    \centering
    \includegraphics[width=1\linewidth]{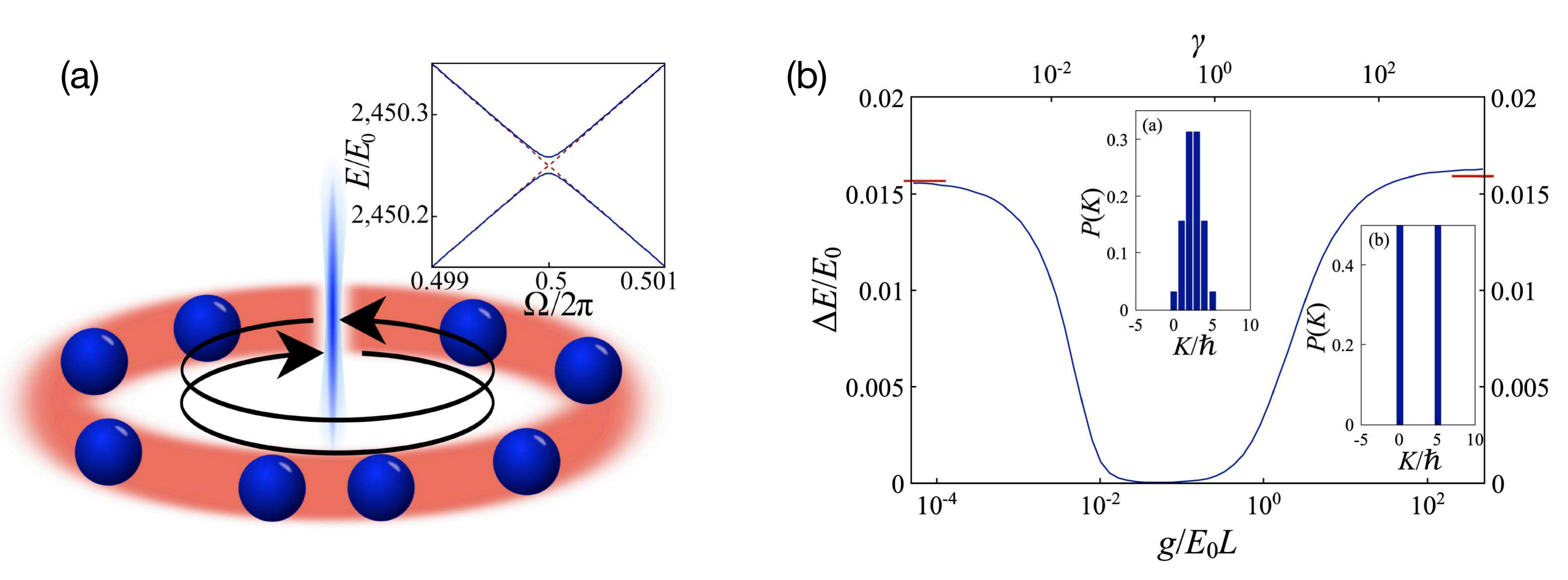}
    \caption{(a) Strongly correlated atoms trapped in a narrow ring with a rotating barrier. The inset shows the energy levels of the supercurrent states with total angular momentum $K = 0$ and $K = N\hbar$, respectively, as a function of the rotational phase  (dashed lines), and the lowest energy levels of N = 99 atoms in the Tonks-Girardeau regime in the presence of a barrier, where a gap opens (solid lines). (b) Energy level splitting between the ground and first excited states at $\Omega = \pi$ as a function of the interaction strength for five atoms in the presence of a barrier. The top horizontal axis shows the Lieb-Liniger parameter. The dashes on the figure margins indicate analytic results for noninteracting ($g = 0$) and strongly interacting atoms ($g = \infty$). The subplots show the distribution of the total angular momentum for (a) $g = 0$ and (b) $g = \infty$, where for the latter the NOON state is clearly visible. Figure adapted from \citep{Hallwood:10}. }
    \label{fig:TonksNoon}
\end{figure}

Beyond the study of fundamental flow effects, the driven dynamics in ring traps can also be used to create macroscopic superposition states. A scheme to create coherent superpositions of annular flow in a system of strongly interacting bosonic atoms in a 1D ring trap was presented in \citep{Hallwood:10} (see Fig.~\ref{fig:TonksNoon}). Here different angular momentum states were coupled by breaking the rotational symmetry using a rotating barrier, and the authors showed how to accelerate the atoms into a superposition state of rotating and non-rotating components. Since the atoms were considered to be in the strongly correlated Tonks–Girardeau regime, this process results in a macroscopically entangled NOON state. However, in order to avoid unwanted excitation during this driving process, especially close to the regions of avoided crossings where neighbouring energy bands come exponentially close to each other, the process requires an adiabatically slow
rotational acceleration of the atoms. Using shortcut-to-adiabaticity and optimal control techniques, \citep{Schloss2016} showed how this process can also be achieved in finite time, thereby making it a realistic proposal.

In the other limit, sudden quenches of the barrier rotation in a ring potential can also create superpositions of rotating and non-rotating states for strongly interacting bosons \citep{Schenke2011}. The ability to create these two components is controlled by the velocity of the barrier, as efficient momentum transfer to the gas can only be achieved when driving the system at integer multiples of the sound velocity of the gas  \citep{Schenke2012}. Exploring drag forces on the transport in strongly correlated systems has also been explored in rotating ring lattices, where the filling ratio of the gas to the number of lattice sites allows switching between superfluid, insulating and supersolid-like phases, which can exhibit many-body stick-slip dynamics when the system is close to commensurability \citep{Mikkelsen2018}.  Similar works have used the MCTDHB method to show how correlation functions can describe the loss of the superfluid fraction through tunneling \citep{Andriati2019}.

Lattices or multi-well systems can also be used to create maximally entangled NOON states of spatial modes. In \citep{Compagno2020} a scheme was proposed to create multi-mode NOON states, $|N000\dots\rangle+|0N00\dots\rangle+|00N0\dots\rangle+\dots$ for applications in multiport interferometry. They considered $N=2$ to $10$ bosons trapped in $M=2$ to $5$ traps which are arranged in a ring geometry, and where the wells have a constant energy off-set with respect to one another. The interactions between the atoms are driven sinusoidally, and it was found that when the driving is resonant with the well offset multi-mode NOON states can be created on short timescales. These states can be characterized experimentally through second-order momentum correlations which are measured via standard time-of-flight measurements used in cold atom systems \citep{bloch2008,pethick2002}.
Another proposal to create NOON states in multi-well systems explores integrability and its breaking in a controlled way \citep{Grun2021}. This approach employs dipolar atoms confined to four sites of an optical superlattice and the insights obtained from integrability allowed for the development of two protocols for the design of NOON states: one probabilistic and the other deterministic. One big advantage of these protocols is that the evolution time does not scale with the total number of particles, opening a potential path for scalability.

Periodic driving of a lattice potential with hard wall boundary conditions that contained a few-body system that had undergone an interaction quench was studied in \citep{mistakidis2015resonant,mistakidis2017mode}. The particles were shown to possess out-of-phase local dipole modes in the outer wells of the lattice and a breathing mode in the center well. Indeed, by tuning the driving frequency resonances between the inter- and intra-well levels, tunnelling dynamics could be observed, which highlights the tunability of the dynamics in these few-body system. The dynamics of lattice trapped few-boson systems after multiple interaction quenches has been explored in \citep{neuhaus2017quantum}. Here the pulsed quenches, consisting of temporal step functions, create interwell tunnelling along with a cradle mode, that describes  dipole-like intrawell oscillation in the outer wells, and a breathing mode. While the breathing mode is strongly dependent on the instantaneous interactions during the quenches, the cradle mode essentially ignores this and persists throughout the dynamics. 

The driving of interacting bosonic impurities immersed in a BEC was explored in \citep{Mukherjee2020}. The trapping potential of the impurities was subject to both pulsed and continuous shaking of its position which induces a plethora of different dynamical features in the system. For instance, pulsed driving below the frequency of the trapping potential causes the impurities to closely follow the motion of the trap during the driving. However, once the driving is terminated the  oscillation amplitude of the impurities decays and they are trapped in the BEC. Driving on resonance with the trap frequency results in more complex dynamics such as impurities escaping and re-entering the BEC and completely decoupling from it after the driving is stopped. For driving frequencies much larger than the trap frequency the impurities remain trapped inside the BEC and exhibit a dispersive behaviour at long times. Pulsed interactions between impurities and the background gas have also been used to explore dynamics going from miscible to immiscible phases \citep{Bougas2021}.

Driving the depths of a double well lattice periodically can break spatial parity and time-reversal symmetry of a system and lead to imbalances in the populations at long times \citep{Chen:20}. These imbalances are a function of the phase of the driving force and of the total number of particles. While the imbalance can be interpreted by comparison with a  driven and non-rigid classical pendulum, the
real-time evolution of the particle population shows significant effects stemming from the quantum many-body correlations. Other ways to break parity dynamically by moving a barrier through a few-body system have also been discussed recently \citep{Nandy_2020} and can be used to study resistivity. 

Periodic driving of few-body systems has also been used to study time crystals as in \citep{Barfknecht2019}, where the two-component gas can be described by an effective spin-chain model, cf.~\ref{CloseToFermi}. Applying periodic spin-flip pulses allows one to realize discrete time-translation-symmetry breaking in the regime of strong inter-species interactions and weak intra-species interactions. Such time crystals are robust even with imperfect pulses and the ability to tune the interactions via Feshbach resonances can allow one to study the melting of the time crystals as the Hamiltonian parameters are changed. Related studies in both few- and many-body setups within cavity systems can be found in \citep{gong2018PhRvL}. 
Periodic driving of an external trapping potential can also be used for delta-kick cooling, with recent extensions being proposed for interacting quantum systems which are scale invariant \citep{Dupays2021}.

\subsection{Transport}

Controlled transport of an interacting many-body system is not an easy task, as interactions can introduce collective effects, correlations and non-trivial excitations which can perturb the system significantly. One way to approach this problem is by trying to expand single particle control techniques to interacting few-body systems, and an example that allows for high-fidelity transport is the technique of spatial adiabatic passage (SAP) \citep{MenchonEnrich:2016}. This method is an analog of the well-known STIRAP technique from atomic physics \citep{Bergmann:2019}, and uses a {\it dark} eigenstate to coherently transfer single particles between two localized spatial states. 

The SAP technique relies on the presence of three neighbouring localised states, between which individually adjustable tunnel couplings exist. Labelling the localised states as $|L\rangle$, $|M\rangle$ and $|R\rangle$ and assuming that their onsite energies are the same, one can describe the  system with the simplified Hamiltonian
\begin{equation}
    \label{eq:H3}
    H(t) = \hbar
	\begin{pmatrix}
		0&\Omega_{LM}(t)&0\\
	   	\Omega_{LM}(t)&0&\Omega_{MR}(t)\\
		0&\Omega_{MR}(t)&0
	\end{pmatrix},
\end{equation}
where $\Omega_{LM}$ and $\Omega_{MR}$ are the time-dependent coupling rates between the states.  Diagonalizing this Hamiltonian yields the so-called {\it dark state} of the form
\begin{equation}
	\ket{D(\theta)} = \cos \theta \ket{L} - \sin \theta \ket{R},
\end{equation}
where $\tan \theta = \Omega_{LM}/\Omega_{MR}$. Since this state has only contributions from $|L\rangle$ and $|R\rangle$, transport between these two states can be achieved by changing the angle $\theta$ from $0$ to $\pi/2$. This process is well explored for single particle systems \citep{MenchonEnrich:2016}.

\begin{figure}[tb]
    \centering
    \includegraphics[width=0.8\linewidth]{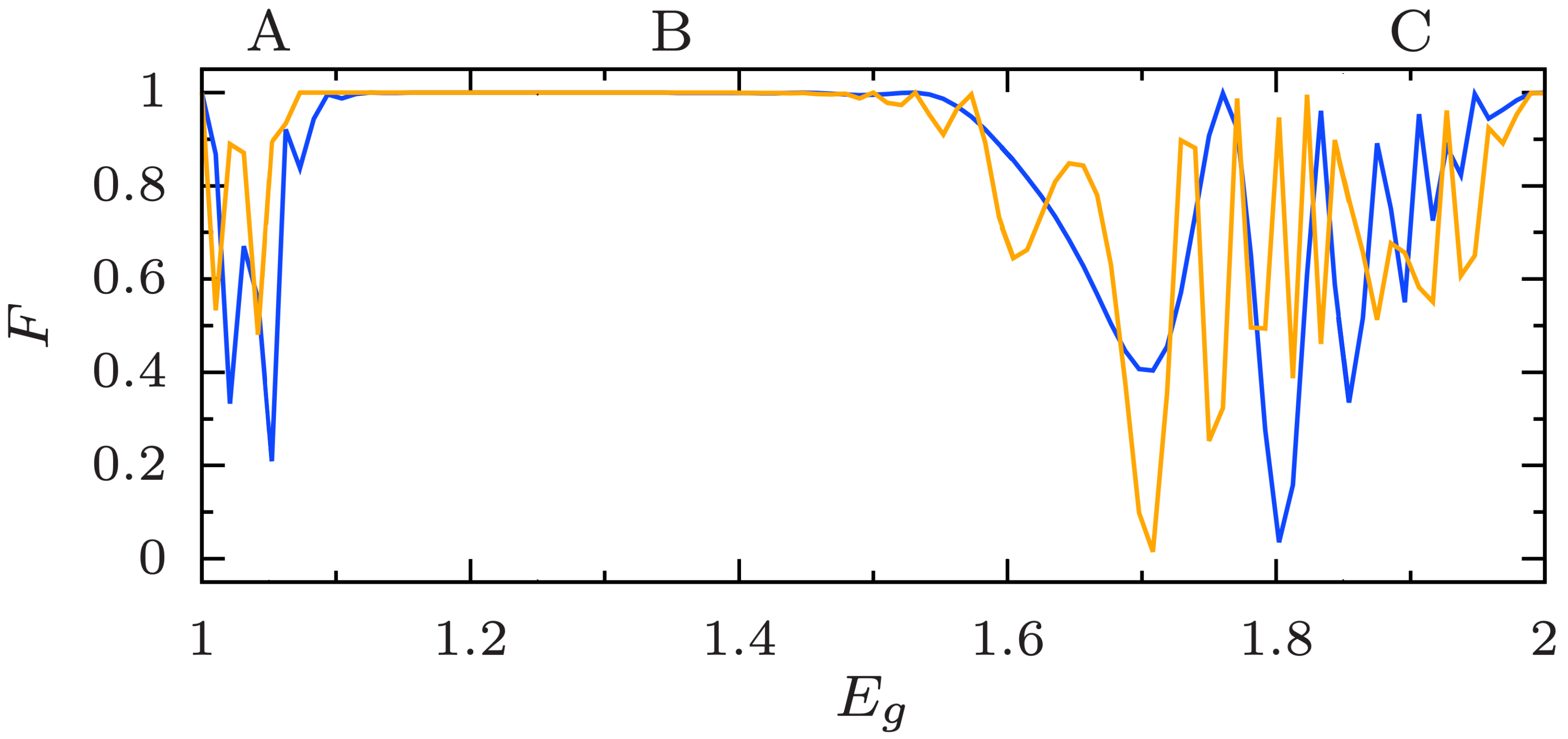}\hfill
    \caption{Final population in state $|R\rangle$ as a function of the ground state energy of the two particle system (in harmonic oscillator units) after the SAP protocol is carried out over a total time $T = 4000$ [blue (dark gray)] or $T = 12 000$ [orange (light gray)] with the system initially in state $|L\rangle$. In the B region the decoupled energy band allows for high-fidelity transfer, whereas in the A and C regions level crossings with the ground and first exited band, respectively, lead to transitions out of the dark state. Figure from \citep{Benseny:2016}}
    \label{fig:2PSAP}
\end{figure}

In many-particle systems the interaction energies need to be included into the Hamiltonian \eqref{eq:H3}, which means that the {\it dark state} solution used for high-fidelity transport is no longer available. This problem was studied in \citep{Benseny:2016} for a few-body bosonic system over the whole range of repulsive interactions.  While the non-interacting and the Tonks-Girardeau limits can obviously be described by single particle states and are therefore amenable to SAP, a large and continuous region of intermediate interactions was also identified in which high transport fidelities can be obtained (see Fig.~\ref{fig:2PSAP}). In this region the system develops a decoupled energy band which possesses a {\it dark state} that uses  two-particle cotunneling. However, when approaching the non-interacting and infinitely interacting limits this energy band overlaps with other energy levels creating energy level crossings which make it difficult to follow the dark-state at all times. 

The SAP scheme also allows to transport only a predefined number of particles from a larger, interacting ensemble \citep{Reshodko:17}, for instance $|L,M,R\rangle=|N,0,0\rangle \rightarrow |N-M,0,M\rangle$. This is done by adjusting the onsite energies so that a {\it dark-state} resonance becomes possible for well defined interaction energies only. Similar work to this, but considering emission into free space from a trapped gas, was carried out in \citep{Lode:14} and \citep{Dobrzyniecki:18escape} and clearly showed how the interaction energies can be used to control the many-body boson tunneling process.

Another wide area of study concerning transport is Josephson tunneling for small systems with interactions. While this problem is well understood for mean-field settings, an exact numerical study by \citep{Sakmann:09} showed that the presence of correlations and the possibility of fragmentation can lead to different and richer dynamics, even for weak interactions. The numerical technique used in this work was the multiconfigurational time-dependent Hartree method for bosons, which allowed the study of systems with up-to 100 particles. In the  strongly correlated regime the Josephson
dynamics among two Luttinger liquids or two Tonks gases coupled head-to-tail through a weak link was studied in \citep{Polo:18}. This work found that the oscillations are intrinsically damped due to a coupling to the longitudinal, collective excitations of the system. It also provided analytical expressions for the natural frequencies and damping rates as a function of the microscopic parameters of the model.

The importance of interband tunneling in small bosonic samples in triple-well potentials was carefully examined in \citep{Cao:11}, where the authors showed that robust tunneling can be achieved in certain interaction regimes. Interestingly this process is, in principle, not limited to small systems, but can be generalized to multiwell systems and even larger cloud sizes. Tuning the interaction strength allows one to control the tunneling between arbitrary spatial and energetic configurations of bosons in a multiwell trap, thereby opening the door for controllable dynamical transport single atoms to specific wells and specific in-well energy levels. In \citep{LU2015} it was also shown that periodic driving of the interactions can allow a single particle to tunnel out of a small ensemble trapped in the middle well.

The correlated tunneling dynamics of an impurity species inside a lattice trapped finite-sized gas of interacting bosons was explored in \citep{keiler2020doping}. For strong interactions between the impurities and the gas an insulating phase emerges, which manifests itself in the localization of the impurity at the edge of the lattice and the creation of a particle-hole pair. After a quench from the insulating to the superfluid phase the impurity tunnels through the gas and can refocus fully on the opposite side of the lattice. Adding a second impurity allows for the creation of impurity-impurity correlations, however stable transport is more difficult to achieve in this case.

\begin{figure}[tb]
    \centering
    \includegraphics[width=1\linewidth]{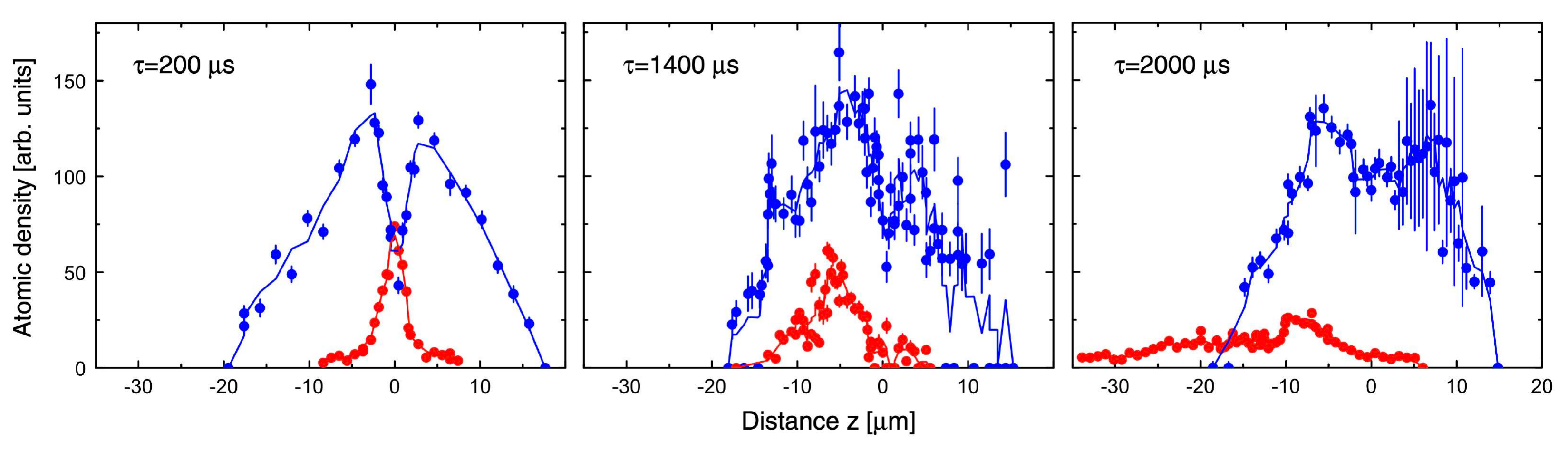}\hfill
    \caption{Density profiles for a small TG gas falling through a bigger sample for different times $\tau$ after the outcoupling. The trapped component is given by the upper blue curve and the impurity component by the lower red curve. Figure from \citep{Palzer:09}.}
    \label{fig:TGTransport}
\end{figure}

The transport of a small group of atoms in the Tonks-Girardeau (TG) limit through a TG gas of a different species was experimentally studied in \citep{Palzer:09} (see Fig.~\ref{fig:TGTransport}) and theoretically analysed in \citep{Rutherford:11}. This situation can be created when a small number of atoms is spin-flipped into an untrapped state inside a larger TG gas sample. Due to the gravitational pull the untrapped sample then falls down through the remaining gas. Such a system shows complex nonequilibrium dynamics that depends on the inter-species interaction strength and even possess a regime of atom blockade, where the free component is trapped by the other matter wave.

However, the interaction in small gas samples can also be used to inhibit transport. A microscopic analogue of source–drain transport with ultracold bosonic atoms in a triple-well potential was discussed in \citep{Schlagheck:10}. Using a time-dependent tilt in a triple-well potential provides a gate voltage and the authors observed that the transport was hampered over large areas of the parameter space by the interaction blockade.

The opposite of transport, i.e.~the phenomenon of localisation, has attracted a lot of interest over the recent years, in particular due to the fact that interactions can be important. This has been dubbed many-body-localisation \citep{nand2015,alet2018} and has become a topic of great interest in cold atoms as well \citep{abanin2019RvMP}. For a small bosonic gas in the presence of spatially correlated disorder due to an optical speckle field it was explored in \citep{Mujal:2019}. Analyzing a broad range of interaction strengths the authors found that a contact potential does not destroy a Poisson level-spacing distribution that was carefully engineered for the non-interacing case, which indicated that localization can occur also in interacting few-body systems in a continuum. This behaviour can also be found in the limit of strong interactions within a spin-chain model of cold atoms in 1D \citep{duncan2017FBS} and, indeed, recent experiments have found evidence for many-body localization for systems of up to eight atoms \citep{Lukin:19}.

\subsection{Quantum control}

While precise control of non-interacting and mean-field quantum systems has been well established using both numerical optimization protocols and analytic shortcuts-to-adiabaticity (STAs) \citep{STAs}, these techniques have recently started to be extended to few-body systems with finite contact interactions. This has practical considerations for the implementation of collisional quantum gates, whereby adiabatic dynamics are needed to ensure high fidelity \citep{Calarco2000} with the trade off being longer gate times. In order to speed up the dynamics while maintaining high fidelity gates, \citep{Jensen2019} devised an optimal control scheme using the GRAPE algorithm to control the merging and subsequent separation of two interacting particles in a time-dependent double well potential. Using the fast quasiadiabatic technique (FAQUAD) \citep{Martinez2015} controlled the tunnelling of two interacting particles in a double well showing co-tunnelling and splitting processes, and also created mesoscopic superposition states of strongly interacting bosons in a ring trap.  

In a similar manner the interaction strength between the particles can be controlled directly by modulating the applied magnetic field, allowing one to quickly ramp from weak to strong interactions and, in the process, create large correlations. To minimize irreversible excitations and fluctuations in the entanglement, \citep{Fogarty2019} derived an approximate STA for the relative motion of two atoms based on a superposition ansatz between the initial and target states, $\psi_i(x)$ and $\psi_f(x)$ respectively. The ansatz is given by
\begin{equation}
    \Phi_c(x,t)=\mathcal{N}(t)\left[ \left(1-\eta(t) \right)\psi_i(x) + \eta(t) \psi_f(x)   \right] e^{i b(t) x^2},
\end{equation}
with $\mathcal{N}(t)$ being the normalization factor and $b(t)$ the so-called chirp which accounts for the dynamical phases. The dynamics is controlled by an arbitrary switching function $\eta(t)$, which is chosen to smoothly change the state from $\psi_i(x)$ to $\psi_f(x)$. The ansatz is used to minimize the effective Lagrangian of the system to ultimately yield the optimal interaction ramp $g(t)$, which gives a big advantage over non-optimized ramps when driving to strong interactions. This technique is robust to fluctuations in $g(t)$ and has been extended to three particle and two component systems \citep{Kahan2019}, while optimal control techniques have been used to efficiently drive interactions in similar fermionic systems \citep{Li2018}.

Interparticle interactions can have detrimental effects on the control of quantum systems. For instance, controlling the dynamics of particles in the bosonic Josephson junction to transport all particles from one well to another can be hindered by the presence of even weak interactions, resulting in large infidelity far from the quantum speed limit. In \citep{Brouzos2015} a compensating control pulse (CCP) to control the zero point energy of each well was used to negate the destructive effects of the interactions and allow the system to follow a geodesic path at the quantum speed limit. This is the fastest theoretically allowed time to transfer from one prescribed state to another in a given Hilbert space and under the evolution of a given class of Hamiltonians. A simple analytic pulse was derived from a two-mode theory $D_{\text{CCP}}(t)=U \cos^2 (Jt)$, where $J$ is the coupling between the two wells and $U$ is the nonlinear interaction strength. This was compared with a numerically optimized chopped random basis algorithm (CRAB) pulse which allows one to control the system when CCP would not work effectively due to constraints on the pulse. The many-body state was described using the MCTDHB method and also compared with the mean-field Gross-Pitaevskii theory to understand many-body effects. The non-optimized CCP pulse drastically improves the fidelity of the transport for both the mean-field and many-body states. However, close to the quantum speed limit differences between these approaches become substantial. In this regime, the CRAB method was shown to only slightly decrease the infidelity for the mean-field description, while it is necessary for optimization of the many-body state when quantum depletion and fragmentation take place resulting in reduction of infidelity up to an order of magnitude when compared to CCP.

%% file: section08.tex
\section{Quantum droplets}\label{sec:Droplets_contact} 

Another intriguing phase of matter is the so-called liquid phase originating from the competition between attractive and repulsive interactions. 
Liquid helium is a prototype example that appears in nature forming self-bound liquid droplets which are dense nanometer-sized strongly interacting clusters of helium atoms~\citep{barranco2006helium,dauxois2006physics}. 
Their underlying binding mechanism indeed stems from the balance between attractive and repulsive forces with the former driving the system to collapse and the latter stabilizing it to a finite size. Quantum droplet formation takes place also in bosonic ultracold atomic mixtures experiencing two-body contact interactions, that are tunable through Feshbach resonances, see also Section~\ref{dipolar_droplets} for the formation of relevant structures in dipolar gases~\cite{bottcher2020new,ferrier2016observation,kadau2016observing}.  
In particular, referring to 3D, droplet nucleation generically appears in the regime where the interspecies attraction is larger than the average intraspecies repulsion. 
Here, according to mean-field theory, the mixture is expected to collapse~\citep{donley2001dynamics,roberts2001controlled} but the involvement of quantum fluctuations compensates this effect providing in turn a crucial component for the stabilization of the system and the subsequent generation of a quantum droplet.  
Therefore, standard mean-field (Gross-Pitaevskii) treatment would not allow the formation of quantum droplets.  
Importantly, the presence of quantum fluctuations yields a minimum of the energy at a non-zero density, which is a necessary condition for the occurrence of a liquid state at zero temperature~\citep{petrov2015quantum,Petrov2016Liquids}. 
To describe this physics, it is common to use the first beyond mean-field energy correction to the mean-field energy, namely the Lee-Huang-Yang (LHY) term~\citep{huang1957quantum,lee1957many}, which was first introduced to describe hard core repulsive bosons. Nevertheless, the relevance of higher-order quantum correlations and few-body effects on these self-bound structures is a highly active research field the state-of-the-art 
of which will be summarized in this review.

\begin{figure}[tb]
    \center
    \includegraphics[width=0.7\textwidth]{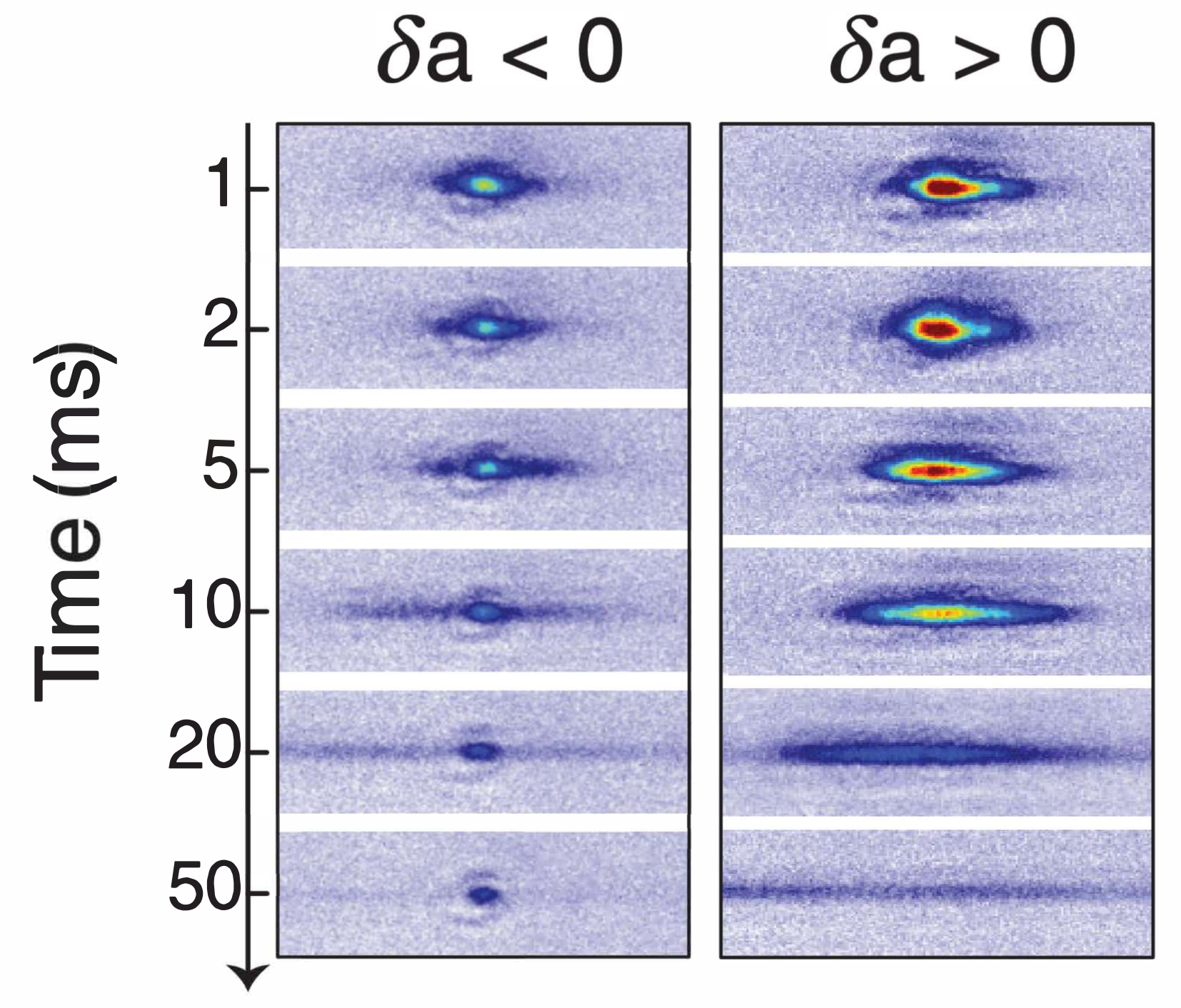}
    \caption{\textit{In-situ} images after releasing a BEC in the quasi-1D optical waveguide for different evolution times. A self-bound state is realized in the attractive regime with $\delta a<0$, whereas for the repulsive region $\delta a>0$ the gas can be seen to expand. 
    Here, $\delta a = a_{\uparrow \downarrow}+\sqrt{a_{\uparrow \uparrow} a_{\downarrow \downarrow}} \propto \delta g$, with $a_{\uparrow \downarrow}$ denoting the interspecies 3D $s$-wave scattering length.  
    Figure from  \citep{cheiney2018bright}
    }
\label{fig:droplets_experiment} 
\end{figure}

Quantum droplet formation was originally theoretically proposed in the seminal paper by Petrov~\citep{petrov2015quantum} and consecutively observed in various 3D ultracold atom experiments, see for example Fig.~\ref{fig:droplets_experiment}. 
The experiments were performed with homonuclear mixtures of $^{39}$K in different hyperfine states both in the presence of an external trap and in free space~\citep{cabrera2018quantum,semeghini2018self,cheiney2018bright,ferioli2019collisions,lavoine2021beyond} as well as in heteronuclear mixtures composed of $^{41}$K and $^{87}$Rb~\citep{d2019observation,burchianti2020dual}. 
More recently, quantum droplets were also realized in heteronuclear mixures of $^{23}$Na and $^{87}$Rb \citep{guo2021lee}. 
These experimental demonstrations motivate a new era of theoretical investigations concentrating, for instance, on the collisional aspects of these states~\citep{ferioli2019collisions} revealing a transition from compressible to incompressible regimes and their relation to non-linear structures~\citep{kartashov2018three}, see also the relevant review of~\citep{luo2021new}.
Deviations from Petrov's theory have been explored by utilizing either variational treatments~\citep{parisi2019liquid,cikojevic2019universality,parisi2020quantum,cikojevic2018ultradilute} or including higher-order energy correction terms in a self-consistent way~\citep{hu2020microscopic,hu2020microscopic1,ota2020beyond,hu2020collective,gu2020phonon}, or considering three-body contact interactions in the absence of two-body ones~\citep{sekino2018quantum,morera2021quantum}. 
A peculiar feature of these quantum objects in higher-dimensions is their self-evaporation, i.e. the fact that in certain parameter regimes they are not able to sustain any collective mode~\citep{petrov2015quantum}. Also, droplets in homonuclear mixtures suffer from significant three-body losses which lead to typical lifetimes being of the order of a few milliseconds~\citep{cabrera2018quantum,semeghini2018self}. 
This behavior was theoretically predicted in~\citep{ferioli2019collisions}. 

However, as heteronuclear mixtures provide promising setups to circumvent this problem in higher-dimensions~\citep{d2019observation}, a result that is theoretically supported~\citep{fort2020self}, the realization of long-lived quantum droplets has become a topic of increasing interest. 
In this sense, 1D geometries where liquid states can indeed also be formed~\citep{petrov2015quantum,Petrov2016Liquids}, see also Fig.~\ref{fig:miscible_transition}, provide ideal platforms for probing the properties of these structures since three-body losses are suppressed~\citep{astrakharchik2006correlation}. 
Also, it is experimentally possible to reach the strongly correlated regime~\citep{stoferle2004,kinoshita2004,haller2009,paredes2004}, which is not easily accessible in higher-dimensions and the role of quantum fluctuations is naturally more enhanced in 1D~\citep{giamarchi2003,cazalilla2011}. 
For the above-mentioned reasons the relevant 1D scenario has received particular attention and it will be the focus of our discussion below.

Before delving into the specific 1D properties of quantum droplets we should emphasize their intrinsic differences when compared to higher-dimensional cases. 
First and foremost, the corresponding beyond mean-field contribution depends strongly on the density of states and therefore on the dimensionality of the system. Indeed, in higher-dimensions the underlying stabilization mechanism stems from the balance of attractive mean-field interactions $\sim n^2$ and repulsive correlations $\sim n^{5/2}$, where $n$ denotes the density of the gas. 
In sharp contrast in the 1D case droplets manifest as a consequence of the competition of attractive quantum fluctuations $\sim n^{3/2}$ and the mean-field repulsion, a result that becomes directly evident from the corresponding energy functional (see also Eq.~(\ref{LHY_energy_general}) below). 
As such, they are expected to form in the interaction region where according to mean-field theory a homogeneous binary mixture is still stable~\citep{Petrov2016Liquids}. 
They are anticipated to also persist for small particle numbers and independently of the sign of the mean-field interaction term contrary to 3D geometries. 
Another interesting feature of 1D droplets is their distinction from bright solitons especially for small particle numbers. This difference manifests itself in the fact that droplets generically do not support internal normal excitation modes~\citep{tylutki2020collective,cappellaro2018collective} and exhibit inelastic collisions~\citep{Petrov2016Liquids,astrakharchik2018dynamics} in contrast to bright solitons.

The simplest framework to describe the stationary and dynamical properties of quantum droplets, irrespective of the dimension, is the so-called generalized Gross-Pitaevksii (GGP) approach. It incorporates the first-order effect of quantum fluctuations within the local density approximation~\citep{petrov2015quantum,astrakharchik2018dynamics},  see below for a detailed discussion.  
As such, it is necessary to initially analyze this concept in order to provide a smooth introduction and a  comprehensive overview to the reader into this relatively new research direction before reporting on results predicted by  \textit{ab-initio} methods which are naturally restricted to the few- and mesoscopic-body regimes. 
Indeed, later on, more elaborated variational treatments such as Quantum Monte Carlo techniques~\citep{parisi2019liquid,cikojevic2019universality,parisi2020quantum,cikojevic2018ultradilute} and ML-MCTDHX~\citep{mistakidis2021formation} have been employed, mainly focusing on relatively small particle numbers and on stationary properties, in order to benchmark the validity of the GGP predictions. 
These studies demonstrated various deviations from the GGP framework attributed to residual beyond LHY correlations for increasing interactions where GGP is expected to fail.  
Particularly, within the strongly interacting regime and for few- and mesoscopic atom numbers it is naturally anticipated that the impact of higher-order correlations beyond the validity of the GGP will be dominant.  
Summarizing, the majority of investigations to date rely on relatively large weakly correlated atomic systems adequately described by the GGP approach and there is a comparatively smaller amount of few-body studies.  
While the latter are the main focus of our presentation below, it is necessary for self-consistency reasons and the readers' understanding to report on the basic properties of the GGP since the droplet concept originates from this approximation.

\subsection{First order quantum correction within the generalized Gross-Pitaevskii framework} 

Let us initially concentrate on a general case of a 1D two-component BEC in free space of length $L$ described by the Hamiltonian of Eq.~(\ref{eq:Hamiltonian_pseudospinor}) introduced in Sec.~\ref{sec:spinor} with $V_A(x)=V_B(y)=0$. 
Particularly, the two-component system is characterized by intracomponent repulsion coefficients $g_1>0$, $g_2>0$ and an intercomponent attraction $g_{12}<0$. 
The mass $m$ of both components is considered here to be the same and $n_i=\abs{\Psi_i}^2$ denotes the density of each component represented by the wave function $\Psi_i$ with $i=1,2$. 
In the following, particle/density balanced setups, $N_1=N_2=N/2$ or $n_1=n_2\equiv n/2$, where additionally $g_1=g_2$ holds, will be referred to as symmetric in order to be discerned from asymmetric settings in which at least one of these conditions is not satisfied. 

The beyond mean-field energy density of the asymmetric system has been derived within second-order perturbation theory in~\citep{Petrov2016Liquids,Lieb1963GS} accounting for quantum fluctuations in the local density approximation.  It is given by 
\begin{equation}
    \mathcal{E}_{1D}=\frac{(\sqrt{g_1} n_1-\sqrt{g_2} n_2)^2}{2}+\frac{g \delta g(\sqrt{g_2} n_1+\sqrt{g_1} n_2)^2}{(g_1+g_2)^2}-\frac{2 \sqrt{m}(g_1 n_1+g_2 n_2)^{3/2}}{3\pi \hbar}, 
    \label{LHY_energy_general} 
\end{equation}
where $\delta g \equiv g_{12}+g$ and $g\equiv \sqrt{g_{1}g_{2}}$, see also Fig.~\ref{fig:miscible_transition}. The first two terms in Eq.~(\ref{LHY_energy_general}) denote the mean-field contribution and  the last one corresponds to the leading-order beyond-mean-field LHY correction that accounts for quantum many-body effects~\citep{Petrov2016Liquids}. 
The correction due to quantum fluctuations in Eq.~(\ref{LHY_energy_general}) is relevant if $\abs{\delta g} \ll g$, i.e., when the mean-field repulsion and attraction are balanced; otherwise, they are expected to be suppressed and one should use the standard Gross-Pitaevskii equation to describe the system.
Since in 1D the beyond-mean-field terms account for a negative correction to the energy functional, in particular $\propto -n^{3/2}$, a positive mean-field imbalance i.e. $\delta g>0$ is required for ensuring an energy minimum. 
Note that the sign of the LHY term strongly depends on the dimensionality of the system. 

In the 3D case the LHY contribution $\propto +n^{5/2}$ is positive and thus 
$\delta g <0$ is needed for producing an energy minimum~\citep{petrov2015quantum}. 
An analysis of the dimensional crossover from 3D to 1D and from 3D to 2D is partially discussed in \citep{zin2018quantum,ilg2018dimensional,luo2021new}. 
However, it should be emphasized that the impact of the LHY term in each crossover is currently a largely unexplored and active research topic.
As such, the formation of 1D quantum droplets stems from the balance of the weakly repulsive mean-field interaction $\delta g >0$ and the LHY term introducing an effective attraction. 
Another intriguing but much less explored regime, irrespectively of the dimension, arises at the mean-field balance point where $\delta g=0$. 
There the mixture is governed only by the LHY contribution since the mean-field interactions are cancelled and the so-called LHY fluid forms. 
This phase was experimentally probed in 3D using either different hyperfine states of $^{39}$K~\citep{skov2021observation} or the different isotopes $^{23}$Na and $^{87}$Rb~\citep{guo2021lee}. Experimentally, $\delta g$ can be tuned~\citep{cabrera2018quantum,semeghini2018self,cheiney2018bright} to be either positive or negative. 
Note that a corresponding discussion for the energy functional and existence of a 1D quantum droplet in a Bose-Fermi mixture can be found in~\citep{rakshit2019self} where it has been argued that the role of quantum fluctuations is negligible for the droplet formation. 

Within the framework of the Bogoliubov theory it has been demonstrated~\citep{petrov2015quantum,Petrov2016Liquids} that the coefficients $g$ and $\delta g$ are related to the so-called hard ``spin'' and soft ``density'' modes, respectively. Thus, the condition $\abs{\delta g} \ll g$ results in the separation of the two involved characteristic spatial scales. 
Indeed, in quantum droplets the variation of their density profile is determined by the ``soft'' healing length $\xi_{-}=\hbar/(\sqrt{2} m c_{-}$) while the beyond mean-field energy originates from distances comparable to the ``hard'' healing length $\xi_{+}=\hbar/(\sqrt{2}m c_{+})$, where $c_{\pm}=\sqrt{(g \pm g_{12})n/m}$ is the speed of the associated Bogoliubov modes~\citep{Petrov2016Liquids}. 
It is evident that $\delta g$ and $g \gg \delta g$ define the speed of the ``soft'' and ``hard'' modes respectively within Bogoliubov theory~\citep{lifshitz1980statistical} while the relatively large scale separation determined by $\xi_{-} \gg \xi_{+}$ justifies the local character of the LHY correction term which allows one to describe the dynamics taking place at timescales $t \gg t_{+}=(2m \xi_{+}^2)/\hbar$.

The Euler-Lagrange variational principle \citep{pitaevskii2016bose} leads to the coupled system of GGP equations of motion for the asymmetric mixture 
\begin{align}
\begin{split}
 i \hbar \frac{\partial \Psi_1}{\partial t}= - \frac{\hbar^2}{2m} \frac{\partial^2 \Psi_1}{\partial x^2}+ (g_1+G g_2) \abs{\Psi_1}^2 \Psi_1&-(1-G)g\abs{\Psi_2}^2 \Psi_1\\&- \frac{g_1 \sqrt{m}}{\pi \hbar} \sqrt{g_1 \abs{\Psi_1}^2+g_2\abs{\Psi_2}^2} \Psi_1,\\ 
 i \hbar \frac{\partial \Psi_2}{\partial t}= - \frac{\hbar^2}{2m} \frac{\partial^2 \Psi_2}{\partial x^2}+ (g_2+G g_1) \abs{\Psi_2}^2 \Psi_2&-(1-G)g\abs{\Psi_1}^2 \Psi_2\\&- \frac{g_1 \sqrt{m}}{\pi \hbar} \sqrt{g_2 \abs{\Psi_1}^2+g_2\abs{\Psi_2}^2} \Psi_2,\label{GGP_asymmetric}
 \end{split}
\end{align}
where the parameter $G=2g \delta g/(g_1+g_2)^2$ measures the deviation from the balance point of mean-field repulsion and attraction. 
Extensions of the GGP equations and the form of the corresponding LHY term have been discussed in 3D when taking into account either mass-imbalanced mixtures \citep{burchianti2020dual,minardi2019effective} or Rabi-coupling between the condensate components \citep{lavoine2021beyond,cappellaro2017equation}.

For a symmetric mixture with $n_1=n_2\equiv n/2$ and $g_{1}=g_{2}=g$, Eq.~(\ref{GGP_asymmetric}) reduces to a single-component GGP describing a Bose liquid \citep{Petrov2016Liquids,astrakharchik2018dynamics} under the assumption that 
the corresponding `wave function' is expressed through the individual components as $\Psi(x)=\Psi_i(x)\sqrt{\sqrt{g_{i}}+\sqrt{g_{\bar{i}}}}/g_{\bar{i}}^{1/4}$, where if $i=1$ ($i=2$) then $\bar{i}=2$ ($\bar{i}=1$).  
This reduced single-component equation admits two fundamental exact solutions, namely the so-called quantum droplet of  finite size having a flat-top density profile and the plane-wave solution with a uniform density. 
Their properties have been extensively discussed in \citep{Petrov2016Liquids,astrakharchik2018dynamics} and,
for completeness, we briefly review the basic characteristics of the quantum droplet solution existing at $\delta g/g >0$ and having the form~\footnote{Here the wave function is expressed in units of $\hbar/\sqrt{mg}$, while the length (time) in terms of $mg/\hbar^2$ ($mg^2/\hbar^3$) as in Ref.~\cite{mithun2020modulational}.} 
\begin{equation}
\Psi(x;t)\propto \frac{A e^{-i \mu t}}{1+B\cosh(\sqrt{-2\mu}x)}.\label{droplet_wfn} 
\end{equation}
Here, the amplitudes are $A=\sqrt{n_0}\mu/\mu_0$ and $B=\sqrt{1-\mu/\mu_0}$, where $\mu$ is chemical potential of the system while $n_0\propto [(\sqrt{8}g)/(3\pi \delta g)]^2$ and $\mu_0\propto-4g/(9\pi^2 \delta g)$ are the equilibrium density and its associated chemical potential. The particle number in terms of the chemical potential of the exact quantum droplet solution is $N(\mu)=n_0\sqrt{-2/\mu_0}[\ln((1+\sqrt{\mu/\mu_0})/(\sqrt{1-\mu/\mu_0}))-\sqrt{\mu/\mu_0}]$.

The solution of Eq.~(\ref{droplet_wfn}) originates from the balance of the cubic self-repulsion and quadratic attraction of the beyond mean-field equation.  
It occurs within the interval $\mu_0<\mu< 0$ and exhibits a flat-top shape for $\mu-\mu_0\ll \abs{\mu_0}$ of size $L \approx (-2 \mu_0)^{-1/2} \ln [1-(\mu/\mu_0)^{-1}]$. 
As a consequence, the size of the droplet diverges for $\mu=\mu_0$ and thus its wave function tends to a delocalized plane-wave solution of maximally uniform density $n=n_0$. 
This implies the incompressibility of the BEC which can be accordingly considered as a fluid. 
Importantly, the full-width-at-half-maximum of the quantum droplet $n(x=0)/2=n_0(\mu/\mu_0)^2(1+\sqrt{1-(\mu/\mu_0)})^{-2}$ decreases for increasing $\mu$ reaching the value $\sim 2.36/\sqrt{\mu_0}$ at $\mu\approx 0.776 \mu_0$ and afterwards as $\mu \to 0^{-}$ it becomes larger and diverges as $\sim 1.71/\sqrt{-\mu}$. 

Finally, let us note that the above-described exact quantum droplet solution exists also in the range $\delta g /g <0$ and features a crossover from a Korteweg-de Vries~\citep{lamb1980elements,hirota1981soliton} $\sim \sech^2(\sqrt{-\mu/2x})$ to a bright-like soliton $\sim \sech(\sqrt{-2\mu}x)$~\citep{mithun2020modulational}. 
Other types of exact solutions of the single-component GGP, corresponding for instance to dark and anti-dark solitons or kink structures, as well as for the asymmetric mixture described by Eq.~(\ref{GGP_asymmetric}) and, for instance, featuring Gaussian-like shapes or expressed in terms of Jacobi elliptic functions, have been addressed in some detail in~\citep{mithun2020modulational}. Their stability properties are still an open issue.

\subsection{Stationary properties of one-dimensional droplets}\label{droplet_GS}

The ground state phase diagram in free space and the characteristic properties of quantum droplets including correlation functions and surface tension have been studied for a symmetric mixture in \citep{parisi2019liquid,parisi2020quantum} using a numerically exact Diffusion Monte Carlo (DMC) method. 
In the large particle number (thermodynamic) limit it was demonstrated that there is a critical ratio $\abs{g_{12}}/g \approx 0.45$ above which the ground state energy of a uniform system exhibits a minimum at a finite density. 
This situation corresponds to the liquid phase which can be either stable or unstable, breaking into droplets above and below the spinodal point, respectively, where the compressibility diverges. 
Below this critical value the energy increases for larger densities, which signals the presence of a gas phase. 
Here, the energy minimum corresponds to half of the binding energy of dimers i.e. $\epsilon_b/2=-\hbar^2/(2m \tilde{a}^2)$ with $\tilde{a}=2\hbar^2/(m\abs{g_{12}})$ being the 1D scattering length of the interspecies contact potential and it is associated with a vanishing density. 
Note also that this value of $\abs{g_{12}}/g$ is in accordance with the prediction of the four-body scattering problem and in particular the crossing point of the effective interaction between dimers from repulsive to attractive values~\citep{pricoupenko2018dimer}. 

Comparing these exact results with the GGP predictions, an adequate agreement has been demonstrated in high density regions but large deviations have been found at small densities where the weak coupling theory is not valid. 
Moreover, by inspecting suitable pair correlation functions it was found that there are interaction regimes of strongly correlated liquid states which lie beyond the applicability of the GGP framework. 
Notice also that based on Bogoliubov theory the phase diagram and the thermodynamic properties of 1D weakly interacting liquids in terms of the intercomponent interaction and the temperature have been discussed in~\citep{de2021thermal}. 
It was shown that the dominant contribution to the thermal effects stems from the excitation of the soft density mode.

\begin{figure}[tb]
    \center
    \includegraphics[width=\linewidth]{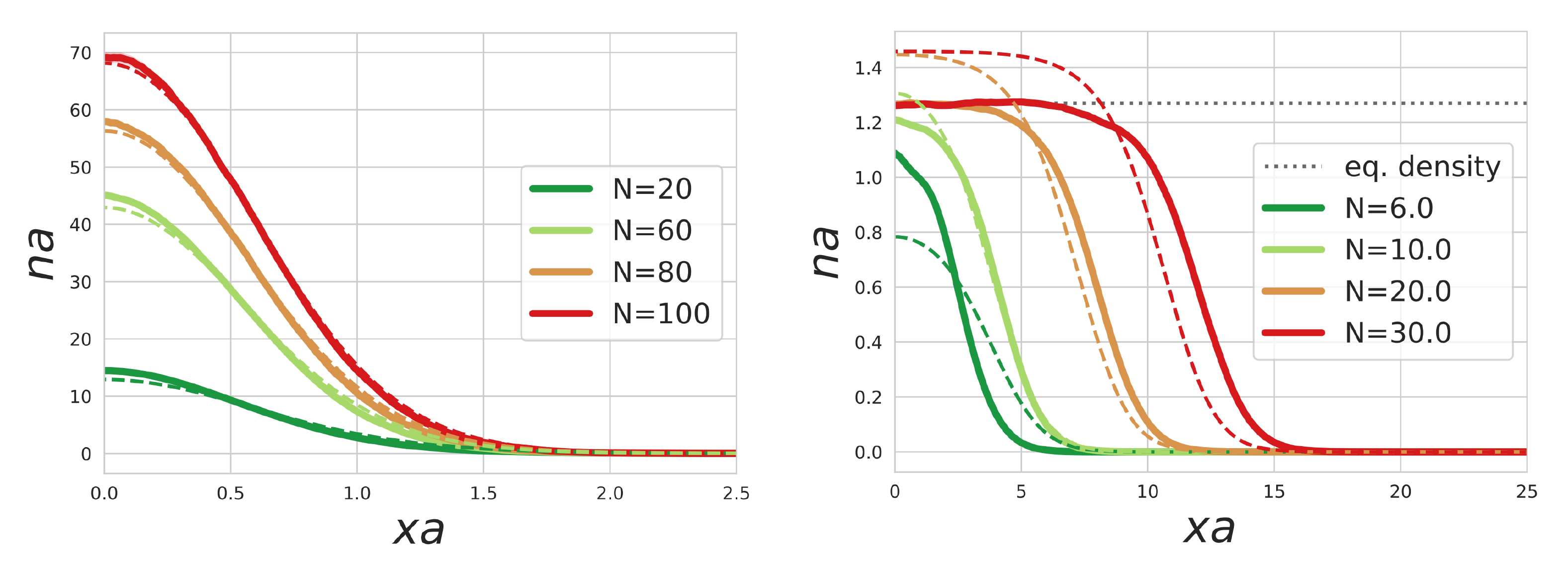}
    \caption{Ground-state density profiles of 1D droplets for varying particle number of a symmetric mixture as predicted within the DMC approach (solid lines). Shown are the cases of $\abs{g_{12}}/g=0.95$ (left panel) and $\abs{g_{12}}/g=0.6$ (right panel). 
    Dashed lines represent the density obtained in the GGP framework, whilst the dotted line corresponds to the equilibrium density of a uniform mixture at $\abs{g_{12}}/g=0.6$ as calculated using DMC. The normalization parameter ${\rm a}=-2 \hbar^2/(mg)$ refers to the 1D intracomponent scattering length. Figure from \citep{parisi2020quantum}.}
\label{fig:droplet_profiles_QMC} 
\end{figure}

Interestingly, a quantitative comparison between the DMC and the GGP approach on droplet properties as captured by their density profiles, surface tension and breathing mode frequency yielded a very good agreement for symmetric mixtures consisting of a finite particle number~\citep{parisi2020quantum}. 
Deviations between the results of the two methods were mainly identified within interaction regions where the GGP theory is not expected to be accurate, i.e. $\abs{g_{12}}\ll g$, see for instance Fig.~\ref{fig:droplet_profiles_QMC}.

We note in passing that for large particle numbers, the effect of the surface tension is important leading to the formation of a self-bound nonuniform configuration. The surface tension is in general contained in the energy density of the system. 
In 1D and for large particle numbers the energy per particle is expressed with respect to the bulk energy $E_B$, which provides the energy per particle of a homogeneous system in the thermodynamic limit, and the surface energy $E_S$ is the first order correction due to the presence of a finite particle number, leading to
\begin{equation}
 \frac{E}{N}\approx E_B+\frac{E_S}{N}+\dots.\label{energy_surface_tension}
\end{equation}
Higher-order correction terms are neglected here, for simplicity. 
The surface tension of the droplet is determined by the ratio of the surface energy $E_S$ to the surface area, which in 1D corresponds to only two points. Therefore this ratio takes the value $E_S/2$. 

\begin{figure}[tb]
    \center
    \includegraphics[width=\linewidth]{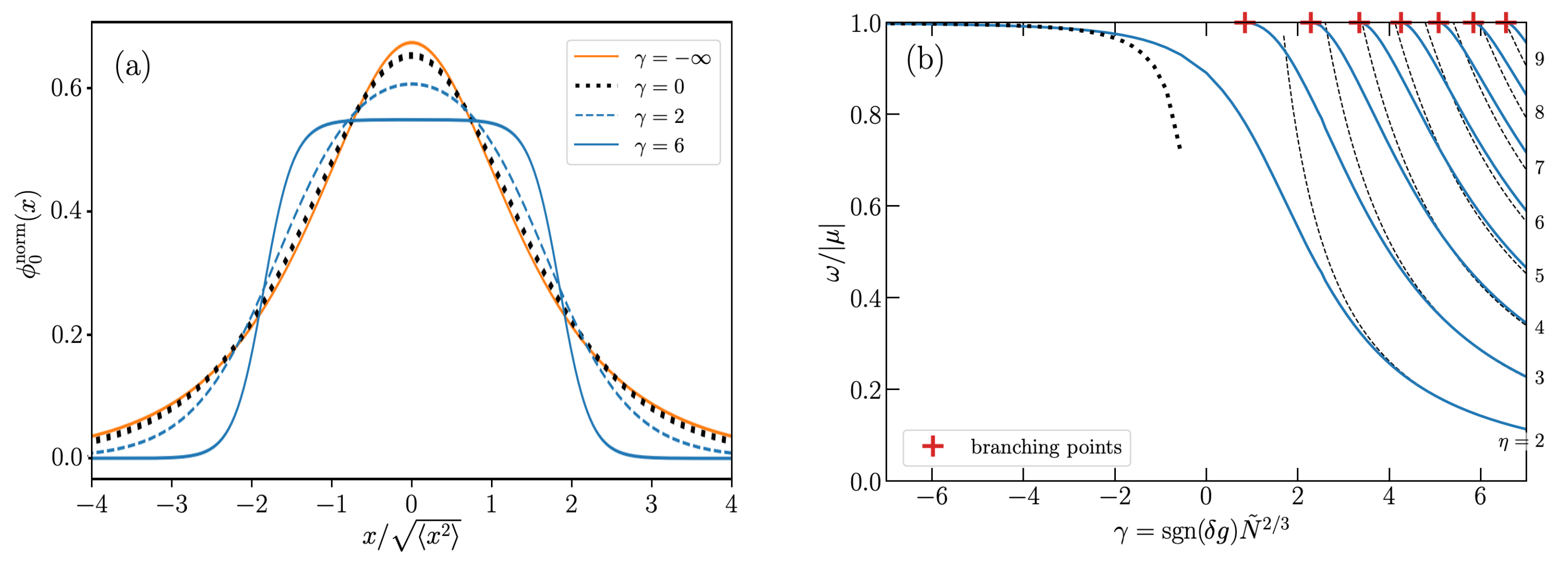}
    \caption{(a) Droplet `wave function'  [Eq.~(\ref{parametrization_droplets})] features a  transition from soliton-like to flat-top configurations when varying $\gamma$. (b) The ratio of discrete Bogoliubov frequencies $\omega_{\eta}$ to the particle-emission threshold $-\mu$ with respect to $\gamma$. Red crosses signify the so-called branching points at which the discrete modes cross $-\mu$ and enter the continuum of excitations. Dashed and dotted curves represent asymptotic approximations of the spectrum and the energy of the bound state respectively. Figure adapted from \citep{tylutki2020collective}. }
\label{fig:droplet_spectrum} 
\end{figure} 

This linear dependence of the energy $E$ on $1/N$ is captured for increasing $N$ both within the DMC approach and the GGP theory. 
However, it has been argued that the GGP underestimates (overestimates) $E$ ($E_S$) for $\abs{g_{12}}/g<0.95$. Regarding the density of the mixture it has been demonstrated~\citep{parisi2020quantum} that, for a constant ratio $\abs{g_{12}}/g$ and increasing particle number, the density of the mixture becomes higher and wider, while in the large $N$ limit the density near the center of the cloud becomes flat reaching the equilibrium value of the uniform system, see Fig.~\ref{fig:droplet_profiles_QMC}. 
Therefore, an important result is that for large $N$, the density of the bulk is independent of the particle number whereas the radius of the cloud continues to expand. 
The GGP framework shows differences compared to the DMC predictions already for $\abs{g_{12}}/g\approx0.95$, which are  attributed to residual beyond LHY quantum fluctuation effects.  
However, for $\abs{g_{12}}/g\ll 1$ these deviations become significant since interparticle correlations are important and lie far beyond the GGP description.

Another important characteristic of the droplets is their excitation spectrum, which has been derived by~\citep{tylutki2020collective,zin2020zero,salasnich2018self} using a generalized (beyond mean-field)  Bogoliubov-de-Gennes linearization analysis. 
In~\citep{tylutki2020collective} the control parameter is the dimensionless constant 
\begin{equation}
    \gamma=\frac{\pi N^{2/3} \delta g}{2g}.
    \label{parametrization_droplets}
\end{equation}  
It allows one to regulate the shape of the droplet, see Fig.~\ref{fig:droplet_spectrum}~(a). 
When decreasing $\gamma$,  the density of the system changes from a flat-top profile ($\gamma \gg 1$) to a Gaussian-like shape as $\gamma$ takes negative values, gradually approaching a soliton-like configuration for $\gamma\to - \infty$. 
The behavior of the mode frequencies as function of $\gamma$ has also been studied. The conclusion is that the lowest non-trivial collective mode, namely the breathing mode, remains bounded. It is therefore, essentially, the only mode that remains supported by the droplet configuration independently of $\gamma$, while the higher-lying modes eventually cross into the continuum for decreasing $\gamma$, see the spectrum in Fig.~\ref{fig:droplet_spectrum}~(b). 
In the vicinity of the aforementioned crossings the corresponding mode is related to a large probability amplitude of finding a particle outside the droplet. 
More precisely, flat-top droplets occurring for $\gamma\gg 1$ and having a large bulk region support plane-wave phonons and resemble constant-density elastic media with free ends. 
Decreasing $\gamma$ results in a reduced bulk and the phonons with the smaller wavelengths approach the continuum faster. 
At the balance point, i.e.~at $\gamma=0$, the system is described by a mean-field theory having only quadratic nonlinearity, which gives rise to a Korteweg-de-Vries type droplet wave function of the form $\sim \cosh^{-2}(\sqrt{-\mu/2}x)$.  
A detailed discussion on the properties of the Korteweg-de-Vries equation and its soliton solutions can be found in~\citep{hirota1981soliton,hirota1972exact,wadati1972exact,dodd1982solitons}. 
Another interesting feature of short-range attractively interacting Bose mixtures is the formation of supersolid-like states in the presence of Rashba-Dresselhaus spin-orbit coupling as it was demonstrated  in~\cite{sachdeva2020self}. Specifically, a first-order phase transition from a self-bound supersolid to a droplet state was identified for varying strength of the Rabi coupling term. 

The breathing frequency of the droplet has been estimated using the GGP and the DMC methods~\citep{parisi2020quantum} and significant deviations between these two approaches have been found only for small particle numbers. 
Keeping $\abs{g_{12}}/g$ fixed, the breathing frequency in the GGP framework increases for larger particle numbers, reaches a maximum at a certain $N$, and then decreases approaching zero for very large $N$. 
As such, it has been argued that the origin of the breathing mode changes from a surface mode triggered by surface tension effects to a bulk density mode due to the bulk compressibility. 

\begin{figure}[tb]
\center
\includegraphics[trim={0cm 7cm 0cm 8cm},clip,width=0.9\textwidth]{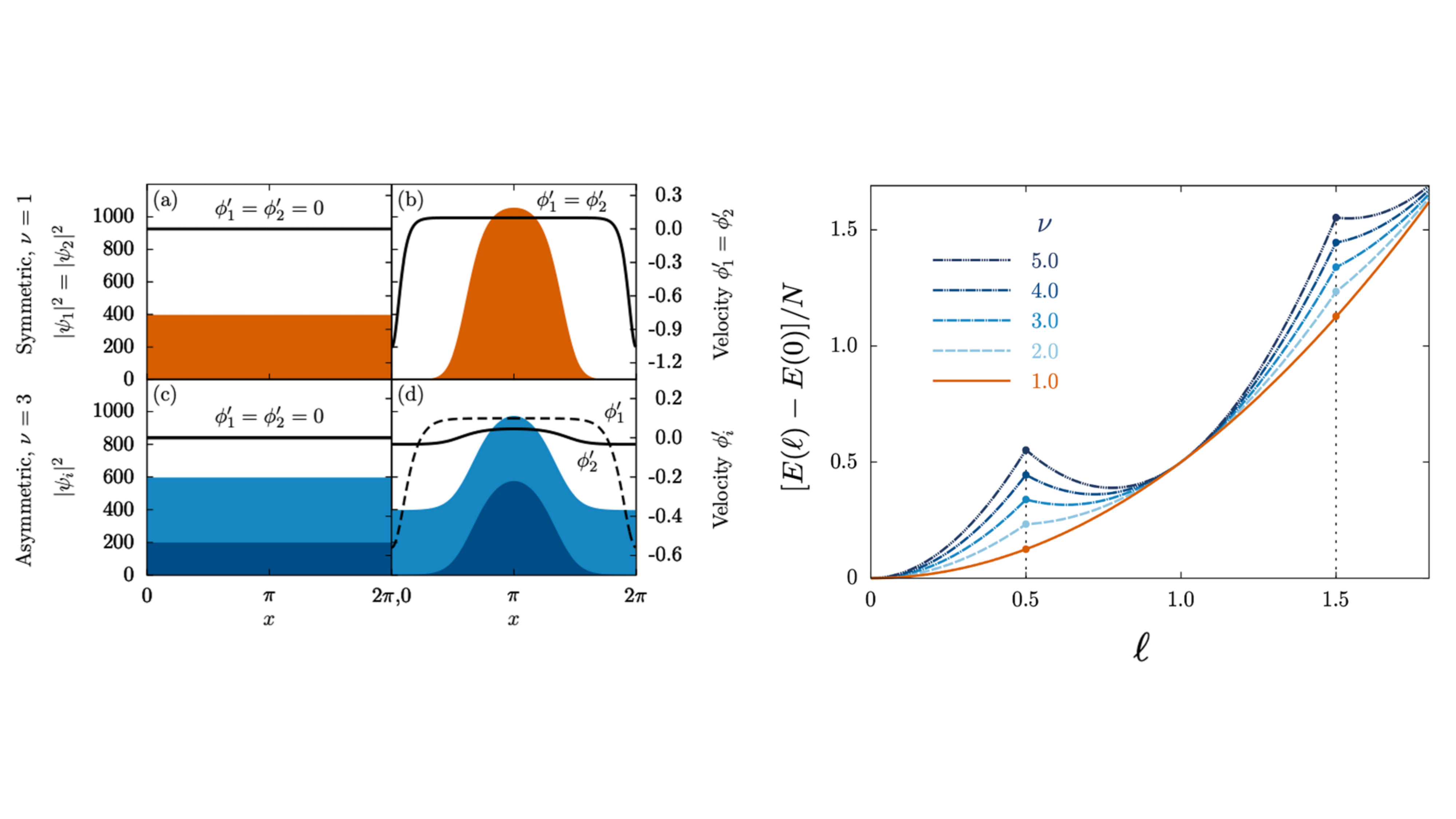}
\caption{(Left panels) Ground state densities (red, blue curves) and velocities $\phi_i'$ (black lines) of the individual components in the rotating frame with frequency $\Omega=0.1$ of an attractively interacting two-component mixture in a 1D ring containing $N=5000$ atoms. 
Panels (a), (b) present the case of a symmetric $\nu=1$ mixture with (a) $g=0.5$ and (b) $g=1.5$.   A transition from a homogeneous to a droplet profile occurs. 
(c), (d) represent the asymmetric $\nu=3$ scenario for the same parameters as in (a), (b). 
In all panels the right axis corresponds to the condensate velocities $\phi_i'$. 
(Right panel) Energy with respect to the angular momentum per particle for different population imbalances $\nu$ (see legend) and interaction $g=1.5$.  
In all cases the quantities presented are in dimensionless units. 
Figure from~\citep{tengstrand2022droplet}.}
\label{fig:mixed_drops} 
\end{figure} 

Having analyzed the droplet properties in free space a remaining question is the impact of the external confining geometry. 
In this context, the stationary droplet configurations and excitation spectrum of a particle imbalanced bosonic mixture featuring equal intracomponent repulsions and experiencing a 1D rotating ring trap have been analyzed within the GGP~\cite{tengstrand2022droplet}. For a symmetric mixture in the ring a larger interaction leads to a transition from a homogeneous to a droplet profile [Fig.~\ref{fig:mixed_drops}(a), (b)]. 
Importantly, it is found that an increasing intracomponent coupling with fixed $\delta g/g$ breaks the translational symmetry of the system. 
It gives rise to a composite configuration for the majority populated species being in a superposition of a droplet and a uniform distribution kept together by the confinement, while the minority component  exhibits a localized droplet profile, see Fig.~\ref{fig:mixed_drops} (c), (d). 
Moreover, the above-described mixed phase exhibits a characteristic behavior of the ground state energy for varying angular momentum per atom $l=(L_1+L_2)/N$ with $L_i$ being the angular momentum of the $i$-th component. 
This is illustrated in the right panel of Fig.~\ref{fig:mixed_drops} where for $\nu=1$ it has a parabolic shape while in the case of $\nu>1$ the distinct parabolas intersect. 
Interestingly, it was shown that the asymmetric system can be modelled as having $N_c$ atoms acquiring angular momentum as a classical rigid body under rotation and $N_v$ ones 
taking angular momentum only in terms of vorticity. In this way, the asymmetric system features properties of a solid and superfluid simultaneously where rigid-body rotation and quantized vorticity coexist. 
This coexisting nature was further confirmed in the response of the system subjected to infinitesimal rotations where non-droplet atoms move in the opposite direction of the rigid body. 
Finally, it should be emphasized that the effect of an external harmonic trap,  especially for few-body and mesoscopic systems,  on the droplet properties and in general the LHY contribution is a subject of ongoing investigations~\citep{mistakidis2021formation,englezos2022correlated}. 

On the other hand, the ground state phase diagram of these structures for symmetric bosonic mixtures ($N_A=N_B\equiv N/2$) in 1D lattice potentials of unit filling has been investigated in \citep{morera2020quantum,morera2021universal} using the DMRG approach, see also Section~\ref{sec:Methods}, in order to simulate the underlying Bose-Hubbard Hamiltonian. Since tight-binding models lie beyond the focus of this review, we provide in the following only some corresponding remarks. 
Various droplet phases were classified by inspecting the behavior of the density as well as the one- and two-body correlation functions. To characterize the droplet properties the dimensionless ratio $r=1+U_{AB}/U>0$ was  introduced such that the balance point of the mean-field interactions occurs at $r=0$, while $U_{AB}$ ($U$) denotes the on-site interspecies (intraspecies) attraction (repulsion). Characteristic phases include the Mott-insulator and the pair superfluid exhibiting exponential and algebraic decay of pair correlations respectively. Also, one encounters a two atomic superfluid state where each species shows quasi-long-range coherence separately and an increasing droplet size which becomes comparable to the system size in the non-interacting limit. Moreover, the structure of the lattice trapped mixture in the vicinity of the balance point has been analyzed in~\cite{morera2021universal} where analogies to an effective dimer model~\cite{kuklov2004commensurate,kuklov2004superfluid} were made. Here, the transition from a gas to a liquid phase has been identified when the effective dimer-dimer interactions~\citep{pricoupenko2018dimer,valiente2009scattering} change from repulsive (gas) to attractive (liquid). 
Note in passing that analytical estimates of the dimer-dimer effective range and scattering length, as well as the tetramer bound state threshold and binding energy can be found in~\citep{pricoupenko2018dimer,hadizadeh2011scaling,greene2017universal}. 
Finally, entering the regime of sufficiently strong dimer-dimer repulsions the fermionization of the dimers forming an effective Tonks-Girardeau state can be seen. 
The impact of the filling factor of the lattice on the density profiles of the phases is discussed in \citep{morera2020quantum}, but a detailed analysis of this issue is still an outstanding question. Further discussion on 1D lattice trapped droplet configurations can be found in the recent review article~\cite{khan2022quantum}.

\subsection{Dynamics of one-dimensional droplets}\label{droplet_dynamics} 

To further understand droplet properties and their potential applications, one studies dynamics, e.g.~by considering collisions between two droplets~\citep{astrakharchik2018dynamics,katsimiga2023interactions} or their spontaneous generation triggered by an instability mechanism~\citep{mithun2020modulational}. 
We note here that the relevant dynamical studies are restricted to only a few (see below) mainly relying on the GGP framework and thus the current understanding, especially the build-up of higher-order correlations, is still far from complete. Below, we provide an overview of most recent dynamical droplet properties as captured by the GGP where quantum fluctuations are crucial and also discuss implications of higher-order quantum corrections whenever applicable. 

The collisional dynamics between two counterpropagating droplets is very interesting since it allows one to distinguish them from bright soliton structures in spite of the fact that droplets have density profiles similar to those of bright solitons. 
Recall that a soliton maintains its shape when moving with constant velocity and also when colliding with another soliton as long as the system remains nearly integrable~\citep{kevrekidis2007emergent}. 
Similarly, droplets whose wave function is described by Eq.~(\ref{droplet_wfn}), exhibit a constant density profile $n(x;t)=\abs{\Psi(x;t)}^2$ in the course of their time-evolution if moving with a constant velocity. 
However, they generally do not preserve their shape when undergoing pairwise collisions~\citep{astrakharchik2018dynamics,katsimiga2023interactions}, a result that clearly differentiates them from bright solitons. 
Let us also remark here another interesting property that concerns the stability of a single droplet when it is exposed to a periodic density modulation $\sim \cos(kx)$ of wavenumber $k$. 
It has been shown that there is a particle dependent critical wavenumber $k_c$ above which the droplet splits into several pieces (\textit{viz.} fragments) propagating in opposite directions while for $k<k_c$ it maintains its shape featuring almost periodic oscillations.

\begin{figure}[tb]
    \center
    \includegraphics[width=0.95\linewidth]{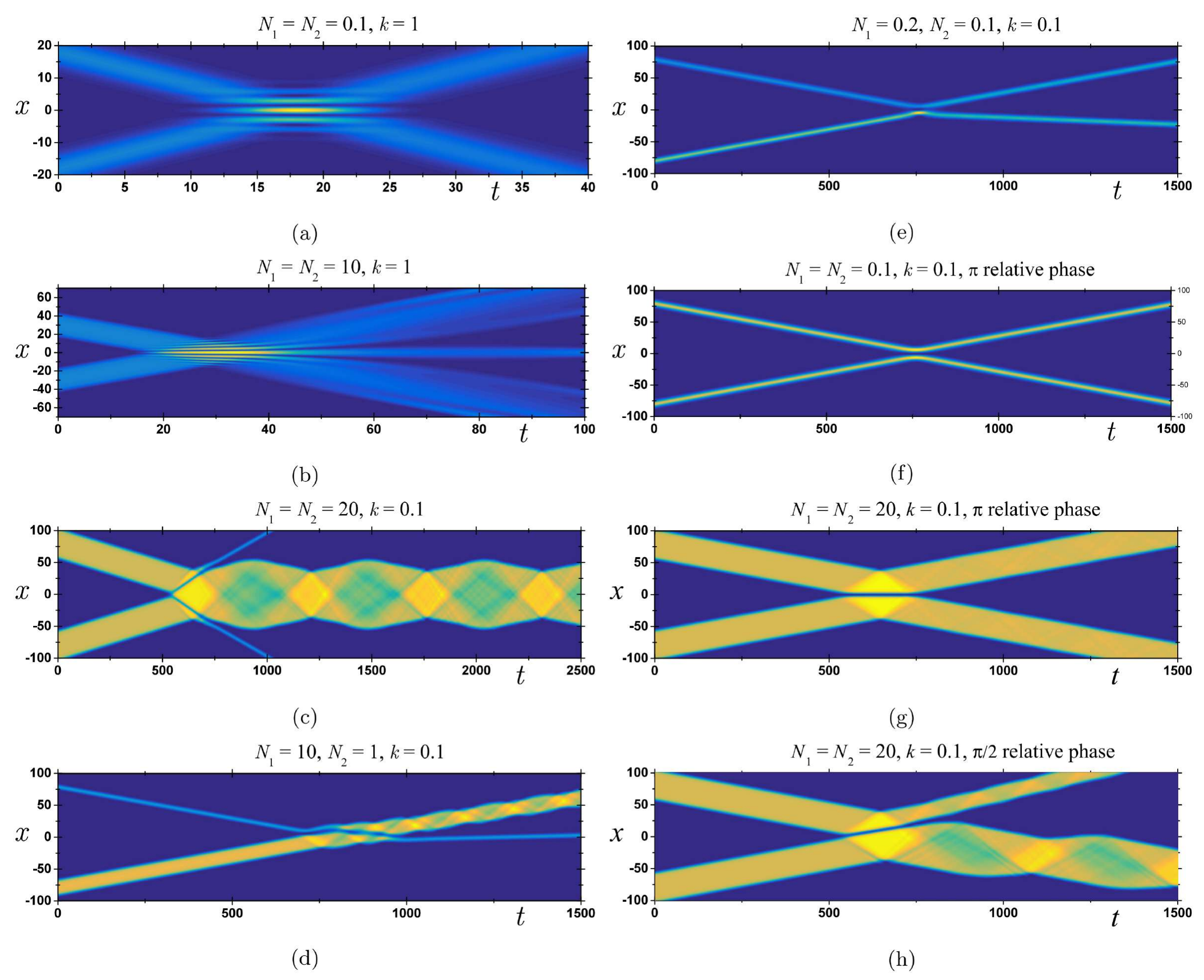}
    \caption{Density plots of two colliding droplets whose initial form is given by Eq.~(\ref{counterprop_droplets}). Different situations for symmetric ($N_1=N_2$) and asymmetric ($N_1 \neq N_2$) mixtures are shown for different incident momenta $k$ (see legends). 
    Unless stated otherwise, the collisions are \textit{in-phase} ($\phi=0$). Figure from \citep{astrakharchik2018dynamics}.}
\label{fig:droplet_collisions} 
\end{figure} 

Two counterpropagating droplets were considered in \citep{astrakharchik2018dynamics} assuming that the initial `wave function' has the form
\begin{equation}
 \Psi(x;0)=e^{ikx/2+\phi} \Psi_1(x+x_0)+e^{-ikx/2} \Psi_2(x-x_0),\label{counterprop_droplets}
\end{equation} 
where $\Psi_1(x)$, $\Psi_2(x)$ are given by Eq.~(\ref{droplet_wfn}), $\pm x_0$ denote the initial locations of the droplets, $\pm k/2$ corresponds to their initial momenta and $\phi$ is the relative phase between them. 
\citep{astrakharchik2018dynamics} demonstrated that the collisional properties of the droplets depend strongly on the considered particle number, their velocity and relative phase. 
As discussed in Section~\ref{droplet_GS}, for small particle numbers the droplet profile has a Gaussian-like shape while for large atom numbers it is of a ``puddle'' type. 
Below, the basic collisional features reported in \citep{astrakharchik2018dynamics} are summarized, see also Fig.~\ref{fig:droplet_collisions}. 

Two small droplets preserve their shapes after the collision and show an interference pattern at the collision point, which is similar to the interplay of coherent matter waves. 
In sharp contrast, two counterpropagating ``puddles'' generally display aspects of inelastic collisions, accompanied by mass-transfer phenomena and leading to a larger number of droplets or a merger of the outgoing ones for fast and slow velocities, respectively. 
Depending on the phase difference it is possible to tune the degree of excitation and the population of the outcoming droplets, especially for initially large droplets. 
On the other hand, the collision of a ``puddle'' and a Gaussian-like droplet results in the excitation of the former performing internal vibrations and an almost un-perturbed shape of the latter. Two particle imbalanced Gaussian-like droplets show an almost elastic collision with their outgoing trajectories being only slightly affected, a result that equally holds for few-body settings~\cite{englezos2022correlated}.

The impact of beyond LHY correlations in the droplet nonequilibrium dynamics of mesoscopic systems  induced by an interspecies interaction quench within the nonperturbative ML-MCTDHX approach was studied in~\citep{mistakidis2021formation}.  
Specifically, a sudden change from weak to strong intercomponent attractions enforces the droplet expansion acquiring flat top signatures during the dynamics. Here, the expansion velocity is found to be reduced as compared to the GGP prediction. 
Following the opposite quench, namely from strong to weak interactions, a breathing motion is triggered whose frequency becomes slightly smaller than in the GGP framework due to residual higher-order correlations~\citep{parisi2020quantum,astrakharchik2018dynamics}. 
In this sense, both the expansion and the breathing dynamics of the droplet provide experimental means for identifying beyond LHY physics. 
Interestingly, two-body correlations build-up in the course of the evolution evidencing that two bosons of the same component experience a bunching at long distances and an anti-bunching at the same location.

\begin{figure}[tb]
    \center
    \includegraphics[width=\linewidth]{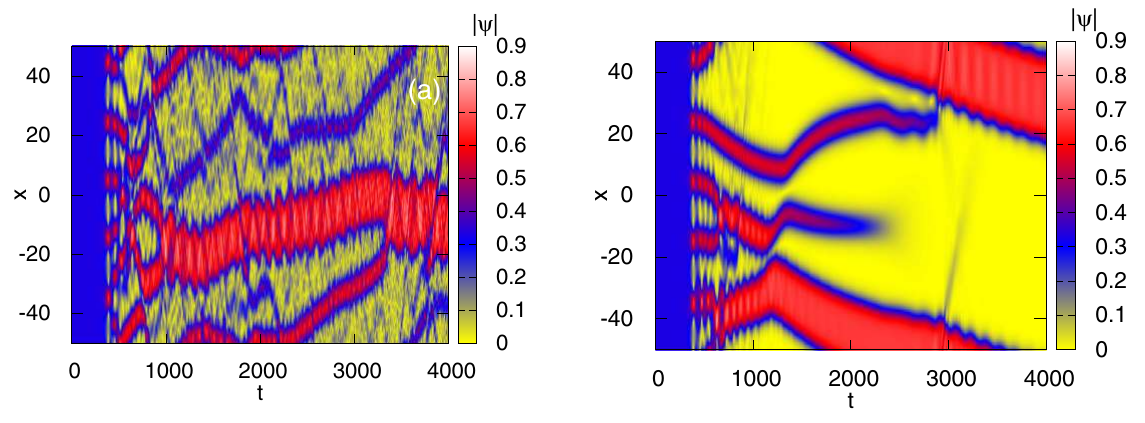}
    \caption{Time-evolution of $\abs{\Psi(x,t)}$ representing an initially perturbed plane-wave solution in the ultracold regime (left panel) and for the case of the presence of dissipation (condensate thermal fraction) with strength $\gamma=0.03$ (right panel). The spontaneous generation of an array of quantum droplets and their subsequent coalescence can be observed. When dissipation is taken into account to emulate thermal effects of the background its fluctuations are minimized. The initial homogeneous density corresponds to $n\approx 0.1$ while $\delta g/g=0.05$.  Figure adapted from \citep{mithun2021statistical}.} 
\label{fig:MI_droplets} 
\end{figure} 

The nucleation of quantum droplets from plane-wave solutions due to the manifestation of the modulation instability\footnote{The modulational instability wave phenomenon is related with the exponential growth
of perturbations due to the presence of nonlinearity and anomalous dispersion. It manifests in various physics areas e.g. ranging from BECs (quantum fluids)~\citep{nguyen2017formation,everitt2017observation}, and waveguides~\citep{tai1986observation} to systems in nonlinear optics~\citep{agrawal2000nonlinear} and charged plasmas~\citep{thornhill1978langmuir}.} has been discussed in~\citep{mithun2020modulational,mithun2021statistical}. 
In particular, let us consider a stationary plane-wave subjected to small perturbations i.e. $\Psi(x;t)=[\sqrt{n}+\delta \Psi(x;t)]e^{-i\mu t}$. Here, $\delta \Psi=\zeta \cos(kx-\Omega t)+i \eta \sin(kx-\Omega t)$ with $\zeta$, $\eta$ being infinitesimal amplitudes, while $k$ and $\Omega$ denote the wavenumber and frequency of the pertubation. 
Performing a linearization analysis with respect to $\delta \Psi$ yields the corresponding generalized  Bogoliubov-de-Gennes equation of motion and leads to the dispersion relation   
\begin{equation}
 \Omega^2=\frac{k^4}{4}+\bigg(\frac{\delta g}{g}n-\frac{\sqrt{n}}{\sqrt{2}\pi}\bigg)k^2.\label{dispersion_MI}
\end{equation} 
The modulation instability occurs when $\Omega$ becomes imaginary, namely when $n<[2 \pi^2 (\delta/g)^2]^{-1}$ or $k^2<4[\sqrt{n}/(\sqrt{2}\pi)-(\delta g/g)n]\equiv k_0^2$. This imaginary contribution becomes maximal for $k_{max}=k_0/\sqrt{2}$. Letting the perturbed plane-wave solution evolve within the single-component GGP framework 
and operating within parameter regions where the modulation instability is expected to take place [see Eq.~(\ref{dispersion_MI})] leads to the emergence of chains of quantum droplets, see the left hand side panel of Fig.~\ref{fig:MI_droplets}. 
Their number reduces during the evolution due to mergers of the colliding droplets and
these observations clearly indicate that the modulation instability is indeed the underlying mechanism for the creation of quantum droplets in this setting. 
This result also holds true for asymmetric mixtures characterized either by a population imbalance or by different intracomponent coupling strengths. 
Finally, in an attempt to account for thermal effects, a phenomenological dissipative term of strength $\gamma$ was considered in the GGP by~\citep{mithun2021statistical}. 
It was shown that in the presence of dissipation the overall generation of droplets is similar and happens on the same timescale as the zero temperature scenario. 
However, in this situation the formation of "large" droplets becomes more prominent and the background fluctuations are minimized, see the right hand side panel of Fig.~\ref{fig:MI_droplets}. 
It should be emphasized at this point that the role of thermal fluctuations and their adequate modelling including their impact on the LHY term is certainly an issue that deserves further study. 

Independently, the quench dynamics of a quasi-1D repulsively interacting binary bosonic mixture was investigated in~\citep{cidrim2021soliton}. 
The quench protocol consisted of a linear ramp of the interactions over a finite time interval and the calculations were based on the GGP framework including also the effect of three-body losses~\citep{semeghini2018self}. It was found that after quenching to weak (strong) intercomponent couplings the arising modulation instability allows for the formation of dark soliton trains (self-bound droplets). 
An analytical prediction of the number of generated solitons was also provided through the unstable eigenmode of the Bogoliubov spectrum. 
In sharp contrast, a quench from the weakly attractive (soliton) regime to strong interactions (droplet region), produces a splitting of the original gas into two bright solitons, but not a train (see also~\citep{cheiney2018bright}).   

Over the past few years it has been demonstrated that quantum droplets feature inelastic collisions and that the modulation instability process provides means for their spontaneous generation. 
Despite the fact that various theoretical questions for the dynamical response of droplet configurations in 1D still remain open, it is worthwhile to comment that these systems constitute fruitful platforms for probing beyond mean-field effects in weakly interacting ultradilute Bose gases~\citep{navon2011dynamics}. They also provide a new arena for generalizing already well studied concepts such as nonlinear excitations~\cite{edmonds2022dark,PhysRevA.107.063308} and impurity physics~\cite{bighin2022impurity}.  
Another future research direction of particular importance is the study of the competition between temperature effects and quantum fluctuations and also the generic modelling of temperature~\citep{mithun2021statistical,de2021thermal}. 
Finally, we note that most of the current investigations have focused on the dynamics of quantum droplets described by the GGP framework. 
However, beyond LHY correlations result in a reduced velocity of expanding droplets and a slightly smaller breathing frequency.
This motivates beyond LHY studies,  especially in the few-body regime,  to reveal higher-order correlation effects in the course of the time evolution.

%% file: Section09.tex
\section{Methodologies and computational approaches } 
\label{sec:Methods}

In what follows we will provide a brief account of some of the most important methods which are frequently used to compute the stationary properties and in particular the nonequilibrium quantum dynamics of few- to many-body trapped ultracold atoms. We start by describing the main features of the well-known Exact Diagonalization method, followed by the celebrated Bethe Ansatz-based methods. As a prime example of an iterative and variational numerical method we then address the Density Matrix Renormalization Group approach. The Multi-Layer Multi-Configuration Time-Dependent Hartree for Atomic Mixtures focuses on the quantum dynamics of ultracold systems with intra- and interspecies interactions. 
Finally, we sketch the mapping onto a spin chains description.

\subsection{Exact diagonalization}

Arguably, the most widespread numerical technique to approach quantum systems with a few degrees of freedom is 
exact diagonalization (ED). ED is a term that applies broadly to any numerical method based on writing the Hamiltonian matrix in a prescribed basis and then diagonalizing the obtained matrix. The ED method is useful if the Hilbert space is sufficiently\footnote{The meaning of the word `sufficiently' here strongly depends on the problem at hand, in particular, on the sparsity of the Hamiltonian matrix.} small. It is widely used to analyze lattice many-body problems (see, e.g.,~\citep{Weisse2008}), and few-body systems. The main advantage of ED is the possibility to control the accuracy of calculations -- one can reduce the error bars in a systematic manner. In addition, one can obtain the full energy spectrum of the Hamiltonian, which gives a basis for simulating time dynamics, see, e.g., recent studies on time dynamics of few-body systems~\citep{Ebert:2016,Yin:2016, Nandy_2020,Nandy_2021}. 

For lattice systems with a finite number of degrees of freedom the Hamiltonian matrix is in principle finite. However, the `exponential curse', i.e., the fact that a Hilbert space will generally grow exponentially with the number of particles, means that in practice it is always necessary to perform some kind of truncation procedure to keep things manageable. A number of general purpose codes 
exist in the open source domain, including the widely used ALPS~\citep{Bauer:2011} and QuSpin packages~\citep{Weinberg2017,Weinberg2019}. 

For continuum systems, which are the focus of this review, the Hilbert space is infinite to begin with, and, to employ ED, one {\it always} works with a truncated basis. One common strategy is to use a non-interacting one-body basis with $n_{\mathrm{max}}$ states.
To ensure a converged result when working with finite bases, it is customary to calculate the spectra, eigenstates, and other properties of the system as a function of the basis size. A key question is the approach to the infinite basis size limit, i.e., how fast will the results converge as the basis size is changed.

For a concrete setting with a finite number of particles there are a number of systematic tricks that can be applied to either speed up convergence or obtain higher accuracy for a given quantity of interest, or both. This is a vast field that features a lot of creativity in `clever' basis choices that accomplish these tasks more efficiently. We will not attempt a full review, but merely point out some results that have been successful in cold atoms, and in particular in the direction of few-body static and dynamic studies. The focus here is on short-range interactions. 

For particles interacting via zero-range potentials in one spatial dimension, it is known that convergence as a function of $n_{\mathrm{max}}$ can be slow. The slow convergence is due to a cusp of the wave function. It leads to a non-analytical behavior, which is hard to reproduce using bases based upon non-interacting models. For example, for eigenfunctions of harmonic oscillator potential, the convergence is expected to be $1/\sqrt{n_{\mathrm{max}}}$~(\citep{Grining_2015}), whereas for a plane wave basis, the convergence should be $1/n_{\mathrm{max}}$~(\citep{Volosniev_2017Flow,Jeszenszki2018ED}). The difference between $1/\sqrt{n_{\mathrm{max}}}$ and $1/n_{\mathrm{max}}$ is related to a different dependence of the maximal one-body energies on $n_{\mathrm{max}}$, which in those cases determine the resolution of non-interacting wave functions\footnote{One can understand this intuitively by noticing that the extend of an eigenstate in a box does not depend on the energy of the state -- it is always given by the size of the box. In contrast, the `size' of an eigenstate in a harmonic trap depends on the energy. This implies that the basis formed from the latter eigenstates gains less resolution by changing $n_{\mathrm{max}}$.}. 
It is worth noting that for attractive interactions, the slow convergence pushes the solution out of reach faster than for repulsive interactions~(see, e.g.,~\citep{Amico2014}). 

To improve convergence, one can introduce an effective interaction instead of the bare one. Ideally, the effective interaction is an interaction term that produces results close to the exact ones already for relatively small values of $n_{\mathrm{max}}$.  One simple empirical method to obtain an effective interaction from the bare one is to rescale the interaction strength as 
\begin{equation}
    g\to\frac{g}{1+g/g_0},
    \label{eq:g_rescale}
\end{equation}
where $g_0$ is an empirical constant that depends on $n_{\mathrm{max}}$, such that $g_0\to\infty$ as $n_{\mathrm{max}}\to\infty$. It was shown that a rescaling of Eq.~\eqref{eq:g_rescale} significantly improves the accuracy of the energies, see~\citep{Ernst2011}. 
The value of $g_0$ can be determined in various ways, for example, one can extract it (more rigorously) from the solution to a two-body problem or (more phenomenologically) from the Tonks-Girardeau limit~\citep{Ernst2011}.
\citep{Jeszenszki2018ED,jeszenszki2019} studied convergence patterns of the latter approach, and showed that Eq.~(\ref{eq:g_rescale}) allows one to improve convergence of the energy of the system from $1/n_{\mathrm{max}}$ to $1/n_{\mathrm{max}}^2$ for a plane-wave basis. These works also introduced a transcorrelated method that can improve the convergence even further (to $1/n_{\mathrm{max}}^3$).
For contact interactions, this method implemented in the open source code NECI\footnote{https://github.com/ghb24/NECI\_STABLE}~\citep{Kai2020} and Rimu.jl\footnote{ https://github.com/joachimbrand/Rimu.jl}
is a similarity transformation of the Hamiltonian that enforces the boundary (cusp) conditions of Eq.~(\ref{eq:bethePeielrsBoundary}). The downside of this approach is that
 the similarity transformed Hamiltonian is non-Hermitian and has three-body terms (that can be avoided). Still, it can be diagonalized with the widely available Arnoldi
iteration, or in the case of larger systems, with full-configuration interaction quantum Monte
Carlo~\citep{Kai2020}. Finally, we note that the method has also been used to study systems in three spatial dimensions~\citep{jeszenszki2020}. 

Another effective interaction for cold atoms can be obtained using the Lee-Suzuki method from nuclear physics~\citep{Christensson2009,Rotureau2013}. For studies of few-body systems in one spatial dimensions, see~\citep{Lindgren2014Fermionization,Dehkharghani2015Bosons,Volosniev_2017Flow}; freely available software is introduced in \citep{Rammelmuller2022Effective}; a comparison to the transcorrelated method described above is presented in~\citep{Rammelmuller2022Magnetic}. The effective interaction in those works is based upon the solution of a two-body problem. In a truncated Hilbert space, the effective interaction can be written as 
\begin{equation}
    V_{\mathrm{eff}}=Q^T E Q-H_0,
\end{equation}
where $E$ is the matrix that contains two-body energies on a diagonal, $Q$ is an orthogonal matrix, and $H_0$ is a non-interacting two-body Hamiltonian. The $Q$ is chosen as $u/\sqrt{u^T u}$, where $u$ is a matrix whose rows contain a projection of eigenvectors on the truncated Hilbert space. The two-body solutions for a ring and harmonic potentials are discussed in~Secs.~\ref{sec:EarlyMod} and~\ref{sec:OneComp}. Using those solutions one can construct an effective potential that incorporates correct two-body information in a truncated Hilbert space, thus speeding up the convergence. 

Finally, we mention that the convergence can be improved even for a Hamiltonian with a bare interaction if one modifies the basis used for diagonalization. For example, one can use a basis with additional variational parameters, e.g., the frequency of the harmonic potential~\citep{KOSCIK2018ED}; a basis that utilizes conserved quantities~\citep{Harshman2012}; a basis based upon random states (explicitly correlated Gaussians) that are not related to a non-interacting problem~\citep{Andersen:16}, see also~\citep{Mitroy2013Gaussians} for a review of that approach. 
Those methods can be very successful. However, they might require a system-dependent pre-selection of suitable basis states and departure from the established numerical techniques, which stall their development. 

\subsection{Bethe ansatz-based methods}
\label{subsec:BetheAnsatzMethods}

\begin{figure}[tb]
  \centering
    \includegraphics[width=0.7\linewidth]{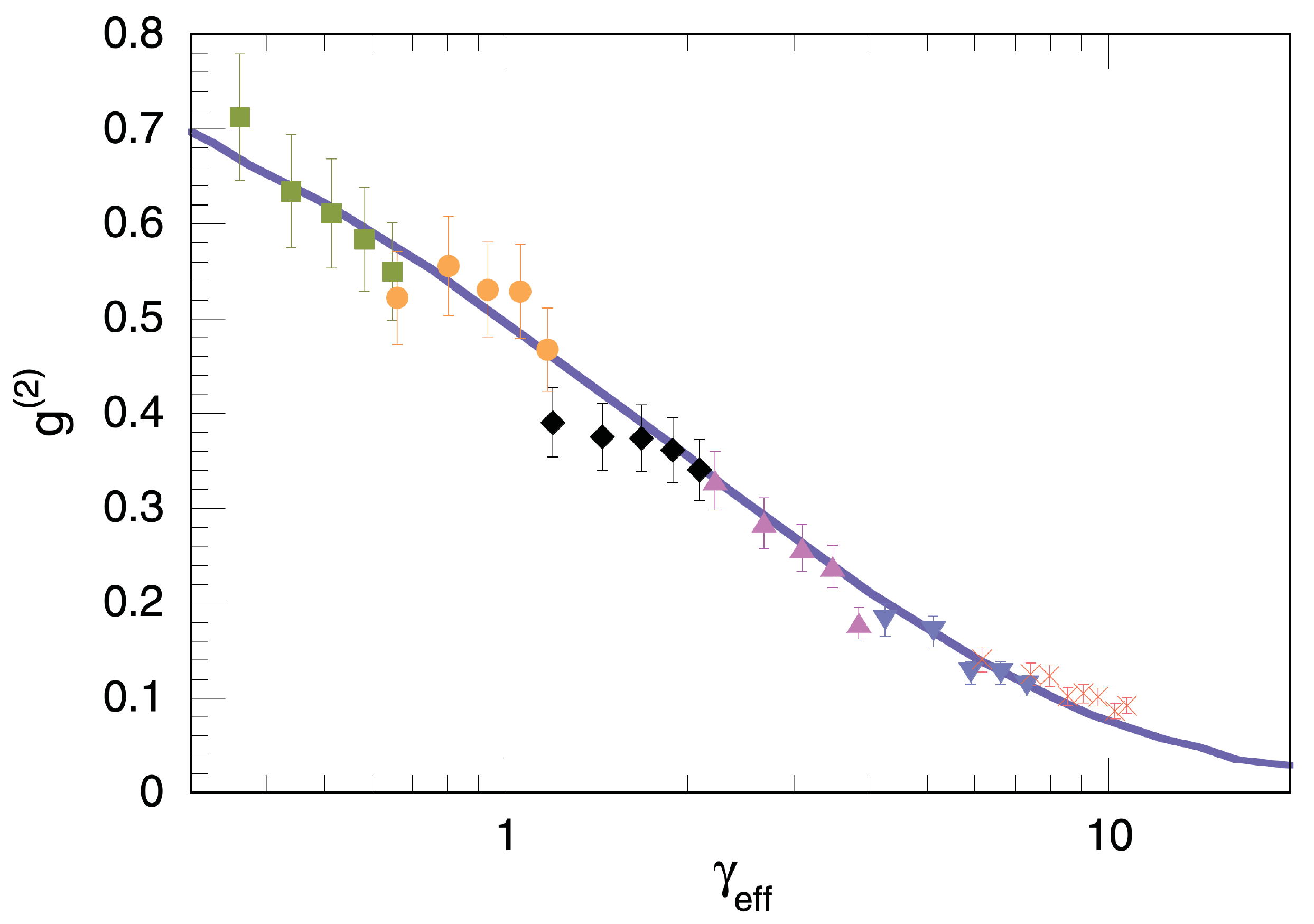}
    \caption{Experimental measurement of the pair correlations in the repulsive 1D Bose gas. The solid curve shows the theoretical prediction obtained using the Bethe ansatz for the Lieb-Liniger model \citep{Gangardt2003Stability}. Figure from \citep{kinoshita2005}.}
\label{fig:2bodycorr}
\end{figure}

In Section \ref{sec:EarlyMod} we presented the form of the Bethe ansatz wave function for a bosonic system and discussed the solutions in the few-body limit. The Lieb-Liniger model, for general $N$ is described by the Hamiltonian
\begin{equation}\label{llH}
    H= - \sum_{i=1}^N\frac{\partial^2}{\partial x_i^2}+2c\sum_{i<j}\delta(x_i-x_j),
\end{equation}
where $\hbar=m=1$ is once more assumed. 
The Bethe ansatz for arbitrary $N$ has been given in Eq. \eqref{coordBA}. Imposing the contact interaction condition on the derivatives, as well as periodic boundary conditions, yields a set of Bethe ansatz equations (BAE) given by
\begin{equation}\label{BAEbosons}
    k_jL=2\pi I_j-(N-1)\pi-2\sum_{l=1}^N\text{arctan}\left(\frac{k_j-k_l}{c}\right),
\end{equation}
where $I_i$ is an integer or half-integer for a total odd or even number of particles, respectively. Solving these equations specifies all the possible values of $k_i$, thus determining the complete wave function. To express a physical solution, these values must all be different from each other. It is worth pointing out here that the solutions strongly depend on the sign of the interaction: for the repulsive case, all solutions are real \citep{Yang1969BosonsTD}, while for the attractive case complex solutions called {\it strings} are allowed \citep{Takahashi1999OneDimensional}. Solving the eigenvalue equation $H \psi = E \psi$ then returns the total energy of the system, which is simply given by
\begin{equation}\label{llEnergy}
    E=\sum_i k_i^2.
\end{equation}
The problem of finding the whole spectrum of the model is thus reduced to solving the BAE, from which the ground state energy, elementary excitations, and correlation functions can be derived. Remarkably, the method allows one to handle both the repulsive and attractive regimes as well as to deal with an arbitrary number of particles and has been successfully employed in a number of applications. A comparison between experimental results obtained for the pair correlations of a 1D Bose gas and the results predicted by the Bethe ansatz approach is shown in Fig.~\ref{fig:2bodycorr}, where a decrease of pair correlations can be observed as the interactions are made stronger.

For the fermionic case, the situation becomes more involved since one has to deal with different components or spins. Thus, the spatial wave function, provided by the coordinate Bethe ansatz, must be combined with a corresponding spin wave function. Yang's solution for the spin wave function in a two-component system of $N$ fermions with $N_\downarrow$ down spins is obtained with the {\it generalized} or Bethe-Yang ansatz, and reads \citep{Yang1967Fermi,Takahashi1999OneDimensional}
\begin{equation}
    \Phi=\sum_R A(R)F_P(\Lambda_{R1},y_1)...F_P(\Lambda_{RN_\downarrow},y_{N_\downarrow}),
\end{equation}
where the $y_i$ are the coordinates of spin-down particles with $i \in \{1,...,N_{\downarrow}\}$ and $R$ are permutations acting on these coordinates. For each permutation, the functions $F(\Lambda,y)$ depend on the {\it spin rapidities} $\Lambda$ and are given by
\begin{equation}
F(\Lambda,y)=\prod_{j=1}^{y-1}\frac{k_j-\Lambda+ic'}{k_{j+1}-\Lambda-ic'}
\end{equation}
where $c'=c/2$. In analogy with the bosonic case, the Yang-Baxter equation allows for the solution of the fermionic problem. The periodic boundary conditions are now satisfied with a set of $N$ equations for the quasimomenta $k$ and rapidities $\Lambda$
\begin{eqnarray}
    k_jL=2\pi I_j-2\sum_{\alpha=1}^{N_\downarrow}\arctan\left(\frac{k_j-\Lambda_\alpha}{c'}\right),
\end{eqnarray}
and
\begin{equation}
    \sum_{j=1}^{N}2 \arctan\left(\frac{\Lambda_\alpha-k_j}{c'}\right)= 2\pi J_\alpha+2\sum_{\beta=1}^{N_\downarrow}\arctan\left(\frac{\Lambda_\alpha-\Lambda_\beta}{c}\right).
\end{equation}
In these expressions, $I_j$ and $J_\alpha$ are integers or half-odd integers, depending on the numbers $N_\uparrow$ and $N_\downarrow$ in each component. The energy expression is the same as Eq.~\eqref{llEnergy}. However, we now have a nested set of the Bethe ansatz equations due to the spin degree of freedom. The spin wave function obtained with the Bethe-Yang ansatz yields the solution of the two-component fermionic problem when coupled with the spatial wave function. It is important to point out that these parts must be combined by taking into consideration the appropriate permutation symmetries, which can be obtained from the Young tableaux \citep{Guan2013Review}. The final wave function must then by antisymmetric upon simultaneous permutation of the coordinates and spins of two particles.

The Bethe ansatz approach thus offers an elegant way to solve a number of 1D models of particles with short range interactions. In addition, it also allows for the solution and discussion of physical properties of other types of models in the context of ultra-cold atoms - see, for example, \citep{Links2003, Santos2006, Duncan2007heteronuclear, Foerster2007Atomic, Rubeni2012, Rubeni2017}. However, the wave functions produced can often become too complex and of little practical use in the calculation of experimentally relevant quantities. This is due to the fact that they grow factorially with the number of particles, making them hard to handle for increasingly large $N$. This characteristic is particularly problematic in a dynamic setting, where the effect of integrability is expected to be more pronounced.

To address these issues, efficient numerical methods based on -- or employing technical aspects related to -- the Bethe Ansatz have been developed and applied to the solutions of integrable many-body problems in recent years. The first of these, known as ABACUS (Algebraic Bethe Ansatz-based Computation of Universal Structure factors) \citep{Caux:2009} is based on utilizing the solutions provided by the Bethe ansatz, along with an optimized scan of the Hilbert space to specifically tackle the problem of calculating the generalized dynamical structure factor, defined as
\begin{equation}
    S^{a\bar{a}}(k,\omega)=\frac{1}{N}
    \sum_{j,j'=1}^{N}
    e^{-ik(j-j')}\int_{-\infty}^{+\infty}dt 
    e^{i\omega t}\langle
    \mathcal{O}^a_j(t)
    \mathcal{O}^{\bar{a}}_{j'}(0)\rangle
\end{equation}
where $\langle \mathcal{O}^a_j(t)\mathcal{O}^{\bar{a}}_{j'}(0)\rangle$ is the ground state expectation value of a given correlation function at sites $j,j'$ and $a,\bar{a}$ denote site-specific indices. Using the Lehmann spectral representation \citep{fetter1998}, the structure factor can be cast as
\begin{equation}\label{abacussf}
    S^{a\bar{a}}(k,\omega)\frac{2\pi}{N}\sum_\mu\vert \langle \lambda^0 \vert \mathcal{O}^a_k\vert \mu\rangle \vert^2 \delta (\omega-E_\mu-E_0),
\end{equation}
where $\vert \lambda^0 \rangle$ and $E_0$ denote the ground state and its correspondent energy. The essential step in obtaining the dynamical structure factor thus becomes calculating the form factors $\langle \lambda^0 \vert \mathcal{O}^a_k\vert \mu\rangle$ and summing over the intermediate eigenstates $\vert \mu\rangle$. The form factors can be expressed analytically using the algebraic Bethe ansatz for different models, such as the 1D Bose gas \citep{Slavnov:1989,Slavnov:1990} and the Heisenberg chain \citep{Kitanine:1999,Kitanine:2000}. The summation over intermediate eigenstates is performed by assuming certain properties of the form factors for large systems and realizing an optimized scan of the Hilbert space. Once the method returns a list of form factors, a dynamical correlation function can be calculated.

\begin{figure}[tb]
  \centering
    \includegraphics[width=\linewidth]{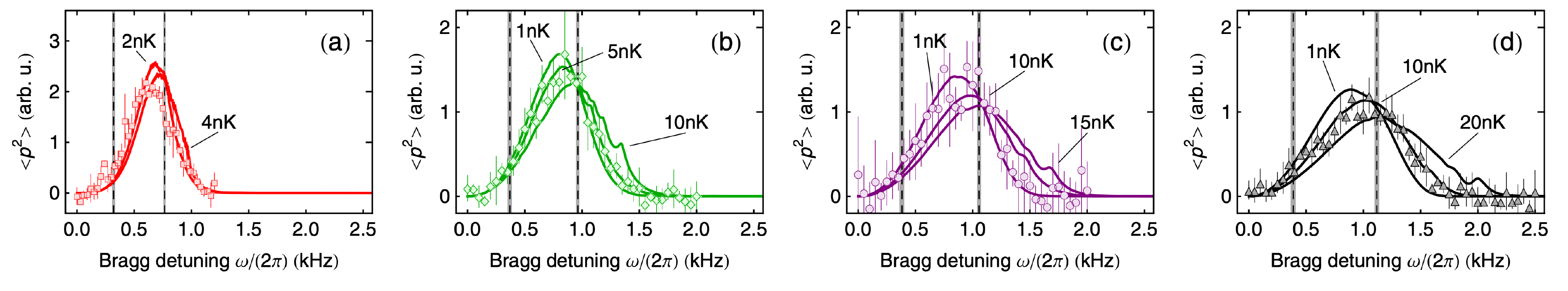}
    \caption{Experimentally measured Bragg-excitation spectra of a 1D Bose gas at finite temperatures. Panels (a)-(d) display the results for increasingly stronger interactions. The solid curves show the averaged theoretical results obtained from the dynamical structure factor calculated with ABACUS. Figure adapted from \citep{Meinert:2015}}.
\label{fig:excitationBoseAbacus}
\end{figure}

Since its development, ABACUS has been used to obtain accurate results for the correlation functions of different 1D integrable many-body models. Applications include the calculation of properties of the 1D Bose gas such as one-body dynamical correlations, \citep{Caux:2007}, correlations of highly excited \citep{Fokkema:2014}, and finite-temperature regimes \citep{Panfil:2014}, as well as the four-spinon structure factor of the Heisenberg model \citep{Caux:2006}. Additionally, the method has been successfully applied in the prediction of experimental results such as the dynamical structure factor \citep{Fabbri:2015} and the excitation spectrum of the 1D Bose gas \citep{Meinert:2015} (see Fig.~\ref{fig:excitationBoseAbacus}).

A notable limitation of the Bethe ansatz is the fact that it is restricted to homogeneous systems, where the integrability condition is preserved. Such a feature can be troublesome because most experimental results with cold atoms depend on the presence of a (usually harmonic) trapping potential. While the local density approximation (LDA) is able to reproduce equilibrium results in the presence of external potentials, that is not the case for dynamics; one particular result that posed a challenge to the theoretical developments of its time is the experiment known as Quantum Newton's Cradle \citep{kinoshita2006}. In this experiment, a many-body Bose gas is initialized in an out-of-equilibrium state, which evolves in time with the particles oscillating and colliding in a trap. Interestingly, the system does not seem to thermalize even after several collisions, meaning that the particles are only allowed to exchange momentum and their momentum distributions do not change in time considerably. Such a feature is a characteristic of integrable systems, where diffraction is absent. However, while usual Bethe ansatz-based techniques are not able to incorporate the presence of the trapping potential in a real-time simulation of dynamics, the standard hydrodynamic approaches could not correctly reproduce the lack of thermalization observed in the experiments.

\begin{figure}[tb]
    \centering
    \includegraphics[width=0.9\linewidth]{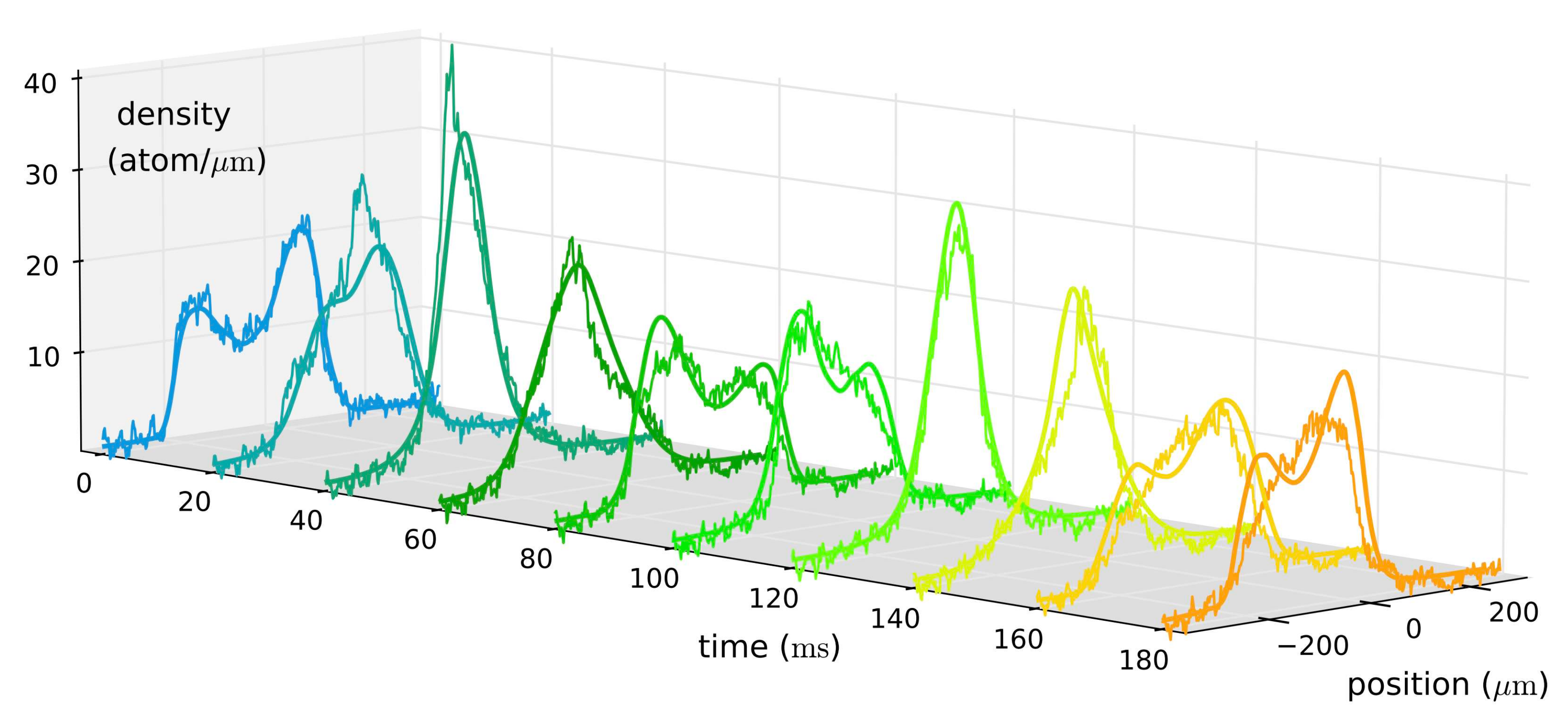}
\caption{Experimental measurement of the expansion of an atomic cloud of ${}^{87}$Rb released from a double-well potential generated on an atom chip. The solid curves show the GHD results. Figure from \citep{Schemmer:2019}.}
\label{fig:AtomChipReleaseGHD}
\end{figure}

These limitations were overcome with the development of a new tool that became known as Generalized Hydrodynamics (GHD) \citep{Bastianello:2022}. It is often described as an emergent hydrodynamic theory used to obtain the large-scale dynamics of integrable systems far from equilibrium, and was originally applied to the study of transport properties of the XXZ chain \citep{Bertini:2016}, as well as the sinh-Gordon and Lieb-Liniger models \citep{Castro-Alvaredo:2016}. The method is built upon principles such as the generalized Gibbs ensemble \citep{Rigol:2007}, which describes the dynamics and relaxation of integrable systems to steady states after a long time evolution, and local entropy maximization, defined utilizing the language of the thermodynamic Bethe ansatz. The goal of GHD is to solve the Euler-scale equations
\begin{equation}
    \partial_t \rho_p(p,x,t)+\partial_x\left[v^{\text{eff}}(p,x,t)\rho_p(p,x,t)\right]=0
\end{equation}
where $\rho_p(p,x,t)$ denotes neither a fluid density in the classical sense nor a density of particles; instead, it describes a density of ``quasiparticles'' moving with an effective velocity $v^{\text{eff}}$. This density thus defines the state of a fluid cell, which depends on momentum, time and position. 
Since its original proposal, GHD has been developed in great detail \citep{DeNardis:2018,Bertini:2018,Alba:2019,DeNardis:2019,Vu:2019,Bastianello:2019,Bertini:2020} and applied to a number of different many-body models \citep{Doyon:2018,Bastianello:2018,Doyon:2019,Urichuk:2019,Mestyan:2019,Nozawa:2020,Bettelheim:2020}, including the Quantum Newton's Cradle problem \citep{Caux:2019} (a full description of the problem, including the occurrence of late-time thermalization is provided in \citep{Bastianello:2020}). The method has been recently experimentally verified with measurements of the time-evolution of weakly \citep{Schemmer:2019} (see Fig.~\ref{fig:AtomChipReleaseGHD}) and strongly \citep{Malvania:2021} atomic gases under trap quenches.

\subsection{Density Matrix Renormalization Group}
\label{subsec:DMRG}

Very few numerical methods for quantum systems are able to address static and dynamic problems outside the realm of integrable models while still being capable of handling fairly large systems. A remarkable exception is the case of the Density Matrix Renormalization Group (DMRG) method \citep{Schollwock:2005,Schollwock:2011}, originally proposed by White in \citep{White:1992, White:1993}. It quickly proved to be one of the most powerful and versatile methods for obtaining the low-energy properties of 1D quantum many-body models defined on lattices.

DMRG is an iterative and variational numerical method. Its core idea is relatively simple: it consists of splitting the length of a discrete system into two blocks (which can initially be of small size), introducing at each step two additional sites, such that the total system is now dubbed a {\it superblock} and then finding the ground state of this superblock with a diagonalization method such as the Lanczos algorithm. 
The resulting set of eigenstates is truncated so that only the states corresponding to the largest eigenvalues of the reduced density matrix for a superblock are maintained. These are then used to build the basis of states corresponding to the new superblocks, which can now be augmented by two additional sites. This process, called infinite-system DMRG, is repeated until the desired system size $L$ is reached. It is often viewed as a warm-up cycle for the finite-system DMRG stage, which starts with the basis found in the infinite-system DMRG stage and continues the growth of the two blocks in the system, but now keeping the total size fixed such that each block grows at the expense of the other. The process of each of the blocks being increased to maximum size as the other shrinks to its minimum size is called a {\it sweep}, and usually several sweeps are required before convergence in the energy of the system is achieved. The resulting states can then be utilized to calculate properties such as energy, correlation, and order parameters. While the main focus of DMRG is obtaining the ground state observables, it can also produce accurate results for the wave function and for low-lying excited states, within the approximations of the method.  

In recent years, DMRG has seen a number of updates and adaptations, including formulations in terms of Matrix Product States (MPS) \citep{McCulloch:2007,Verstraete:2008,Schollwock:2011,Wall:2012} and the development of related methods capable of treating dynamics, such as the time-evolving block decimation (TEBD) algorithm \citep{Vidal:2003,Vidal:2004} and the adaptive time-dependent DMRG \citep{Daley:2004,Schollwock:2006}. 
These tools have been widely employed in the study of discrete many-body quantum systems, especially in the treatment of Heisenberg-type spin chains \citep{White:1993spin,Hallberg:1995,Gobert:2005,Fuhringer:2008} and Hubbard-like models \citep{Daul:1996,Kuhner:1998,Kollath:2005,Kollath:2007,Ejima:2011} and has been useful in the understanding of important phenomena both in the theory of condensed matter and cold atoms.  More importantly, DMRG can also be employed in the study of static and dynamic properties of continuous systems. In the low-density regime, a Bose-Hubbard model can be mapped into the Lieb-Liniger model \citep{Haldane1980}, and tools tailored for discrete systems can be applied. For instance, the dynamics of excitations in the two-component Bose gas can be described in the superfluid regime with the introduction of a small lattice spacing parameter that is related to the hopping term and the mass. Such an approach has proven crucial in the study of phenomena such as spin-charge decoupling in two-component bosons, even when compared to the hydrodynamic approximation \citep{Kleine:2008a, Kleine:2008b}.  

While originally devised for studying the low-energy properties of 1D systems, DMRG has also been used to approach problems in higher dimensions \citep{Nishino:1995,White:1996,White:1998,Stoudenmire:2012,orus2014}, although this is considerably more difficult due to the scaling with system size beyond 1D. More recently, a number of finite temperature DMRG studies have also been performed \citep{Verstraete:2004,Zwolak:2004,Feiguin:2005,White:2009,Karrasch:2012,Karrasch:2013}. 
A generalization of MPS to continuous systems (cMPS) has proven to be very successful in predicting the properties of the Lieb-Liniger gas \citep{verstraete2010c} and other works exploring the properties of this technique have followed
\citep{haegeman_2013,cuevas2018c}. 
For further information and for hands-on implementations, we refer to the different open-source DMRG-based libraries that are currently available \citep{Bauer:2011,Jaschke:2018,Itensor:2020}.

\subsection{Multi-Layer Multi-Configuration Time-Dependent Hartree Method for Atomic Mixtures} 
\label{subsec:ML-MCTDHA}

Another way to approach the quantum many-body problem is to use wave function propagation methods that address the exponential scaling of the Hilbert space with the particle number by
expanding the many-body state in terms of a dynamically optimized truncated basis. 
The Multi-Configuration Time-Dependent Hartree (MCTDH) family of methods~\citep{meyer1990multi,beck2000multiconfiguration,lode2020colloquium} is such an approach for the {\it ab-initio} treatment of multi-dimensional time-dependent problems. 
A more elaborated discussion on the historical development of this family of methods can be found in~\cite{lode2020colloquium}, while more recent progress have been addressed e.g. in~\citep{wodraszka2016using,wodraszka2017systematically,leveque2017time,kohler2019dynamical,larsson2017dynamical}. 
Notable examples where these methods have been successfully applied include tunneling processes~\citep{Sakmann:09,lode2012interacting,schurer2016impact}, collective 
excitations~\citep{Schmitz2013}, quench-induced dynamics~\citep{mistakidis2015negative,mistakidis2014interaction,neuhaus2017quantum}, periodic driving of lattice-trapped atoms~\citep{mistakidis2015resonant,mistakidis2017mode}, the effect of long-range 
e.g. dipolar interactions~\citep{chatterjee2020detecting,chatterjee2019correlations,bera2020relaxation}, quantum non-linear excitations~\citep{streltsov2011swift,kronke2015many,katsimiga2018many,katsimiga2017many,beinke2018enhanced} and atom-ion systems~\citep{schurer2015capture}. 

\begin{figure}[tb]
    \center
    \includegraphics[width=0.8\linewidth]{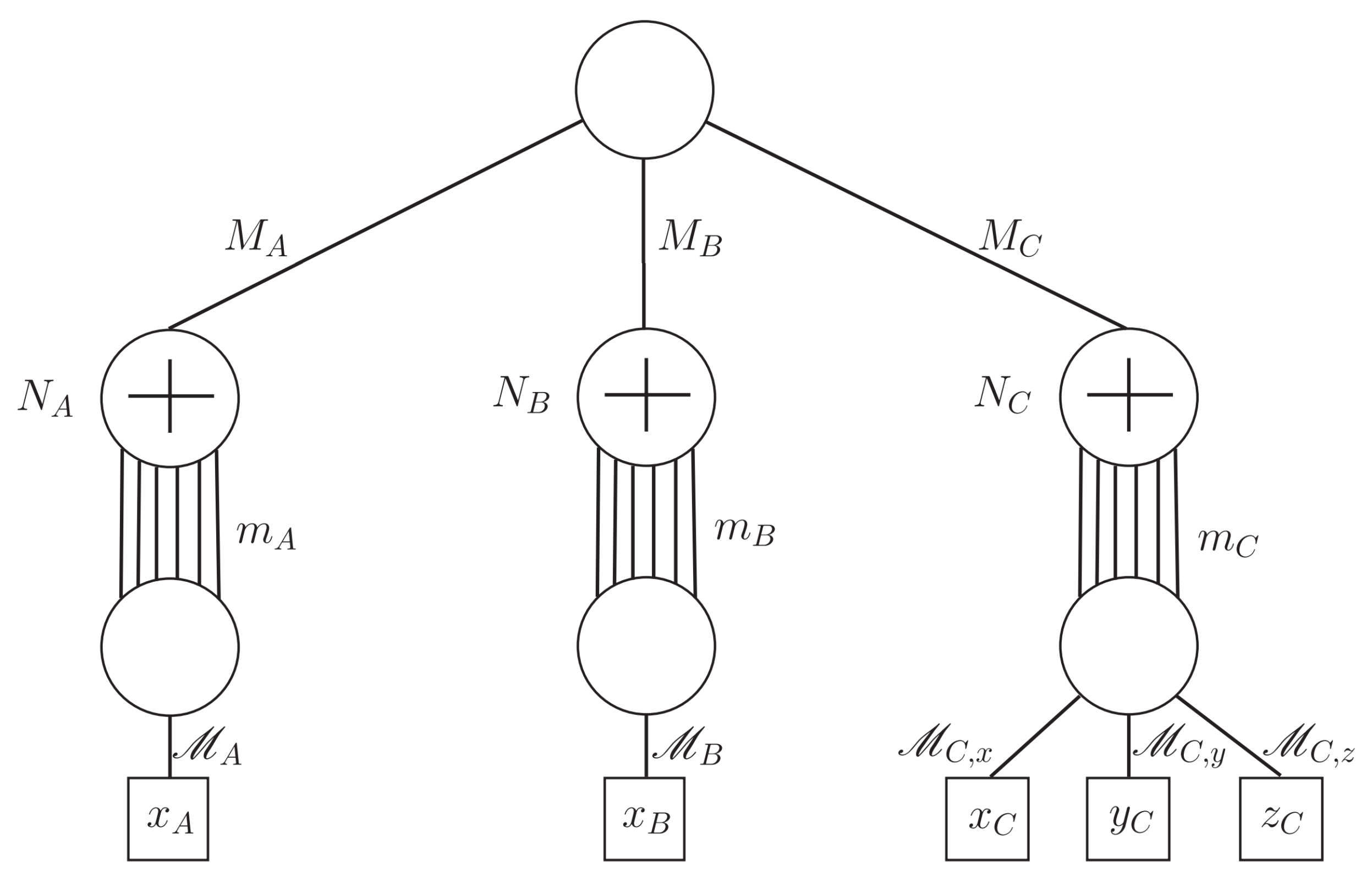}
    \caption{Schematic representation of a three-layer tree diagram containing (from top to bottom) the top layer, the species layer, and the particle layer. The system corresponds to a three-species bosonic mixture with species $A$, $B$, and $C$. The number of species (single-particle) functions are $M_A$, $M_B$, and $M_C$ ($m_A$, $m_B$, and $m_C$). The primitive degrees-of-freedom $x_A$, $x_B$ and ($x_C$, $y_C$, $z_C$)  are marked by the squares and have dimension $\mathcal{M_A}$, $\mathcal{M_B}$ and ($\mathcal{M}_{C,x}$, $\mathcal{M}_{C,y}$, $\mathcal{M}_{C,z}$). The "+" symbol within the species nodes designates the fact that the species are bosonic. Figure from~\citep{cao2013multi}.}
\label{fig:tree_structure_MLB} 
\end{figure} 

An important extension was the integration of the multi-layer (ML) scheme into the wave function ansatz~\citep{manthe2008multilayer,vendrell2011multilayer,wang2003multilayer} that can be adapted to system specific intra- and intercomponent correlations (see Fig.~\ref{fig:tree_structure_MLB}). 
This led to the development of ML-MCTDHX~\cite{cao2017unified,cao2013multi,kronke2013non} allowing to treat cold atomic mixtures by utilizing dynamically optimized truncated basis states for the involved components to solve the time-dependent many-body 
Schr{\"o}dinger equation while capturing the emergent correlation effects using a computationally feasible basis size. 
This way it is possible to treat mesoscopic 
Bose-Bose~\citep{cao2013multi,kronke2013non}, Bose-Fermi and Fermi-Fermi mixtures including 
spinor components~\citep{koutentakis2019probing} and different spatial dimensions while still handling multiple internal components \citep{bolsinger2017beyond,cao2017unified}. 
The characteristic structure of the wave function expansion at the different layers for a three-component bosonic mixture is provided in Fig.~\ref{fig:tree_structure_MLB}. 
Recent applications include, but are not limited to, tunneling of mixtures~\citep{keiler2020doping,theel2020entanglement}, phase-separation processes~\citep{mistakidis2018correlation,erdmann2019phase}, Bose~\citep{mistakidis2020pump} and Fermi~\citep{Mistakidis2019FermiPolarons} polaron dynamics, quantum dark-bright solitons~\citep{katsimiga2017dark}, mesoscopic charged molecules~\citep{schurer2017unraveling} and spinor BECs~\citep{mittal2020many}. 

In general, an $S$-component ultracold atomic mixture is a composite system residing in the Hilbert space 
$\mathcal{H}^{AB,\dots,S}=\mathcal{H}^A\otimes\mathcal{H}^B\otimes\dots\otimes\mathcal{H}^S$, where $\mathcal{H}^{\sigma}$ denotes
the Hilbert space of the $\sigma=A,B,\dots,S$ component. 
To account for the intercomponent correlations of the mixture the many-body wave function can be expanded in terms of a 
set of $k=1,2,\dots,D_{\sigma}$ distinct species functions $\Psi^{\sigma}_k (\vec x^{\sigma};t)$ for each $\sigma$-component, see also Fig.~\ref{fig:tree_structure_MLB}, as  
\begin{equation}
\begin{split}
    \Psi(\vec x^A,\dots,\vec x^S;t) = \sum_{i_{A},i_{B},\dots,i_{S}=1}^{D_A,D_B,\dots,D_S} A_{i_{A},i_{B},\dots,i_{S}}(t) 
    \Psi^A_{i_{A}} (\vec x^A;t)\otimes\Psi^B_{i_{B}} (\vec x^B;t)\otimes \dots \otimes \Psi^S_{i_{S}} (\vec x^S;t). 
\end{split}
\label{decomposition_S_species}
\end{equation}
In this expression the $\vec x^{\sigma}=(x_1^{\sigma},x_2^{\sigma},\dots,x_{N_{\sigma}}^{\sigma})$ are the spatial coordinates and 
the $A_{i_{A},i_{B},\dots,i_{S}}(t)$ are the time-dependent expansion coefficients offering a measure of the intercomponent correlations. 
Also, the functions $\{\Psi_k^{\sigma}\}$ form an orthonormal set of $N_{\sigma}$-body wave functions in a subspace of $\mathcal{H}^{\sigma}$.

In the case of a binary mixture ($S>2$), we can set $D_A=D_B\equiv D$ without loss of generality since according to the Schmidt decomposition for bipartite systems~\citep{horodecki2009quantum} the maximal strength of intercomponent correlations depends solely on $\min\{D_A,D_B\}$. 
As such, the ansatz in Eq.~\eqref{decomposition_S_species} reduces to a truncated Schmidt decomposition of rank $D$, namely 
$\Psi(\vec x^A,\vec x^B;t) = \sum_{k=1}^D \sqrt{ \lambda_k(t)}~\Psi^A_k (\vec x^A;t) \Psi^B_k (\vec x^B;t)$ with $\sqrt{\lambda_k(t)}$ being the Schmidt weights. 
These are referred to as the natural species populations of the $k$-th species function $\Psi^{\sigma}_k$ of the $\sigma$-component since they correspond to the eigenvalues of the species' reduced density matrix 
\begin{align}
\rho^{N_{\sigma}} (\vec{x}^{\sigma}, \vec{x}'^{\sigma};t)=\int dx^{\sigma'}_{1} dx^{\sigma'}_{2}\dots dx^{\sigma'}_{N_{\sigma'}} \Psi^*(\vec{x}^{\sigma}, 
\vec{x}^{\sigma'};t) \Psi(\vec{x}'^{\sigma},\vec{x}^{\sigma'};t),
\end{align}
where $\sigma \neq \sigma'$. 
They quantify the presence of intercomponent correlations and entanglement, and 
if multiple eigenvalues of $\rho^{N_{\sigma}}$ are macroscopically occupied then the system is considered species 
entangled~\citep{roncaglia2014bipartite}; otherwise it is disentangled. 
In this sense, the deviation $1-\lambda_1(t)$ estimates the degree of system entanglement. 

Next, in order to include intracomponent correlations into the many-body wave function ansatz each species function 
$\Psi^{\sigma}_k(\vec x^{\sigma};t)$ is expressed as a linear superposition of time-dependent number-states $|\vec{n} (t) \rangle^{\sigma}$ with 
time-dependent expansion coefficients $B^{\sigma}_{k;\vec{n}}(t)$. 
Each number state $|\vec{n} (t) \rangle^{\sigma}$ corresponds to a permanent for bosons and a determinant for fermions.  
It is constructed upon $d_{\sigma}$ time-dependent variationally optimized single-particle functions (SPFs) 
$\left|\phi_z^{\sigma} (t) \right\rangle$, with $z=1,2,\dots,d_{\sigma}$ and occupation numbers 
$\vec{n}=(n_1,\dots,n_{d_{\sigma}})$. 
The occupations $n_i$ are integers, and in particular $n_i\in \{ 0, 1 \}$ for fermions, 
and determine the number of particles in a certain SPF while satisfying the restriction $\sum_{i=1}^{d_{\sigma}} n_i=N_{\sigma}$. 
Furthermore, the SPFs are expanded with respect to a time-independent primitive basis of dimension $M_{pr}$, which could, for example, be a discrete variable representation.

In order to address the time-evolution of the ($N_A+N_B+\dots+N_S$)-body wave function $\Psi(\vec x^A,\vec x^B\dots,\vec x^S;t)$ obeying the Schr\"odinger equation under the influence of the Hamiltonian $H$, one can numerically determine the ML-MCTDHX equations of motion~\citep{cao2017unified} based upon a variational principle such as the 
Lagrangian~\citep{broeckhove1988equivalence}, McLachlan~\citep{mclachlan1964variational} or 
the Dirac-Frenkel~\citep{dirac1930note,frenkel1934wave} 
principle for the above-discussed generalized wave function ansatz. 
As a case example, referring to a binary Bose-Fermi mixture of species B and F, these equations consist of a set of $D^2$ ordinary 
(linear) differential equations of motion for the coefficients $\lambda_k(t)$, coupled to a set of $D[\binom{N_B+d_B-1}{d_B-1 }+ \binom{d_F}{N_F }]$ non-linear integrodifferential  equations for the species functions, and 
$d_B+d_F$ non-linear integrodifferential equations for the SPFs~\citep{cao2017unified}. The main control parameters of the ML-MCTDHX are $D_{\sigma}$ and $d_\sigma$, and 
it is possible to use the scheme in different orders of approximation. 
As such when choosing $d_\sigma=M_{pr}$ and $D_{\sigma}=\binom{N_{\sigma}+d_{\sigma}-1}{d_{\sigma}-1}$ for bosons [$D_{\sigma}=\binom{d_{\sigma}}{ N_{\sigma}}$ for fermions], the wave function ansatz is of full configuration-interaction type thus covering the entire Hilbert space of the system. 
Neglecting all correlations i.e. setting $D_{\sigma}=d_\sigma=1$ for bosons ($d_\sigma=N_\sigma$ for fermions), the ML-MCTDHX equations reduce to the set of coupled Gross-Pitaevskii (Hartree-Fock) equations of motion~\citep{pitaevskii2016bose}. For concrete cases with systems 
that do not feature very strong correlations, it is typically 
sufficient to consider a limited number of species functions and orbitals, and hence
to obtain a very efficient representation of the many-body state.

A major challenge for many-body simulations is to achieve a desirable degree of convergence. Within the ML-MCTDHX framework one has to carefully inspect the truncation order of the total 
Hilbert space which is determined by the numerical configuration space $C=(\{D_{\sigma}\};\{d_{\sigma}\};M_{pr})$. This is done by systematically 
increasing $D_{\sigma}$, $d_\sigma$, and $M_{pr}$, and comparing the results for the observables of interest. Additionally, in limited cases where there exist analytical estimations for certain observables, e.g.~the center-of-mass variance~\citep{klaiman2015variance,alon2019variance}, they can be employed to test the convergence of the ML-MCTDHX simulations~\citep{katsimiga2017dark}. 
Finally, we note that the predictions of ML-MCTDHX have been benchmarked against various numerical data, e.g., for two particles in a harmonic oscillator~\citep{gwak2018benchmarking}, for the generalized Gross-Pitaevskii framework with the Lee-Huang-Yang contribution~\citep{mistakidis2021formation}, for systems with impurities~\citep{koutentakis2021pattern,brauneis2022artificial}, and for spin-chain models~\citep{koutentakis2019probing}.
In general, ML-MCTDHX behaves as most other {\it ab-initio} methods and the degree of convergence in general depends on the observable of interest and the system under consideration. 
Indeed, lower order quantities such as the energy of the system or densities display a relatively faster convergence as compared to higher-order observables like the $N_{\sigma}$-body reduced density matrix.

\subsection{Spin-chain mapping}
\label{sec09SC}

As discussed in Section~\ref{CloseToFermi}, strongly interacting particles with two or more internal degrees of freedom can be mapped onto a spin-chain Hamiltonian whose coupling coefficients are determined by the external trapping geometry. This subsection illustrates the mapping for a two-component system following~\citep{Volosniev2015Dynamics}, see also~\citep{Deuretzbacher2014Mapping}. For the sake of discussion, it is assumed that $g_{A}=g_{B}$, allowing for the following form of the Hamiltonian from Eq.~(\ref{eq:Hamiltonian_pseudospinor})
\begin{align}
    H=&\sum_{n=1}^{N_A} H_0(x_n)+\sum_{n=1}^{N_B} H_0(y_n) \nonumber \\
    &+\kappa g\sum_{n<m}^{N_A}\delta(x_n-x_m)  +\kappa g\sum_{n<m}^{N_B}\delta(y_n-y_m)+g\sum_{n=1}^{N_A}\sum_{m=1}^{N_B}\delta(x_n-y_m), 
    \label{SCHam}
\end{align}
where $H_0(x)=-\frac{\hbar^2}{2m}\frac{\partial^2}{\partial x^2}+V(x)$ is the single particle Hamiltonian. In the impenetrable limit ($1/g=0$), the eigenstates of the system can be written as
\begin{equation}\label{SCansatz}
\Psi=\sum_{k=1}^{L(N_A,N_B)}a_k P_k\Phi_F(\{x_i,y_j\}),
\end{equation} 
where $\Phi_F(\{x_i,y_j\})$ describes the state of $N=N_A+N_B$ spinless fermions for the ordering of particles $x_1<...<x_{N_A}<y_1<...<y_{N_B}$. $P_k$ is a permutation operator acting on the coordinates. There are 
\begin{equation}
L(N_A,N_B)=\frac{(N_A+N_B)!}{N_A!N_B!}
\end{equation}
combinations, and thus independent coefficients $a_k$, which lead to $L(N_A,N_B)$ degenerate states at $1/g=0$.

The states in the vicinity of $1/g=0$ should extremize the slope of the energy near the fermionization limit, which is obtained from the Hellmann-Feynman theorem,
\begin{eqnarray}\label{SChf}
\frac{\partial E}{\partial g}=\frac{\sum_{n=1}^{N_A}\sum_{m=1}^{N_B}\langle\delta (x_n-y_m)\rangle+\kappa \sum_{n<m}^{N_A}\langle\delta (x_n-x_m)\rangle +\kappa \sum_{n<m}^{N_B}\langle\delta (y_n-y_m)\rangle}{\langle\Psi|\Psi\rangle},
\end{eqnarray}
where $\langle\delta (x)\rangle\equiv\langle\Psi\vert\delta (x)\vert\Psi\rangle$.
The numerator in this expression can be calculated using the boundary conditions due to the zero-range interaction. For particles of the same type, the conditions are given as
\begin{eqnarray}
\left(\frac{\partial\Psi}{\partial x_{i}}-\frac{\partial\Psi}{\partial x_{i'}}\right)\Bigg\rvert_{x_{i}-x_{i'}=0^-}^{x_{i}-x_{i'}=0^+}=2\frac{\kappa g m}{\hbar^2} \Psi(x_{i}=x_{i'}), \nonumber \\
\left(\frac{\partial\Psi}{\partial y_{i}}-\frac{\partial\Psi}{\partial y_{i'}}\right)\Bigg\rvert_{y_{i}-y_{i'}=0^-}^{y_{i}-y_{i'}=0^+}=2\frac{\kappa g m}{\hbar^2} \Psi(y_{i}=y_{i'}), 
\label{SCcond1}
\end{eqnarray}
and for the interaction between $A$ and $B$ particles one has
\begin{equation}\label{SCcond2}
\left(\frac{\partial\Psi}{\partial x_{i}}-\frac{\partial\Psi}{\partial y_{j}}\right)\Bigg\rvert_{x_{i}-y_{j}=0^-}^{x_{i}-y_{j}=0^+}=2\frac{g m}{\hbar^2} \Psi(x_{ i}=y_{j}).
\end{equation}
Equations~\eqref{SChf}, \eqref{SCcond1} and~\eqref{SCcond2} with the wave function proposed in Eq.~\eqref{SCansatz} lead to the expression for the energy in the vicinity of $1/g=0$ 
\begin{eqnarray}\label{SCfunc}
E=E_F-\frac{1}{g}\frac{\sum_{i=1}^{N-1}\alpha_i \mathcal{A}_i}{\sum_{k=1}^{L(N_A,N_B)}a_{k}^2},
\end{eqnarray}
where $E_F$ is the eigenenergy of a system of $N_A+N_B$ spinless fermions, and 
\begin{equation}
    \mathcal{A}_i=\sum_{k=1}^{L(N_A-1,N_B-1)}(a_{i|k}^{AB}-a_{i|k}^{BA})^2+\frac{2}{\kappa}\sum_{k=1}^{L(N_A-2,N_B)}(a_{i|k}^{AA})^2+\frac{2}{\kappa}\sum_{k=1}^{L(N_A,N_B-2)}(a_{i|k}^{BB})^2.
\end{equation}
The coefficient $a_{i|k}^{s s'}$ corresponds to $a_k$ of Eq.~\eqref{SCansatz} that multiplies $\Phi_F$ with $s$ particle at the position $i$ and $s'$ particle at the position $i+1$. There is no need to specify the order of other particles, in fact the sum over $k$ is the sum over the ordering of all other particles. The coefficients $\alpha_i$ depend only on the shape of the trapping potential and on the total number of particles $N=N_A+N_B$. In particular, they are the same for bosonic and fermionic systems and are given by 
\begin{equation}\label{SCgeo}
    \alpha_i=\frac{\hbar^4}{m^2}\frac{\int_{x_1<x_2\cdots<x_{N-1}}dx_1\cdots dx_{N-1}\Big|\frac{\partial \Phi_F(x_1,\cdots,x_i,\cdots,x_N)}{\partial x_N}\Big|^2_{x_N=x_i}}{\int_{x_1<x_2\cdots<x_{N-1}}dx_1\cdots dx_N |\Phi_F(x_1,\cdots,x_i,\cdots,x_N)|^2}.
\end{equation}

The mapping of  a strongly interacting system onto a spin chain is established upon observation that the energy functional in Eq.~\eqref{SCfunc} can be obtained with the wave-function ansatz
\begin{equation}
\Psi_s=\sum_{k=1}^{L(N_A,N_B)} a_k P_k |\uparrow_1...\uparrow_{N_A}\downarrow_1...\downarrow_{N_B}\rangle
\end{equation}
for the spin-chain Hamiltonian
\begin{equation}\label{SCxxz}
H_s =\left(E_F-\frac{\kappa+1}{g\kappa}\sum_{i=1}^{N_A+N_B-1}\alpha_i \right)\mathbf{I}+\frac{1}{g}\sum_{i=1}^{N_A+N_B-1}\alpha_i \left[\mathbf{P}_{i,i+1}-\frac{1}{\kappa}\sigma_z^{i}\sigma_z^{i+1}\right],
\end{equation}
where $\mathbf{P}_{i,i+1}$ is the operator that swaps the spins at the $i$th and ($i+1$)th sites, $\sigma_z^{i}$ is the third Pauli matrix that acts on the site $i$ and $\mathbf{I}$ is the identity operator. This correspondence implies that to find the set $\{a_k\}$ that minimizes the energy in Eq.~\eqref{SCfunc}, it is enough to find the eigenstates of the Hamiltonian $H_s$.

The non-trivial part of applying the mapping is the computation of the $\alpha_i$. Given the external potential $V$, it is typically easy to compute a set of $N=N_A+N_B$ one-body eigenstates and then construct the Slater determinant, $\Phi_F$, from these $N$ states. However, standard techniques for multi-dimensional integration allow one to calculate the multiple integral in Eq.~\eqref{SCgeo} only for $N\lesssim15$~\citep{Volosniev2014StrongInteractions,Levinsen2015Ansatz,Yang:2015,Li2016multibranch}. It is hard, if not impossible, to extend multi-dimensional integration to a large number of particles ($N\simeq 15$), due to the curse of dimensionality, see, e.g.,~\citep{Hinrichs2012} and references therein. 

A method to tackle systems with $N>15$ is highly desirable to study a number of research questions, such as few- to many-body transition, and finite-size effects in mesoscopic cold-atom experiments. 
Luckily, Eq.~\eqref{SCgeo} can be rewritten as a sum of 1D integrals~\citep{LOFT2016_Conan}, which allows one to calculate the $\alpha_i$ for systems with $N\lesssim 35-60$~\citep{LOFT2016_Conan,Deuretzbacher2016Numerics}.  An open source code CONAN (Coefficients of 1D $N$-Atom Networks) for such calculations is also available \citep{LOFT2016_Conan}.

To go to even larger systems (e.g., with $N\simeq 60$), one could use the existing approximate expressions of $\alpha_i$ for different confining potentials. For large systems in external traps that change weakly on the length scale given by the density of spinless fermions, the local density approximation~\citep{Deuretzbacher2014Mapping} can provide one with an accurate expression
\begin{equation}
    \alpha_i=\frac{\hbar^4\pi^2 n_{TF}^3(z_i)}{3m^2},
    \label{eq:alpha_LDA}
\end{equation}
where $n_{TF}$ is the Thomas-Fermi density profile, and $z_{i}$ is the center of mass of the $i$th and the ($i+1$)th particle. The  expression in Eq.~\eqref{eq:alpha_LDA} can be anticipated from the results obtained for homogeneous quantum wires~\citep{Matveev2004wire}. Its accuracy has been checked for various trapping geometries~\citep{Deuretzbacher2014Mapping,Marchukov2016FewBody,Li2016multibranch}, where it was shown that Eq.~\eqref{eq:alpha_LDA} is qualitatively correct even for a handful of particles, provided that the trap changes weakly on the length scale given by the density of the gas. Although, quantitative differences between the exact values and the ones obtained with Eq.~\eqref{eq:alpha_LDA} might lead to noticeable effects in time dynamics for a small numbers of particles~\citep{Marchukov2016FewBody}, Eq.~\eqref{eq:alpha_LDA} is a powerful starting point for understanding the relationship between the trapping potential and the coefficients $\alpha_k$. In reverse-engineering applications, Eq.~\eqref{eq:alpha_LDA} provides an initial guess for trapping potentials that may lead to perfect state transfer~\citep{Loft2016Transfer} or that may realize desired entanglement Hamiltonians~\citep{Barfknecht2021Entanglement}.

For a harmonic trap, the values of the $\alpha_i$ have been approximated by~\citep{Levinsen2015Ansatz} 
\begin{equation}
    \alpha_i\simeq C_N i(N-i),
    \label{eq:alphai_levinsen}
\end{equation}
where the coefficient $C_{N}$ depends only on the total number of particles. For a given system with $N$ fixed, it sets the overall energy scale; its value is not important for calculations of particle-particle correlations. The values of $C_N$ including its asymptotic expansion are presented in~\citep{Massignan2015Magnetism}.
Equation~(\ref{eq:alphai_levinsen}) is not exact, but it is surprisingly accurate~\citep{Levinsen2015Ansatz,Yang:2015,Laird2017SUN}. Its simplicity might imply some hidden symmetries for strongly interacting systems in harmonic traps. 

For a system in a box potential\footnote{\citep{Duncan2017SpinImpurity} observed similar physics in an optical lattice potential for certain fillings.} that is Bethe-ansatz solvable, one expects a Bethe-ansatz solvable spin chain. This expectation is supported by indications that the exchange coefficients do not depend on the position. This was seen numerically for the ground~\citep{Marchukov2016FewBody} and excited states \citep{Barfknecht2021Current}. 
For the ground state, this observation can be supported by the Bethe-ansatz results~\citep{GuanSpinChain2008,Pan2017ExactOrdering}. Assuming that the coefficients are indeed identical, one can show that\footnote{Note that this expression could be used to derive  Eq.~(\ref{eq:alpha_LDA}).} $\alpha_i=2\hbar^2 E_F/(m L)$~\citep{Volosniev2017MassImbalance,Barfknecht2021Current}. This simple form of the coefficients suggests a path for engineering integrable spin models without lattices. One can use it to design multicomponent, in particular, SU(N) models, whose properties are known for fermions~\citep{Yang1967Fermi,Takahashi1999OneDimensional,Guan2013Review} and are less explored for bosons. Since the Bose systems have additional interactions, they can be used to study integrable and non-intergrable parameter regimes in a single experimental set-up opening possibilities to study transitions between them, see, e.g.,~\citep{Hao2009TwoComponent,Massignan2015Magnetism}.

%% file: Section10.tex
\section{Beyond contact interactions: an extended cold-atom laboratory}\label{sec:beyond} 

 Physical phenomena discussed in the previous sections are based on the assumption that two-body interactions can be modelled using delta-function (or contact) potentials, $g\delta(x_1-x_2)$ -- the standard approach for cold-atom systems in 1D, for a review see~\citep{YUROVSKY2008}. Unlike higher dimensions, see~\citep{BraatenHammer2006}, the zero-range two-body interaction leads to a well-defined many-body problem in 1D. Therefore, from a purely mathematical point of view, there is no need for theoreticians to consider other (more fundamental) interactions. However, from the physical standpoint, the potential, $g\delta(x)$, has its limitations, and cannot be used to describe a number of important phenomena. For example, delta-function interactions cannot bind spinless fermions (due to the Pauli principle) or lead to 
 Borromean and Brunnian systems\footnote{Borromean (Brunnian) systems are bound states of more than two particles in which two-body (all) subsystems are unbound, see, e.g.,~\citep{Baas2012Brunnian}. In 3D, many properties of such states can be explored using zero-range potentials assuming a three-body parameter~\citep{jensen2004, BraatenHammer2006}. In 1D and 2D, finite-range potentials are necessary for Borromean and Brunninan systems since any attractive potential supports a two-body bound state~\citep{LandauQM}.}. These observations imply that delta-function potentials cannot be used to model physics of, for instance, unconventional superconductors or superfluids [such as classic helium-3 systems~\citep{Leggett1972pwave}], or of exotic systems in which few-body settings play the role of building blocks. 
 
Studies of particles interacting via other-than-delta-function potentials are therefore necessary. This necessity has motivated recent theoretical and experimental studies of cold atoms with beyond contact interactions.
In our opinion, few-body physics of these systems will be the topic of many future works, and we find it necessary to set the stage for them by providing a succinct description of some rapidly developing topics lying beyond the main focus of the present review. As such, this section is different from the previous ones. It provides an outlook into the future of the field by listing directions explored by (mainly) time-independent methods and approaches that pave the way for studies of dynamics.

\subsection{Finite-range corrections in atom-atom interactions}\label{fin_range_corrections} 

In cold-atom physics, atom-atom interactions are usually considered to be of short range character. The validity of this approximation has been established for many systems by analyzing various experimental data with theoretical models based upon $g\delta(x)$, where $g$ then describes only some integral information about the actual potential, see, e.g.,~Fig.~\ref{fig:2bodycorr}. It turns out, however, that finite-range corrections are important for narrow Fano-Feshbach resonances, which are closed-channel dominated~\citep{friedrich2006theoretical,greene2017universal} yielding an energy-dependent scattering length~\citep{chin2010feshbach}.

A similar scenario may arise for collisions in the presence of quasi-1D set-ups, where the transverse spatial directions are tightly confined as compared to the axial one (i.e., $\omega_x, \omega_y \gg \omega_z$). 
Then, it might be beneficial to incorporate finite-range corrections into the coupling constant $g$ regardless of the type of Fano-Feshbach resonances that are employed to tune the two-body atomic interactions. To model two-body physics of such systems, one can for example use the energy-dependent scattering length
\begin{equation}
    a(k)\simeq \frac{a}{1-\frac{1}{2}k^2 a r_e},\label{energy_dep_scat}
\end{equation}
where $k$ is the two-body momentum in the relative two-body frame given by the relation $k=\sqrt{2 \mu E/\hbar^2}$ with $\mu$ indicating the reduced mass of the particles and $E$ their total collisional energy; $r_e$ is the effective range correction and $a$ denotes the energy-independent scattering length parameter.

In quasi-1D trapping potentials the relation~(\ref{energy_dep_scat}) was used in \citep{Naidon2007EffectiveRange,YUROVSKY2008} (see~\citep{Blume2002pseudopotential} for relevant 3D discussions) to construct models that allow for the description of resonant phenomena, such as confinement-induced resonances~\citep{YUROVSKY2008,dunjko2011confinement}. Note that many-boson problems with energy-dependent coupling coefficients require a modification of the commonly used numerical approaches, see, e.g.,~\citep{Fu2003Beyond,Collin2007Energy,Cappellaro2017finiterange,Tononi2018}. This is one of the reasons why the many-body (and consequently few-body) physics of these systems is not comprehensively explored.

\subsection{Odd-wave interactions}

Simulations of `unconventional' many-body physics in 1D may be based upon 3D
$p$-wave physics, which leads to interactions between two spin-polarized fermions. 
The strength of the interactions can be tuned using external fields, see,
e.g.,~\citep{Granger2004pwave,Peng2014pwave}.

\begin{figure}[tb]
    \begin{center}
    \includegraphics[width=0.8\textwidth]{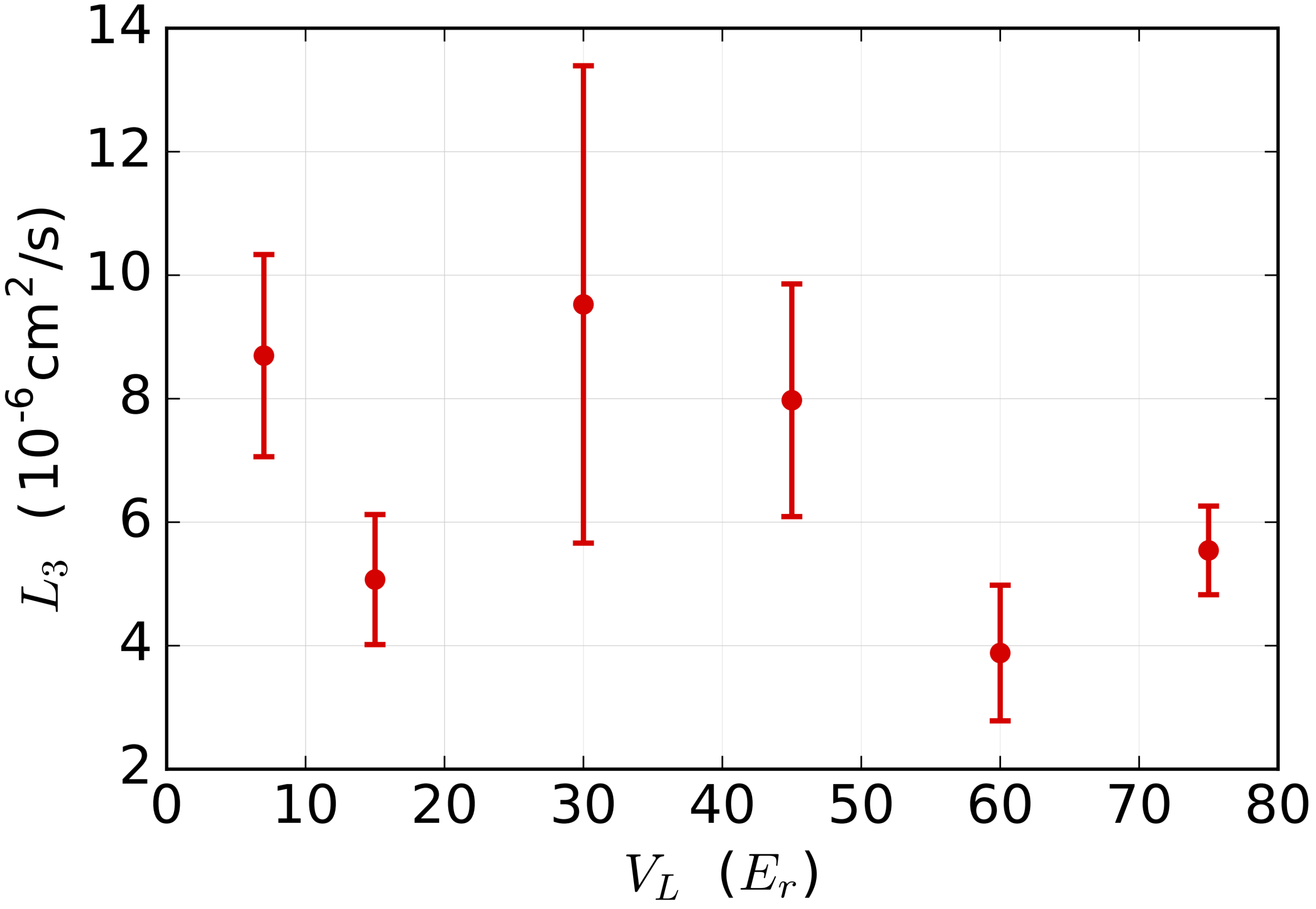}
    \end{center}
    \caption{Peak three-body loss coefficient for fermions interacting via a $p$-wave resonance as a function of of the lattice depth. Figure from supplemental material of~\citep{Chang2020Loss}.}
\label{fig:rec_p_wave}
\end{figure}

Severe atom losses complicate studies close to $p$-wave resonances in 3D, see, e.g.,
\citep{Regal2003pwave,Luciuk2016pwave}. However, recent theoretical and experimental
work~\citep{gunter2005, zhou2017pwave, Kurlov2017Two_body, Chang2020Loss} may provide a background for
future experimental realization of strongly interacting 1D systems with odd-wave
interactions. As an example, Figure~\ref{fig:rec_p_wave} presents experimental results for the
dependence of the peak three-body loss coefficient, $L_3$, with respect to the lattice depth, $V_L$, in $^6$Li
\citep{Chang2020Loss}. Only a minor dependence on the confinement
strength is visible, and \citep{Chang2020Loss} suggest that suppressing loss in
heavier fermions such as $^{40}$K, $^{161}$Dy and $^{167}$Er might be even more promising.
Those systems may then allow one to study few- and many-body statics and dynamics with odd-wave interactions.

To describe systems with odd-wave interactions, one requires
a beyond-contact potential, since the term $g\delta(x)$ acts only in the {\it even-wave channel} due to the symmetry
$\delta(-x)=\delta(x)$. 
A possible route to describe odd-wave interactions is via the pseudopotential
\begin{equation}
    v(x)=g_P\delta'(x)\hat O_{\pm}, \qquad \hat O_{\pm} f(x)=\frac{1}{2}\left( f'(0+)+f'(0-) \right),  
\end{equation}
which defines the odd-wave boundary conditions for the wave function \citep{girardeau2004,GIRARDEAU20043}. 
Note that this zero-range approximation to odd-wave potentials in 1D requires regularization~\citep{valiente2015effective,Cui2016pwave}, which is similar to zero-range $s$-wave problems in 3D~(\citep{BraatenHammer2006}). 
A theoretical approach that includes beyond-zero-range physics and does not require a regularization can be based upon the finite-range potential~\citep{Veksler2014Model}:
\begin{equation}
 v(x)= g_0\delta(x)+g_1[\delta(x-d)+\delta(x+d)],
 \label{eq:interaction_Veksler}
\end{equation}
where $g_0,g_1$ and $d$ are parameters that determine the strength and the range of the interactions.

Experimental progress in engineering odd-wave interactions has motivated theoreticians to study stability and phase diagrams of many-body systems with odd-wave interactions~\citep{Yajiang2007pwave, Imambekov2010pwave, Cui2018pwave}. Theoretical tools for such studies are the Bethe ansatz, which can be used under certain conditions, in particular when the Bose-Fermi mapping is applicable~\citep{Cheon1999FermiBose,valiente2020bose,valiente2020uni}.
To tackle many-body problems that cannot be solved by the Bethe ansatz, one can rely on numerical methods discussed in Sec.~\ref{sec:Methods} as well as on variational approaches, see, e.g., ~\citep{Kocik_2020}. For strongly interacting systems one can employ effective spin-chain formulations~\citep{Hu2016pwave}, and it is possible to use a mapping of a continuous model onto a lattice representation~\citep{Schmidt2010pwave}. Several such approaches inspired by effective field theory~\citep{Hammer2000Effective} and few-body routes to the physics of liquids have been introduced recently~\citep{valiente2016few,valiente2020bose,morera2021quantum}. The physics of the beyond Lieb-Liniger model that corresponds to the interaction in Eq.~(\ref{eq:interaction_Veksler}) is discussed in~\citep{Veksler2016BeyondLL,Jachymski2017BeyondLL}, see also~\cite{Syrwid2022many_molecule} for relevant dynamics of a two-component ultracold Fermi gas confined in a 1D harmonic trap. 

Generalizations of the $s$-wave confinement induced resonances  to higher partial wave two-body interactions are considered in~\citep{Giannakeaspra2011, Giannakeaspra2012,hesspra2015}. Specifically, it is demonstrated that the inclusion of higher orbital angular momentum, i.e., $\ell>0$, leads to a series of confinement-induced resonances coupled to each other due to the trapping potential. The energy dependence of such resonances enjoys several universal attributes, such that for deep enough interatomic potentials only four $\ell$-wave confinement-induced resonances are relevant, i.e., two for identical bosons and two for identical fermions~\citep{hesspra2014}.

The $\ell$-wave confinement-induced resonances can be important in dipolar gases~\citep{Giannakeasprl2013} or in ultracold gases in the vicinity of high-$\ell$ Fano-Feshbach resonances. 
In particular, recent experiments report the existence of high-partial wave Fano-Feshbach resonances where open and closed channels are dominated by $\ell\geq2$ angular momentum in binary mixtures of $^{87}$Rb--$^{85}$Rb atoms \citep{cuiObservationBroadWave2017}.
Furthermore, the experimental observation of the first Bose gas near a $d$-wave shape resonance introduced alternative possibilities where $\ell$-wave confinement-induced resonances can play a pivotal role~\citep{yaoDegenerateBoseGases2019}. 

We note in passing what happens when dealing with spin-polarized Fermi gases as it is one of the main motivations for studying odd-wave potentials. In this case, the leading interaction terms stem from a $p$-wave
channel as it was confirmed in cold-atom gases relatively early \citep{gunter2005}. 
Low-dimensional Fermi gases were used to the study spin-imbalanced systems 
\citep{liao2010}, effectively demonstrating that a cold-atom setup can be used to simulate pairing phenomena related to 
the so-called FFLO pairing phase and its 
stabilization by confining a system to a low-dimensional 
geometry, for a more elaborated discussion on this topic see~\citep{Guan2013Review}.

\subsection{Dipole-dipole interactions}
\label{Sec:LongRange}

A conceptually different approach to simulate beyond-zero-range physics is based on long-range interactions\footnote{In the context of the present review, long range means that the potential has a tail which decays slower than the tail of the van der Waals potential, which characterizes the long-range behavior of the interaction among neutral atoms.}. Note that the 1D geometry allows one to solve analytically problems with certain long-range potentials, notably with $1/x^2$ potentials~\citep{Calogero1971inverse}. Those potentials can describe cold-atom experiments only in a relatively crude approximation, and will not be discussed here. Instead, the focus will be on the experimentally more relevant case of dipole-dipole potentials that occur naturally in systems of cold molecules or atoms with large magnetic moment (e.g., Cr, Dy, Er)~\citep{Lahaye2009Review,Baranov2012Review}.  

The systems discussed in this subsection can be used to engineer long-range physics, which enriches quantum simulators, see, e.g., the review by~\cite{defenu2021long}.  For example, magnetic atoms \citep{kotochigova2014,norcia2021} used to produce BEC of dipoles \citep{griesmaier2005,lu2011,aikawa2012,petter2019} can realize 
supersolid states of matter \citep{bottcher2019,sohmen2021}. 
A supersolid refers to a peculiar state of matter where a frictionless flow of a superfluid and crystal periodic density modulations of a solid coexist~\citep{boninsegni2012colloquium,leggett1970can,yukalov2020saga}. 
The initial experimental realization of supersolids in ultracold atom systems were reported in BECs coupled to light fields created by optical cavities~\citep{leonard2017monitoring} or featuring spin-orbit coupling~\citep{li2017stripe}.  We also note that cold molecules can have a very large dipole moment, and may lead to other exciting states of matter. However, some further experimental development is needed as molecular systems typically suffer from losses more significant that atomic ones, see~\citep{Bohn2017Review} for a recent review of the topic.

Systems with dipole-dipole interactions do not have Bethe ansatz solutions in a continuum, and the $1/x^3$ divergence at the origin complicates calculations in much of the standard numerical machinery\footnote{
It is worth noting that in the context of models that can be solved by the Bethe ansatz methods, there is a family of integrable models that describe dipolar atoms in multi-well systems \citep{Tonel2015,Ymai2017} for different physical applications \citep{Tonel2020,Wilsmann2018,Grun2021,Grun2020}. Also, a variational Bethe Ansatz approach for dipolar 1D bosons, based on analytic expressions for the ground state energy and structure factor of the Lieb-Liniger model, has been discussed \citep{Palo2020}.
From the experimental side, the far-from-equilibrium dynamics of a strongly interacting nearly-integrable system as it is tuned away from integrability has been investigated in a dipolar quantum Newton's cradle consisting of highly magnetic dysprosium atoms \citep{Tang2018}. Dysprosium atoms have also been used to create strongly correlated prethermal states by topological pumping to stabilize a strongly attractive dipolar system in 1D \citep{kao2021}. }. Luckily, in quasi-1D geometries a dipole-dipole potential is regularized at short distances~\citep{Sinha2007dipoles, Deuretzbacher2010dipoles,Zinner2011Dipoles} and can be written in the form
\begin{equation}
    V(x)=\frac{D^2}{l^3} \left[\frac{1}{4}F\left(\frac{x}{l}\right) -\frac{2}{3}\delta\left(\frac{x}{l}\right)\right],
\end{equation}
where $D$ is the dipole moment, $l$ is the length associated with the confinement, and the bounded function $F(u)$ has the expression
\begin{equation}
    F(u)=-2u+\sqrt{2\pi}(1+u^2)e^{-\frac{u^2}{2}}\mathrm{Erfc}(u/\sqrt{2}).
\end{equation}
Therefore, the dipole-dipole interaction under confinement is determined by the delta-function potential at $x=0$. Everywhere else it is given by the function $F(x/l)$, which decays as $1/|x|^3$ in the limit $|x|\to\infty$. 

At the level of few-body systems, the interaction given by $V$ allows one to study (quasi)-crystallization of a harmonically trapped system, in particular the transition from a fermionized state to a Wigner-like crystal in the ground state properties~\citep{Deuretzbacher2010dipoles,Zollner2011Dipoles,Koscik2017Dipoles,Bera2019Sorting} [see~\citep{Cremon2010Wigner} for relevant studies in two dimensions]. An external lattice enables to effectively tune this transition, and study the bottom-up assembly of lattice models with next-to-nearest-neighbor interactions. 

\begin{figure}[tb]
    \center
    \includegraphics[width=0.8\linewidth]{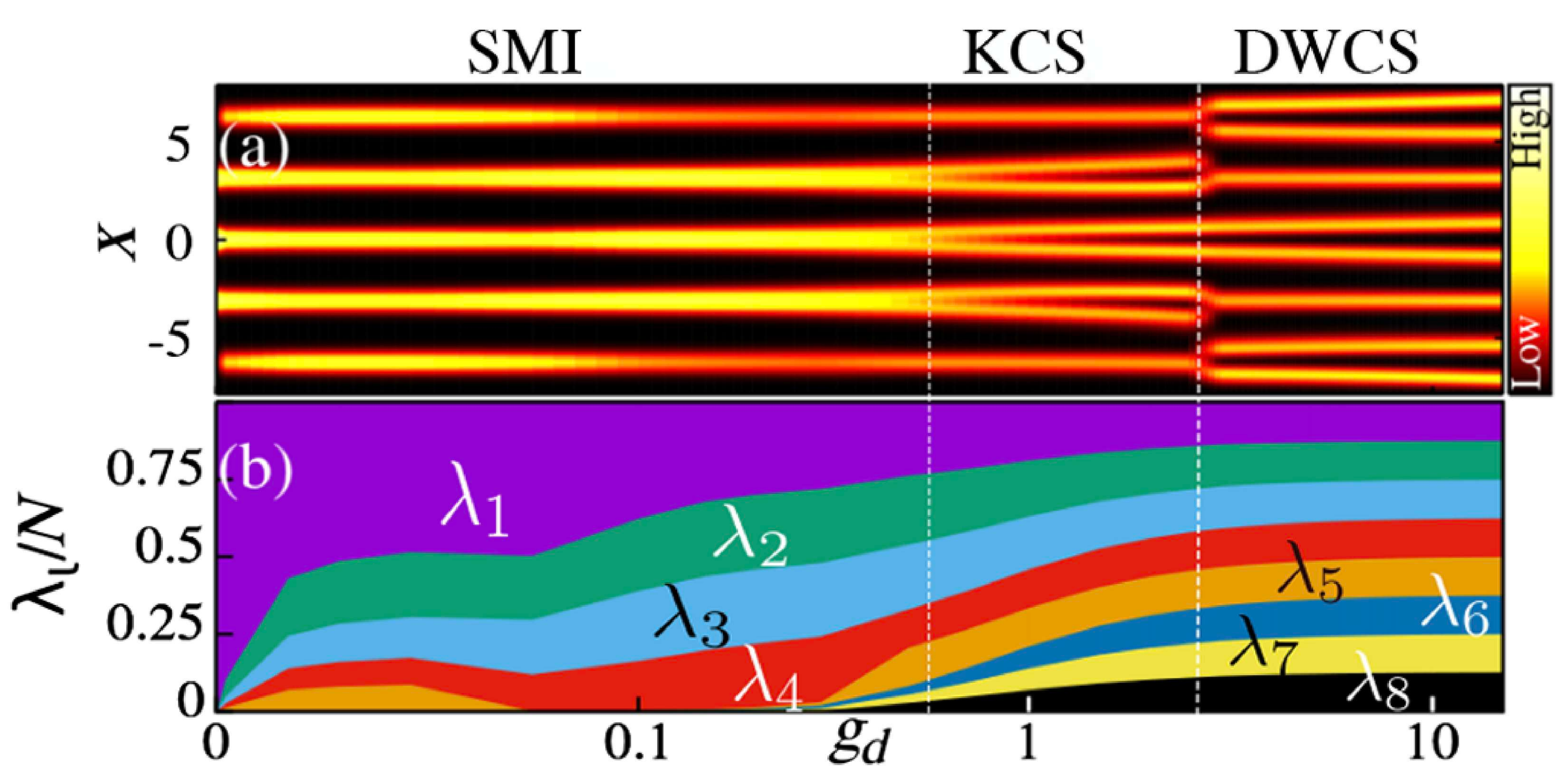}
    \caption{One-body density as function of the dipolar coupling $g_d$.  For $g_d \leq 0.8$ the system forms an SMI state, whose density is mainly concentrated in the three central wells (superfluid fraction). A larger $g_d$ results in the KCS phase characterised by a two hump structure of the density profile in the doubly occupied wells. An even stronger dipolar strength leads to the formation of DWCS with a density configuration alternating between single and double occupations. (b) Natural populations $\lambda_k$, $k=1,2,\dots,8$ with respect to $g_d$. In the SF the $\lambda_1$ is dominant while an increasing $g_d$ gives rise to fragmentation which becomes substantial beyond the KCS state. In all cases, $N=5$ bosons are trapped in a five well lattice potential with barrier height $V_0=8$. All quantities are given in terms of the recoil energy. Figure is from~\citep{chatterjee2020detecting}.}
\label{fig:dipole_bosons_lattice} 
\end{figure} 

The ground-state phase diagram of a few dipolar bosons confined in both commensurate~\citep{chatterjee2013ultracold} and incommensurate~\citep{chatterjee2013ultracold,chatterjee2018order,chatterjee2019correlations,chatterjee2020detecting} finite size 1D optical lattices has been explored relying on the MCTDHB method, see Section~\ref{subsec:ML-MCTDHA}. 
It has been shown that various states -- the few-body precursors of the phases in the thermodynamic limit -- emerge for increasing dipolar strength, see also \citep{pizzardo2016quantum} for the creation of cat-states in a double-well. 
Their existence is a consequence of the competition between the kinetic, potential, and the involved interaction energies.  
These states range from a pure superfluid (SF) for weak dipole-dipole interactions\footnote{The employed interaction potential has the form $V_d=g_c\delta(x_i-x_j)+g_d/(\abs{x_i-x_j}^3+\epsilon)$ with $\epsilon$ being the short-scale cutoff introduced by the transversal confinement i.e. $\epsilon=a_{\perp}^3$~\citep{Deuretzbacher2010dipoles,deuretzbacher2013erratum,Sinha2007dipoles} regularizing the divergence at small separations $\abs{x_i-x_j}\leq a_{\perp}$. Here, $a_{\perp}$ is the transveral length scale. Note also that for electric (magnetic) dipoles $g_d=\mu_m^2/(4 \pi \epsilon_0)$ ($g_d=\mu_m^2 \mu_0/(4 \pi)$) where $\mu_m$ refers to the dipole moment and $\epsilon_0$ ($\mu_0$) is the vacuum permittivity (permeability).}, $g_d$, which turns into a Mott-insulator coexisting with a superfluid (SMI) for larger couplings as can be seen in Fig.~\ref{fig:dipole_bosons_lattice}(a). 
Here, due to the hard-wall boundaries the superfluid fraction resides in the central wells in order to minimize the kinetic energy. 
A further increase of the dipolar interaction results in the so-called kinetic crystal state (KCS)\footnote{Note that the KCS phase does not exist for larger system sizes and only the DWCS persists.}, where the bosons avoid each other and the density exhibits a two-hump structure in the doubly occupied wells. The latter correspond to the central sites due to the kinetic energy. 
For even stronger dipole-dipole couplings, where the interaction energy overcomes the kinetic energy, a density wave ordered structure appears, in which next-to-the-nearest neighboring wells are doubly occupied and their density feature a two-hump structure.
The resultant phase is called a density-wave crystal state (DWCS). It is characterized by substantial two-body anti-correlations within each lattice site. 
This behavior is general for crystal phases~\citep{chatterjee2013ultracold,chatterjee2018order,chatterjee2019correlations}. 
As a further illustration of the many-body nature of these systems, Fig.~\ref{fig:dipole_bosons_lattice}(b) shows the fragmentation parameter $\Delta=\sum_k(\lambda_k/N)^2$ ($N$ is the atom number), which serves as a mesoscopic order parameter characterizing these phases. In this expression, $\lambda_k$ denotes the $k$-th eigenvalue of the reduced single-particle density matrix~\citep{sakmann2008reduced}. 
It is apparent that the formation of a crystal-like state is associated with a population $\Delta=1/N$ of the eigenvalues, a result that indicates the strongly correlated origin of these states, see also \citep{fischer2015condensate} for the dependence of the degree of fragmentation in long-range settings on the spatial dimension. 
An important point here is that in contrast to the SF to Mott transition, which is inherently related to the presence of an external lattice potential, crystalization is a result of the dominant effect of the dipolar interactions and occurs irrespectively of the boundary conditions and the one-body potential~\citep{arkhipov2005ground,astrakharchik2008ground,Zollner2011Dipoles,Bera2019Sorting}. 
Interestingly, an experimental protocol has been proposed for the detection of the above-described phases \citep{chatterjee2020detecting}, based on the particle number distributions  obtained through single-shot absorption images~\citep{bakr2009quantum,sherson2010single}. 

The multi-band tunneling dynamics of two dipolar bosons in a 1D triple well is studied in~\citep{chatterjee2013ultracold}\footnote{For the dynamics of two magnetic atoms with a focus on a single well, see~\cite{Suchorowski2022}.}.  In particular, the authors connect transitions between bands to the two boson energy spectrum, which features several avoided-crossings between eigenstates of different bands.
The bosons are initially prepared in their ground state in the most left well of the triple well and are then left to evolve experiencing a fixed dipole-dipole coupling lying close to the aforementioned resonances. 
Depending on the value of the dipolar strength, being below or above the location of the resonance, a different dynamical response is observed. 
It ranges from pair tunneling to multifrequency tunneling events whose period is substantially impacted by the long-range coupling. The characteristics of the dynamics are monitored via the population of each well and the pair probability in the course of the evolution while even the participating multi-band number states can be identified due to the few-body nature of the system. 
It is worthwhile noting that, in the Floquet picture, a directed selective tunneling pathway of ($N-1$) atoms was found to be excited for dipolar bosons initialized in the middle well of a high-frequency-shaken triple well potential~\citep{lu2014directed,luo2015directed}.   
For dipolar bosons in a double-well potential, the dominant tunneling mechanisms remain qualitatively the same as for contact interactions. They range from Josephson oscillations to macroscopic quantum self trapping as it has been discussed e.g. in \citep{haldar2018impact,roy2020quantum,mazzarella2013two}. For completeness, it should be mentioned that estimations of the lowest breathing mode in strongly attractive harmonically trapped dipolar gases providing signatures of a super-Tonks-Girardeau phase can be found in~\cite{astrakharchik2008super}.  

\begin{figure}[tb]
    \begin{center}
    \includegraphics[width=0.6\linewidth]{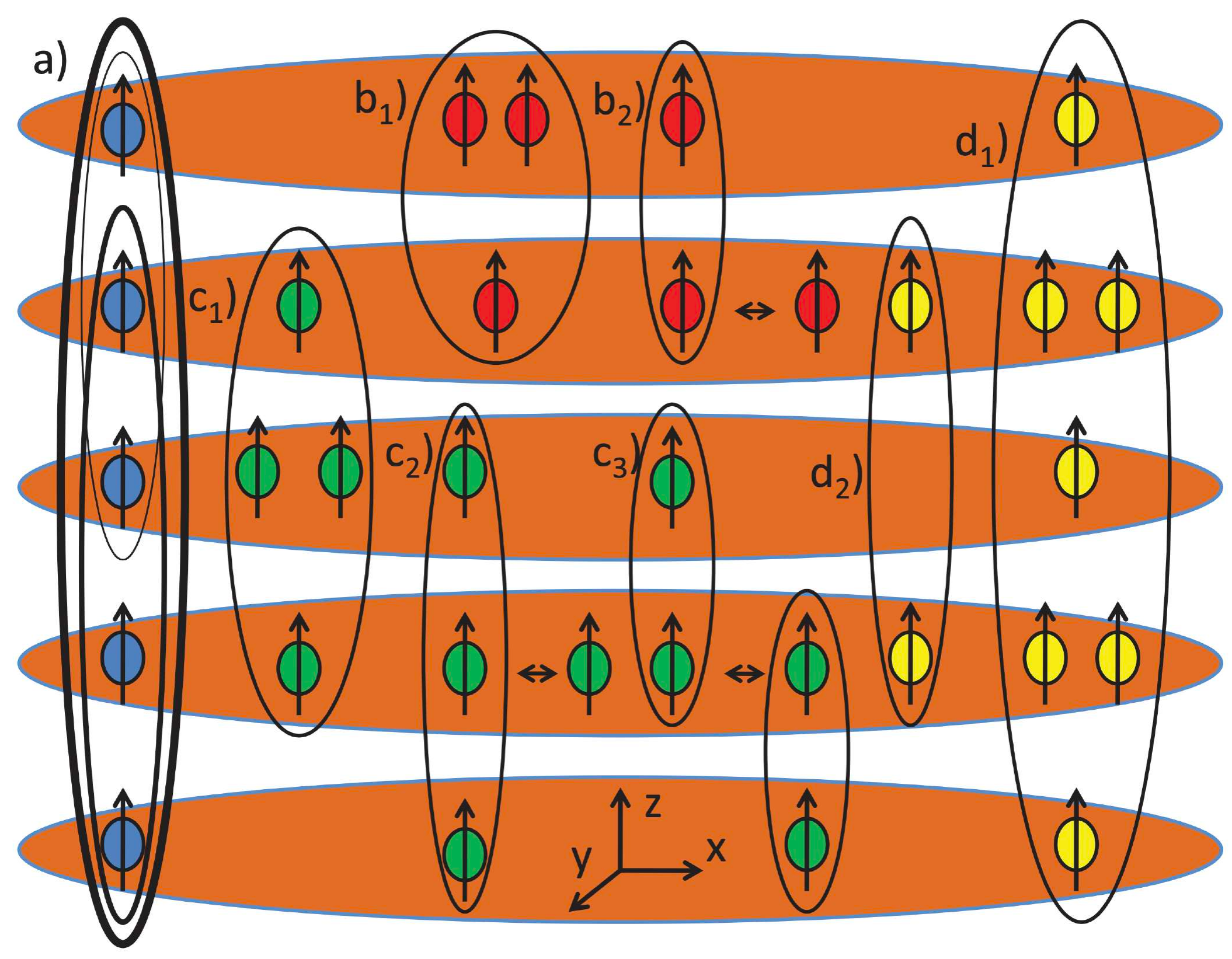}
    \end{center}
    \caption{Illustration of different complexes that can form in layered geometries with long-range 
    dipolar interacting particles. The complexes are relevant for both 1D tubes and 2D layers. Notice that the dipole moments are perpendicularly oriented in these examples. Figure from~\citep{volosniev2012dipoles}}
\label{fig:dipcomplex}
\end{figure} 

Beyond a single-tube geometry, the long-range dipolar interaction motivates studies of coupling between tubes. The inter-tube coupling is overall attractive and supports a two-body bound state~\citep{Zinner2011Dipoles,Volosniev2013Dipoles}, which may lead to the existence of few-body bound states between tubes, and, hence, a new type of quantum matter where 
few-body bound states play the role of building blocks \citep{Wang2006Polar,Armstrong2020Clusters}. There are many types of clusters of few dipoles. Chains can include only one particle per layer, but more complicated structures contain more that a single bound particle from one layer, and can create a bifurcated many-particle bound state in layered systems. Such structures can be created by tilting the direction of dipoles~\citep{Wunsh2011Dipoles,Volosniev2013Dipoles}, and probed for instance by detecting scattered light~\citep{Mekhov2013Probing}. 
An illustration of various complexes that could be possible with long-range dipolar interactions is shown in Figure~\ref{fig:dipcomplex}. Note that while the figure illustrates the case of 2D layers and perpendicularly oriented 
dipole moments\footnote{This system can also be used to simulate ultradilute quantum liquids~\citep{Guijarro2022Quantum}.}, the same complexes are imaginable with 1D tubular geometries. In particular, if the orientations of the 
dipole moments are tilted with respect to the planes/tubes, the relative contributions of attraction and 
repulsion can be changed \citep{Zinner2011Dipoles}. 
\begin{figure}[tb]
    \center
    \includegraphics[width=0.8\linewidth]{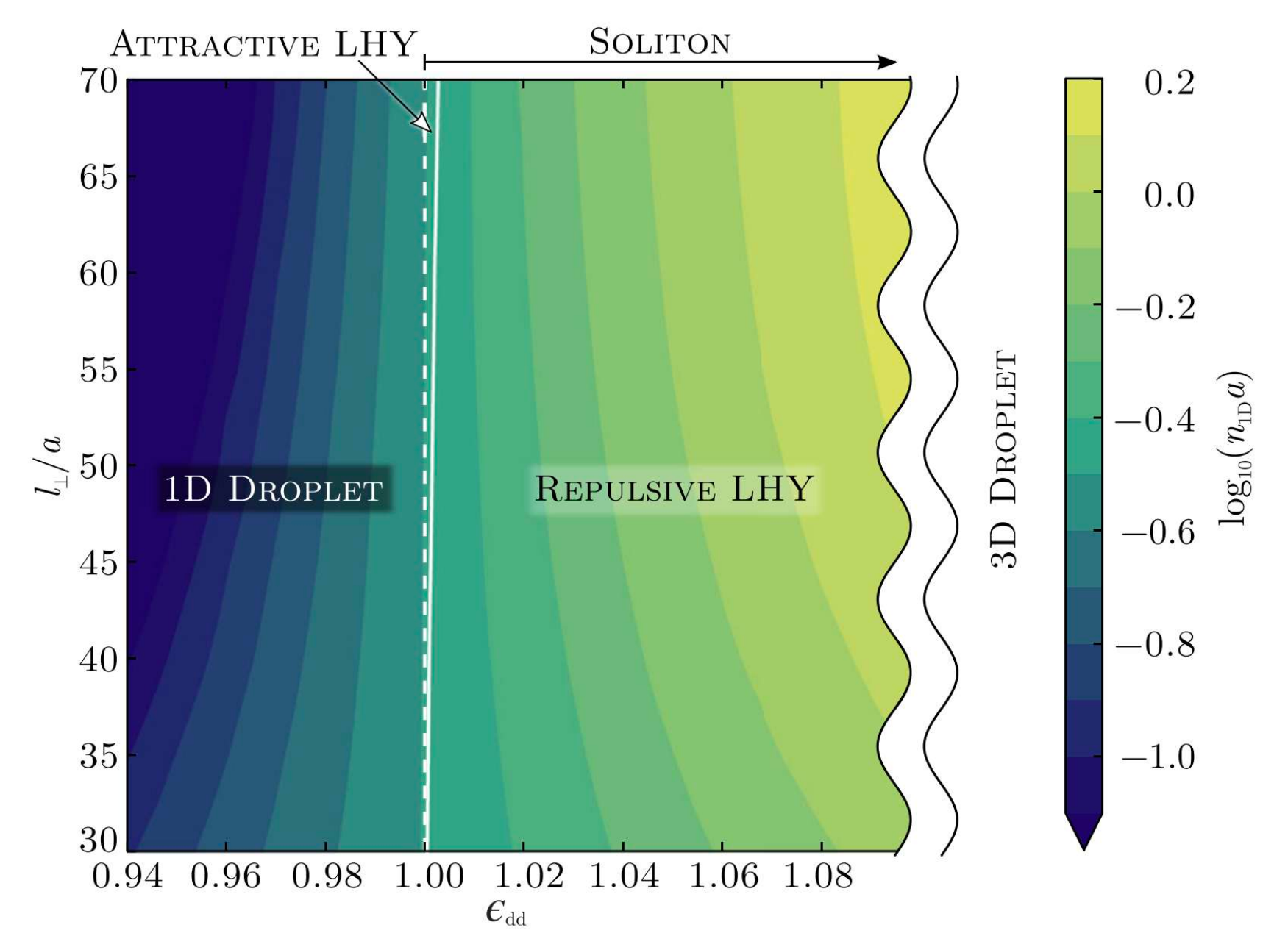}
    \caption{The peak density in a gas of $N=5000$ atoms with respect to $l_{\perp}/a$ and $\epsilon_{dd}$ quantifying the regimes of attractive and repulsive LHY contribution in 1D. For $\epsilon_{dd}>1$ a smooth crossover from droplets to solitons is realized. The white solid line marks the region where the LHY contribution is on-average repulsive. Here $l_{\perp}=\sqrt{\hbar/(M \omega_{\perp})}$ is the transversal oscillator length with frequency $\omega_{\perp}$. Figure from \citep{edler2017quantum}}
\label{fig:dipolar_phase} 
\end{figure}

\subsection{Dipolar droplets}\label{dipolar_droplets}

As discussed in Section~\ref{sec:Droplets_contact}, quantum droplets may appear in interacting Bose-Bose mixtures with short-range interactions due to the presence of quantum fluctuations. In these settings the liquid phase occurs as a result of the competition between the repulsive intracomponent interactions and the attractive intercomponent coupling. 
In the same manner droplet structures may emerge in single-component dipolar bosonic systems due to the interplay between the long-range dipole-dipole attraction and the short-range repulsion \citep{bottcher2020new}. 
In fact, an ensemble of confined attractive dipoles is typically unstable and its stability can be achieved by the counteraction of strong repulsive contact interaction that prevents the collapse of such settings. 
However, in the case of an overall mean-field attraction the system is stabilized only by the presence of quantum fluctuations. 
Dipolar quantum droplets were observed experimentally for the first time using $^{164}$Dy~\citep{ferrier2016observation,kadau2016observing} soon after their theoretical prediction in Bose mixtures \citep{petrov2015quantum}. 
Since then they have been realized using several different atomic species including $^{133}$Er  \citep{chomaz2016quantum},  $^{164}$Dy \citep{schmitt2016self,ferrier2016liquid} and $^{162}$Dy \citep{bottcher2019dilute}. 
One of the conclusions drawn from these experiments is that beyond mean-field effects might be more important in dipolar condensates than in those with short-range interactions.

\begin{figure}[tb]
    \center
    \includegraphics[width=0.7\textwidth]{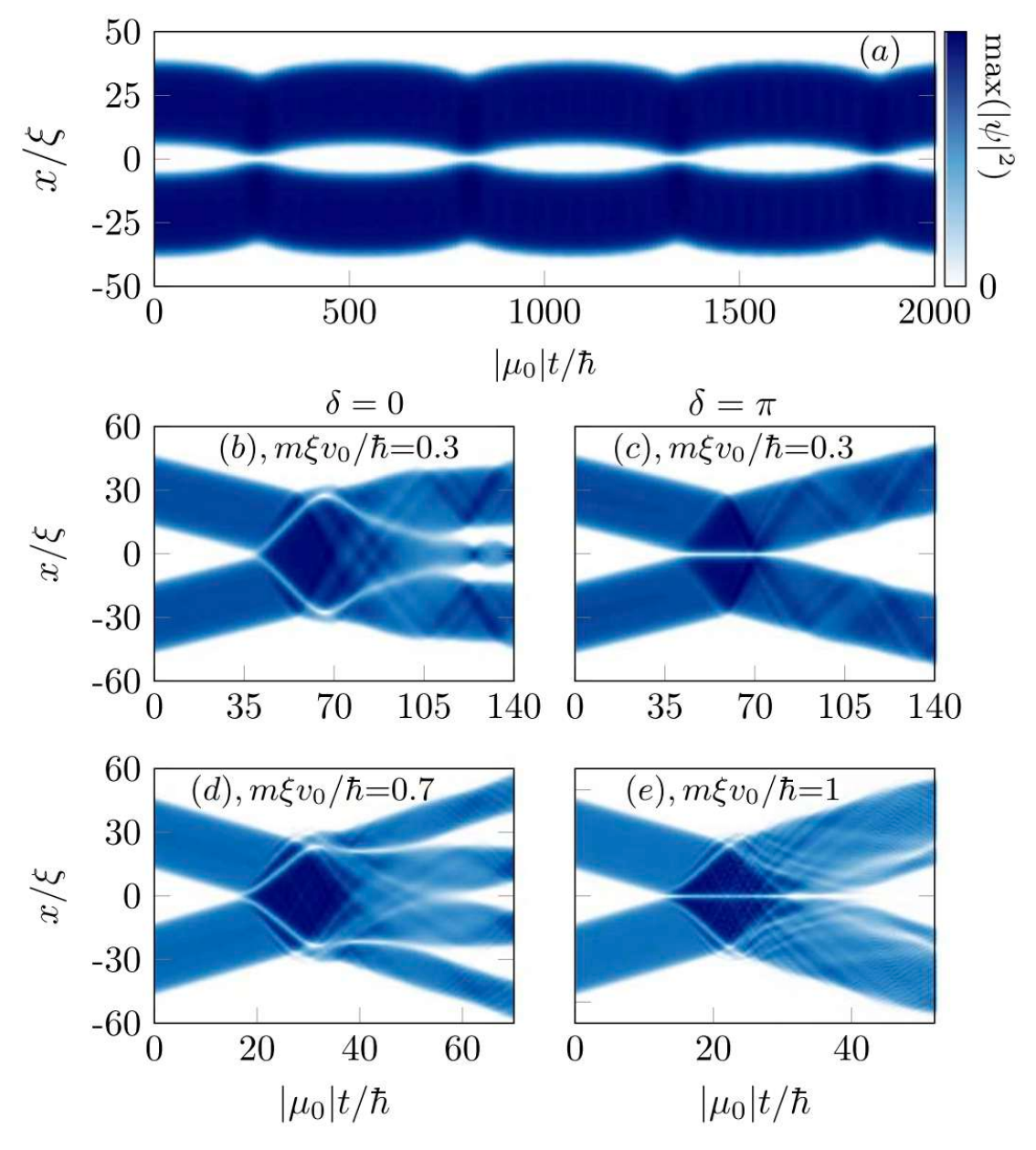}
    \caption{Representative dipolar droplet collisions. 
Panel (a) depicts the propagation of a long-lived droplet dimer while panels (b)-(e) present in ($\delta=0$) and out-of-phase ($\delta=\pi$) dynamics for various velocities (see legends). The initial velocity is $u_0$, $\xi(\epsilon_{dd})$ is the dipolar healing length and $m$ the atom mass. Figure from~\citep{edmonds2020quantum}.}
\label{fig:dipolar_drop_collision} 
\end{figure}

It has been demonstrated \citep{edler2017quantum} that the momentum dependence of the dipolar interactions in quasi-1D geometries~\citep{Sinha2007dipoles} leads to beyond mean-field corrections that differ significantly from those in non-dipolar counterparts. 
In particular, besides the difference in the density dependence of the quantum fluctuations in the presence and absence of long-range interactions also their sign can be changed due to the crucial role of the transversal directions in 1D dipolar gases. For a relevant characterization of the emergent phases see Fig.~\ref{fig:dipolar_phase}. An important control parameter for entering different phases in these gases is the relative ratio between the dipolar and the $s$-wave interaction strengths, namely $\epsilon_{dd}=\mu_0 \mu_m^2/( 6 \pi l_{\perp}^2 g_{1D}$). Here, the effective 1D interaction strength is related to the 3D one via $g_{1D}= g_{3D}/(2 \pi l_{\perp}^2)$ with the transversal length scale being $l_{\perp}=\sqrt{\hbar/(M \omega_{\perp})}$, while $\mu_0$ is the vacuum permeability and $\mu_m$ the dipole moment.  
Specifically, for dipolar bosons the density dependence of the LHY term is $\sim -n_{1D}$ as long as $n_{1D}a \to 0$, while for increasing $n_{1D}a$ it departs from this linear behavior~\citep{edler2017quantum}. Here, $n_{1D}$ denotes the gas peak 1D density and $a$ represents the 3D $s$-wave scattering length. Recall that in Bose mixtures the LHY scales as $\sim n_{1D}^{1/2}$. 
It should be clarified at this point that this behavior is in sharp contrast to the 3D scenario\footnote{Similarly, three-body losses exhibit a distinct density dependence for quasi-1D dipolar and non-dipolar Bose gases \citep{edler2017quantum}, a behavior that is not encountered in 3D.} where the density dependence of the corresponding LHY contribution in both dipolar gases and Bose-Bose mixtures is $\sim n_{3D}^{3/2}$~\citep{bottcher2020new,petrov2015quantum}. 
Moreover, it was found in~\citep{smith2021quantum,bisset2021quantum} that more complex phases forming miscible and immiscible droplet states can be achieved in binary magnetic gases, namely two-component dipolar bosonic gases. 
The excitation spectrum of self bound dipolar droplets using Bogoliubov-de-Gennes linearization analysis has been examined in~\citep{bottcher2020new,baillie2017collective,lima2012beyond}. 
A main result was the identification of modifications e.g. in the frequencies of the collective modes caused by the inclusion of quantum fluctuations in the presence of dipolar interactions. 
Furthermore, by monitoring the behavior of the low-lying excitation modes it was found that for increasing atom number the droplets transit from being compressible to incompressible.  

The collisional properties of two counterpropagating 1D dipolar droplets with the constant velocities $\pm v_0$ and the phase difference $\delta$ are explored in~\citep{edmonds2020quantum}. It is shown that the droplets undergo inelastic collisions, that their post-collision shapes and sizes depend crucially on $\delta$, and that excitations build up on the droplets in the form of sound waves. 
It is also shown that, for a fixed phase difference, varying the initial velocity gives rise to a rich droplet response. Characteristic examples are presented in Fig.~\ref{fig:dipolar_drop_collision}. 
One can readily see that two initially out-of-phase droplets produce a long-lived droplet dimer. 
Furthermore, by considering different finite initial velocities one concludes that in-phase collisions lead to the
droplet fission~\citep{dingwall2018non} associated with the generation of several droplet states after collision [see Fig.~\ref{fig:dipolar_drop_collision} (b), (d)]. 
Turning to out-of-phase collisions, one observes the formation of two droplets [see Fig.~\ref{fig:dipolar_drop_collision} (c), (e)] accompanied by the emission of sound waves, which are more pronounced for larger initial velocities. 
Finally, the rotational properties of toroidal dipolar Bose gases have been discussed in~\cite{tengstrand2021persistent} explicating that supersolids are able to sustain persistent currents and show hysteretic response which is affected by the background superfluid fraction. 
This proposal suggests a complementary signature of supersolidity that should be experimentally accessible.

\subsection{Atom-ion interactions}
\label{subsec:atom_ion}
In recent years, there has been a new push to combine cold atoms with trapped ions in order to 
leverage the advantages of both platforms, see~\citep{tomza2019} and references therein.
In the context of many of the previous sections concerned with various types of impurities 
in backgrounds, we could be tempted to consider the atom-ion setup with the typical picture of a single ion in a gas of atoms with the ion playing the role of an impurity. However, the long-range nature of the atom-ion interactions may complicate the usual 
polaron picture and requires a more careful treatment \citep{cote2002,Astrakharchik2021Ion,christensen2021charged}.

The atom-ion interaction of a trapped ion immersed in a cold atomic gas falls off with 
the inverse distance to the fourth power, although as with the dipolar interactions discussed 
above, the short-range part of the potential needs regularization (see, e.g.,~\citep{krych2015}). 
Once this is taken into account, there are many opportunities to study novel physics in 
the atom-ion setups, including collision dynamics \citep{idziaszek2009}, charge transfer \citep{cote2000}, and cold chemistry, see~\citep{Rios2021Cold} and references therein. 

Already at the level of the static properties, one encounters marked differences
due to the atom-ion interactions. The long-range attraction 
between the neutral atoms and the ion leads to a number of new bound states, which  will be possible in
the gas \citep{Goold:10}. This implies that applicability of the standard mean-field theories such as the Gross-Pitaevski approach 
must be re-evaluated~\citep{massignan2005}. At the level of dynamics, the atom-ion 
systems present a number of opportunities as quantum simulators of models of 
condensed-matter systems. As noted by \citep{bissbort2013}, a crystal of trapped
ions immersed in a cold atomic gas features close analogies to classic solid-state systems, and, thus, may provide a very accurate quantum simulator for 
both structure and transport properties. Furthermore, as recently pointed out, 
ions interacting with cold atoms in tailored lattice setups may be used to 
simulate different types of lattice gauge theories~\citep{Dehkharghani2017LGT} 
and Tomonaga-Luttinger liquids~\citep{michelsen2019}.

In closing this section, we note that the combination of optical tweezers and atom-ion 
systems has been recently proposed as a road to engineer spin-spin 
interactions in more general quantum spin models~\citep{Arias2021} with possible applications in quantum
simulation tasks or for quantum computing purposes.
For more information on different aspects of both experimental and theoretical studies of 
atom-ion systems, we refer to the review by~\citep{tomza2019}.

\section{Summary and future prospectives}

In this review we have discussed the physics of low-dimensional ultracold quantum gases in the regime of few to many atoms, with the aim of providing a comprehensive but still manageable account of the vast body of existing literature. 
It is our hope that the review will be useful and serve as a guide through the recent developments for both theoreticians and experimentalists interested in this broad field of research and also inspire future endeavors. 
Given that recent experimental advances represent a major motivation for the research in this field, we have mainly focused on systems in harmonic traps and pointed to the connections to homogeneous systems and lattice geometries in the most relevant cases. 
Our focus has been on few-body systems, but with a keen eye on the connection to many-body physics throughout the sections.  In the following we will summarize some of the insights that have been gained and how they fit into the current state and future development of the field.  

Fundamentally, cold-atom experiments have renewed the interest in mathematical models of one-dimensional interacting particles, like the Lieb-Liniger model,
which can offer an understanding of various physical concepts, such as integrability. This availability of a detailed theoretical basis, which includes exact solutions for many problems across a wide range of interaction regimes, but also different dimensionalities, allows for experimental and advanced theoretical developments to be readily benchmarked and tested even in a many-body context. 

The possibility to analyze small systems fully quantum mechanically is the key towards unraveling the behaviour of complex systems beyond the exactly solvable models. Already single-component small systems 
 show the emerging physics stemming from interactions, thus, providing excellent test cases for approximate methods that can identify which effects and properties are important and dominant. By calculating properties of few-body systems on current computational hardware, one can study complex non-equilibrium dynamics and the role of interactions and statistics. Single-component systems are likely to reveal more interesting phenomena in the future and also pave the way for understanding multi-component gases. 

A key case in point are gases with two components, where there are both intra- and intercomponent interactions to consider, giving rise to interesting few-body ground state configurations. Here, for instance, the miscible and immiscible phases share a many-body counterpart but the boundary of this second-order phase transition is shifted towards larger intercomponent interactions in the few-body regime.  It is also possible to observe the emergence of the strongly correlated phases in the Tonks-Girardeau and composite fermionization limit.  Turning to the dynamics of these gases, we have highlighted the interaction-dependence of collective breathing modes from a few-body  perspective, an investigation that can readily provide insights for larger strongly interacting  bosonic systems. The dynamical phase separation process of mesoscopic bosonic gases revealed the formation of domain-wall structures among the spontaneously nucleated filamentary density patterns for quenches towards the immiscible phase and the formation of dark- anti-dark solitary waves following quenches to the miscible regime. These processes also take place in many-body systems. Another generic mechanism that occurs both in the few- and many-body regimes of binary systems is inter-species energy transfer triggered by finite intercomponent couplings.  Such processes are anticipated to be relevant also in more complicated multicomponent Bose gases.   Finally, tunneling dynamics is a key example where few-body systems provide a bottom-up approach to understand processes arising in larger systems since they enable monitoring of microscopic mechanisms. In this context, few-body settings show more rich tunneling features as compared to many-body setups, e.g. correlated pair tunneling channels,  while in turn also experiencing the same dynamical responses such as Josephson oscillations, macroscopic quantum self-trapping, coherent quantum tunneling as well as collapse and revival events. A similar correspondence between few- and many-body systems occurs also in the tunneling dynamics of single-component quantum gases. In the case of mass-imbalanced mixtures, the heavy component acts as a “material barrier” as seen by the lighter component, and this is a mechanism that one can observe going all the way from few- up to the many-body regime. 

Arguably,  the simplest 
two-component system is a small Bose gas with an impurity atom. For these systems, the few-body point of view provides novel insights on many-body dressing, and sheds light onto two problems of modern condensed matter physics. 
 The first problem is the emergence of many-body properties of the bath from the underlying two-body correlations. Here, the impurity acts as a tool that enables measurement in a laboratory. 
 The second problem concerns   
the limits of applicability of the polaron concept. Here, the impurity is not simply a probing device. Its existence is actually necessary for the definition of the problem and appearance of a quasiparticle.  For Fermi gases, it was shown both experimentally and theoretically that a handful of particles are enough to screen the impurity in its ground state. This implies that one can study a few-body problem to understand the self-energy of the impurity and other relevant properties. Bose gases are highly compressible, and therefore (in general) one needs more bosons than fermions to screen the impurity. Still, many properties of 1D Bose polarons can be understood by studying systems with around 20 to 30 particles. This has been demonstrated in a number of theoretical works, but experimental confirmation is still an outstanding goal.  Ultracold-atom setups appear promising in this context in both weakly (Fr{\"o}hlich polaron) and strongly interacting limits.  

Beyond a two-component system, we considered spin-1 Bose gases, and again exposed several aspects of the interplay between few- and many-body physics, involving a multitude of intriguing states related to first- and second-order phase transitions for both ferromagnetic and anti-ferromagnetic gases. These phases arise due to the presence of inter-particle and spin-spin interactions and are naturally absent in two-component gases. To provide an intuitive description for the generation of these states it is natural to start with a somewhat simpler mean-field framework and subsequently argue about their persistence when correlations are taken into account. In particular, as we discussed, few-body analogs of these phases exist and the impact of correlations is most pronounced across the respective second-order phase transitions disfavoring states where all spin components are occupied and therefore shrinking the region of existence of the antiferromagnetic and broken-axisymmetry phases. Correlation processes are naturally enhanced for few-body systems and for increasing spin-spin interactions. In the dynamics of these spinor gases spin-mixing processes are mostly inevitable irrespective of the size of the bosonic ensemble, while few-body systems provide invaluable insights regarding the microscopic nature of these processes due to the numerically tractable number of involved configurations. 

As a final example of the few- to many-body crossover, we discussed the physics of quantum droplets appearing within the attractive intercomponent coupling regime of short-range interacting mixtures. The existence of these configurations is inherently related to the presence of quantum fluctuations that are commonly taken into account through the first-order Lee-Huang-Yang (LHY) correction to mean-field theory. Here we must note that there are significant differences for one- and higher dimensions. The quantum droplet concept was originally developed in the context of an extended Gross-Pitaevskii approach which we have outlined for pedagogical reasons, since it provides an instructive framework for discussing results from ab-initio methods accounting for higher-order correlations. Concretely, an important phenomenon is the transition from a flat-top density configuration to a Gaussian type occurring for increasing intercomponent repulsion. Importantly, this occurs both in the few- and many-body cases. Particularly, it was argued that bosonic samples of twenty atoms already suffice to produce flat-top density signatures, while ensembles containing fewer atoms favor progressively more spatially localized structures. Beyond LHY, correlations become more prominent for stronger attractions and generically become important in the strongly interacting and few-body regimes. We expect the same to hold for the dynamics of these configurations, but note that this is as yet an under-explored area.   
To date, deviations from the LHY description in the breathing frequency and the expansion velocity of droplets have been reported. 
In this spirit, mechanisms to sustain excitations processes, pattern formation, and impurity physics constitute some of the important directions for future research in this field.  
Experimental demonstration of these few-body droplet states and especially droplets in 1D remain still elusive.

Droplet configurations also emerge in the presence of dipolar interactions and they again exist exclusively due to quantum fluctuations. The latter becomes more prominent for dipolar bosons in 1D as compared to their short-range counterparts. The formation of droplets in this context originates from the interplay of long- and short-range interactions, and a corresponding crossover from droplets to soliton entities can be realized via tuning the relative interaction ratio.  
The overall structure of the generated crystals depends typically on the atom number and the trap aspect ratio, while the few-to-many-body crossover in these systems is still an open question. 
Near the balance point of interactions, other states of matter dubbed supersolids can be formed  combining the crystal arrangement of a solid and the frictionless flow of a superfluid. Their few-body analogs could provide insights into the participating state configurations and correlation patterns; a statement that equally holds for the relevant mixture settings.  
Dipolar few atom systems embedded in multiwell geometries are found to favor higher-band excitation processes. 
These give rise to complex strongly correlated ground state phases such as density-wave crystals. 
They can also lead to a richer nonequilibrium dynamical response e.g. involving interband tunneling channels, while maintaining standard pathways like Josephson oscillations and macroscopic quantum self-trapping. As such, they constitute promising platforms to study interaction-induced resonant excitation mechanisms as well as trigger pattern formation in small and mesoscopic gases by exploiting, for instance, the knowledge of the energy spectrum.  

These examples demonstrate both the power of the few-body approach in cold atoms, and how it can be leveraged to enrich our understanding of many-body systems. 
Many static properties of systems of interest are well understood by now, and a large number of exact or approximate analytical methods exist that allow one to build up some solid physical intuitions. In contrast, the corresponding non-equilibrium quantum dynamics can often be analyzed only numerically, and even in cases susceptible to exact treatments the inclusion of all correlations is only possible within certain limits. This should, however, not obscure the fact that recent progress in methodology has pushed the frontiers enormously for systems in which all correlations can in principle be calculated.
Indeed, the work we have summarized in the review reveals that (i) an in-depth analysis of the non-equilibrium dynamics in these controllable quantum systems can deliver interesting insights and that (ii) ultracold atoms may provide a highly valuable and flexible laboratory to study correlated dynamical processes induced by a quench, continuous driving, or a suitable time-dependent protocol. 
A comparison of experimentally obtained data with state-of-the-art numerical results and advanced analytical models will keep leading to a deeper understanding of non-equilibrium processes in the future, including the physics of other interesting effects such as dissipation and decoherence. It might even pave the way to prepare novel metastable states with unseen properties or even functionalities.

Let us finally indicate a number of further research directions for many of the systems discussed in this review. This is by no means an exhaustive list of topics, but merely a selection mainly based on our own interests. It is our humble attempt to create an outlook into the future from our limited perspectives and the order in which we present our points should not be taken as a prioritization.

To further understand the few- to many-body crossover, a new and promising route is to investigate the impact of physical effects in spinor gases in different correlation regimes. More generally, multicomponent systems in different interaction regimes have recently been shown to hold great potential to create up-to-now unseen quantum many-body states~\citep{Keller:22,Keller:23}, for which
fundamental questions about the on-demand generation of spin-wave excitations, triggering controllable spin-mixing processes and the impact of correlations on the excitation spectrum exist. The latter has recently been explored for many-body systems~\citep{lavoine2021beyond}, and it would be 
interesting to observe how these features will develop as one explores 
the crossover down to few-body systems.
Furthermore, in the context of spinor gases,  
unraveling the emergence of static phases originating from the cooperation of spin and spatial degrees-of-freedom as well as the competition between intraband and interband tunneling dynamics in optical lattices are topics worth pursuing. 

Going beyond simple short-range interactions, we believe that it would be fruitful 
to explore 
the impact of correlations on the interplay between contact and long-range interactions during non-equilibrium dynamics. One particular aspect here is how few-body mixtures can be used to design complex pattern formation by triggering specific instabilities and excitation mechanisms~\citep{chomaz2022dipolar}.
More generally, the utilization of the 
presence of quantum fluctuations in quantum liquid phases (droplets) as a means to explore the fate of non-linear structures or the emergence of quasi-particles ranging from polarons to magnons in systems where short- and long-range interactions are competing is an intriguing problem.

Another challenging direction that has a long-standing 
history in both few- and many-body physics, is the competition between
quantum fluctuations and temperature effects. From our point of view there is a need to further develop novel effective models as well as numerical techniques in order to elucidate this interplay. In conjunction with 
the additional degrees-of-freedom that a multicomponent setting possesses, 
one may use few-body setups to elucidate new applications within 
quantum information and quantum thermodynamics 
\citep{Fogarty:20}, as well as the connection between information scrambling and work statistics \citep{Mikkelsen:20} to the spread of correlations in small systems. Additionally, it would be worth exploring how issues of topological physics in finite interacting 
systems can impact the physics and the interplays discussed above, also in 1D \citep{Reshodko:19}. Here we again stress that multicomponent setups are particularly interesting as internal degrees-of-freedom can be reinterpreted as synthetic dimensions as the particle number in the system grows. This poses the interesting question of how this behavior evolves as a function of system size.
    
Another key question for low-dimensional systems is whether they are integrable or not. 
To further comprehend and explore the transition from integrability to a non-integrable situation in which chaos may arise, machine learning shows promise in identifying this crossover \citep{Kharkov2020Chaos,Huber2021}. A major topic of interest for this discussion is "how many interacting particles are needed to bring a quantum system to the regime of strong chaos?"~\citep{schiulaz2018few,zisling2021many,Fogarty2021}, especially in multi-component settings~\citep{Anhtai:23,daug2023many}. This question is 
an important aspect to explore in few-body physics as it has implications 
on the study of transport and quantum control, among others. 
It also has the potential to aid the design and control quantum states and 
devices, and can open new paths as a foundation for next-generation 
quantum technologies, through the use of methods developed around the notion of quantum integrability and its breaking~\citep{Wilsmann2018, Grun2021, Grun2020}. 
    
However, above all, we must stress again the unstoppable and always prodigious progress on the experimental side that will keep being a motivator for new and advanced work also on the theory side. As one example, it is worth mentioning the recently obtained control for allowing time-resolved detection of even single atoms with high spatial resolution \citep{guan2019}. These kind of advanced techniques  can open up new directions and shed new light also on old results. They are part of the perpetual exchange of advances and ideas between experimental and theoretical developments, that has been a constant and fundamental pillar of supporting the fast progress in this area.

The advances mentioned above are some of the insights and results that we believe 
may pave the way for future studies. These will address research questions related to thermalization, quantum hydrodynamics, state engineering, chaos and integrability and in general the management of correlations. 
In turn, these directions of research may become starting points for developments in quantum technologies.  We hope that our review will provide a useful stepping stone for the corresponding theoretical developments and serve as a fruitful guide for experimental advances.

\section*{ACKNOWLEDGMENTS} 
This review could not have been written without the many fruitful discussions and great collaborations with colleagues throughout the years, there are too many to mention. Here we acknowledge conversations regarding the context of the review with Joachim Brand, Fabian Brauneis, Adolfo del Campo, Alberto Cappellaro, Panagiotis Giannakeas, Tommaso Macr\'i, Oleksandr Marchukov, Lukas Rammelm{\"u}ller and Manuel Valiente.  
S. I. M. acknowledges support from the NSF through a grant for ITAMP at Harvard University.  T.F. acknowledges support from JSPS KAKENHI-21K13856 and T.F. and Th.B. acknowledge support from the Okinawa Institute for Science and Technology Graduate University, and JST Grant Number JPMJPF2221. A.F. and R. E. B. acknowledge support from CNPq (Conselho Nacional de Desenvolvimento Científico e Tecnológico) - Edital Universal 406563/2021-7.
A.~G.~V. acknowledges
support by European Union's Horizon 2020 research and innovation programme under the
Marie Sk{\l}odowska-Curie Grant Agreement No. 754411.
P.~S. is supported by the Cluster of Excellence
'Advanced Imaging of Matter' of the Deutsche Forschungsgemeinschaft (DFG) - EXC2056 - project ID 390715994. N.~T.~Z. is partially supported by the Independent Research Fund Denmark.